%% file: tobins-thesis.tex
\DocumentMetadata{ 
	pdfstandard = a-2b,
	pdfversion  = 1.7,
	lang		= en-US,
}

\documentclass[twoside]{mitthesis} 
\usepackage{listings}

\usepackage[version=4]{mhchem}

\usepackage{lipsum}
\IfPackageAtLeastTF{lipsum}{2021/09/20}{\setlipsum{auto-lang=false}}{}

\usepackage{booktabs}

\usepackage{array}

\usepackage{tcolorbox} 

\usepackage{epigraph} 
\usepackage{framed}
\usepackage[textsize=tiny]{todonotes}

\usepackage{import}

\usepackage{booktabs}       

\usepackage{amsmath,amssymb}
\usepackage[ruled,linesnumbered]{algorithm2e}
\usepackage{adjustbox}

\usepackage{microtype}

\usepackage{enumitem}

\usepackage{tikz}
\usetikzlibrary{arrows, arrows.meta, positioning, automata, calc}

\RequirePackage[numbers,sort&compress]{natbib}
 
\apptocmd{\bibliography}{\addcontentsline{toc}{chapter}{\protect\textbf{\bibname}}}{}{}

\renewcommand{\bibname}{References}

\input{custom-enviroments}

\usepackage{parskip}
\hypersetup{%
	pdfsubject={Private, Verifiable, and Auditable AI Systems},
	pdfkeywords={Private, Verifiable, Auditable, AI Systems, Artificial Intelligence, LLMs, Massachusetts Institute of Technology, MIT},
}

\begin{document}


\title{Private, Verifiable, and Auditable AI Systems}



\Author{Tobin South}{MIT Media Lab}[M. Phil Applied Mathematics, Adelaide University, 2021][B.Math.Comp.Sci, Adelaide University, 2018]

\Degree{Doctor of Philosophy}{Program in Media Arts and Sciences, School of Architecture and Planning}

\Supervisor{Alex `Sandy' Pentland}{Toshiba Professor of Media Arts and Sciences at MIT}

 \Acceptor{Tod Machover}{Academic Head, Program in Media Arts and Sciences}{}

\DegreeDate{May}{2025}

\ThesisDate{April 25th, 2025}

\maketitle




\begin{abstract}
	\import{content/}{abstract.tex}
\end{abstract}

\import{content/}{reviewers} 

\import{content/}{acknowledgments}



\tableofcontents
\listoffigures



\import{chapter-1/}{chapter-1} 
\import{chapter-2/}{chapter-2} 
\import{chapter-3/}{chapter-3} 
\import{chapter-4/}{chapter-4} 
\import{chapter-5/}{chapter-5} 
\import{content/}{conclusion}










\bibliographystyle{plainnat}
\bibliography{chapter-1/roadmap, chapter-1/report_risks, chapter-1/mystuff, chapter-2/zktax, chapter-2/verifyevals, chapter-3/nlrRAG, chapter-3/PRAG, chapter-3/community, chapter-4/auth-delegation, chapter-5/c5}

\end{document}

%% file: custom-enviroments.tex
\newenvironment{paperabstract}{
  \begin{center}
    \bfseries Project Abstract
  \end{center}
  \begin{quote}
}{
  \end{quote}
}


%% file: content/abstract.tex
%
%





\noindent
The growing societal reliance on artificial intelligence necessitates robust frameworks for ensuring its security, accountability, and trustworthiness. This thesis addresses the complex interplay between privacy, verifiability, and auditability in modern AI, particularly in foundation models. It argues that technical solutions that integrate these elements are critical for responsible AI innovation. Drawing from international policy contributions and technical research to identify key risks in the AI pipeline, this work introduces novel technical solutions for critical privacy and verifiability challenges. Specifically, the research introduces techniques for enabling verifiable and auditable claims about AI systems using zero-knowledge cryptography; utilizing secure multi-party computation and trusted execution environments for auditable, confidential deployment of large language models and information retrieval; and implementing enhanced delegation mechanisms, credentialing systems, and access controls to secure interactions with autonomous and multi-agent AI systems. Synthesizing these technical advancements, this dissertation presents a cohesive perspective on balancing privacy, verifiability, and auditability in foundation model-based AI systems, offering practical blueprints for system designers and informing policy discussions on AI safety and governance.

%% file: content/reviewers.tex

\newpage 
\thispagestyle{empty} 

\begin{center} 
    {\large\bfseries Private, Verifiable, and Auditable AI Systems} 

    \vspace{1em} 

    by

    \vspace{1em}

    \ExplSyntaxOn
    {\seq_item:cn { g_author_name_seq } { 1 }}
    \ExplSyntaxOff

\end{center}

\vspace*{5em} 

\noindent
This dissertation/thesis has been reviewed and approved by the following committee members:

\vspace{3em} 

\noindent 
\ExplSyntaxOn
\textbf{\seq_item:cn { g_supervisor_name_seq } { 1 }}\\ 
\seq_item:cn { g_supervisor_title_seq } { 1 }\\ 
\ExplSyntaxOff


\vspace{5em} 

\noindent
\textbf{E. Glen Weyl}\\ 
Research Lead at Microsoft Research Special Projects\\    
Senior Advisor, GETTING-Plurality, Ash Center, Harvard

\vspace{7em} 

\noindent
\textbf{Michiel Bakker}\\ 
Professor at MIT Sloan and MIT Institute for Data, Systems, and Society\\ Senior Research Scientist at Google DeepMind \\


\vfill 


%% file: content/acknowledgments.tex


\chapter*{Acknowledgments}
\pdfbookmark[0]{Acknowledgments}{acknowledgments}

{\setlength{\parskip}{0pt}

First and foremost, this thesis owes its existence to Robert, without whom my PhD would have been unacceptably boring. To the many hours spent in the office scheming, and the many more hours spent traveling on absurd adventures, I owe much of the joy of this PhD to your friendship. I'll be retaining you as general counsel for all future adventures.

To Sandy for the wisdom and guidance, even when I didn't realize where the nudge was going--and for taking on one last PhD student.

To Hope for teaching me how to thrive in America, and to Nic for inconveniently timed yet immeasurably life-enriching calls. 
To Thomas for sparse calls but spectacular adventures and shared dreams. 
To Claire for showing me that New England is for more than work. 

To Zivvy for showing that life is better when it's a little crazy, and to Morgan Frank for getting lunch with me on a Spring day in Adelaide, your support unlocked everything that came after.
To Lewis and Matt, who, above all others on the list, taught me what it meant to do science and ask the right questions. 

To Longpre et al. for teaching me the art of coauthorship and to Wilson and Hardjono for providing adult supervision in the lab. To Kush for desk exchanges and Naana for being human.
To Calvin and Habib for believing I could be a VC, still yet to be proven right, and to Sarah for teaching me how it works.

To Glen for unlocking more unexpected opportunities for real connection and growth than I could have hoped for, and to Michiel for a jovial wisdom that's hard to find. To Shrey for figuring out what plurality means with me, and to Jules for showing me what a stylish yet smart PhD looks like. To John W for creating events punctuating my time at MIT like nothing else.

To everyone at the MIT Media Lab and Camberville community for making me feel at home, for which there are too many names to mention. In ranked order by appearances in my spreadsheet, Marcus, Al (and the Calvin fam), Suyash, Maisy, Brian (and the Ashdown squad), Mohsen, Berke, Keeley (the third spice weasel), Paris, Eyal, Alfonso, Celia, Jason, Jiaxin, Joost, John Werner, Oriana, and so many more.

To all my interns and UROPs (and Zac), for waking up for my early morning Zoom calls and visiting me in Boston.

And finally to my Mum, who is unfailingly supportive in everything I do (and cries every time I leave Australia), and my Dad, whom I become a little more like every day (for better or worse).
} 

%% file: chapter-1/chapter-1.tex
\chapter{Risks and opportunities for privacy and security in general-purpose AI}\label{chapter:1}

\epigraph{``Trust, but verify.''}{\textit{Russian Proverb (popularized by Ronald Reagan)}}

Artificial intelligence is no longer the stuff of science fiction speculation; it is rapidly becoming the invisible architecture scaffolding our modern world. From the algorithms curating our news feeds and guiding our financial decisions to the fast-growing tool for answering our daily questions and helping us complete work tasks. This integration promises unprecedented efficiency, innovation, and convenience. Yet, as its influence grows, so too does the complexity—and opacity—of the systems that power it.

With the advent of powerful foundation models and large language models (LLMs) these concerns are amplified. How can we trust the outputs of systems whose internal workings are often opaque? How do we protect the sensitive data they process? How can we verify the claims made about these systems—-claims regarding their capabilities, their fairness, or their adherence to safety protocols? When something inevitably goes wrong, how do we establish accountability in a chain of algorithmic decision-making?

These are not just technical puzzles; they are fundamental questions at the heart of our societal and economic stability in a future underpinned by AI.Without robust frameworks for security, confidentiality, and accountability, we risk building our future on foundations of sand (in both the literal silicon wafer and figurative sense), vulnerable to unforeseen failures, malicious exploitation, and an erosion of public trust that could derail progress altogether.

This thesis confronts this challenge. It opens with a critical question: How can we engineer end-to-end security within the sprawling ecosystems of large-scale AI, and why is achieving this not just beneficial, but absolutely essential for the AI-reliant society we are becoming? This is not a question answerable by technology alone, nor by policy alone. It demands a synthesis between the technical realities of AI development and the societal imperatives articulated in emerging international policy discussions.

Drawing from both these domains, this research undertakes a systematic exploration of the multifaceted risks that haunt the AI pipeline--from data collection and model training to deployment and real-world interaction. We dissect the vulnerabilities that threaten the security of the systems themselves, the confidentiality of the data they handle, and the accountability of the outcomes they produce.

But identifying risks is only the first step. The core contribution of this work lies in charting a path forward through pragmatic, novel technical solutions. This thesis argues that the bedrock for a safe, trustworthy, and ultimately prosperous AI future rests upon the deliberate and sophisticated balance of three crucial pillars: \emph{privacy}, \emph{verifiability}, and \emph{auditability}. These are not optional add-ons but foundational design principles.

\section{How this thesis is structured}

\emph{In many chapters and important sections of the thesis, an italicized note will be added at the top of the text to highlight relevant published academic papers that are included in the chapter as well as key concepts and ideas that are presented in the chapter. Look out for these notes to help navigate the thesis.}

This opening \autoref{chapter:1} serves as an introduction to the thesis, outlining the key concepts and risks that will be explored, before presenting two forward-looking pieces on the risks to privacy posed by AI and the roadmap towards building solutions to address these risks. The first is drawn from the `Risks to Privacy' and `Methods for Privacy' sections of the First International AI Safety Report \cite{bengio2025internationalaisafetyreport} that I had the honor of authoring, which highlighted, at a global level, the challenges to privacy that are presented by the creation, use, and downstream application of AI systems. The second draws from a paper titled `A Roadmap for End-to-End Privacy and Security in Generative AI' \cite{South2024Roadmap}, published by MIT Press as part of the MIT President's call for work on critical issues in Generative AI. This project brought together cryptographers and AI experts from across MIT to highlight the importance of privacy and auditability, and present a technical roadmap towards end-to-end security.
Like all papers included in this thesis, the original manuscript content has been edited for clarity, conciseness, and coherence with the rest of the thesis.

\autoref{chapter:2} explores the use of zero-knowledge cryptography, specifically zkSNARKs, to make verifiable claims about AI models across their lifecycle without revealing sensitive information like model weights. It presents methods for creating verifiable evaluations of model performance and fairness \cite{SouthVeriableEvaluations} and discusses how these principles can be applied to ensure transparency in other areas, such as verifying the provenance of training data. It draws on my work on verifiable evaluations of model performance and fairness \cite{SouthVeriableEvaluations} and zero-knowledge data attestations \cite{zkTax}.

\autoref{chapter:3} focuses on the privacy and auditability challenges related to Retrieval Augmented Generation (RAG)--the process of LLMs retrieving external information to answer queries. It introduces technical solutions using cryptographic methods like multi-party computation (MPC) \cite{Zyskind2023} and trusted execution environments (TEEs) \cite{southsecure} to enable secure and auditable RAG systems, ensuring private data remains protected while allowing for updates and verification.

\autoref{chapter:4} tackles the security implications of increasingly autonomous AI agents that can act on behalf of users. It proposes frameworks for authenticated delegation \cite{south2025authenticated}, allowing users to securely grant and restrict permissions for AI agents. The chapter explores how to extend existing authentication protocols (like OAuth 2.0 and OpenID Connect) and define clear access controls to ensure agent actions are authorized, auditable, and accountable. It also touches upon verifying the human origin behind agent actions using concepts like personhood credentials \cite{Adler2024k}.

\autoref{chapter:5} synthesizes the technical contributions presented throughout the thesis. It connects the concepts of privacy, verifiability, and auditability, showing how the proposed cryptographic tools and frameworks (zkSNARKs, TEEs, MPC, authenticated delegation) can be combined to build end-to-end secure AI systems. It revisits the core thesis questions, outlines a path toward safer AI, and discusses the practical implications for system designers and policymakers.

Collectively, this work offers a distinct perspective: that through the rigorous application of technical methods, particularly those grounded in security and cryptography, we can construct foundation model-based AI systems that reconcile the often-competing demands of privacy, verifiability, and auditability. It is a perspective grounded in the belief that engineering trust is not only possible but paramount, as we continue our journey across the algorithmic tightrope towards an increasingly intelligent future.

\subimport{./}{risks.tex}

\subimport{./}{roadmap.tex}

%% file: chapter-1/risks.tex
\section{Risks to privacy from AI systems} \label{c1:sec:privacy-risks}    

\emph{This section is based on the `Risks to Privacy' section of the First International AI Safety Report \cite{bengio2025internationalaisafetyreport} with additional edits to clarify concepts in this thesis.}

Artificial intelligence (AI) has rapidly evolved from a niche scientific pursuit into a transformative technology impacting countless aspects of modern life. At its most fundamental level, all AI operates on an input-to-output basis: it processes data provided to it (input) to generate a result (output). However, the capabilities and applications of AI vary significantly. While many AI tools are specialized for one specific function, a new and powerful category has emerged: general-purpose AI.

General-purpose AI distinguishes itself not by performing a single task well, but by its remarkable versatility. These models or systems demonstrate the ability to handle a diverse range of cognitive tasks, such as summarizing lengthy texts, generating original images from descriptions, translating languages, or writing functional computer code. This flexibility often stems from underlying technologies like large language models (LLMs), sometimes referred to as `foundation models' or `generative AI'. Terms that are used interchangeably in this work.

To navigate the complexities of this field, it is crucial to differentiate between two core concepts: AI models and AI systems. An AI model can be thought of as the foundational mathematical engine -- the raw, trained intelligence that powers AI applications. It represents the core capability, developed through extensive training on data. An AI system, on the other hand, is the practical application built around one or more AI models. It integrates the model(s) with other necessary components (like user interfaces, data processing pipelines, and safety mechanisms) to create a tool designed to be useful to humans in a specific way. For instance, the widely known ChatGPT application is an AI system; the powerful engine driving its conversational abilities, such as GPT-4, is the underlying AI model.

Understanding this interplay between versatile AI models and the integrated AI systems built upon them is essential for appreciating the capabilities, applications, and potential implications -- including the range of risks -- associated with the powerful general-purpose AI technologies shaping our world. 

\subsection{What are privacy, auditability, and verifiability?}
Central to building safe and reliable AI are the concepts of privacy, auditability, and verifiability, which together form the bedrock for responsible AI development and deployment.

\emph{Privacy} is a notably complex and multifaceted term \cite{Nissenbaum2009c, GPASInternationalEnforcementCooperationWorkingGroup2023k, Solove2025r}. It can refer to (a) the broad right allowing individuals to control their personal information and decide who can access it; (b) specific technical mechanisms designed to reduce the likelihood of data leakage (such as differential privacy); or (c) the ability to control how an AI system uses one's data or to interact with the system confidentially. This section will delineate between risks related to training data, system use, and intentional harm below. In general, this thesis focuses on use risks, as this is a rapidly growing area of concern (given the widespread deployment of AI), yet remains less explored from a technical security perspective compared to others. Whereas training data risks and techniques like differential privacy have a longer history of research, and intentional harm risks are a primary focus of the AI alignment and safety community, use risks bridge the fields of cybersecurity and cryptography with the challenges of large foundation models to ensure confidentiality during system interaction.

Separately, \emph{auditability} is a critical component for understanding how AI systems are used and deployed, ensuring they behave as expected, and enabling traceability when harm occurs. Closely related to concepts like logging, monitoring, transparency, and oversight, this is a key property that allows for the confirmation of claims about AI systems, the rectification of harms, and the establishment of accountability. Privacy can sometimes stand in tension with auditability, but technology can play a crucial role in reconciling these objectives.

To these ends, \emph{verifiability} is the capability to confirm that specific properties (such as privacy guarantees or audit logs) or other claims about an AI system hold true, often using technical methods. Verifiability of confidentiality allows one to ascertain whether third-party deployers of AI or external data sources are maintaining privacy without needing to blindly trust the third party. Verifiability for auditability allows one to confirm that the claims and audit trails generated for an AI system are accurate and complete, even when privacy-preserving techniques are employed.

These needs stem from tangible risks: deploying AI at scale often requires systematic observability and auditability (frequently for regulatory compliance), while complete confidentiality of user data is also often necessary (and sometimes mandated by different regulations). This thesis explores these trade-offs and presents technical solutions designed to address these interconnected challenges.

\subsection{Privacy risks}
AI systems rely on and can process vast amounts of personal data, posing significant privacy risks. Such risks include loss of data confidentiality for people whose data was used to train AI systems, loss of transparency and control over how data-driven decisions are made, unauthorized use or processing of personal data \cite{EuropeanDataProtectionBoard2024j}, and new forms of privacy-related abuse that AI systems could enable. These risks are already present with existing AI tools but are exacerbated by the increased scale of training, capacity for information processing, and ease of use presented by AI. These privacy risks and definitions also do not capture the concepts in tension with privacy, such as auditability, verifiability, transparency, and utility. These concepts will be explored in more detail later.

For purposes of international policy discussion, the risks to privacy from General-purpose AI can be broadly categorized into: 
\begin{enumerate}
\item Training Risks: risks related to training and the collection of data (especially sensitive data), 
\item Use Risks: risks related to AI systems' handling of sensitive information during use, and
\item Intentional Harm Risks: risks that malicious actors will apply AI to harm individual privacy. 
\end{enumerate}

\begin{figure}[htbp]
    \centering
    \includegraphics[width=0.8\textwidth]{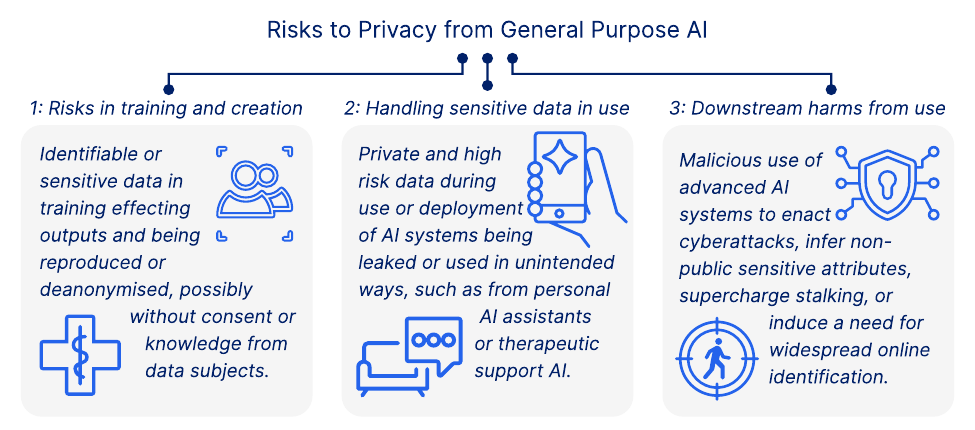}
    \caption{Risks to privacy from AI fall into three risk groups: (1) risks associated with training on sensitive data, (2) risks related to handling sensitive information during the use of AI, and (3) risks from malicious actors applying AI to compromise individual privacy. In general, this thesis will focus on use risks.}
    \label{fig:privacy-risks}
\end{figure}

\paragraph{AI systems may expose their training data.} The training of AI models generally requires large amounts of data. Academic studies have shown that some of this training data may be memorized by AI models \cite{Carlini2022m,Chen2023v}, enabling users to infer information about individuals whose data was collected \cite{Shokri2017m,Fredrikson2015i,Duan2024q} or to even reconstruct entire training examples \cite{Carlini2021u,Carlini2023k,Shi2023s,Lukas2023a}. However, definitions of memorization vary, so it is challenging to make concrete claims about the harms that might arise from memorization \cite{Carlini2022m}. Many systems are trained on publicly available data containing personal information without the knowledge or consent of the individuals it pertains to, in addition to training on proprietary web content owned by media distributors \cite{Longpre2024e,GPASInternationalEnforcementCooperationWorkingGroup2023k}. This extends to cases where one person posts personal information about another person online – for example, Facebook posts including pictures and information about a person's peers or friends without explicit consent from those peers. In specific domains, training on sensitive data (such as medical or financial data) is often necessary to improve performance in that domain, but could result in serious privacy leaks. These risks can be reduced – for example, existing medical AI systems such as Google's Gemini-Med \cite{Saab2024m} are only trained on anonymized or pseudonymized public patient data – but more research is needed to assess the risks associated with this. Privacy-preserving training approaches or synthetic data may help address this.

\paragraph{Information used during the application of AI can be leaked, such as private data used to personalize responses.} AI models do not have knowledge of current affairs occurring after their training or knowledge of private information not included in the training data. To address this, it is common practice to provide relevant contextualizing information to AI systems during usage through the so-called `Retrieval Augmented Generation' (RAG) \cite{Lewis2020l,Karpukhin2020j,Ram2023w}. This process can also allow for personalized responses using private personal data, for example, with personal assistant AIs on phones \cite{GeminiTeam2023a,Gunter2024a}. It can also be used to include external information, such as web search results \cite{Nakano2021n}, in the context used to provide a response. These can be combined; for example, a healthcare AI support tool may include or access sensitive medical records about an individual and then search the web or medical databases for relevant information before providing a response to support a clinician. While the use of on-device private data can make AI more useful, it can create additional risks of leaking this data. Risks of information leakage to third parties increase substantially when data (or insights from the data) leave a device \cite{Arora2023d,Zyskind2024j}, although cybersecurity approaches can minimize these risks \cite{UKNationalCyberSecurityCentre2023e}. In practice, balancing privacy, user transparency, and consumer utility in this context is a difficult challenge; technical approaches to balance this exist, but it is also important to find policy approaches that safeguard rights, enable transparency, and create trust for data sharing to promote innovation. 

\paragraph{AI systems could enable increased privacy abuse by malicious actors.} There are many scenarios relevant to privacy risk in which malicious users may exploit AI's increased information processing capabilities. For example, fine-grained internet-wide search capabilities, such as powerful reverse image search or forms of writing style detection, allow individuals to be identified and tracked across online platforms, and sensitive personal characteristics can be inferred \cite{Kosinski2013b,Weidinger2021g} (such as gender, races, medical conditions, or personal preferences), further eroding individual privacy \cite{Staab2023t}. LLMs can enable more efficient and effective searches for sensitive information in data. Detection, redaction, or sanitization of personally identifiable information alone is insufficient to fully mitigate inference of sensitive personal content: many user attributes, such as detailed sexual preferences or specific drug use habits, can often still be found from `redacted' data \cite{Mireshghallah2024q}, although AI systems may also be useful in supporting the monitoring and removal of sensitive information online. These risks can arise across many contexts and may result in broad unauthorized processing of personal data. This includes risks associated with the ability of AI systems to infer private information based on model inputs \cite{Weidinger2021g,Gabriel2024c}. Beyond analysis and search, AI content generated using private data, such as non-consensual deepfakes, can be used to manipulate or harm individuals. This raises concerns about the harm caused by the malicious use of personal data and the erosion of trust in online content.
\\

The increased prominence and capabilities of AI have led to its increased use in sensitive contexts and subsequent scrutiny of its possible violations of privacy laws. AI is now more common in contexts with sensitive data, such as personal devices with smart assistants \cite{GeminiTeam2023a,Gunter2024a} and healthcare \cite{Lamb2024v}. Privacy harms from training on sensitive data may not become realized for an extended period after training since the time between the collection or use of data and the subsequent deployment of an AI system may be substantial. Regulators are increasingly enforcing privacy laws to protect consumers from companies that use AI without privacy controls or safeguards \cite{FederalTradeCommission2024n,FederalTradeCommission2023x}. Meanwhile, new modalities of interactions with AI create new risks to privacy. For example, high-quality video generation models \cite{Ho2022u} may be capable of memorizing video information (such as faces of students in live-streamed classrooms) or of being used to exploit privacy by reasoning over video data \cite{RekaTeam2024r} or through speaker identification \cite{OpenAI2024d} (for example, using AI to watch individuals and automatically take notes on their behavior). Other concerns about privacy from downstream consequences of AI have also emerged. For example, in the future, there may be a need to differentiate humans from capable AI online, which could make mass identification and subsequent online surveillance more likely \cite{Adler2024k}.

%% file: chapter-1/roadmap.tex
\section{A roadmap for end-to-end privacy and security in generative AI}

\noindent
\emph{Learning from the risks above, this section outlines the technical solutions that can be used to address training and use risks, and how to understand the difference between privacy and verifiability. Remember that generative AI here is a stand-in for any modern general-purpose foundation model-based AI system.}

End-to-end security and privacy are increasingly urgent as generative AI is deployed in organizations and integrated into our daily lives. Many partial solutions exist across the research and practice landscape, yet none provide an end-to-end solution for generative AI. These technologies address different threats or attackers and aim to satisfy varied security guarantees. Complex supply chains and computation pipelines used in training and deploying foundation models require combining a plurality of technical solutions to address security and privacy concerns. This roadmap provides a framework for clarifying the different security challenges encountered in AI and organizing the existing defense technologies to address them. First, we define the security goals we want these systems to achieve; contrasting privacy (which implies that no sensitive information, such as private training data, proprietary model weights, or sensitive user inputs, is leaked) and verifiability (which makes it possible to confirm the integrity of the computational steps in the generative AI pipeline). We examine how these goals apply to two types of attackers: an internal attacker who can interfere along the AI pipeline and an external attacker who can only see outputs. We show how different existing technologies for verifiability and privacy sit in this framework, where they can be used, and how they can be swapped or combined to achieve end-to-end security. No single technology will solve the security woes of AI. A modular and hybrid solution that combines existing and new innovations in cryptography and privacy can pave the way for a secure future for AI.

\begin{figure}[htbp]
    \centering
    \includegraphics[width=0.8\linewidth]{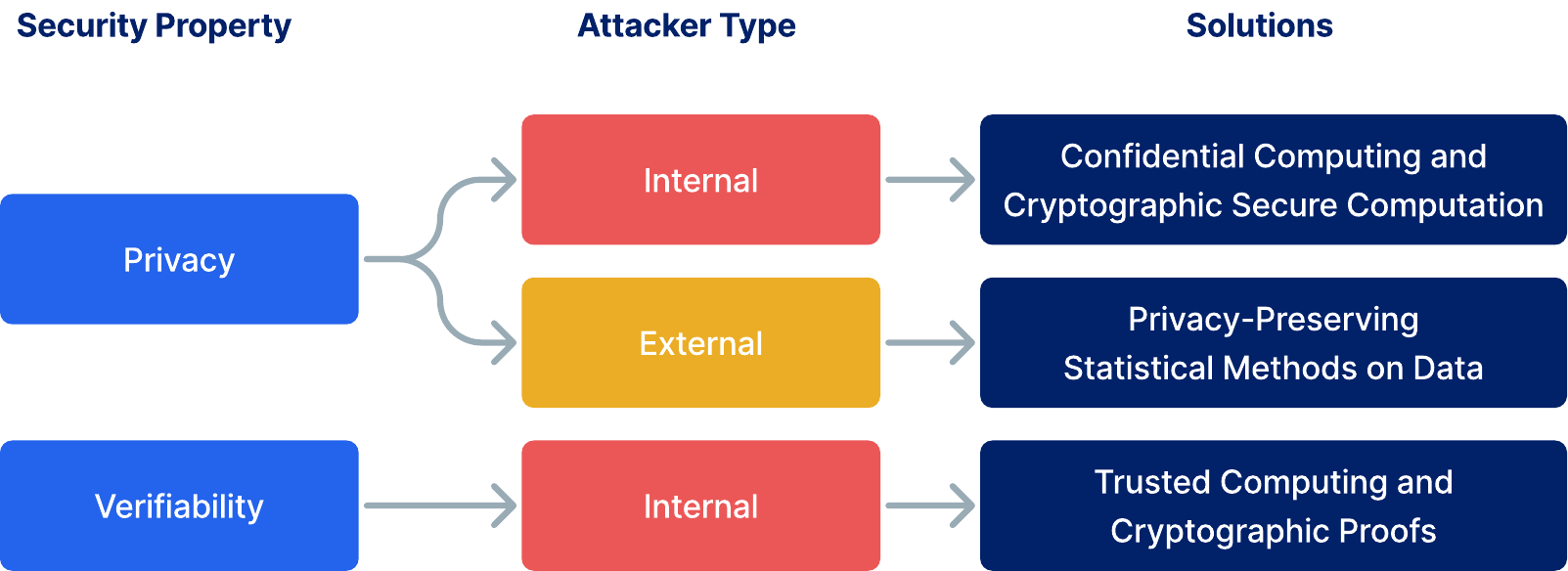}
    \caption{A high-level overview of the relationship between security goals, attacker types, and the solutions that can be used.}
    \label{fig:e2e-framework}
\end{figure}

\subsection{Why Generative AI Needs
Security}\label{sec:why-generative-ai-needs-security}

There is an increasing need for technologies to be end-to-end secure and handle user data privately in the twenty-first century; generative AI is no exception. The new capabilities of generative AI hold immense promise, yet unbridled, they present privacy and security risks to user inputs, training data contributions, model weights, and usage patterns. This has led to a status quo of models trained on undisclosed collections of scraped public content hosted by model providers that see all user behavior, retain the right to train future models on that user data, and provide no real security guarantees. Deployed AI models face significant privacy challenges. If users include sensitive information in their prompts (the input to a model), that data can leak via data breaches or when the model's usage data is used to train future AI systems. When model developers choose to ignore content privacy risks during training, they risk causing substantial downstream harm to individuals whose private information may be inadvertently exposed or misused by the AI model. These risks and limitations ultimately hold back the deployment of this technology. End users are concerned about how their data will be used, given the history of large tech companies exploiting their data; enterprises are wary of the risks created by uploading intellectual property or commercially sensitive data; and society at large is worried about how these risks interact with vulnerable populations such as children. Legislation such as the General Data Protection Regulation, California Consumer Privacy Act, and European Union AI Act have identified these concerns and created rules around the use of consumer data, explicitly calling out high-risk contexts in which privacy is critical \cite{EUAIACT}. If generative AI is to have widespread adoption, especially in high-risk contexts such as education and healthcare, new solutions for privacy and security in generative AI will be needed. Building this private toolkit is not just a matter of complying with emerging regulations but an opportunity to look forward to how we want private user data to be handled in a world where more and more of society is mediated through the lens of AI. Privacy can slow progress in the short term, yet it is critical to the technology's long-term growth. Building user trust, both from the content provider and end-user sides, is a necessary step in the full-scale rollout of generative AI technologies. Such a toolkit will not be static; new research questions will be asked and answered as AI models and the ways we use them evolve rapidly. This section provides a brief survey of existing solutions, a framework of how these tools fit into the broader picture of privacy and security for generative AI, and a call to action for researchers to fill in the missing gaps.

\subsection{The Supply Chain of AI and Its Actors}\label{sec:the-supply-chain-of-ai-and-its-actors}

\begin{figure}[ht]
    \centering
    \includegraphics[width=0.9\linewidth]{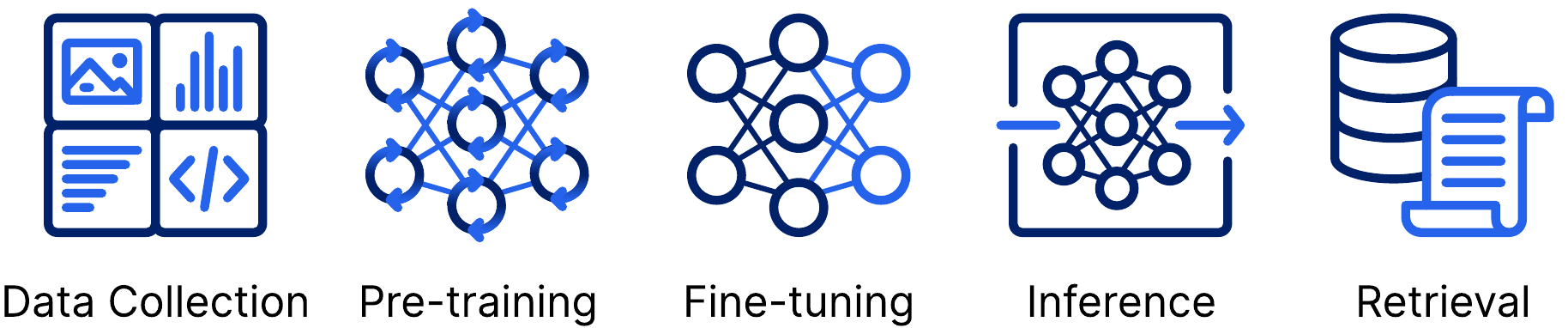}
    \caption{The key general computational components of the generative AI pipeline from data collection through to model inference and use. While many other components exist, these are the key building blocks for an end-to-end secure generative AI system.}
    \label{fig:e2e-pipeline}
\end{figure}

To understand where security matters for generative AI, we need to look at the computational pipeline and supply chain it relies on. For simplicity, let's boil down the pipeline of AI into five (not strictly ordered) steps. Generative AI starts with a `large-scale data collection' of web-scale pretraining data, domain-specific pretraining data, and other control datasets such as instruction finetuning data. This data is then used in the `pretraining of large foundation models' to learn the general purpose and emergent capabilities. These large models then undergo `finetuning' to become useful for a specific application. This can include instruction finetuning, reinforcement learning from human feedback, learning to take actions, or task-specific finetuning. This step requires separate training data, but the lessons and requirements from pretraining data collection still apply. Once the model is fully trained, it needs to be provided to users (in which users could be enterprises using AI agents, consumers using chatbots, or any other AI application). While there are many instances in which models are deployed on a local machine, the large scale of these models and their onerous compute hardware requirements result in many AI models being deployed in the cloud. In such an instance, inference refers to the model receiving a prompt, repeatedly generating output tokens (words), and sending the response back to a user. In addition to this inference, there is often a second step in which the AI model will request, or need to be provided, an additional piece of input. This paradigm of `retrieval augmented generation' (RAG) accesses databases of information external to the AI model and pulls that information into the context window (the content the AI can see) \cite{Lewis2020}. These steps towards the use of AI are performed by many different participants, each of which may have sensitive data to be kept private or computations that need to be trusted, and each is an opportunity for added security. Consider the following participants in the end-to-end supply chain of AI and where their concerns might lay: 

\begin{itemize} 
  \item   data contributors, who generate and provide training data for   foundation models for use in pretraining or finetuning 
  \item   model creators, who use such data to create and train AI models 
  \item   model deployers, who take pretrained models and host their inference   and general use for users 
  \item   end users, who receive the outputs of generative AI and use the models 
  \item   external databases, which may be accessed by the model during runtime   for RAG (this includes web browsing or other document   retrievals) \cite{Nakano2021n}. 
\end{itemize} 

These distinctions are amorphous. There are many instances in which two parties are the same, such as the common cases in which model creators are also model deployers (see OpenAI), model providers also provide the external database (such as Google Gemini using Google Search), or data contributors are also end users. Further, many of these categories can encompass multiple parties. An AI model can be trained by one company and finetuned by another (such as through reinforcement learning from human feedback as a service), in which both parties are lumped into model creators. Similarly, a large model may have been hosted for inference by a cloud provider but accessed through a third-party app that handles user inputs and model outputs (e.g., any software wrapping around a generative AI API); not only are both deployers here with respect to the personal end user but the third-party app is also an end user with respect to the cloud model inference host. This web of privacy and security dependencies can become complex quickly, but is underpinned by a set of relationships that can be individually addressed via specific security and privacy solutions between each step.

\subsection{Defining security in AI}\label{sec:defining-security-in-generative-ai} 
The concept of security is multifaceted, encompassing an extremely broad range of security properties for a system. To organize the different security objectives in generative AI that we discuss in this section, we condense the `threats' against AI systems into two principal attacker models, each calling for different defense strategies. We then define privacy and verifiability, the two specific security properties we want our generative AI systems to satisfy. 

\begin{figure}[h]
    \centering
    \includegraphics[width=\linewidth]{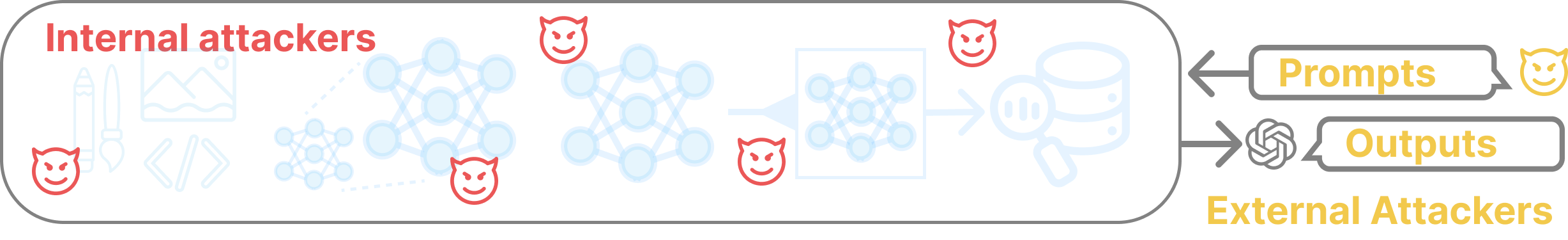}
    \caption{Two attacker models: internal and external.}
    \label{fig:e2e-attackers}
\end{figure}

\subsubsection{Attacker Models}\label{sec:attacker-models}

 An `attacker' model tries to describe the capabilities of a potential malicious actor who is trying to bypass the system's security. While many details are obfuscated here, at a high level, there are two types of entities we're worried about. 

\paragraph{Internal Attacker}\label{sec:internal-attacker}

 First, we address an attacker inside the computational pipeline. An organization or individual hosting, running, or controlling one of the computational steps might be ill-motivated. Such an attacker may wish to extract sensitive data or tamper with the computation itself. Consider the case of a malicious large language model provider that wants to access user input, a database provider that wants to know what AI is being used for, or an inference service wishing to deploy smaller models to reduce costs. 

\paragraph{External Attacker}\label{sec:external-attacker}

 Our second attacker only has black box access to the model. That means it has control over the prompts sent to the model and can observe the model responses. Here, the attacker can try to extract sensitive information such as training data or model weights. This encompasses practically any end user of an AI system trying to extract private or proprietary information about the model and the training data by prompting the model. 

\subsubsection{Security Properties}\label{sec:security-properties}

Noting the above threats, let's turn our attention to the security goals we actually want to achieve. At a broad level, the two properties we want AI systems to uphold are privacy and verifiability. 

\begin{figure}[h]
    \centering
    \includegraphics[width=\linewidth]{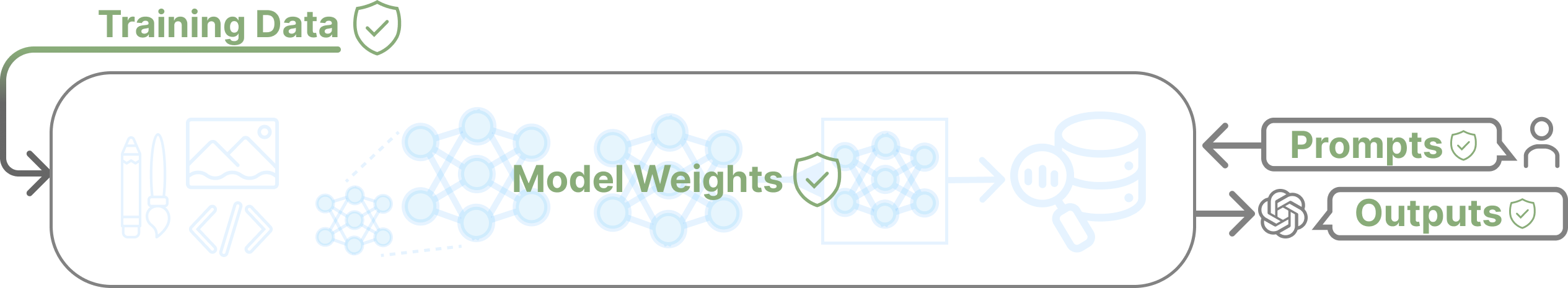}
    \caption{What are we protecting? Key classes of sensitive data in generative AI.}
    \label{fig:e2e-sensitive-data}
\end{figure}

\paragraph{Privacy}\label{sec:privacy}
 A system enforces privacy if it does not leak sensitive information to a potential attacker. We focus on three types of data to help us frame how to organize different existing threats and defense mechanisms. 
 
\begin{enumerate}
\item
  \textbf{Private training data:} The training of generative models
  requires huge amounts of data, for both pretraining and finetuning,
  which in some instances contains proprietary data or sensitive
  information regarding individuals. If a language model memorizes this
  data, it can be leaked to an end user. Similarly, at the inference
  stage, the model may query a database holding sensitive data, which
  could, in turn, be leaked to an end user. There are also instances in
  which private training data should be kept secret from an internal
  attacker. Many regulations mandate that personal or sensitive data
  (such as medical records) never leave the device or custodian server
  in their raw form, and only aggregate and privacy-preserved insights
  can be shared.
\item
  \textbf{Proprietary model information:} In many cases, model weights
  are the most valuable intellectual property across the pipeline.
  Keeping model weights or model architecture private not only from the
  public (an external attacker) but also from hackers or malicious
  actors within the supply chain (internal attackers) is paramount.
  There are also more complex instances of privacy here. Consider the
  case in which a benchmarking provider has private data (that they wish
  to keep private so that no one can train on it and game the
  benchmark), and a model provider wishes to keep their model weights
  private. Without cryptographic solutions or a trusted third party,
  these two are at an impasse in maintaining the privacy of their
  proprietary data while performing the benchmark on a private model.
\item
  \textbf{Sensitive user input:} When interacting with the generative
  model, a user might share sensitive information through their prompts,
  especially when the model is used in a sensitive context such as
  personal health or company record management. It is critical that this
  information isn't leaked to an external party or the public; equally
  important is that companies minimize how much these user records are
  transmitted to other parties in the generative AI supply chain.
\end{enumerate}

\paragraph{Verifiability}\label{sec:verifiability}

Verifiability is the ability to create guarantees and verify that computational steps of the AI pipeline have been run and done so correctly. Such a `computation step' can be extremely broad and can range from verifying that model inference was performed on the proper model to verifying a computational claim that the training data did not contain \emph{New York Times} URLs. Such a guarantee allows a user or actor in the supply chain to obtain a verifiable proof of the integrity of what has occurred inside an opaque system. This is useful when an end user wants to verify which model it is remotely accessing or if a company in the supply chain wants confidence in the computations performed by its suppliers. 

\subsubsection{How These Fit Together}\label{sec:how-these-fit-together}

These threats to generative AI systems are not competing goals. Some of the technical solutions we see below sometimes address several of these threats at once and often complement each other. Indeed, a key goal of many security solutions is to both provide privacy and allow a user to verify that the operations on the private information have succeeded.

\paragraph{What We're Not Focusing On}\label{sec:what-were-not-focusing-on-non-goals}

While the roles of privacy and auditability in security are already broad, it is critical to point out the large number of security goals we do not address here. The security properties above do not include risks associated with model behavior, including biased model outputs, effective guardrails on outputs or use, and dual-use capabilities. This is similar to how transport layer security (think of the secure `s' in `https') doesn't concern itself with the \emph{content} of the webpages it secures, only that it is secure from attackers. Some innate properties of models, such as their ability to be compromised by adversarial prompts or their tendency to hallucinate, are not part of this discussion.

Further, we don't address the risks created by the supply chain of code and training data. This data can provide undesirable knowledge (how to make a bomb) or be an attack vector for data poisoning \cite{Goldwasser2022}. While some of the tools we describe (e.g., differential privacy or zero-knowledge data proofs) can help address this issue, we leave the effects of data on the model outside the scope here. Supply chains of other inputs (open model weights and inference code) can also be a mechanism for attacks, but are not addressed via the tools laid out here.

These risks should be taken into account when designing end-to-end generative AI systems but cannot be addressed using the cryptographic and confidential computing mechanisms we present in this section.

\subsection{How Existing Solutions Can Address These
Threats}\label{sec:how-existing-solutions-can-address-these-threats}

\begin{figure}[h]
    \centering
    \includegraphics[width=\linewidth]{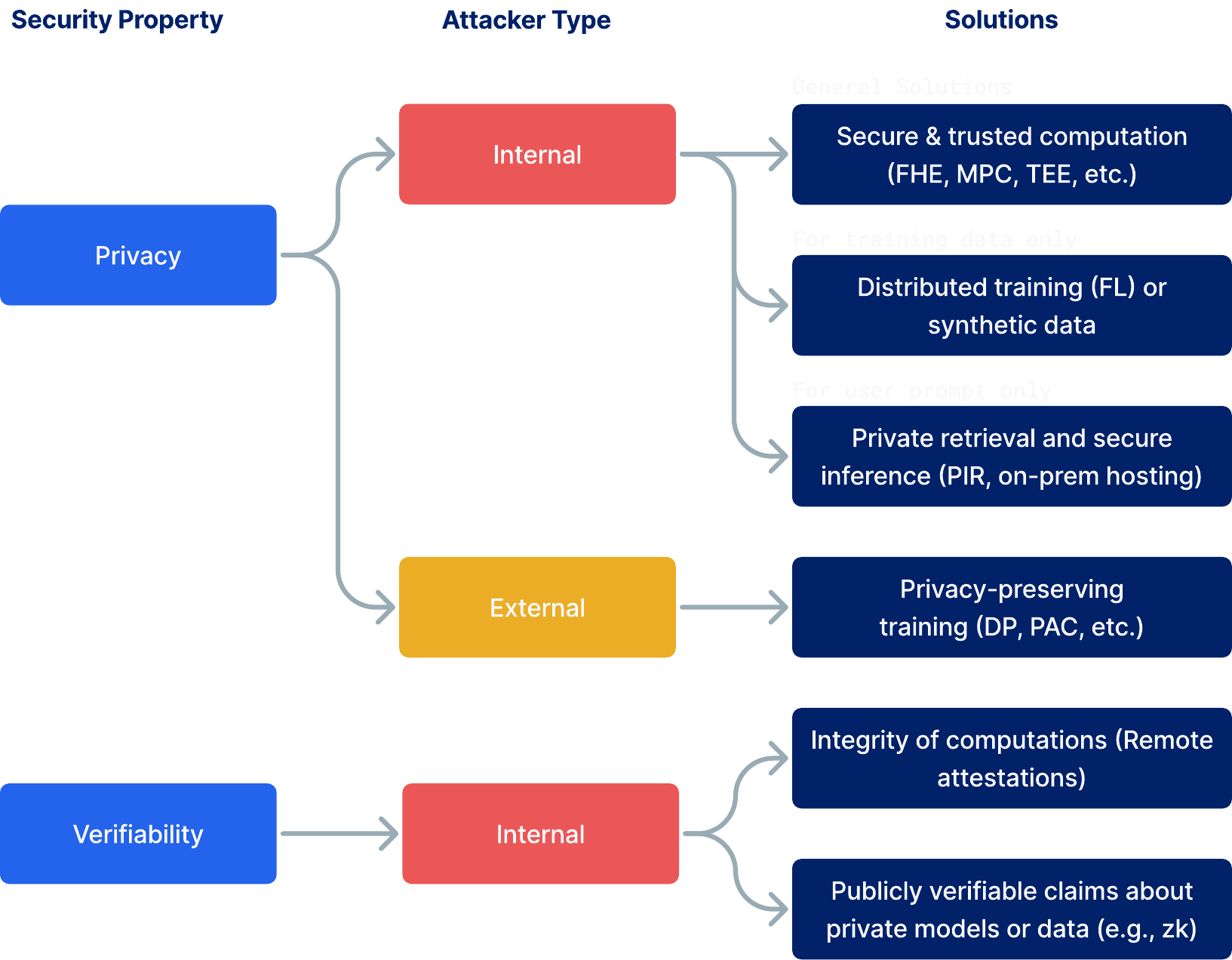}
    \caption{A summary of how classes of solutions exist in the framework presented.}
    \label{fig:e2e-solutions}
\end{figure}

There is a vast array of open-source technologies, commercial solutions, and new research that address the challenges of security in generative AI. For the most part, these tools will fit in the above framework of security guarantees and what they're defending from. For now, let's elaborate in more technical detail on some of these technologies to address the above concerns.

\subsubsection{Privacy, from an Internal
Attacker}\label{sec:privacy-from-an-internal-attacker}

\paragraph{General Purpose
Solutions}\label{sec:general-purpose-solutions} These solutions prevent the leakage of sensitive data by keeping it encrypted during transit and usage or by isolating the sensitive computation from other pieces of software at the hardware level. Each solution presents a different tradeoff between performance, security or trust assumptions, and usability, providing different generic primitives for secure computation that can address privacy concerns against an internal attacker.

\begin{itemize}
\item
\textbf{Homomorphic encryption (HE):} HE enables computations directly on encrypted data without requiring decryption, ensuring confidentiality throughout processing. HE usually relies on public key encryption schemes, which means that anyone can encrypt data using a public key, and several parties can provide different inputs for HE, but only the parties holding the private key can decrypt outputs or intermediate states. Unfortunately, HE has a significant performance overhead, sometimes orders of magnitude above basic computation, and implementing HE for the diverse operations involved in the pipeline can be quite complex. As a result, in practice, HE has been mostly used for inference, protecting the user's private prompts \cite{GiladBachrach2016}. Nevertheless, HE could also be used for model training, in which HE permits learning from encrypted datasets, promoting privacy-aware collaboration \cite{Hesamifard2017}.

\item
\textbf{Multi-party computation (MPC):} MPC makes it possible to split trust among several parties to perform sensitive computations securely.\footnote{\textbf{As a quick note on HE versus MPC:} HE is often described in the context of a single user holding sensitive data and the decryption key. This is sufficient in the simple case of secure inference, when the model is known to the service provider. In more elaborate cases, a threshold HE or a multi-key HE solution can be seen as a specific MPC solution. In these cases, no one party holds the decryption key, and instead multiple keys (or shares of a single key) are needed to decrypt the data.} This can allow generative AI inference to occur across multiple servers without any one server seeing the model weights or input data \cite{Mohassel2017}. MPC also makes it possible for different parties to each contribute their own private inputs without revealing those inputs to other parties. This can allow setups in which a user and a server can collaboratively run inference on the user's private prompts and the server's private model without revealing those secrets to each other. Like other cryptographic techniques, MPC can be extremely slow, especially when servers must communicate intermediate values to one another. There have been attempts to optimize the systems for machine learning, which has allowed small transformers to be run with MPC \cite{Knott2021,Li2022}.

\item
\textbf{Confidential computing:} Confidential computing, leveraging trusted execution environments or other secure hardware enclaves, allows a user to execute code on a remote server, for instance, in a public cloud, without trusting the software stack or the cloud provider. They rely on hardware primitives that isolate the sensitive computation from other pieces of software and allow for bare-metal performance with limited overheads. They are applicable across the computational pipeline and have recently been extended to GPUs (although challenges to deploying large language models still exist). These hardware solutions are exciting but lack the mathematical security guarantees provided by the cryptographic solutions above.

\item
\textbf{On-premise hosting}: Hosting sensitive parts of the process on premise can address many of the concerns around data privacy for the owner of the sensitive data by not involving external parties. For instance, training could happen on the local servers of the data contributor to protect training data, while inference could happen on premise of the model provider to protect model weights. While this privacy solution is simple and easy to understand, it is often infeasible. First, in modern pipelines, the owners of the different sensitive data might be different, and no clear party can be trusted to handle training data, model weights, and user inputs. Furthermore, large models can be hard to deploy, requiring expensive and difficult-to-manage infrastructure that is easier to find and maintain in public clouds. Nevertheless, on-premise computation can be a valuable and simple tool to protect computations in generative AI from attackers.

\end{itemize}
\subsubsection{Solutions Specific to Securing Training
Data}\label{sec:solutions-specific-to-securing-training-data}

\begin{itemize}
\item
\textbf{Federated learning}: Federated learning (or its related alternative, split learning) is a training method that distributes the training of a model across each data contributor, allowing models to be trained without raw data leaving the contributor's device or local server \cite{Yin2021}. This can slow down or limit the fidelity of training but avoids the risk of exposing raw data to a potential internal attacker, like a malicious centralized server, while allowing patterns to be learned across diverse data sources.

\item
\textbf{Synthetic data:} Synthetic data in machine learning refers to artificially generated data designed to emulate real-world data without being collected from real-world observations \cite{Cheng2021}. It is created to replicate the statistical properties and patterns of real data but is often an incomplete representation. Synthetic data is commonly used when real data is limited, expensive, or difficult to obtain or when privacy concerns prohibit the use of real data. By generating and transmitting only synthetic data, other actors in the training pipeline cannot see the sensitive original data. 
\end{itemize}

\subsubsection{Solutions Specific to Securing User
Input}\label{sec:solutions-specific-to-securing-user-input}

\begin{itemize}
\item
\textbf{Local execution:} Perhaps the most obvious solution to ensuring the privacy of user inputs is to store all data and execute all model inferences locally on the user's device. There are many contexts in which this is a useful privacy solution that is both simple to implement and easy for consumers to understand. However, many models, especially the most powerful ones, are extremely hard to run locally due to their size. This makes many use cases, especially those involving phones, very hard without cloud computing.

\item
\textbf{Private information retrieval (PIR):} Putting aside training and inference, many modern generative AI workflows leverage external databases or web searches to generate relevant content. This ecosystem of retrieval augmented generation often leaks information about how the generative AI system is being used to external databases and tools. PIR can be critical in protecting this information from leaking to external parties. PIR can be done using various technologies, including searchable symmetric encryption, private set intersection, or MPC, and has been shown to be useful in generative AI contexts for web search and database retrieval \cite{Henzinger2023,Zyskind2023}.
\end{itemize}

\subsubsection{Privacy, from an External
Attacker}\label{sec:privacy-from-an-external-attacker}

So far, all approaches have focused on protecting input privacy throughout the computation itself. Still, they did not capture any leakage that may have occurred from the model's output to an external attacker. Thus, we need other approaches that specifically focus on protecting against an attacker who intentionally extracts data by repeatedly querying the model and examining the outputs.

\begin{itemize}
\item
\textbf{Differential privacy (DP) in model learning (and beyond):} DP is a statistical method that ensures the privacy of training data from model memorization by injecting noise during training in order to maintain a given `privacy budget.'\cite{Cummings2024} This is the first of these solutions to address the privacy of training data from end users directly and can, to an extent, allow for the training of large models on private data at either pretraining or finetuning.\cite{Yu2021} This can be combined with tools such as federated learning but can also reduce the performance gain or knowledge improvement that a piece of data could contribute.

\item
\textbf{Other statistical noising approaches:} Recent work has presented other approaches to adding privacy during training, such as PAC (Probably Approximately Correct) Privacy, Pufferfish privacy, and others\cite{Xiao2023,Kifer2014}. These approaches take alternative formulations on how to minimize the memorization and regurgitation of training data while still learning generalized information from the data.

\item
\textbf{Noise introduced by other methods:} Federated learning and the synthetic data above can also introduce limitations on memorization and the leaking of individual data points, which, beyond limiting the transmission of raw data, can help avoid leaking sensitive information to an end user. Unfortunately, these methods usually do not provide any formal guarantees alone, and strong privacy for training data can only be achieved by combining them with the ones presented above.

\item
\textbf{Using privacy on prompts and retrieved documents:} While the above privacy methods are most commonly used to minimize information leakage during training, they can sometimes also be used during inference time. Since both prompts and retrieved documents can be represented numerically in embeddings, the above privacy tools can be applied to limit the ability of an external attacker to see retrieved data or original prompts (or, separately and related to the internal attack, could be used to obfuscate the prompt from the inference provider at the cost of accuracy).
\end{itemize}

\subsubsection{Verifiability}

Verifiability enables checking that no one is manipulating the AI pipeline and allows statements about generative AI that cannot be faked. These verifiable statements and guarantees of integrity may be seen by internal or external parties and protect against attackers who are manipulating the internal pipeline, not the external prompts.

\begin{itemize}
\item
\textbf{Remote attestation of confidential computing:} Many of the privacy tools outlined above also allow a deployer or user to verify that privacy has been maintained and computation occurred correctly. This ability to check on the process is possible in everything from trusted platform modules to MPC, but its implementation varies. This verification itself rarely slows down computation significantly but is often associated with the heavy cryptographic tools that already create overhead.

\item
\textbf{Zero-knowledge model proofs:} The rapidly growing field of zero-knowledge machine learning allows proofs of model inference to be run that allow a third party to verify the inference occurs while maintaining privacy over model weights or inference inputs\cite{Sun2024}. A key aspect of this is having confirmation that the correct model was being run when inference is external \cite{Kang2022}. This can, in turn, enable powerful additional guarantees like verifiable evaluations of model performance without exposing model weights or benchmarking data to end users or the public eye \cite{SouthVeriableEvaluations}. Proving is, however, very slow compared to standard inference.

\item
\textbf{Attestations about data:} The public or other parties in a supply chain may often have questions about training data which can, at times, contradict our privacy requirements. Using methods such as Merkle trees or smaller zero-knowledge proofs, parties can share attestations (verifiable statements) about the data they are using. This can allow for verifiable data provenance or confirmation that sensitive data was excluded (such as through an AI Bill of Materials).
\end{itemize}

\subsection{How These Solutions Can Be
Combined}\label{sec:how-these-solutions-can-be-combined}

Building and deploying generative AI systems require enforcing security at each step along the supply chain. Guaranteeing robust privacy and verifiability for an end-to-end system will require composing the above solutions as building blocks to cover each corresponding threat, as no one approach covers the entire attack surface.

Solutions that address different security properties or attacker models can be combined to cover as many threats as needed, while ones addressing the same risk can be used at different steps of the pipeline. Often, methods addressing the same type of threat can be compared and offer a tradeoff between performance, security, and usability. For instance, fully homomorphic encryption offers strong security relying on well-studied cryptographic assumptions, but performance overheads make it impractical in most cases. On the other hand, multiparty computation relaxes assumptions on trust by requiring some servers to be trusted but offers better performance.

To illustrate how to compose these different solutions, let's examine a real-life example. Consider the canonical example of using AI in healthcare, where privacy is paramount, and generative AI has huge potential for improved diagnosis and personalized treatment. First, large foundation models require huge quantities of training data, but not all data needs the same privacy constraints. Data collection and pretraining could start using open public data to build a base capability within the model. Such training could be attached to zero-knowledge attestations about training data to enable the public to audit inputs verifiably. Pretraining could then shift to data owned by hospitals that has sensitive attributes or personally identifiable information; for this, we could use federated learning to make sure no central server learns information about the patients but combine it with differential privacy to guarantee that the final model will not leak patient private data to a malicious user. This model could now undergo fine-tuning for instruction following or alignment. These approaches so far help keep training data private from internal and external attackers by ensuring that sensitive data isn't exposed to a cloud provider or regurgitated to an end user.

Next, deploying this model could be done in various ways to fit a wide class of privacy guarantees. For general questions, standard cloud hosting will suffice with traffic encryption to avoid web snooping. However, for sensitive queries involving patient data, we may want to turn to a solution that ensures the privacy of user prompts with regard to the cloud provider. One such example is hosting the model in a trusted execution environment such that even the model provider cannot access or see what user prompts are being passed to it.

This model usage can, in turn, be augmented by assessing up-to-date medical data from databases. For simple augmentation, such as local patient data, it might suffice to have a locally hosted database with the model hosting or the client; for external data, such as new clinical practices for external vendors, private information retrieval could be used, ensuring that the external database doesn't see the sensitive content of prompts.

This example is all possible using existing privacy and security technologies available today. It demonstrates how such a patchwork of solutions can be combined to provide real privacy guarantees across the computational pipeline of AI. In many of the above instances, more advanced technologies could replace existing ones. For example, with the improvement of HE for machine learning, the inference step could be entirely performed under FHE. While a wide array of new approaches can be expected over time, this framework in which different aspects of the generative AI life cycle need different solutions will remain true.

\subsection{Where Does the Future
Lie?}\label{sec:where-does-the-future-lie}
The world will see an increasing demand for privacy from consumers and enterprises when generative AI is evermore integrated into our lives and used in sensitive contexts. To address this, end-to-end security will be required in many settings. As we see above, this will require combining technologies to balance speed, security, and usability. Even in cases in which simpler solutions, such as locally hosted models, become popular, there will still be a need for specific technical instantiations of privacy and security to enable these models to reach their fullest potential.

Regulation will further drive this. AI regulations increasingly require developers and deployers to consider the role of security and privacy in AI systems. The EU AI Act highlights the importance of security, requiring high-risk AI systems (which includes systems deployed in education, employment, and emotional recognition) to be ``resilient against attempts by unauthorized third parties to alter their use, outputs or performance by exploiting system vulnerabilities'' (Art. 15), and where AI systems are evaluated on sensitive personal data, these systems must include ``state-of-the-art security and privacy-preserving measures'' (Art. 10) to safeguard the personal data. Similarly, President Biden's Executive Order on the Safe, Secure, and Trustworthy Development and Use of Artificial Intelligence calls for AI systems to be ``resilient against misuse or dangerous modifications.'' These statements are echoed across regulations around the world, and their compliance obligations will create a potent demand for privacy and security for AI systems.

\begin{figure}[htbp]
  \centering
  \includegraphics[width=0.57\linewidth]{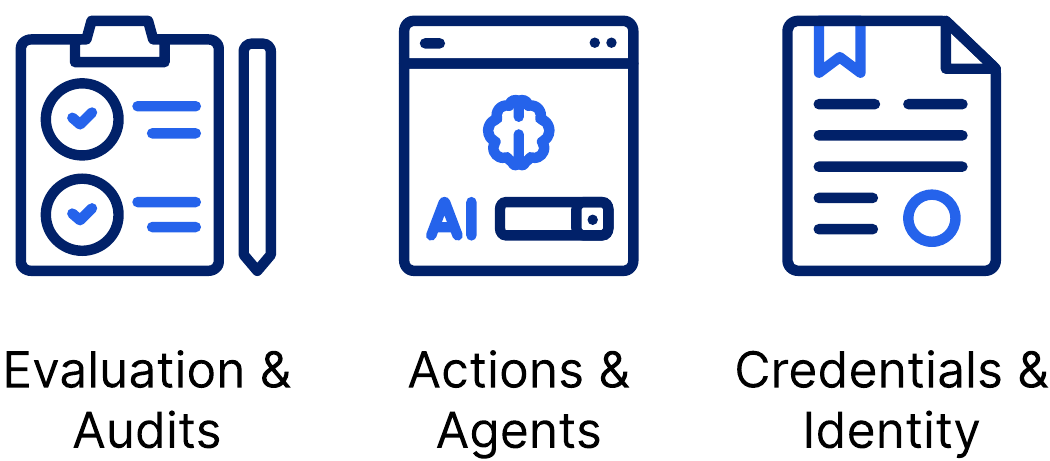}
  \caption{Three key elements of the AI ecosystem that we didn't discuss in this section, but will return to throughout this thesis: evaluations, agentic behavior, and systems credentials. While not necessary for end-to-end security in generative AI, these elements are critical for the safe and trustworthy deployment of AI.}
  \label{fig:not-in-e2e}
\end{figure}

Agentic behavior can follow naturally from this existing patchwork approach. Much discussion and effort have gone into developing AI systems that can take action in the world (agents) built upon the utility provided by foundation models. As foundation models develop new capabilities and modalities, cryptographic and privacy tooling needs only be developed for these new applications and then integrated into the existing stack of software designed for the trustworthy use of AI systems. 

Still, more research is needed across this computational pipeline and supply chain. While much of the existing work in security for AI is production-ready (and many tools are widely used), other aspects still require scaling, speed improvements, production testing, or continued deep research. 

%% file: chapter-2/chapter-2.tex
\chapter{Verifiable claims about models and data}\label{chapter:2}
\epigraph{``zkSNARKs may be to cryptography what transformers are to AI: a general-purpose technology that is so powerful that it will completely steamroll a whole bunch of application-specific techniques for a whole bunch of problems developed in the decades prior.''}{\textit{Vitalik Buterin, co-founder of Ethereum}}

\noindent
\emph{Following from the roadmap to privacy and auditability in the previous chapter, this chapter dives deeper into a specific technology--zkSNARKs--to examine the range of applications it can enable. This chapter draws on three papers I co-authored. It will show how we can build a verifiable approach to AI model evaluation, and will then explain how these principles can be extended to other parts of the AI supply chain, such as through training data attestations.}

One of the hardest problems to solve in the AI development and deployment process is balancing the needs for privacy and auditability. This is a theme that will come up repeatedly in this thesis, and this chapter is the first to provide a concrete technical solution.

This chapter explores how we can use a specific cryptographic technology, zkSNARKs, to enable verifiable claims about models and data. What this means is that we can make statements about AI systems and their evaluations, deployments, and data, in such a way that we can mathematically verify with certainty that the statement is true, without revealing any additional information beyond the truth of the statement itself. Model weights, training data, and other sensitive information can be verified without being revealed.

As the epigraph suggests, zkSNARKs represent a paradigm shift in verifiable computation. Their core magic lies in the ability to prove that a specific computation was performed correctly---yielding a particular result---without revealing any of the secret inputs involved in that computation. Imagine being able to prove a model achieved 95\% accuracy on a benchmark dataset without revealing the model's architecture or weights, or proving that a dataset used for training excluded certain types of sensitive information without revealing the data itself. This capability directly addresses the opacity and trust deficits inherent in much of contemporary AI development and deployment.

Essentially, zkSNARKs allow someone (a prover) to convince anyone else (a verifier) that a statement about a computation is true, without revealing the secret inputs used in that computation. The name of zkSNARKs itself provides a hint as to what they are:
\begin{itemize}
    \item \textbf{Zero-Knowledge}: The prover can prove the statement without revealing any additional information.
    \item \textbf{Succinct}: The proof is small and can be verified quickly.
    \item \textbf{Non-Interactive}: The prover can generate the proof without the verifier needing to be online.
    \item \textbf{Arguments of Knowledge}: The proof is such that the verifier can be convinced of the truth of the statement without having to know the entire statement itself.
\end{itemize}

Here, we move beyond the theoretical promise to explore the practical engineering required to apply zkSNARKs within the complex AI lifecycle. This chapter demonstrates how this technology can serve as a cornerstone for building systems that balance the pillars of privacy and verifiability identified previously. We will investigate three key applications: first, establishing verifiable evaluations of complete machine learning models, allowing third parties to confirm performance or fairness metrics without direct model access; second, tackling the significant computational overhead of zkSNARKs by exploring verifiable computation for parts of AI systems, enabling pragmatic trade-offs for large-scale models; and third, extending these principles beyond models to create verifiable attestations about data itself, enabling trustworthy data provenance and usage claims throughout the AI supply chain. Through these explorations, this chapter aims to substantiate the claim that cryptographic verification is not just a theoretical possibility, but an increasingly viable engineering solution for building more trustworthy AI.

\subimport{./}{verifyevals.tex}

\subimport{./}{partialzk.tex}

\subimport{./}{zktax.tex}

%% file: chapter-2/verifyevals.tex
\clearpage
\section{Verifiable evaluations of machine learning models using zkSNARKs}\label{sec:verifyevals}


\begin{paperabstract}
In a world of increasingly closed-source commercial machine learning models, model evaluations from developers must be taken at face value. These benchmark results---whether over task accuracy, bias evaluations, or safety checks---are traditionally impossible to verify by a model end-user without the costly or impossible process of re-performing the benchmark on black-box model outputs. This project presents a method of verifiable model evaluation using model inference through zkSNARKs. The resulting zero-knowledge computational proofs of model outputs over datasets can be packaged into verifiable evaluation attestations showing that models with fixed private weights achieve stated performance or fairness metrics over public inputs. We present a flexible proving system that enables verifiable attestations to be performed on any standard neural network model with varying compute requirements. For the first time, we demonstrate this across a sample of real-world models and highlight key challenges and design solutions. This presents a new transparency paradigm in the verifiable evaluation of private models. 
\end{paperabstract}

Model transparency, bias checking, and result reproducibility are at odds with the creation and use of closed-source machine learning (ML) models. It is common practice for researchers to release model weights and architectures to provide experimental reproducibility, foster innovation, and iteration, and facilitate model auditing of biases. However, the drive towards commercialization of models by industry (and, in the case of extremely large language models, the concern over the safety of open-source models) has led to the increasing practice of keeping model weights private \citep{karpur2023securing,brown2020language}.

\begin{figure}[th]
    \centering
    \includegraphics[width=\textwidth]{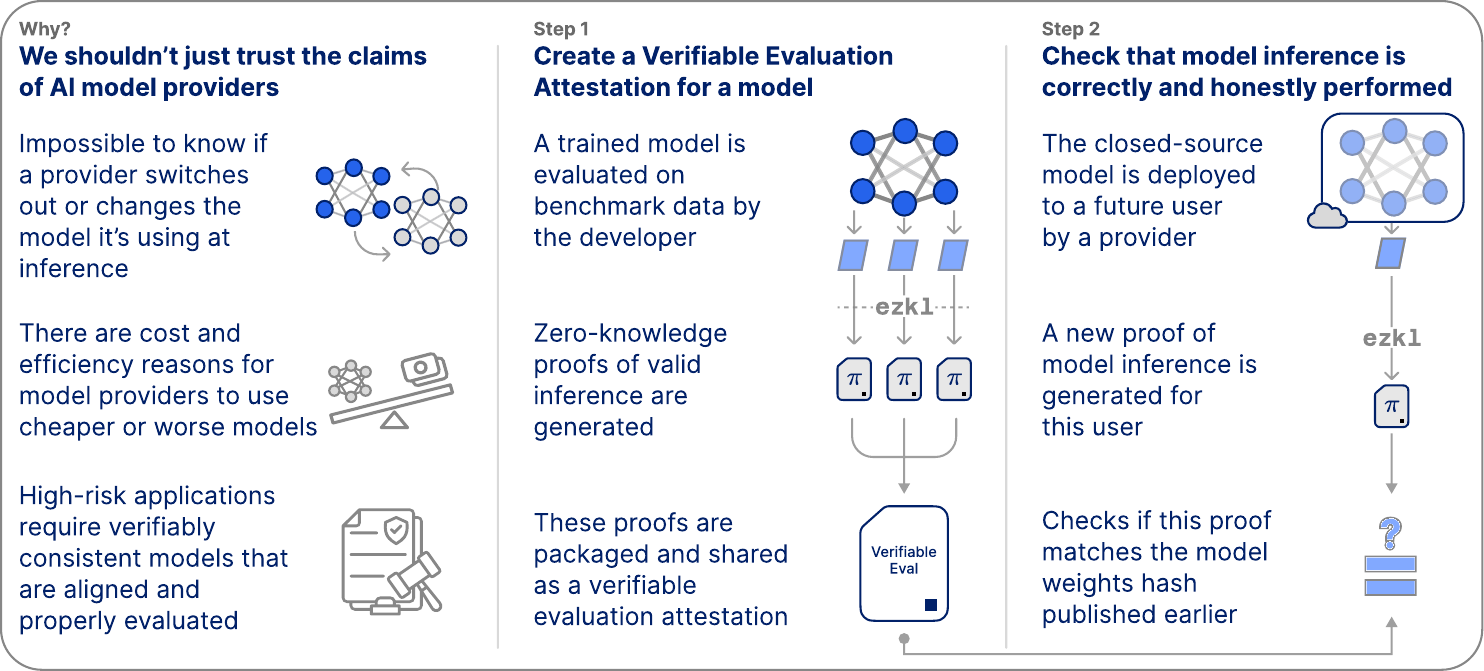}
    \caption{A high-level overview of the motivations and system design, which is augmented by the flexible ezkl proving system that can handle any ML model.}
\end{figure}

Keeping model weights private (regardless of whether the model architecture is public) limits the ability of external observers to examine the model and its performance properties. This presents two key concerns. Firstly, it can be hard to verify any claims that are made about a model's performance, either in the form of scientific evaluation results or commercial marketing. Secondly, performance characteristics that were not intended or tested for are hard to examine. Instances of racial and gendered bias have been found post-hoc in production ML models~\citep{buolamwini2018gender}. These unwanted model performance characteristics, such as bias, come in many forms and are often not initially tested for, leaving it to the public and interested parties to determine the characteristics of a model.

Tools such as algorithmic audits~\citep{buolamwini2018gender, raji2019actionable,kroll2015accountable} can partially allow auditing via API, but come at a significant expense and are not possible when models are not publicly facing. Many high-risk and customer-impacting models exist without public APIs, such as resume screeners~\citep{bogen2018help} or models used internally by law enforcement or governments. Even cases where models are public can benefit from verifiable evaluations, given the technically challenging and expensive process of running these models.

While a model creator or provider can always state claims about a specific model's performance on an evaluation task or set of characteristics (often in the forms of documentation or model cards such as \citet{mitchell2019model}), this is insufficient to verify a model being used has a performance characteristic. To verify the claimed capability, an end user must be able to confirm that the model in use was run on the benchmark in question, that it performs at a desired level, and that the same model is being used at a later date. This problem is not purely academic, as some scholars have raised concerns that OpenAI's models have degraded over time~\citep{chen2023chatgpt}.

\paragraph{The problem.} Users of AI systems have a vested interest in knowing about the performance capabilities of these systems. At the same time, high-performance AI systems may be more expensive to provide, or systems that perform well on certain benchmarks may underperform elsewhere. This creates incentives for model providers to provide users with a different model than advertised. Users without technical sophistication or computational resources struggle to confirm that the models they are using meet advertised benchmarks. While the EU AI Act and the US Executive Order on AI include transparency obligations for model providers to share technical details about the model \citep{us2023executive, eu2023aiact}, enforcing compliance with this requirement is extremely challenging in practice (especially as the market for AI becomes increasingly fragmented). We sketch a technical safeguard against this practical problem by allowing AI system users to verify the properties of the model they are using without requiring them to run their own tests or trust model providers.

\paragraph{Contributions.} To address these challenges, this work presents the first end-to-end demonstration of using zero-knowledge machine learning to generate evaluation proofs on arbitrary ML models. 
In doing so, this work contributes:

\begin{itemize}
\item A framework for general-purpose verifiable attestation of neural networks and other models across a variety of datasets and tasks using fully succinct non-interactive proofs;
\item Built upon a new proving approach that maps from ONNX model formats to proof circuits, flexible to any ML model.
\item A challenge-based model for checking model inference matches performance-attested model weights.
\item Analysis of computational costs and complexity of model evaluation, with an eye to how speed and scalability can be addressed via a `predict, then prove' strategy. 
\item A demonstration of evaluation across traditional non-neural ML models, multi-layer perceptions, convolutional neural networks, long short-term memory networks, and small transformers. 
\item A working implementation for public use with arbitrary models using the ezkl toolkit.
\end{itemize}

\subsection{Background and Related Works}
This project seeks to combine the extensive literature in accountable algorithms \citep{kroll2015accountable,kim2017auditing,desai2017trust}, ML reproducibility \citep{Beam2020ChallengesTT,semmelrock2023reproducibility,Albertoni2023ReproducibilityOM,Belz2023MissingIU,10.5555/3546258.3546422,10.1001/jama.2019.20866,10.1145/3576915.3623130,gundersen2022machine} and AI fairness \citep{Muthukumar2018UnderstandingUG,Mehrabi2019ASO,CorbettDavies2018TheMA,May2019OnMS,Raji2019ActionableAI,Krkkinen2021FairFaceFA,Liu2023FairCompassOF,Parraga2023FairnessID} with the growing field of zero-knowledge machine learning \citep{Feng2021ZENAO,Liu2021zkCNNZK,Lee2020vCNNVC,kang2022scaling}, in a pragmatic way. Previous work has typically focused on model-specific proofs, unlike our flexible proving system. Recently, speed improvements have enabled proofs for small language models \citep{sun2024zkllm, chen2024zkml} of similar size to our work, prompting questions of verifiable evaluation of models. While previous works briefly discuss this \cite{kang2022scaling, chen2024zkml}, we flesh out the practical constraints and design choices involved in the verifiable evaluation of models and provide possible solutions to scalability beyond faster chips and future research. A broader and more comprehensive background and related work is available in \autoref{sec:eval-background}.

\subsection{Problem Definition and Threat Model} \label{sec:problemdef}
A model provider has a private model, $f(\cdot, W)$, with weights $W$. The provider wants to make a claim, $C$, about the model's performance over a labeled dataset $\{(x_i, y_i) | i \in {0, \ldots, N}\}$. Traditionally, the provider would publish the model outputs $\{(x_i, f(x_i)) | i \in {0, \ldots, N}\}$ or an aggregation of those outputs (e.g., $ \frac{1}{N} \sum_0^N \left|y_i-f(x_i)\right|$) (e.g., a benchmark). If a model provider is a bad actor, they can simply produce fake outputs or aggregation statistics (i.e., lie about the model performance). 

Here, we want the ability to verify these published facts without needing the full weights. To do so, we publish the results and an attached proof of inference $\pi$ containing the model weights hash $H(W)$. The output then becomes $\{(x_i, f(x_i), \pi, H(W)) | i \in {0, \ldots, N}\}$ or an aggregated claim such as $ (\Pi, H(W), \frac{1}{N} \sum_0^N \left|y_i-f(x_i)\right|)$ where $\Pi$ is a meta-proof that verifies and attests to all included inferences being valid and the aggregation step being correctly computed. It must be possible to verify that the inferences are valid and that the model weight hash matches those inferences.

This is sufficient for scientific result confirmation but insufficient for models being deployed in the real world. To confirm the future use of a model is done by the same model with matching performance characteristics, a proof of inference must be done. This can be achieved by sharing with an end user the result of the inference on their input $x^*$ as well as a proof of inference and the model weights hash $(x^*, f(x^*), \pi, H(W))$. The data that is fed into this proof is called the `witness.' This allows the end user to confirm the model they access matches the model of the performance claim by verifying that the proof is valid and the corresponding model weight hash matches the original claim. A full threat model is available in \autoref{sec:threatmodel}.

\subsubsection{Example Uses}
To further identify the need for this system, we outline two simple real-world threats where verifiable evaluations are needed. First, consider the case of a new model architecture with private weights being published in an academic context, where the authors claim a high performance on a benchmark. As the model weights are private, model users and auditors cannot verify that the model performs at the described levels. The public needs to verify that a model \emph{exists} such that its execution over a dataset produces the benchmark result. 
The verifiable evaluation performance attestation demonstrates that the authors have a set of model weights that can achieve such performance on an evaluation. Even though weights are kept private, the general architecture of the model (e.g., whether it is a CNN or transformer) will be implicitly leaked in the current proof system (although future work could obfuscate some elements of this such as the number of layers in a CNN at the cost of proof size).

Second, consider the case of a model being used in production, via a publicly facing API or for internal use only. As above, a model consumer wants to know that there exists a model that performs well and has the described characteristics (such as not exhibiting racial bias on test sets). However, the model consumer also wants to know that this well-performant model is the one being used during inference. This motivates the model weight hash verification during inference time. 

\paragraph{Scalability in production}
Performing zero-knowledge proof of inference is a computationally expensive process compared to standard ML inference. Previous work treats the steps of inference and proving as intertwined \citep{kang2022scaling, chen2024zkml}; when they don't need to be. The zkSNARK computation operates on a witness, $(x^*, f(x^*), H(W))$, which must be generated at inference time with very low compute cost, but the proof itself can come later. This, in essence, leverages the fact that zkSNARKs prove that a witness and a circuit combine to produce valid outputs.

In high-risk contexts, an inference proof can be generated and checked for every run \emph{before} using model outputs (as suggested in previous work). However, it is possible to use a `predict, then prove' strategy, in which the model inference is provided immediately, but the proof (possibly computed in parallel) is provided at a delay after the proof process completes. This delay could be seconds, minutes, or more and still provide useful finality that doesn't slow down the delivery of the result. This is especially relevant in legal and regulatory regimes where model performance needs to remain auditable over time, including contexts such as using FDA-approved models in medical settings or model vendors used in high-risk settings (akin to `Trust, but Verify' \citep{desai2017trust}). 

Alternatively, model consumers can choose, for a cost, to challenge the model provider to present a proof of inference and the model weight hash for an input of concern. This guarantees only this inference, not all previous inferences, but can be leveraged strategically with a much lower total burden. If challenges are done randomly, a probabilistic guarantee of the model is created; or challenges can be done when model outputs are suspicious. Conditional on randomness in inference and reasonable accuracy bounds due to quantization, an audit can be performed post hoc on an inference output pair so long as a verifiable evaluation attestation and witness exist. 
A longer discussion of challenge-based model auditing can be found in  \autoref{sec:challenge}.


\subsection{System Design}

\begin{figure*}[thpb]
    \centering
    \includegraphics[width=0.9\textwidth]{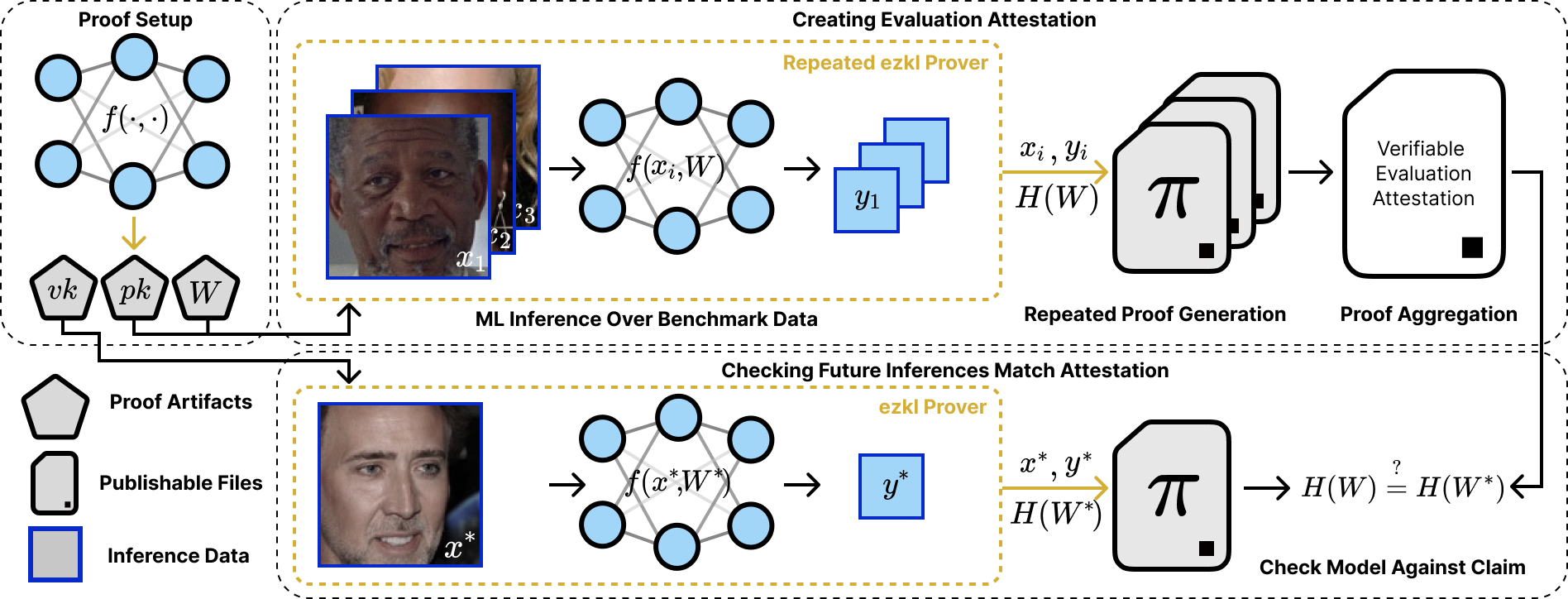}
    \caption{System diagram of verifiable ML evaluation using the zkSNARK ezkl toolkit. A model can be compiled into a proving key ($pk$) and verification key ($vk$) which can be used to generate repeated inference proofs over a dataset ($\pi$), which can then be aggregated into a verifiable evaluation attestation. Using the same proving and verification keys, any future inference of a model can be checked to confirm a model with the same model weight hash, $H(W)$, was used to generate the output. Inference data can be arbitrary.
    }
    \label{fig:fullsystem}
\end{figure*}

To generate and audit verifiable evaluation attestations, the system follows a broad four-step process shown in \autoref{fig:fullsystem}. 
First, a model of interest that has been trained and prepared for deployment is set up for inference. The model is converted into a standard model format (ONNX) and a circuit corresponding to the internal operations of the inference is created. This circuit generation process can be calibrated according to accuracy vs. resource tradeoffs and generates a large proving key, $pk$, and a verification key, $vk$, for the proof setup. This proof setup takes the form $\text{Setup}(1^{\lambda}, W, f) \rightarrow (pk,vk)$ using a standard security parameter $\lambda$, the model weights, and architecture to produce a commitment to the execution of a circuit. This circuit corresponds to the sequence of operations used in the inference of the model. This work is the first to present a flexible setup that works for any ML model in ONNX format to be converted to a proof circuit, made possible by creating a large number of interoperable proof arguments to support the diversity of model architectures; technical details of how this was achieved are in \autoref{sec:setup}.


Second, an evaluation dataset is selected to confirm performance or check for bias. Inference over the dataset is performed using standard practices to produce a set of input files containing input-output pairs $(x,y)$. Dataset choice is critical and can be dynamic (e.g., red teaming) as explored in \autoref{ch2:sec:datasetchoice}.

Third, for each data pair, a witness file is generated (including some quantization) for the $(x,y)$ pair, and a computational inference proof is created as a zkSNARK. This proof takes $W$ as a secret input value to the circuit and $x$ as a public input value and uses the previous commitment $pk$ to generate the proof $\pi$. This $\text{Prove}(pk, W, x, y) \rightarrow \pi \supset \left\{H(W), y\right\} $ is a computationally expensive step, with more details in \autoref{sec:datasetinference}.

These proof files can be aggregated into a verifiable evaluation attestation, including artifacts showing model performance and consistent model weight hashes. This can either be done by naively bundling the inference proofs or through an additional aggregation circuit that checks if all the proofs are valid and combines all the outputs into one. This additional setup comes with important additional privacy considerations for the model results and is elaborated in \autoref{sec:aggregation}. This claim can be publicly published to the world without the risk of leaking model weights.

At a later time, a model user or auditor who sees the result of a model inference $(x^*, y^*)$ can challenge the model provider to confirm which model was used to generate this pair. At an increased cost over standard inference (levied on either the auditor or provider), the provider holding the private model weights can be asked to prove using a new zkSNARK that such an input-output pair can be generated within accuracy bounds from a model with a matching circuit and model weight hash. By performing $\text{Prove}(pk, W, x^*, y^*) \rightarrow \pi^* $ using the same $W$ and $pk$, the provider can prove that they can generate $y^*$ using the model with $H(W)$, and hence the model with the same benchmark properties from the verifiable evaluation attestation. If a proof cannot be generated with matching hashes, the challenge has been failed. More discussion of this challenge-based approach can be found in \autoref{sec:challenge}.

\subsubsection{Aggregation of proofs}\label{sec:aggregation}

Once a collection of inference proofs has been generated, there is a question of how to package them for publication and sharing. The simplest naive solution is to zip up the proofs, which include the witness outputs, and a verification key, allowing any future examiner to calculate performance measures. This is computationally efficient and allows for flexible measures of performance to be made. However, this naive bundle exposes the test data outputs to unwanted use (such as training a copycat model based on these outputs \citep{hinton2015distilling}). Even though this is a reduced risk on test benchmarks, some model providers would prefer not to release inference.

Having the full test data inference results available also affords the possibility of a post-hoc audit on other properties contained in that data (e.g., a test set might only be intended for showing performance, but a careful audit of response performance across classes could reveal biases in the performance across classes). One such example might be sharing overall facial recognition accuracy, of which a finer grain analysis would show that the performance is much larger for white and male faces~\citep{buolamwini2018gender}. Since proof files and verification keys are very small, these naive bundles of inference proofs remain small and fast to verify even for large dataset sizes, with the vast majority of the work in proof generation.

A secondary approach would be to perform a vanilla aggregation circuit which checks if all the proofs are valid and combines all the outputs into one. This has similar downsides to compressing the collection but now comes with the additional cost of the aggregation circuit, which can become quite large. This zero-knowledge aggregation step makes sense when posting model inferences to a blockchain, but less so when sharing a performance attestation. We provide an implementation of this through the aggregation toolkit in ezkl.

A third approach uses such an aggregation circuit but explicitly creates a custom Halo2 circuit to aggregate just the result we want (e.g., the accuracy score, confusion matrix). This allows for a separate privacy regime to be applied, where data outputs are kept private, and only a verifiable accuracy metric is shared. This additional step adds complexity both in the form of additional computational requirements and the need to audit accuracy calculation code, but adds a significant layer of privacy. This additional zero-knowledge proof (ZKP) approach would need to be updated for different model types and outputs (e.g., models with varying input output formats, models with different accuracy measures, or benchmarks requiring complex generation of evaluation like the HumanEval example in \autoref{sec:examplemodels}).

\subsection{Experimental Results}

To demonstrate the flexibility and utility of the verifiable model attestations, we perform two groups of experiments. First, we show an evaluation of example models from a range of modalities, ranging from bias checking of facial detection via standard convolutional neural networks to limited inference of a language model. Second, we perform inference on a series of network architectures at varying model sizes to estimate total memory and compute time requirements, so as to estimate total costs of benchmarking. Code to rerun experiments and use the system can be found on Github.

\subsubsection{Example Verifiable Model Attestations}\label{sec:examplemodels}

While general in design, the specifics of generating a verifiable evaluation attestation vary across model types. Any model that can be expressed as a computational graph in ONNX can be evaluated, and a sample of example models is shown in \autoref{tab:examplemodels}.

The simplest such example of an evaluation would be a multi-layer perceptron (MLP) benchmarking on a small dataset such as MNIST \citep{deng2012mnist}. 
An inference proof is generated on each flattened image from the held-out test split to show its mapping from the input to a number. Bundling these inference proofs with the model weight hash and the ground truth values allows an inspector to evaluate the model accuracy by simply summing across the witness outputs and the ground truths in the attestation. Given the openness of the dataset, there is no need to privatize the ground truths. However, the naive bundling of the inference proofs for an attestation allows inspectors to see exact model outputs. Instead, a simple zero-knowledge proof (ZKP) could be made to verify each inference with the verification key, confirm $H(W)$ remains constant, and perform the simple $\tilde{y}_i \mathrel{\overset{\scriptscriptstyle ?}{\scriptstyle =}} y_i$ comparison across all the test data to produce a verifiable evaluation attestation without releasing individual data. In this case, the information in the naive bundle, including the ground truths are provided as witnesses to the proof with model outputs as private inputs and ground truths as public inputs. The full attestation with all test data proofs and a verification key is less than 100MB.


Other small models can behave exactly the same. A benchmark over a CNN applied to MNIST can be done exactly the same (with different performance characteristics as explored below). Models such as linear regression, SVM, or random forest can all be benchmarked in this approach. While these are not natively convertible to ONNX, there are a variety of tools to convert standard sklearn models into tensor-based models for use on ONNX such as hummingbird \citep{nakandala2019compiling} and sk2torch. We use a small dataset \citep{bennett1992robust} to demonstrate their use due to their small size.

Models can be more than classifiers. Consider a variational autoencoder (VAE) \citep{kingma2014auto} trained on the CelebA faces dataset \citep{liu2015deep}. While full inference of VAE can be proved, the internal reparameterization can create minor challenges for proofs due to the random generation. Instead we prove a partial execution of the VAE showing decoding from the latent space. Here the witness inputs are points in the latent space to be passed into the decoder and a naive bundle will contain these points with proven image outputs from the latent space. This provides an interesting approach to allowing for verifiable statements about the latent space. So long as the space has been sufficiently sampled (such as via a grid across the space centered around the mean) an inspector could see if any undesirable outputs are generated or inspect to see the degree to which certain properties of the latent space (such as race in facial images) are clustered. 

This generative approach can be applied to small language models as well. Autoregressive models such as an LSTM~\citep{sak2014long} or a decoder-only transformer~\citep{radford2019language} require a proof for each inference (e.g., each token they generate). This makes the computational cost of generating a long sequence of tokens quite high. It's worth noting, however, that these proofs can be parallelized. Since the witness can be generated preemptively before proving, the full sequence of tokens can be generated from normal inference to create a series of witnesses that autoregressively show the next tokens being generated, which can then each be proven in parallel. We demonstrate small language models (nanoGPT, designed to replicate GPT-2) running on a sample of HumanEval~\citep{chen2021evaluating}. These language models can be pretrained as the verifiable benchmarking attestation only requires inference.

We assume a public tokenizer and pass pre-tokenized sequences of tokens from the prompt into the model to generate the next token. For simple benchmarking datasets with a single token answer from multiple choice reasoning (such as MMLU \citep{hendryckstest2021}) this can be straightforward to benchmark and package. However, the space of LLMs has opened more rich and complex benchmarks such as HumanEval~\citep{chen2021evaluating} which requires generating longer strings as code and then evaluating the execution that code. Creating a naive bundle as a verifiable evaluation attestation is simple, requiring only that all token generation steps be put together such that an inspector can confirm that the code was generated by the model with the hash $H(W)$. Generating a zero-knowledge proof to privatize the generated code while attesting to its performance is much harder. This would require not only verifying each inference using the verification key, but also doing so in the correct order ensuring coherence of the previous tokens in the witness, and verifying the correct execution of the code and its output.

These approaches can be extended to a variety of models, such as other autogressive models or new language model architecture, diffusion models which behave much like VAE decoders, or sequence to sequence models such as Wav2Vec~\citep{baevski2020wav2vec} or Whisper~\citep{fedus2022switch} which apply the public tokenizer assumption to audio processing and generate a sequence output as witness. Larger models, such as a mixture of experts, are also possible, with model size being the only constraint. As we see below, proof time and resource requirements grow dramatically with large models. 

\begin{table}
\centering
\caption{Total resource and time requirements for generating an inference proof for example models.
}
\label{tab:examplemodels}
\begin{tabular}{lccccclcc}
\small\bfseries\begin{tabular}{@{}c@{}}Model\end{tabular} & \small\bfseries\begin{tabular}{@{}c@{}}Model\\Params\\($\times 10^3$)\end{tabular} & \small\bfseries\begin{tabular}{@{}c@{}}Model\\Flops\\($\times 10^3$)\end{tabular} & \small\bfseries\begin{tabular}{@{}c@{}}{$\mathbf{n_{con}}$}\\($\times 10^3$)\end{tabular} & \small\bfseries\begin{tabular}{@{}c@{}}Prove\\Time (s)\end{tabular} & \small\bfseries\begin{tabular}{@{}c@{}}Verify\\Time (s)\end{tabular} & \small\bfseries\begin{tabular}{@{}c@{}}Proof\\Size\end{tabular} & \small\bfseries\begin{tabular}{@{}c@{}}PK\\Size\end{tabular} & \small\bfseries\begin{tabular}{@{}c@{}}VK\\Size\end{tabular} \\
\hline
Regression         &              0.03&             --& 0.062& 0.1& 0.01&13K& 715K& 1.7K\\
SVM                &              0.03&             --& 0.626& 0.3& 0.02&23K& 16M& 2.5K\\
Random Forest      &              0.08&             --& 3.627& 2.9& 0.02&26K& 276M& 2.7K\\
MLP                & 3.6& 3.5& 1.920& 0.3& 0.02&21K& 14M& 2.3K\\
Small CNN          & 19.8& 68.6& 35.84& 3.1& 0.03&15K& 390M& 1.8K\\
VAE (decoder)      & 1065& 12582& 2016.6& 142& 0.42&1.9M& 16G& 2.5K\\
LSTM               & 29& 950& 495.7& 35& 0.10&41K& 4.1G& 2.5K\\
nanoGPT & 250& 51396&                   9398.9&              2781&                   2.69&0.7M&         219G&         4.2K\\
\end{tabular}
\end{table}

\subsubsection{Increasing Costs from Increasing Size}\label{sec:increasingsize}

\begin{figure}[htbp]
    \centering
    \includegraphics[width=\linewidth]{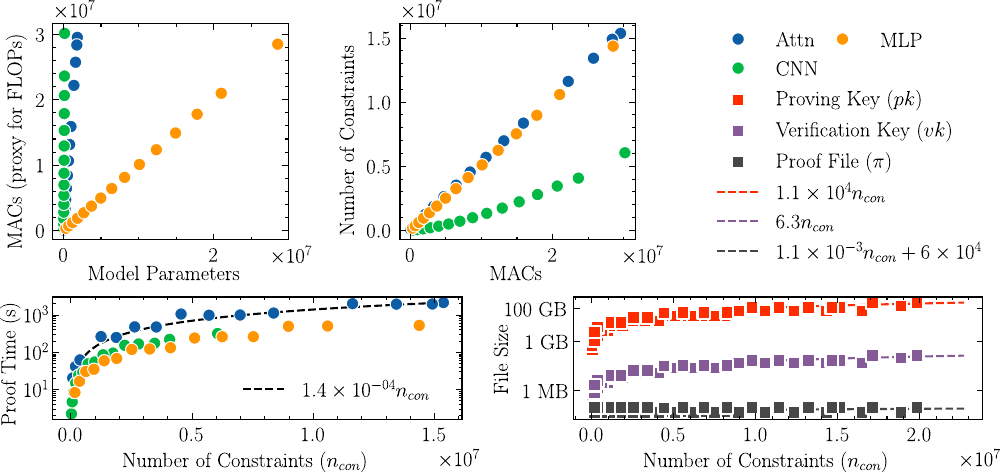}
    \caption{Time and RAM requirements for model proofs with increasing model sizes across multi-layered perceptions (MLP), convolutional neural networks (CNN), and attention-based transformers (Attn). Model requirements scale linearly with the number of constraints, driven by the number of operations used in a model inference. }\label{fig:scaling}
\end{figure}

Increasing numbers of model parameters can lead to dramatically increased performance~\citep{Kaplan2020ScalingLF}. These larger models also require greater resource requirements for inference. The same is true when performing inference of models inside zkSNARKs. 

Increasing the number of operations a model performs---often measured in multiply-accumulate operations (MACs) or floating-point operations (FLOPs)---increases linearly the number of constraints that need to be proved. While the total number of parameters in a model helps determine the number of operations, this is not a direct relationship. Different model architectures use parameters differently, and hence some models (such as CNNs) have a far greater number of operations per parameter than others (such as linear layers in deep neural networks). This relationship can be seen in \autoref{fig:scaling}. 
The CNNs have a lower scaling ratio as optimizations have been made to remove irrelevant operations (such as kernels operating on zero-values) which are included in pytorch operation counts but not in the number of constraints.

To generate an inference proof, the prover must make a KZG commitment~\citep{kzg10} to all the operations needed for inference requiring an SRS file \citep{nikolaenko2022powers} (see \autoref{sec:trustedsetup}). 
These files and the resulting proving keys they enable can become very large on the order of multiple gigabytes for reasonably sized models to hundreds of gigabytes for extremely large models. In general, this large proving key size and the memory requirements it puts on a proving system are the limitations on model size. 

While these keys can be quite large, the burden is on the prover. The resulting proof sizes and verification keys are extremely small. These too scale linearly with model operations but at a far slower rate, allowing for a large number of proof files and their verification key to be shared without a large file size burden. In turn, the verification time is also extremely quick. 

\subsection{Discussion \& Limitations}

While both flexible and powerful, this system is still extremely slow. As a result, verifiable evaluations on datasets are expensive. This motivates key classes of questions for discussion. The first centers around how to estimate current evaluation costs and what future research may provide to reduce them, discussed below.
Second, given limited resources important choices must be made over what benchmarks to run, a key question given the challenges in evaluating models. 
Third, a technical discussion of how speed improvements could be achieved is left for \autoref{sec:speedup}.
Finally, we turn attention to private evaluations. These limitations restrict the current use of the system from frontier models, but future work may enable this. Security limitations related to the use of zkSNARKs are outlined in the threat model in \autoref{sec:threatmodel}, as well as the audit limitations of a challenge-based model.

\subsubsection{Size and Cost Analysis}
For a given model and dataset, the total time it would take to perform inference proofs over the dataset and aggregate it can be roughly approximated. Each specific model architecture has a different relationship between its parameters and the number of operations it performs during inference. However, once a model is profiled for the number of operations (and hence the number of constraints), which can be done extremely efficiently, estimating the total cost to benchmark the model on a given dataset is easy. 
The proof time complexity for a given model is determined by the number of constraints needed to prove. For a given number of constraints, $n_{con}$, the proof rounds up to the nearest SRS file row count, which provides a proof with time complexity of $O\left( 2^{\lceil{\log_2(n_{con})}\rceil}\right)$ which for practical purposes is approximately linear $O(n_{con})$. While the relationship between parameters, operations (MACs), and constraints is usually linear, stating the complexity as $O(n_{parameters})$ masks the varied scaling rate between parameters and constraints across different model types. 

During experimentation, we found proving key size ($pk$) to be the largest limiting factor, even though it grows $O(n_{con})$, as large models require larger memory space to hold the proving key, increasing the size and cost of the compute cluster used.  
To estimate the total compute cost needed after determining the number of constraints for a model, the estimated proving time can be multiplied by the size of datasets to get total runtime, which can be cost calculated for whatever compute cluster has sufficient RAM to hold the proving key. 
Beyond the construction of the verifiable evaluation attestation, these cost estimates are useful for designing costs and incentive structures for model challenges. If challenging a model is costly, either the challenger or provider (or perhaps the loser of the challenger) will need to bear the cost. Having a publicly known cost for model inference makes this process fairer.

\subsubsection{Dataset Selection Choice}\label{ch2:sec:datasetchoice}

A key motivation for this system is the desire to bring transparency, accountability, and robustness to model benchmarking and its importance for fairness and bias checking. This fundamentally relies on our ability to have robust benchmarks to assess model qualities. 
Evaluating ML models is both critically important and often challenging \citep{Liao2021AreWL,HariniFramework}, especially in the case of NLP \citep{jones1995evaluating,kiela2021dynabench,bowman2021will,raji2021ai} and the general purpose usage of foundation models \citep{liang2022holistic,chang2024survey}. 

Even in cases when benchmarking is done well, there are many edge cases that are often missing (e.g., intersectional bias \citep{lalor2022benchmarkingIntersectional,NEURIPS2021_1531beb7,foulds2020intersectional,tan2019assessing} or out-of-domain usage). Measuring and mitigating these issues is, and will continue to be, a challenge in the ML field \citep{blodgett2020language, shah2019predictive, solaiman2023evaluating, bender2021dangers}. While recent work has proposed new more comprehensive benchmarks \citep{liang2022holistic}, further work will be needed for more robust and complete benchmarks; and addressing the challenges of bias in models is more complicated, requiring changes to values and norms around how AI is built and deployed \citep{Schwartz2022TowardsAS, HariniFramework, kroll2015accountable}. 
This is underscored by the push towards prioritizing AI safety \citep{us2023executive}, which includes comprehensive model evaluations. Tasks such as red-teaming model performance or other alignment checks could also be included in a verifiable evaluation attestation; however, for now, a verifiable evaluation attestation is only as useful as the evaluation itself.

\subsubsection{Private Benchmarks}

In the system design used here, the \emph{choice} of benchmark has been made public. While the exact outputs of the model over a benchmark can be hidden through aggregation (see \autoref{sec:aggregation} for more details), the choice of the dataset is fundamental to knowing what the model performance attests to. However, one could imagine a scenario where a model provider wants to prove their performance over a private benchmark dataset. This is immediately possible by making model inputs and outputs private after aggregation of inference proofs, but it requires additional work to justify, as these results are information-free in a vacuum.

Two models could be compared on private benchmarks, where a hash of a benchmark dataset is shared without ever sharing the model. This could allow a model provider to show improvements in performance over time with new models without ever sharing the benchmark that they use. Similarly, we could have benchmarks for general foundation model performance that are not released to the internet to avoid training on the test, which could position such a system run by a company or regulator as a gold standard of performance measures.
Alternatively, you may want to have a benchmark on a private dataset but with additional abilities to prove facts about the dataset. For example, a verifiable evaluation attestation could be made, including a dataset hash, and various facts about the dataset and its properties could be shared via separate zkSNARKs that include the dataset hash.

\subsection{Detailed Threat Model} \label{sec:threatmodel}
Based on the setup outlined in \autoref{sec:problemdef}, we can more clearly define a trust model for these verifiable evaluation attestations. We consider only two participants here: the model provider and the model user, where we assume the possibility that the model provider is adversarial. While other actors in the AI supply chain exist (e.g., model developers training weights or external auditors who verify on behalf of a user), we focus on those hosting the model and viewing its outputs. 

The model provider, as the adversary, has one of two goals. The first (threat model \#1) is to make a public claim that a model exists with a performance characteristic that is not true (e.g., I have a model that gets 100\% on ImageNet). The second and more dangerous (threat model \#2) is the provider changes what model is being served to an end user from inside their black box model hosting. Here, you can assume the provider is hosting a model in the cloud and serving a user over standard via API (as is standard).

For threat model \#1, only a closed-source model makes sense, as any widely accessible open-source model could be replicated and checked. For threat model \#2, the model provider can be serving either an open-source or closed-source model, but we focus primarily on the closed-source model here. It is still possible to change which model is being served when it is open-source, and it is an important problem to verify the model in use. While this system supports this, the privacy guarantees for model weights create unnecessary overhead when a simple computation trace would suffice. The ezkl toolkit does have an option for removing model weight privacy and hashing, which can speed up proofs and still enable any ML model to be used (unlike existing work). Instead, we focus on closed-weight models, where the privacy of the weights must be maintained while preventing the adversary from switching models at inference.  

For threat model \#1, there are many ways an adversary can achieve its goal without attached proof, the most obvious of which is to state a benchmark performance number alone. For threat model \#2, the complexity of the attack is similarly low; a model host simply needs to point the API endpoint at a similar enough model in the hosting stack that could plausibly behave close to the original. There are many reasons a usually trustworthy model provider may do this, from reducing inference costs with smaller models to shifting to models with more favorable performance characteristics.
To help imagine this, consider a Volkswagen scenario for an AI model, where a vendor fine-tunes a model with safety and bias in mind at the cost of performance, reports these safety results, and then serves the model that was not fine-tuned to get better performance. 

Threat model \#2 requires both a claim of model performance and a new proof during inference to check the model weight hash matches. Under our threat model, we only create a guarantee that the model provided to you to produce an output has the same set of model weights as the model of the verifiable evaluation attestation. 

The goal of this work is to remove the need for the public or an end user to trust the model provider. The zkSNARKs enable verification that computational work with a model with weights $H(W)$ occurred, that it produced a given benchmark, and that it was used for a specific inference that is challenged. This requires no specific hardware assumptions and draws from the assumptions baked into zkSNARKs (see below).

\subsubsection{Proving later}
We need proof to guarantee that a model will be served. However, proving is slower than regular inference. As a result, real-time verification is unlikely. Instead, we propose two models (outlined in greater detail in \autoref{sec:challenge}). Either some model inferences are challenged (either when performance degrades or at a regular random interval), or all inferences are challenged. Both of these acts leverage the fact that a witness file (the $(x,y,H(W))$ input to the proof) can be generated at almost no cost during inference, and a proof can be done after the fact on this witness. This ever so slightly weakens the guarantee for threat model \#2. Specifically, this creates the guarantee that the provider has a model with $H(W)$ that can produce $y$ from $x$, not that it necessarily served it earlier. However, if proofs are generated for enough diverse data in production, then we asymptotically have the guarantee that the provider is serving a model with exactly the same properties as the model with weights $W$, which for practical purposes is a reasonable guarantee that they are serving the claimed model. 

\subsubsection{Security Properties}
The security properties of this system derive from the in-built properties provided by zkSNARKs and apply across both threat models. Specifically, the zkSNARKs provide guarantees on: correctness, that the attestation and inference check accurately represent a set of computations that relate to the inputs and outputs of the circuit $(x,y,H(W))$; soundness, that the outputs of the zkSNARK cannot be generated without knowledge of the weights to generate the hash and the output even with a dishonest party; confidentiality of the model weights, which are kept privacy even when at attestation is shared publicly; integrity, ensuring that the model weights or inputs cannot be modified by a dishonest party during a proof; non-repudiation, ensuring that once proofs are published, the model provider cannot deny the authenticity of the published results; succinctness, that proofs and attestations are small in size; and non-interactivity, that a verifier does not need to interact with the prover during verification.

\subsubsection{Security Assumptions}
Similarly, the zkSNARK approach (built atop the halo2 proving system) comes with the same security assumptions of zkSNARKs themselves with additional carveouts. For the cryptographic assumptions of the prover, we point to \citet{groth2018updatable} and \citet{halo2book}. The key one to highlight is the need for a trusted setup to provide input randomness into the zkSNARK, via a structured reference string (SRS) (or common reference string). This challenge has been faced by a variety of zkSNARK projects, and consequently, a number of solutions exist.

\label{sec:trustedsetup}
\paragraph{Trusted setup} To use a zk-SNARK, a Structured Reference String (SRS), must be created to provide a public set of parameters. The creation of these parameters also generates `toxic waste,' which can reveal the random inputs to the model and remove the security and privacy provided by the SNARK. The ezkl proving system, drawing from the original halo2, uses the Perpetual Powers of Tau ceremony~\citep{nikolaenko2024powers}, which has many ongoing contributors generating an SRS, such that as long as a single contributor can be trusted to throw away their toxic waste, the SRS is secure.

\subsubsection{Security Limitations}
While we leverage the security limitation of zkSNARKs as a foundation, we introduce two key additional limitations here. The first is that the guarantees provided, that a model exists, is being used, and has properties, are tied to a specific set of evaluations. As noted in \autoref{ch2:sec:datasetchoice}, evaluations are often fraught. A model provider could train a model specific to game a set of benchmarks and evaluations and then faithfully serve you this model. The model would verify and pass all checks but be fundamentally bad as the model developer `gamed' the system by overfitting on the benchmarks or evaluations. The second is that the slow proof times limit the utility in synchronous contexts for large models. While we propose the `predict, then prove' approach as a workaround (explained in the main text and elaborated below), it minimizes the real-time security of the system.


\subsubsection{Challenge-based Production Model Audits}
\label{sec:challenge}

Any inference of a model that produces a visible input-output pair $(x^*, y^*)$ can be challenged. This includes regularly scheduled audits where an inference step is accompanied by an inference proof challenge, as well as post-hoc audits, where an inference that seems suspicious can be checked against an existing verifiable evaluation attestation. The latter approach limits the ability of the model provider to anticipate a challenge, creating incentives to always use the claimed model. A model user could choose to challenge a provider randomly, creating a predictable likelihood of bad actors getting caught, or selectively in cases where a user is dissatisfied with performance.

The key limitation is that each verifiable inference is vastly more expensive than standard inference. In cases where audits are legally mandated (such as in high risk contexts), a model provider could bear this cost. However, in production deployment models, the provider has an incentive to perform as few verifiable inferences as possible. As such, the model user could pay an additional fee for auditing to account for the cost. If an audit passes, the user is satisfied. However, to make auditing worthwhile in cases of concern, a reward system could be put in place, where a user is rewarded if an audit fails. This could be done using escrow or smart contracts, or via simple terms of service in API usage deals. In the case of a random audit with probability $p$ at cost $c$ per verifiable inference, a reward of $\frac{c}{p}$ would cover the costs in expectation for the user with an untrusted model provider.

Further, even a new verifiable evaluation attestation can be created post-hoc. If concerns are raised about an aspect of model performance (such as a revealed bias on a subclass of data \citep{buolamwini2018gender}), a new verifiable evaluation attestation can be created to examine the issue. This process is extremely costly, matching the cost levels of pre-emptive evaluation. An approach where a user challenges the performance of the model is possible, where a user provides a new test set to run on (as suggested by \citet{kang2022scaling}) requiring both a larger upfront cost and subsequent reward incentives. Such larger audits might be requested by civil liberty organizations or occur due to new standards (such as new bias requirements or evaluations in medical ML settings).

\subsection{Extended Background and Related Works} \label{sec:eval-background}
The field of secure inference of ML models has grown rapidly from specialized interactive proof protocols~\citep{Ghodsi2017SafetyNetsVE} to more general purpose approaches, such as through multi-party computation~\citep{Knott2021CrypTenSM, Mishra2019DE} and ML inference using homomorphic encryption~\citep{Lou2021HEMETAH, juvekar2018gazelle}. These approaches, while preserving privacy of inputs, fail to provide publicly verifiable proof that ML inference occurred correctly.

Instead, much attention has turned to the inference of ML models inside zkSNARKs (Zero-Knowledge Succinct Non-Interactive Argument of Knowledge). Previous work in zero-knowledge machine learning (zkML)~\citep{Feng2021ZENAO} drew on older slower proving systems~\citep{Groth2016}, or using model architecture specific optimizations~\citep{Liu2021zkCNNZK, Lee2020vCNNVC} (such as tailoring for convolution neural networks). These limitations were addressed by \citet{kang2022scaling}, which followed ezkl by leveraging Plonkish arithmetization through the Halo2 \citep{halo2book} proving system. Recent work has also shown that inference proofs of specific LLMs are possible with custom cuda circuits \cite{sun2024zkllm}.

This work builds upon and heavily leverages the ezkl\footnote{\url{https://github.com/zkonduit/ezkl}} toolkit, which itself is underpinned by the Halo2 proving system. This toolkit is continually under development, targeting the deployment of verifiable execution of ML models on-chain in Web3 contexts. 
Although there is a related growing body of literature on verifiable training \citep{Sun2023zkDLEZ, garg2023experimenting, Jia2021ProofofLearningDA}, we do not use or address the issue of training here. Since the release of the first version of this manuscript, recent work has scaled zkSNARKs for small LLMs using the same underlying proving system (halo2) \citep{chen2024zkml} and ezkl itself \citep{ganescu2024trust}.

The idea of using zkML as a tool to create verifiable accuracy claims emerges naturally as an extension from previous zkML work. Previous works mention verifiable accuracy claims as applications of their zkML proof approaches, with \citet{kang2022scaling, chen2024zkml}, but do not explain the evaluation-focused threat models, aggregation options, or benchmarking system.
\citet{Weng2022pvCNNPA} takes this idea and extends it in a constrained multi-stakeholder environment context for CNNs using collaborative inference and homomorphic encryption, and recent work has shown similar trustless fairness benchmarking approaches using logistic regressions \citep{Tang2023PrivacyPreservingAT}.
This work differentiates from the above as the first to provide a generalizable framework for neural network benchmarking, going much further in identifying constraints to provide a practical framework for deployment, as well as discussing choices of evaluation data across context and model type. 

The idea of building accountability into algorithms like this dates back to \citet{kroll2015accountable}, highlighting the important role of cryptography in creating accountable approaches to computer science, a mandate we build upon here. Continuous and post-hoc auditing to ensure these systems live up to their claims \citep{kim2017auditing} and synchrony between the design of accountable systems and their regulation are essential for this to succeed \citep{desai2017trust}. 
In general this work builds on the broad and burgeoning fields of ML reproducibility \citep{Beam2020ChallengesTT,semmelrock2023reproducibility,Albertoni2023ReproducibilityOM,Belz2023MissingIU,10.5555/3546258.3546422,10.1001/jama.2019.20866,10.1145/3576915.3623130,gundersen2022machine} and AI fairness \citep{Muthukumar2018UnderstandingUG,Mehrabi2019ASO,CorbettDavies2018TheMA,May2019OnMS,Raji2019ActionableAI,Krkkinen2021FairFaceFA,Liu2023FairCompassOF,Parraga2023FairnessID}. 
While this literature seeks to ensure unbiased, reproducible, and fair ML models, it often does not address the practical risks that model providers will create unbiased and reproducible models and then later choose to deploy lower quality models due to cost, compute, or performance incentives.
We seek to pragmatically address this issue to give users the ability to verify that the model they are using actually meets relevant benchmarks of bias and fairness. 

\subsection{System Details}

\subsubsection{Flexible Model Setup}\label{sec:setup}

One of the key features is the flexible nature of this proof system. The ezkl toolkit, which underpins the system and is a contribution from the authors, takes an ONNX file as input, an open standard for saving ML models that can be accessed from standard ML libraries.

Each operation in the model's computational graph is represented by one of the ONNX operation types, and each operation will be converted into a proof constraint. While simple in concept, implementation details are critical to achieve efficiency. For example, most operations in the model can be reduced to Einstein summations to minimize the number of constraint implementations needed~\citep{danteHoneyBlog}.

These constraints are constructed into a large proving table for Halo2~\citep{halo2book}. Many further optimizations occur here to achieve speed improvements which are detailed elsewhere. Notably, the ability to create parallel region layouts using cycles, the use of fixed-column lookups to represent nonlinear functions, and the choice to perform operations using fixed-point arithmetic for speed~\citep{danteHoneyBlog}.

While generating this circuit for proving there are many calibration choices to be made around quantization and scale. These can be summarized as tradeoffs between accuracy and the quantity of resources needed for the proof. In this work, calibration for resources is preferred for large model sizes.

Beyond the proof circuit for inference of the model operations, we introduce an additional set of actions in the circuit to calculate a proof of the hash of the model weights (via ZKG hashing per \citet{ezkl2023zerooverheadhashing}). This results in a final proof that contains both the input-output values (which are made public in the proof circuit) and the model weight hash. 

The outputs of this step are the compiled circuit, a large proving key, and a small verification key. \textbf{This is the first public, open-source, working implementation of a zkSNARK proving system that can be used for any type of ML model.}

\subsubsection{Proof System and Arguments} \label{sec:proofsystem}

The flexibility of the ezkl proof system stems from the library's reduction of any zk-equivalent of an ML operation into some combination of four arguments: 
\begin{enumerate}
\item A cumulative dot product argument. 
\item A cumulative sum/product argument. 
\item An elementwise addition/multiplication/subtraction argument. 
\item A lookup argument, used to represent and constrain non-linear functions within the zero-knowledge-circuit. 
\end{enumerate}

These ``arguments'' are used to enforce and guarantee that the result of a computation is indeed from a particular agreed-upon computation graph (such as a neural network).

\paragraph{Cumulative Arguments} \label{sec:cumarg}

Arguments 1. and 2. are constructed in a similar fashion. In the case of the cumulative sum/product consider a vector $\mathbf{x}$ of length $N$. To create a set of constraints within our zk-circuit we create a new vector $\mathbf{m}$ of length $N-1$ and set the following elementwise constraints: 
\begin{equation}
    x_i \circ m_{i} = m_{i + 1} \ \forall i \in 1..N, \ \text{where} \ m_0 = 0.
\end{equation}

$\circ$ is the addition operator in the case of the cumulative sum and the multiplication operator in the case of the cumulative product. The final element of the vector $\mathbf{m}$ then represents the cumulative sum or product, and the constraints above are enforced within the circuit using Halo2 \textit{selectors}. 

The dot product argument is constructed in a very similar fashion. We now have a second input vector $\mathbf{y}$, also of length $N$ and constrain the following: 
\begin{equation}
x_i \circ y_i + m_{i} = m_{i + 1} \ \forall i \in 1..N, \ \text{where} \ m_0 = 0.
\end{equation}

As before, the final element of the vector $\mathbf{m}$ then represents the dot product. 

\paragraph{Elementwise Arguments} \label{sec:elemarg}

Argument 3. is constructed without leveraging the intermediate calculations of \ref{sec:cumarg}. Consider two vectors $\mathbf{x}$ and $\mathbf{y}$, both of length $N$. We constrain the resulting output $\mathbf{m}$ via the following: 
\begin{equation}
x_i \circ y_i = m_{i} \ \forall i \in 1..N, 
\end{equation}

\paragraph{Lookup Arguments} \label{sec:lookuparg}

The ezkl system leverages Halo2 as a proving backend with a few modifications. Most significantly, it changes the original lookup argument, typically used to constraint non-linear functions like $\mathbf{ReLU}$, to the more efficient logUP lookup argument \citep{habock2022multivariate}.

\paragraph{More Complex Arguments} \label{sec:complexarg}

More complex arguments such as those for the $\mathbf{min}$ and $\mathbf{max}$ functions, can be constructed as combinations of the above arguments. 
For instance the $\mathbf{max}$ argument can be constructed as follows: 

\begin{enumerate}
    \item Calculate the claimed $$m=\text{max}(x),$$ and instantiate a lookup table $\mathbf{a}$ which corresponds to the $\mathbf{ReLU}$ element-wise operation.
\item Constrain $\mathbf{w} = \mathbf{x} - (m - 1)$.
\item Use lookup $\mathbf{a}$ on  $\mathbf{w}$, this is equivalent to clipping negative values: $\mathbf{y}=\mathbf{ReLU}(w).$
\item Constrain the values $\mathbf{y}$ to be equal to 0 or 1, i.e. use a selector that enforces that $$y_i*(y_i - 1)=0, \qquad \forall i \in 1\dots N.$$
Any non-clipped values should be equal to 1 as at most we are subtracting the max.
This demonstrates that the there is no value of $\mathbf{x}$ that is larger than the claimed maximum of $\mathbf{x}$.
\item We have now constrained $m$ to be larger than any value of  $\mathbf{x}$, we must now demonstrate that \textit{at least one} value of  $\mathbf{x}$ is equal to $m$, i.e that $m$ is an element of  $\mathbf{x}$. We do this by constructing the argument $$\mathbf{x} = \mathbf{ReLU}(1 - \sum_i y_i) = 0.$$  
Note that $$\sum_i y_i = 0 \iff z = 1$$ and thus no values of the witness are equal to $$\text{max}(x).$$ 
Conversely, if $$\sum_i y_i >= 1 \iff z=0$$ and thus least one value is equal to 1.  
\end{enumerate}

\subsubsection{Dataset Inference} \label{sec:datasetinference}

The most computationally expensive and decision-intensive step is the inference of the ML model over the benchmarking dataset to generate the zkSNARKs. One extremely key choice is the selection of a relevant dataset, an issue we leave for \autoref{ch2:sec:datasetchoice}. Instead, we focus here on the mechanics of how the inferences can be done.

While zero-knowledge ML is often described in terms of the inference occurring inside the proof generation, it can more aptly be described as the proof verifying that an inference was performed; a distinction that has important practical implications.

Firstly, the model can be run over the dataset inputs as it would in any other inference context to generate input-output pairs, $(x_i,y_i)$. These pairs are then quantized per the setup choice above to produce the witness inputs,  $(\tilde{x}_i, \tilde{y}_i)$. When accuracy is sacrificed, the quantized input-output pair may be different from the original values by a few percent (this can be calibrated). Importantly, as far as the test set is concerned, this accuracy trade-off can be examined \emph{before} performing the expensive proof step. The proof step will verify that the witness could be authentically generated using the model weights (which are treated as private inputs for the proof). Hence, iterative calibration on the benchmarking test set to ensure witness data has the performance characteristics is important.

For each of these witness files, a proof is generated using the proving key and the circuit. This is the slowest part of the stack and scales linearly with the size of the test set. For each witness file, $(\tilde{x}_i, \tilde{y}_i, H(W))$, a proof file will be generated $\Pi_i$. These proof files are each very small (on the order of kilobytes) and can be individually verified using the verification key or aggregated.

\paragraph{Posting Performance Claims}
Other work has discussed the use of specific billboards for sharing attestations \citep{Tang2023PrivacyPreservingAT}, but the portable nature of the verifiable evaluation attestations is that they can be shared or hosted anywhere and copied repeatedly without additional risk. As such, posting performance claims to Github, Papers with Code, Huggingface, IPFS, another third-party public-facing hosting service, or even one's own website is sufficient. Ideally, this posting would then be mirrored elsewhere for accountability. 

\paragraph{Hardware}
All experiments run here can be done on commodity hardware. They were initially all done on a 10-core Intel Xeon Processor E5-2687W with 1T of RAM. Some parts of the proving stack (including those using Halo2) are parallelized across CPU cores. The key hardware constraint is the RAM size for storing the proving key, hence the choice of this machine. GPU acceleration is possible in parts of the proving system (and has been done end-to-end for LLMs with custom Cuda circuits recently \citep{sun2024zkllm}), but no experiments here used a GPU. Experiments in the final version of this project were run on a cloud compute cluster provided by ezkl to customers, with similar hardware to the above, but was able to achieve 20\% faster proof times than the Xeon above.

\subsubsection{Scalability \& Future Speed Improvements} \label{sec:speedup}

One of the main constraints on the application of this system (especially with regard to large foundation models) is the speed with which proofs can be generated. This work builds on the ezkl toolkit, which is constantly undergoing speed improvements through optimizations to circuits for inference operations and engineering improvements. Future work will improve these speeds through proof splitting and parallelization \citep{ezkl2023splitting}, GPU acceleration \citep{ezkl2023gpu, derei2023accelerating}, or using alternative underlying proof systems \citep{kothapalli2022nova, boneh2020halo, setty2020spartan}.
Other approaches such as cqlin from \citet{cryptoeprint:2023/393} show promise for ML, while other unpublished work has performed further optimizations \citep{kang_2023_tensorplonk}. GPUs acceleration through implementing attention circuits in cuda has proven effective in creating significant speed improvements for LLMs \citep{sun2024zkllm} and recent work has shown the inference of small LLMs in zkSNARKS \citep{ganescu2024trust, chen2024zkml}.

Interestingly, proof splitting may prove extremely exciting as future work. As we see in \autoref{sec:increasingsize}, the time and resource complexity of models is sublinear. As a result, splitting an AI model into chunks (e.g., each attention block) and completing a proof for each chuck should provide a lower overall computational cost. It's possible that even the largest models could be chunked into reasonable sized parts that could be proved with current hardware resources. 

Further, it is possible to optimize the design of models for more efficient inference in zkSNARKs \citep{Jha2023DeepReShapeRN}. Similarly, choices can be made during benchmarking design, such as model inference batching, which can have small speed improvements at the cost of larger proof requirements. 

As we see improvements in the speed of underlying proving systems and their hardware, the sublinear growth of proving time means that foundation models (which are increasingly performant at small sizes) will be commercially viable at scale.

\subsection*{Project Conclusion}

We present a novel method for verifiable performance and bias benchmarking of ML models using zkSNARKs. This approach addresses the critical issue of verifying model performance claims in environments where model weights are kept private, which is increasingly common in commercial ML applications. This helps ensure transparency and accountability in model evaluations, particularly in high-stakes scenarios where model reliability and fairness are paramount. The system packages together repeated model inference proofs to demonstrate the accuracy of models either through simple bundling of small proofs and verification files or through meta-proofs of performance over model inference proofs. The system's flexibility was demonstrated across a range of ML models ranging from small perceptrons and regression models to medium-sized transformers. Leveraging a `predict, then prove' approach to serving results and proofs combined with a user challenge model of auditing responses reduces the computational costs in production, and shifts compute demands to model trainers. This is the first practical implementation of a verifiable evaluation for arbitrary ML systems, maintaining model weight confidentiality while ensuring model integrity. In doing so, this lays the groundwork for a more transparent and verifiable future for ML model evaluations.

%% file: chapter-2/partialzk.tex
\clearpage
\section{Verifiable computation of partial AI systems}

\noindent
\emph{The previous project on verifiable model evaluations of AI systems suffered from one large flaw: speed. This is a fundamental limitation when dealing with large models and slow proving systems. To address this, a new project was started with Carl-Leander Henneking titled ``Partially private, optimistic, distributed, and verifiable machine learning inference'', which leverages the open source nature of many AI models, and uses partial proving of relevant layers of models to increase speed while maintaining verifiability. For conciseness, this project has been rewritten for this thesis.}

\begin{paperabstract}
 Performing AI model inference inside zkSNARKs is an exciting tool for verifiable and private AI inference. However, the large number of parameters in modern models and the subsequent large number of computational steps to perform inference often make it infeasible to perform full model inference inside a zkSNARK due to computational complexity and time constraints. This project explores the idea of selectively proving certain sub-computations in a model, such as the final classifier head in a deep neural network, or specific layers in a LoRA-tuned model, allowing for a balance between privacy and verifiability.
\end{paperabstract}

Zero-knowledge (ZK) proofs, particularly zkSNARKs, provide strong cryptographic guarantees for verifiable and privacy-preserving AI inference. However, a well-documented limitation of zkSNARKs is their computational cost, which scales significantly with model complexity and parameter count. This is because for each inference step in a model, every operation must be included in the circuit. Models have billions of multiplication and addition operations, to express model weights and biases, in addition to a large number of non-linear operations, which must be encoded as lookup tables or approximated. 

Full-model verification using ZK proofs is often infeasible for large-scale architectures, such as transformer-based models with billions of parameters. The setup and proving phases introduce substantial time and memory overhead, making it challenging to integrate zkSNARK-based verification into real-world AI systems with strict latency and resource constraints.

A promising alternative to full-model verification is the selective application of ZK proofs to targeted portions of a model. Rather than proving the correctness of inference across the entire network, zkSNARKs can be applied to specific layers or model components, such as the final classifier head in a deep neural network or layers involved in a fine-tuned adaptation, such as those used in LoRA (Low-Rank Adaptation) tuning~\cite{hu2021lora}. This selective proof strategy significantly reduces computational overhead while preserving key benefits of ZK verifiability, ensuring that critical computations remain both private and verifiable without requiring the entire model to be incorporated into a zkSNARK circuit.

\begin{figure}[htbp]
    \centering
    \includegraphics[width=0.8\textwidth]{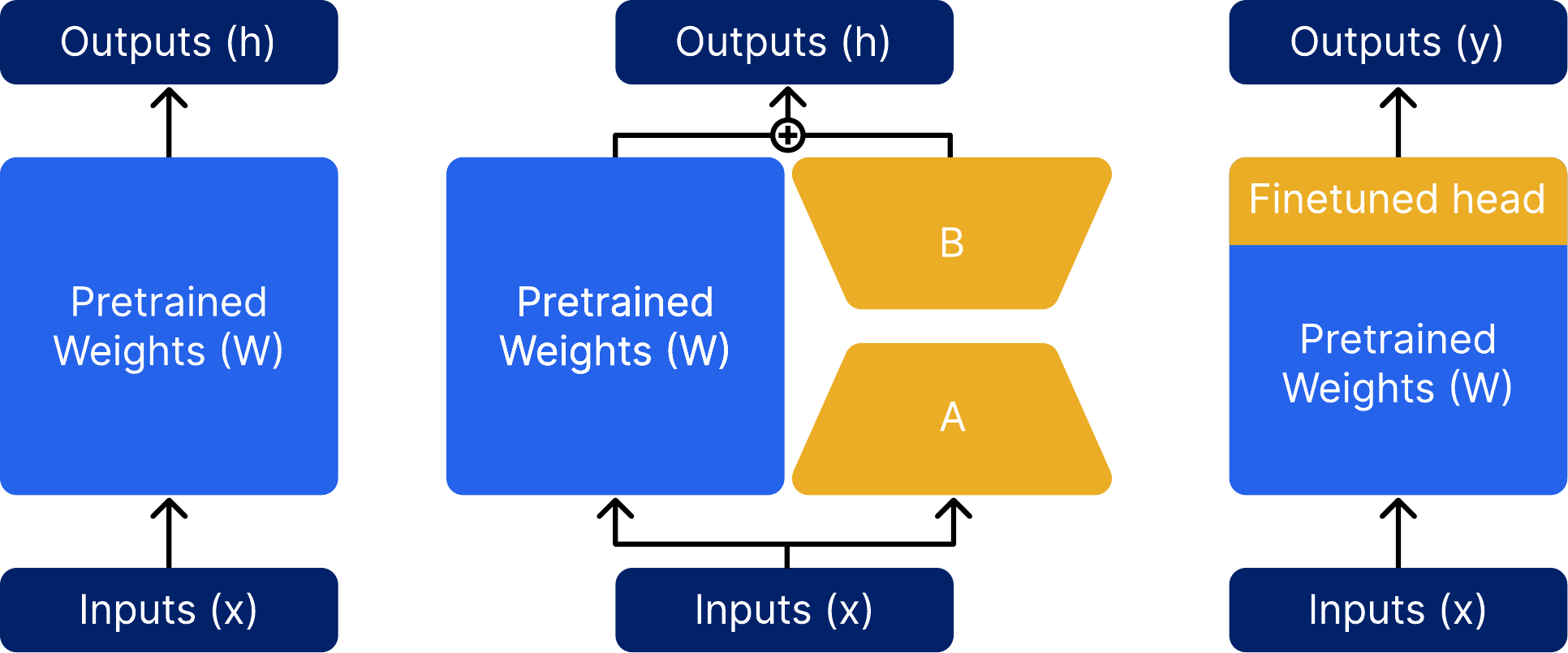}
    \caption{Different approaches to fine-tuning model. Left: a full model, which requires proving computation over all the parameters (Wx). Middle: a LoRA-tuned model, which requires proving only the LoRA-tuned parameters (BAx). Right: a model with a fine-tuned head (note that this is done at the full model level, not the attention layer level), which requires proving only the fine-tuned head computational (such as a set of linear transformations to produce a classifier from a pretrained public model).}
    \label{c2:fig:partial-zk}
\end{figure}

As outlined in \autoref{c2:fig:partial-zk}, we can selectively prove only the relevant portions of a model. To explain, let's return to asking why we want to verify a model. For open weight models with public inputs, there is few reasons to have verifiable computation through zkSNARKs, as it would be cheaper to rerun the inference computation that generate a proof. The key value in zkSNARKs comes from the ability to verify computations on private data, such as private model weights.

If all the model weights (W) are private, then you must run a full model computation. However, if only part of the model's weights are private, then you can selectively prove these with zero-knowledge of the private weights, and use intermediate public inputs and outputs to rerun any needed computations for the public portions of the model. 

In the case of a LoRA-tuned model, we can selectively prove the LoRA-tuned weights side computation on each layer (e.g., prove $BAx$) which can then be combined with the public weights (W). This preserves the private of the changed weights at minimal additional computational cost. For a fine-tuned head (e.g., to produce a classifier from a pretrained public model), we can selectively prove the head computation (e.g., $HWx$).

A key threat to consider here is that model weights can be reconstructed from the public input-output pairs if there are a sufficient number of pairs across the full distribution of inputs. While retraining weights is hard, its difficulty (and demand for data) scales with the number of parameters being tuned. In the same way that LoRA makes fine-tuning more efficient, and proof computations faster, it also makes this threat more feasible. 

\paragraph{Conclusion}
By leveraging this partial zkSNARK approach, AI systems can maintain verifiability where it matters most—such as in sensitive decision-making tasks—while avoiding prohibitive costs associated with full-model verification. This method also allows for modular verification in distributed ML settings, where different nodes can prove only their assigned model segments. Additionally, this enables more practical deployment of zk-verifiable AI in resource-constrained environments, such as edge devices or decentralized networks, by ensuring that only critical computations are subject to cryptographic validation. As advancements in proving systems continue to improve efficiency, the integration of selective zkSNARK-based verification will become an increasingly viable approach for securing AI inference at scale.

%% file: chapter-2/zktax.tex
\clearpage
\section{Portable data using zkSNARKs}\label{sec:zktax}

\noindent
\emph{So far this chapter has focused on the use of zkSNARKs to enable verifiable claims about AI models. However, zkSNARKs can be used to enable a wide range of applications. This section is based on a project in collaboration with Alex Berke, Robert Mahari, Kent Larson, \& Alex Pentland titled `zkTax: A Pragmatic Way to Support Zero-Knowledge Tax Disclosures'. It has been reframed towards the general case of portable data using zkSNARKs, and draws from Data Provenance Initiative I contributed to and co-authored with, led by Shayne Longpre and Robert Mahari. In general, it speaks to the fact that many elements of the AI supply chain can be made more transparent and verifiable through the use of zkSNARKs.}

\begin{paperabstract}
    zkSNARKs can enable the sharing of statements about data without revealing the data itself. This is an extremely powerful primitive that can be used to enable a wide range of applications. In particular, we present a framework for taking in data statements and providing provable redactions of the data. 
    This has general applications across privacy-preserving training data attestations, provable logging, and more pragmatic applications such as zero-knowledge tax disclosures. However, in each instance, the claims are underpinned by a root of trust. The zkSNARK can only make claims about the transformations of data, not the raw authenticity of the data itself in most cases, especially when the data can be synthetically generated or changed. This work explores this tension and presents a general framework for using zkSNARKs to enable portable data.
\end{paperabstract}

Organizations and individuals alike frequently need to share limited or computed information about underlying data without revealing it in full. In corporate deal-making, for example, it may be necessary to disclose certain financial metrics without exposing the complete financial record. Similarly, in governance, public officials may want to exhibit specific aspects of their finances to validate their claims of transparency, yet they may hesitate to reveal personal identifiers in their tax returns. Beyond finance and governance, the rapid growth of AI has created new demands for verifying aspects of training data---such as confirming the percentage of copyrighted material---without exposing entire datasets. These scenarios all pose a common challenge: How can data be selectively disclosed or claims be verified \emph{while maintaining privacy}?

We propose a cryptographic framework that relies on a \emph{Trusted Data Source} (TDS) to sign data, combined with zero-knowledge proof tools to achieve a redact-and-prove workflow. This workflow allows data holders to generate verifiable disclosures—ranging from simple selective releases of certain fields to more complex statements about relationships within the data—while protecting all other sensitive details. The resulting proofs, once created, are easily shareable and can be publicly verified by anyone using the TDS's known public key.

\subsubsection{Motivating Examples}
\paragraph{Selective tax disclosures.} Tax returns are typically replete with sensitive personal information, yet they can be vital for demonstrating income, tax obligations, or other financial metrics in negotiations, loan applications, or public office vetting. Tax authorities commonly serve as a natural TDS in this domain. If the authority or an authorized preparer signs the taxpayer's complete return, the taxpayer can later produce partial disclosures—along with proofs verifying authenticity. This concept forms the foundation of what we refer to as \emph{zkTax}.

\paragraph{AI model pretraining data.} Amid rising concerns about whether AI models are trained on copyrighted material or biased datasets, a TDS (perhaps an AI consortium, a government regulator, or the model developer itself) can sign a manifest of the training data. Subsequent claims---such as ``Less than 10\% of this data is copyrighted" or ``We used only publicly licensed text from domain X''---can be proven without requiring the entire raw dataset to be shared. As AI regulation evolves, having a mechanism to cryptographically verify compliance with copyright or privacy standards may become indispensable.

\subsection{Background and Related Work}\label{sec:background}
\paragraph{Redactable signatures and zero-knowledge proofs.} Redactable signature schemes (RSS)~\cite{sanders2020efficient} and content extractable signatures (CES)~\cite{steinfeld2002content} allow partial redaction of signed documents while maintaining verifiability. However, these approaches often specialize in structured subsets of the original data. Zero-knowledge proofs (ZKPs), on the other hand, offer more flexible ways to demonstrate statements about data without revealing the data itself. ZKPs have found applications in financial privacy~\cite{zcash,provisions} and identity verification. We combine these approaches, leveraging standard public key cryptography with ZK circuits to produce public verifiability alongside minimal disclosure.

\paragraph{Trusted digital infrastructure.} Progress in public key infrastructures (PKIs) and government modernization efforts highlight the increasing feasibility of a TDS. For instance, Mexico's PKI for tax filings~\cite{mexicoPKI} and Estonia's e-government ecosystem both rely on cryptographic signing of user data. Open Banking initiatives~\cite{openbanking} provide further impetus for the standardization of signed financial data. While these programs are focused on practical implementations in financial and e-government contexts, the concept of a TDS can be generalized beyond finance.

\paragraph{Data verification in AI.} Current discussions around AI governance have underlined the need for verifiable statements about training data provenance. Proposals for robust model auditing~\cite{SouthVeriableEvaluations} and dataset documentation reflect a broader trend towards accountability. Our framework addresses precisely this challenge by offering a cryptographic method to confirm claims about pretraining corpora without revealing proprietary data.

\subsection{System Goals and Architecture}\label{sec:goals}
\subsubsection{Core Objectives}
We outline the essential requirements for a broad TDS-based system:
\begin{itemize}
\item \textbf{Establish a Root of Trust:} A recognized entity signs data so that any subsequent disclosures can be verified as authentic.
\item \textbf{Redact-and-Prove Mechanism:} Users selectively hide portions of the data or prove more advanced statements (e.g., aggregations, thresholds) without revealing raw inputs.
\item \textbf{Public Verifiability:} Anyone with access to the TDS public key can verify whether the redacted (or otherwise computed) data is consistent with the TDS-signed original.
\item \textbf{Privacy Preservation:} Sensitive or proprietary details remain undisclosed, aligning with data minimization principles.
\item \textbf{Extensibility:} The system can be easily adapted to new data structures and advanced verification logic.
\end{itemize}

\subsubsection{Three-Service Model}

\begin{enumerate}
\item \textbf{Trusted Data Source (TDS):} The authority or entity that provides signed, machine-readable data. For example, a tax authority for financial returns or an AI dataset provider for a corpus manifest. The TDS uses a known public key to sign a hash of the data, returning $(x, S)$, where $S$ is a signature on $H(x)$.
\item \textbf{Redact \& Prove Service:} A user interacts with a zero-knowledge circuit to remove (redact) selected fields from the TDS-signed data, or to transform it (e.g., computing sums, thresholds). The circuit checks the validity of the signature and ensures the disclosed data is consistent with the hidden original. The output is a verifiable proof $\pi_{x'}$ plus the redacted data $x'$.
\item \textbf{Verification Service:} Any third party, given the TDS public key, the redacted data, and the proof, runs a verification algorithm to confirm validity. If the proof checks out, the third party knows the redacted data or statements indeed come from the TDS-signed original.
\end{enumerate}

\begin{figure}[htbp]
    \centering
    \includegraphics[width=0.8\linewidth]{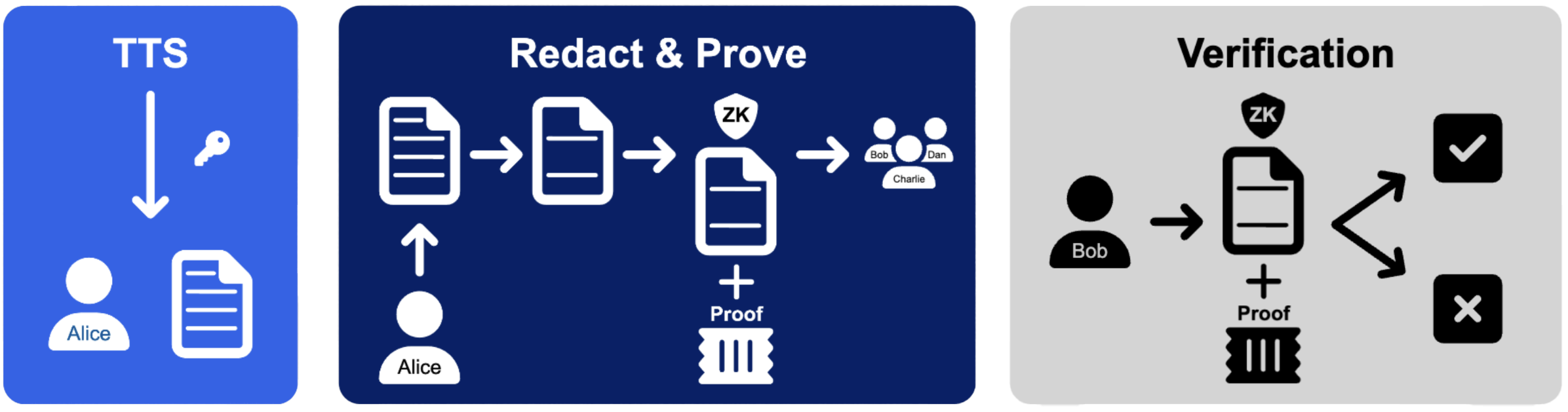}
    \caption{Diagram of the three services in the redact-and-prove system. (Left) An individual, Alice, retrieves data from a Trusted Data Source (TDS), which signs the data with a signature that can be verified using a public key. (Middle) Alice brings the signed data to the Redact \& Prove service, where she selects how the data should be redacted, generates a redacted version, and obtains proof that the original data was signed by the TDS. Alice can make the proof and redacted data public, and anyone can verify its authenticity using the Verify Service. \label{fig:diagram_3services}}
\end{figure}

\subsubsection{Example Workflow: zkTax}
\begin{enumerate}
    \item A tax authority (TDS) signs a complete tax return $x$ using its private key, yielding $S=\texttt{sign}(\textit{sk}, H(x))$. The taxpayer receives $(x, S)$.
    \item The taxpayer uses Redact \& Prove to hide sensitive fields and optionally prove claims like “My reported charitable contributions exceed \$X.” A ZK circuit ensures that $H(x)$ matches the TDS signature and that any disclosed items match the original data.
    \item The taxpayer sends $(x', \pi_{x'}, S)$ (redacted data and proof) to a verifier, who checks correctness using the Verification Service and the TDS's public key. If valid, the verifier is assured $x'$ is consistent with the actual tax return.
\end{enumerate}

\subsection{Extension: AI Pretraining Data Transparency}\label{sec:ai}
As large-scale AI models raise questions around copyright and bias, we propose applying the TDS framework to pretraining data. Suppose an entity—e.g., a model developer or an AI consortium—serves as the TDS, providing a signed manifest of the training corpus in a structured format (e.g., CSV). Once signed, the Redact \& Prove Service can create zero-knowledge proofs about properties of the training data, such as \emph{its size, overlap with copyrighted texts, or compliance with certain content guidelines}. Examples:
\begin{itemize}
\item \textbf{Copyright compliance:} Prove that a given fraction of the data is in the public domain without exposing the entire dataset.
\item \textbf{Bias analysis:} Demonstrate that less than some percentage of the data consists of certain high-risk or sensitive categories.
\item \textbf{Regulatory checks:} Provide cryptographically grounded assurance that the data follows certain licensing agreements or legal standards.
\end{itemize}

\subsection{Implementation Considerations}\label{sec:implementation}
\paragraph{Efficiency and practicality.} Constructing large ZK circuits can be computationally expensive. Implementations may employ succinct proof systems (e.g., Groth16, PLONK, Halo2) to minimize overhead. To remain accessible, the Redact \& Prove Service can be designed as a user-friendly web or desktop application, compiling proof artifacts locally so that sensitive data never leaves the user's control.

\paragraph{Modular design.} The system is most easily extended when the TDS simply signs the hash of an entire dataset and publishes its public key. Different Redact \& Prove solutions can then be developed by private organizations or open-source communities, enabling multiple specialized proof circuits.

\paragraph{Security assumptions.} As with any public key scheme, the security relies on the TDS's private key remaining uncompromised. Furthermore, trust also depends on the TDS accurately representing the original data. If the TDS itself publishes incorrect or incomplete data, the proofs will be valid relative to that original but not necessarily reflect real-world truth.

\subsection*{Project Conclusion}\label{sec:conclusion}
In this section, we have introduced a flexible redact-and-prove framework in which a \emph{Trusted Data Source} signs a dataset, enabling end users to create verifiable disclosures—either partial or aggregated—without exposing sensitive details. While our motivating example is tax data (zkTax), the underlying architecture readily extends to other domains that require verifiable yet private proofs. Emerging concerns in AI around copyright, licensing, and dataset governance illustrate the necessity of such a system, as it combines cryptographic guarantees with fine-grained disclosure policies.

By leveraging progress in zero-knowledge cryptography, open data standards, and robust PKI infrastructures, we envision a future in which organizations and individuals can confidently share verifiable claims about sensitive data in a privacy-preserving manner. Future work includes exploring domain-specific enhancements, improving proof efficiency, and standardizing these approaches for broader adoption, especially in rapidly evolving regulatory contexts.

%% file: chapter-3/chapter-3.tex
\chapter{Private retrieval augmented generation for auditable and updatable LLMs}\label{chapter:3}
\epigraph{``All information looks like noise until you break the code.''}{\textit{Neal Stephenson's `Snow Crash' (1992)}}

\noindent
\emph{Moving beyond just verifiability, this chapter examines the role of privacy and auditability in a key element of AI systems: retrieval. In essence, this addresses the privacy, auditability, and verifiability requirements of knowledge management tools in AI systems beyond core LLM memorization. Specifically, this chapter starts by looking at how RAG can unlock transparency and auditability from a legal lens, but it must be balanced with privacy risks. This chapter then explores two technical approaches to mitigating privacy risks: a novel approach to using MPC for robust privacy across multiple servers with a shared index, and an approach using trusted execution environments (TEEs) for collective knowledge management.}

Where \autoref{chapter:2} focused primarily on using zkSNARKs to achieve the \emph{verifiability} of computations and data properties, often while preserving privacy, this chapter shifts attention to the dynamic process of information access within AI systems. Retrieval Augmented Generation (RAG) introduces distinct challenges and opportunities related to the pillars identified in \autoref{chapter:1}. While zkSNARKs might verify \emph{that} a retrieval process followed certain rules, the act of querying and retrieving external data itself demands robust solutions for privacy, especially when dealing with sensitive information. Concurrently, RAG offers a unique pathway towards enhancing \emph{auditability}, by potentially grounding model outputs in specific, identifiable external sources---a different form of transparency compared to verifying the internal computational integrity of the model itself.

The limitations of static, pretrained models---their inability to access real-time information, incorporate domain-specific private knowledge, or reliably avoid hallucination---have positioned RAG as a critical architectural pattern for deploying truly useful AI systems. By dynamically injecting relevant context into the model's processing window, RAG allows LLMs to operate on current data, answer questions about proprietary enterprise knowledge, and provide personalized responses based on user data. This very dynamism, however, creates a fundamental tension: the grounding capability that makes RAG essential for utility and provides a natural audit trail simultaneously opens significant privacy vulnerabilities if the accessed data or the user's queries are sensitive. This chapter delves into technical mechanisms designed specifically to navigate this crucial trade-off.

Even the most advanced pretrained LLMs lack knowledge of recent events or private information. This explains many complaints seen about hallucinations and the inability of models to perform useful tasks. Fortunately, many ways to solve this draw from lessons in `in-context learning' (the ability of models to flexibly adapt to new situations by using examples or relevant contextual information) and `retrieval augmented generation' (RAG) (the ability of models to retrieve relevant information to answer a question, often from an external database or the web). These techniques are a key tool for unlocking the full potential of LLMs in AI systems, as the economic and personal utility of AI systems is often bottlenecked by context.

While powerful, these approaches introduce a range of risks. Here, we will primarily focus on the privacy risks. Anytime an AI system accesses sensitive information, privacy risks are introduced. Sensitive information can be leaked downstream in outputs, and the implicit act of searching for information via an external system can be used to infer the original private queries.

This chapter unfolds in three main sections. First, we will delve into the conceptual and legal arguments for why RAG architectures inherently offer a powerful mechanism for achieving transparency and auditability by design, drawing from early work exploring these benefits in the context of emerging legal requirements. Building on this foundation, the chapter then pivots to tackle the privacy challenges by presenting two distinct technical approaches. The second section introduces Private Retrieval Augmented Generation (PRAG), a novel system leveraging multi-party computation (MPC) to enable secure querying of distributed, private databases without revealing query content or database details to any single party. Finally, the third section explores an alternative paradigm using hardware-based security, examining how Trusted Execution Environments (TEEs) can create secure enclaves for confidential data pooling, management, and RAG, offering end-to-end confidentiality for sensitive AI workloads. 
    
\import{./}{nlrRAG.tex}

\import{./}{PRAG.tex}

\import{./}{communitytrans.tex}

%% file: chapter-3/nlrRAG.tex
\clearpage
\section{Transparency and auditability by design through retrieval augmented generation}

\noindent
\emph{To begin, let's explore the role of information retrieval in transforming an AI model, which lacks context and can hallucinate, into an AI system that is reliable, transparent, and auditable. 
This section is an updated and contextualized version of a piece coauthored with Robert Mahari published in Network Law Review's Computational Legal Futures edition, titled `Transparency by Design for Large Language Models' in May 2023. It was published on the heels of ChatGPT when RAG was just coming to the fore as a key way of contextualizing AI responses.}

\begin{paperabstract}
Large language models (LLMs) present unique challenges in terms of privacy and transparency, emphasizing the critical need for auditable and updatable systems. Auditing refers to identifying data records utilized by LLMs to generate specific outputs, which is essential for complying with regulations that require transparency in automated decision-making processes. Updatability, the ability to modify or delete data records within the system, is crucial for fulfilling the rights granted by privacy laws like GDPR and California's Consumer Privacy Act (CCPA), including the right to rectification and erasure of personal data. While achieving full explainability of LLMs remains elusive, integrating auditability and updatability into their design could significantly enhance transparency, compliance, and user trust. Thus, a legally sound design approach for LLM systems should prioritize the incorporation of these features to ensure adherence to global privacy standards and user rights.
\end{paperabstract}

Large Language Models -- like ChatGPT, Bard, and Claude -- are akin to modern day oracles: they provide impressively useful outputs without revealing their reasoning. It remains extremely difficult to understand why exactly a large language model has created a specific output, and this issue of explainability continues to attract widespread attention from academics, regulators, and practitioners. A humbler desire is to know what data was used to generate model outputs and to have the ability to modify this input data. This type of transparency matters both from an individual privacy and a business perspective. Individuals have an interest -- and, under some privacy regulations, a right -- to modify or delete data that is stored about them. Meanwhile, organizations that leverage LLMs need to ensure that outputs are based on up-to-date information. 

Building on recent work at MIT, we outline a new proposal for a type of LLM-powered data trust, called the Community Transformer. We explore how this technical architecture provides ways to track what data is used by an LLM and to modify this underlying data in ways that increase privacy and transparency.

\subsection{Updatability and auditability for privacy}

We distinguish between two types of input data for LLMs. The first is pre-training data which is used to create a general-purpose model (putting the `P' in Generative Pre-trained Transformer, GPT~\cite{Radford2018ImprovingLU}). Pre-training datasets are typically extremely large and composed of data scraped from the web and sophisticated additional datasets such as Reinforcement Learning from Human Feedback (RLHF)~\cite{RLHF} and alignment data~\cite{Kenton2021AlignmentOL}. The second type is task-specific data, which is used to tailor a general-purpose LLM to a specific task. A large quantity of task-specific data (although orders of magnitude less than for pre-training) can be used to fine-tune a model -- that is, to update the model weights for a specific application. RLHF and related methods such as value alignment~\cite{Bakker2022FinetuningLM} differ from traditional fine-tuning but broadly fit into this category with their data-hungry processes and permanent model weight updates. Alternatively, typically with lower upfront costs, a small amount of relevant task-specific data may be identified and included as additional information each time a model is used. This second approach, typically referred to as information retrieval, can be as simple as including relevant text or data in a model prompt.

While pre-training data~\cite{washingtonpostInsideSecret} may have privacy or business implications, it merely provides the model with a general ability to perform language tasks and so errors in the pre-training data are generally benign (unless these errors are systematic and give rise to biases in the model). By contrast, errors or omissions in task-specific data have more significant consequences, both because they more directly impact outputs and because there is generally far less of this data available. How exactly pre-training and task-specific data are used also has significant implications for privacy and transparency.

We use the term auditability to refer to the ability to identify what records were used by a machine learning system to generate a specific output. Auditability is a necessary but insufficient condition for explainability, which refers to the ability to understand the computational pathway through which a model produces a specific output or set of outputs. The European Union's General Data Protection Regulation (GDPR) creates an obligation for data processors to share the `logic' involved in reaching an automated decision. Legal scholars have debated~\cite{Selbst2017MeaningfulIA} whether this provision amounts to a right to explainability, however, regardless of whether data subjects are entitled to an explanation it is clear that the drafters of the GDPR intended for automated decisions to be made transparent. As such, auditability – giving users the ability to know which inputs led to a certain output – is an important step towards the spirit of privacy regulation and can form the basis on which individuals may choose to exercise their right to rectify incorrect information. From an organizational perspective, auditability is also a key requirement for LLMs. Organizations leveraging LLMs may wish to understand what inputs gave rise to certain outputs for quality control or compliance purposes. Data audits can also help organizations identify records that give rise to erroneous or outdated outputs to be removed or updated. 

By contrast to auditability, we use the term updatability to refer to the ability to modify or delete data records that are part of a machine learning system. The GDPR and California's Consumer Privacy Act (CCPA) both grant individuals rights amounting to a right to updatability. Namely, the GDPR grants individuals a right to rectification (a right to correct inaccurate information) and the right to be forgotten (a right to erase personal data), while the CCPA grants analogous rights to correction and deletion. While privacy regulations guarantee updatability for individuals, organizations that rely on machine learning enabled decision support systems also require the ability to modify the data that these systems rely upon. For example, an organization that uses LLMs to generate government filings might need to update the template every year. Modifying pre-training data, or data that has been used to fine-tune a model, is much more challenging and costly than simply deleting a record. The simplest approach is to modify the data and then retrain the entire model, however, doing so for the giant LLMs that have become the industry standard is costly. Machine unlearning~\cite{MachineUnlearn} can reduce the cost, but it still demands substantial computational resources and predominantly focuses on data deletion rather than correction. 

Auditability and updatability go hand-in-hand as the former can be used to can reveal errors or omissions to be updated. However, auditability can also be valuable by itself because it increases the transparency of LLMs. For example, auditability can provide insight into how an automated decision was reached and thus provide the basis for an appeal. 

\subsection{LLM Background}

The most basic objective of an LLM is to predict the most likely word to follow a given text input. It is through this simple task that the emergent flexible capabilities of LLMs arise~\cite{GPT4}. During the initial training phase of LLMs, the models ingest their training data, converting it into model weights through an iterative learning process. No record is kept of which training data contributes to what model weight updates. Although the LLM may retain fragments of the training data, this phenomenon is merely an unintended consequence of the learning algorithm attempting to complete the task of next-word prediction.

When utilizing LLMs for downstream applications, it becomes necessary to incorporate task-specific data. As outlined above, there are two primary methods for incorporating this additional information: fine-tuning a pre-trained general-purpose model or providing the model with task-specific data as part of a prompt.

Fine-tuning involves continuing the learning process of the LLM pre-training and adjusting or editing~\cite{hu2021lora} the model weights to optimize for performance on a specific task using the task-specific data. This approach was considered the standard method until relatively recently. However, the latest generation of LLMs offer a significant advantage due to their ability to process extensive amounts of data within a prompt without requiring fine-tuning. As a result, LLMs can adapt to various tasks more effectively, enhancing their overall utility in diverse contexts.

This ability also makes it possible to design LLMs to precisely identify which records contributed to the generation of a particular output and to update the relevant records without the risk of erroneously retaining obsolete information. These developments present a valuable opportunity to create mechanisms for updatability and auditability that comply with privacy regulations and better cater to real-world needs. 

\subsection{A new frontier of prompts \& context }
The new capabilities of the latest generation of LLMs have enabled new approaches to using these models. The rise of prompt engineering via interfaces such as ChatGPT has demonstrated the value that can be extracted from flexible pre-trained models by providing clear instructions and reference pieces of text. Including these additional pieces of textual information in the so-called `context window' that the LLM can parse allows it to make changes to the text or answer questions about it.

Extensions of this, often referred to as information retrieval or knowledge augmentation, draw on specific sets of data to be passed into the LLM during inference time, often based on search queries generated automatically from a user prompt. LLMs are used to identify search queries that would assist in answering the user prompt, which are subsequently used to search the web or specific databases (such as organizational databases) and the retrieved information can be included in the context window for answering the original prompt. This is all done without needing to fine-tune the models on the specific datasets. In a general sense, this is the methodology that is used to power the Bing and WebGPT~\cite{Nakano2021n} experiences.

\subsection{RAG and Auditability}

Explainability, the ability to determine how a machine learning algorithm arrived at a decision, has gained significant academic attention and is widely regarded as a key goal for AI. However, the increasing complexity of large machine learning models has made this goal difficult to achieve. While the Retrieval-Augmented Generation (RAG) architecture does not offer full explainability, it does provide an auditable record of the specific pieces of external data retrieved by the model to generate a given output.

Although RAG design enhances the transparency of language models, it does not fully explain the decision-making process for two important reasons. First, auditability in this context refers only to tracking which inputs the model had access to; it does not explain how those inputs were transformed into the final decision. Second, users can audit what external data was used in generating an output, but they cannot audit how the model's pre-training data influenced the decision. Consequently, errors or biases in the pre-training data could impact outputs independently of the retrieved external data. Nevertheless, auditability provides a significant step toward improving transparency in large language models (LLMs).

Auditability can be applied both ex-post and ex-ante. Ex-post, it helps identify incomplete or erroneous data that contributed to unsatisfactory or incorrect outputs. Ex-ante, it allows users to verify the relevant data being retrieved and offers the opportunity to exclude sensitive data before it is processed.

From a practical perspective, auditability holds great value. It benefits individuals by allowing them to know which records were used in making a particular decision, thus narrowing the scope of data that needs to be reviewed for errors. This, in turn, facilitates appealing decisions made by automated systems. Auditability also aligns with GDPR requirements for automated decision-making. For organizations, it helps ensure that decisions are based on correct data, as when outdated data is accidentally retrieved for reports, allowing such errors to be rectified through the auditable nature of RAG systems.

\subsection{RAG and Updatability}

Within the Retrieval-Augmented Generation architecture, three distinct sources of data contribute to model outputs: personal private data, retrieved external data, and pre-training data. Each data type is managed by separate processes, leading to varying degrees of updatability.

Personal private data, controlled by individuals, is not permanently retained, making it easily updatable. Users have complete control over their private data and can modify or delete it as needed.

External data, retrieved from community or organizational databases, is administered by those managing the RAG system. In principle, this data is also updatable, but practical challenges may arise. In centralized settings, such as businesses, data updates are straightforward. However, in decentralized settings, like municipalities or other public sectors, updating community data may require additional steps, such as contacting an administrator or even requiring a community vote. Balancing updatability and data security is crucial, as unrestricted modifications may introduce risks.

The most significant challenge to updatability lies with pre-training data, typically controlled by third-party developers. Updating pre-training data requires retraining models, which is resource-intensive and often impractical. While initiatives exist to allow individuals to check if their data is included in training datasets and request its removal, the ultimate control over training data rests with developers. Retraining models to incorporate these updates is rarely feasible due to high costs. Fortunately, LLMs are designed to rely on retrieved data in the context window, rather than memorized information from pre-training data. This reduces the risk of erroneous pre-training data affecting outputs, as long as the necessary information is provided in the prompt. However, concerns over the potential retention of incorrect pre-training data persist, particularly as such data is scraped from vast, unregulated online sources. To minimize this risk, LLMs, including those employing RAG, should rely on externally retrieved information to form outputs, rather than memorized knowledge, thereby also mitigating the risk of model hallucinations.

Ultimately, RAG architecture offers individuals and organizations significant flexibility in controlling the data involved in generating outputs. This flexibility greatly enhances the ability to maintain control over the information influencing machine learning models. Moreover, updatability is a key requirement under privacy regulations and will likely be essential for the widespread commercial use of LLMs.

\subsection*{Project Conclusion}

LLMs are often considered 'black box' systems, raising concerns about explainability and privacy. While achieving full explainability remains a challenge, Retrieval-Augmented Generation (RAG) architectures address these concerns by offering built-in auditability and updatability features. Updatability is essential for privacy, allowing users to modify or remove data, while auditability provides insight into the data influencing a given output. Together, these features enhance transparency and facilitate practical deployments of LLMs. Additionally, RAG systems offer communities a path to forming data trusts, enabling responsible use of LLMs in resource-constrained environments. By increasing transparency and control, this architecture promotes the responsible use of machine learning models in various domains.

%% file: chapter-3/PRAG.tex
\clearpage
\section{Private retrieval augmented generation}

\noindent
\emph{It's clear that Retrieval Augmented Generation (RAG) is a powerful tool in an AI system to provide auditable and contextually relevant responses. This utility can, however, come at the cost of privacy. This section seeks to address this need with a privacy-preserving approach to RAG using multi-party computation (MPC), based on a technical paper I coauthored with Guy Zuskind, Robert Mahari, and Alex Pentland titled `Private Retrieval Augmented Generation'.}

\begin{paperabstract}
While the flexible capabilities of large language models (LLMs) allow them to answer a range of queries based on existing learned knowledge, information retrieval to augment generation is an important tool to allow LLMs to answer questions on information not included in pre-training data. Such private information is increasingly being generated in a wide array of distributed contexts by organizations and individuals. Performing such information retrieval using neural embeddings of queries and documents always leaks information about queries and database content unless both are stored locally. We present Private Retrieval Augmented Generation (PRAG), an approach that uses multi-party computation (MPC) to securely transmit queries to a distributed set of servers containing a privately constructed database to return top-k and approximate top-k documents. This is a first-of-its-kind approach to dense information retrieval that ensures no server observes a client's query or can see the database content. The approach introduces a novel MPC-friendly protocol for inverted file approximate search (IVF) that allows for fast document search over distributed and private data in sublinear communication complexity. This project presents new avenues through which data for use in LLMs can be accessed and used without needing to centralize or forgo privacy.
\end{paperabstract}

Heavily pre-trained and fine-tuned Large Language Models (LLMs) have demonstrated exceptional performance on zero-shot~\cite{Kojima2022LargeLM} and few-shot tasks~\cite{brown2020language}. The ability of these models to generalize, combined with their costly pretraining, has shifted the focus from training ad-hoc models to perform specific tasks to utilizing these general-purpose foundational models for a wide variety of use-cases~\cite{Eloundou2023GPTsAG, GPT4}. These pre-trained models lack knowledge of private contexts or recent events.

To provide these LLMs with up-to-date or relevant information, methods such as Retrieval Augmented Generation (RAG)~\cite{Lewis2020, Karpukhin2020DensePR, Mao2020GenerationAugmentedRF} are used to include external information into a generation process without needing fine-tuning on new data. This process allows LLMs to first query an external data source, retrieve relevant information (with respect to a given prompt), and then use both the prompt and the retrieved data as input to the inference phase of the LLM.

Similar to the problem of federated learning~\cite{Kairouz2019AdvancesAO}, it is valuable to aggregate sensitive data from multiple (perhaps many) data owners. To do that, each party should be able to guarantee that their own private data remains private even when it is utilized. On the other hand, model users should be able to query this data from many data owners without needing to share what questions they are asking.

In this work we argue that LLMs require a new model for sharing data for AI tasks. Compared to federated learning, which focuses on the training phase, LLMs should focus on the (i) retrieval phase, and (ii) inference phase. Guaranteeing privacy of \textit{both} the query and any private documents residing in the retrieval database requires that both phases utilize privacy-preserving techniques and are chained together.

Alas, to the best of our knowledge all existing works only tackle the LLM inference problem \cite{Li2022, PUMA, southsecure, Mo2020DarkneTZTM}, but provide no secure solution when retrieval is involved. In this work, we close this gap by introducing Private Retrieval Augmented Generation (PRAG). PRAG allows users to privately search a database, which in itself is private, then send the augmented query privately to any secure (or otherwise trusted) LLM, creating an end-to-end secure solution.


\textbf{Our approach and contributions. } In this project, we propose Private Retrieval Augmented Generation (PRAG), a secure approach to augment neural information retrieval that hides both query vectors and the retrieval database. We use a retrieval database split across a set of servers, and we ensure data remains private by using secure multi-party computation (MPC) techniques. To the best of our knowledge, we are the first to consider the problem of secure distributed retrieval in the context of LLMs, and more broadly, are the first to propose a solution for private similarity search that can protect both the query and a secret-shared (or encrypted) database. This approach can be deployed with any standard neural information retrieval (IR) embedding model to augment distance calculations (e.g., cosine, dot, euclidean) and top-k retrieval over federated vector stores, scaling to medium-size databases with very little accuracy loss (99\% accuracy on real data). 

We further scale the approach to much larger databases using an approximate k-nearest-neighbors approach inside MPC, replicating the accuracy of the state of the art in approximate retrieval using a first-of-its-kind inverted files index inside MPC, providing significant speed improvements for retrieval. Our approach provides both theoretical and empirical improvements of value. We achieve constant communication on the client's side and \emph{sublinear} communication on the servers' side –– the bottleneck in MPC approaches. This work is the first IR approach to work across more than two servers with minimal additional costs. We further present a `leaky' version of the protocol that allows for partial privacy of queries under a privacy budget with significant improvements to speed. 

We evaluate PRAG across a range of data distributions, both real and synthetic, to show it broadly maintains the performance characteristics of non-secure IR approaches. 
We provide a pytorch-native implementation of our system using the Crypten MPC engine.



\begin{figure*}
    \centering
    \includegraphics[width=\textwidth]{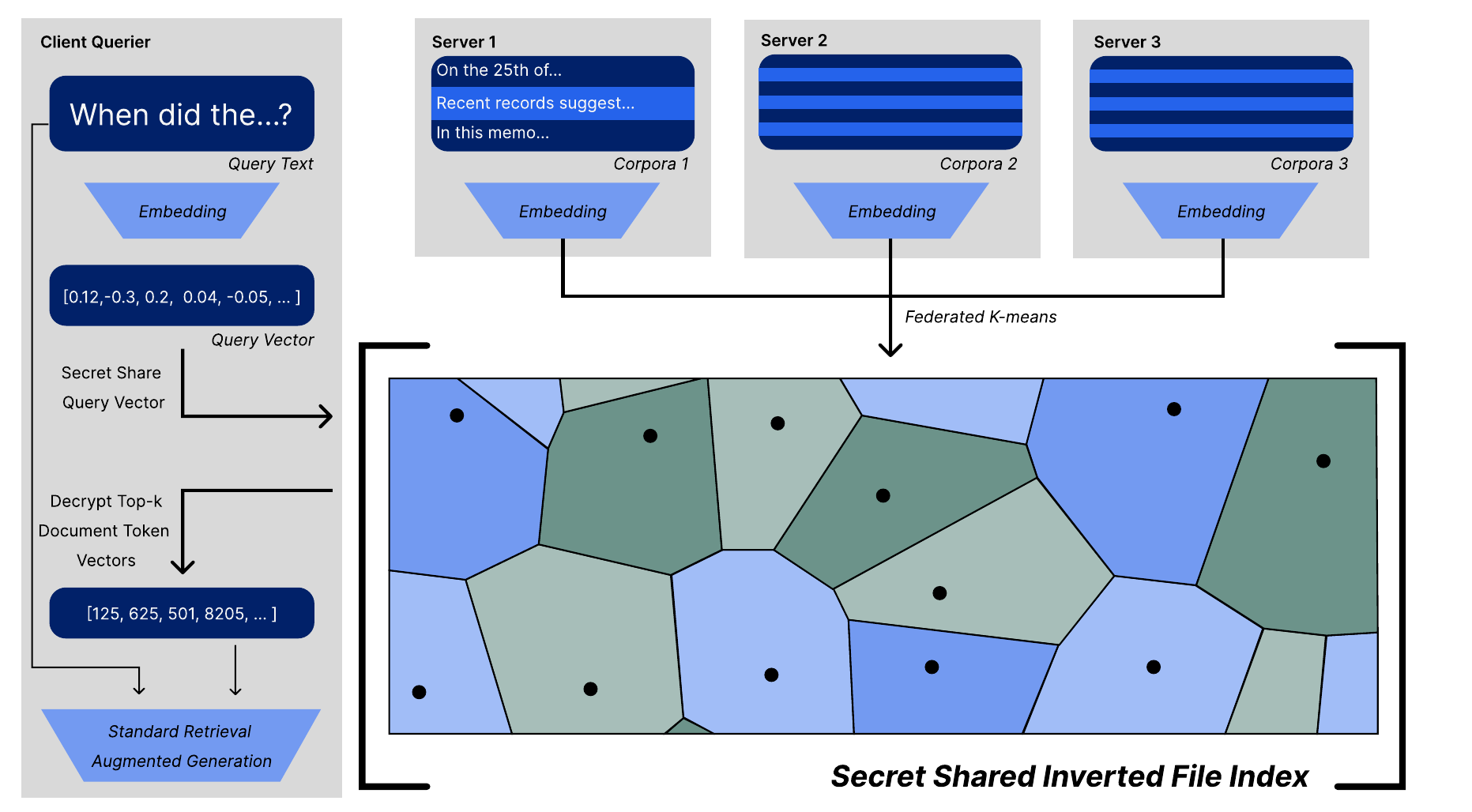}
    \caption{Overview of PRAG architecture using a distributed, secret-shared inverted file index (IVF), for retrieving document token vectors closely matching a privately-generated query vector in LLM-based question answering.}
\label{fig:pragdiagram}
\end{figure*}

\subsection{Methods}
In this section, we present the Private Retrieval Augment Generation (PRAG) framework. The method builds from secret sharing and MPC friendly exact top-k calculations to a new MPC design of an inverted file index for efficient approximate top-k calculation. A visual high-level overview of this design and its usage with a client LLM querier is shown in \autoref{fig:pragdiagram}.

\subsubsection{Overview and Trust Model}
Although a wide array of approaches exist for training document embedding models and augmenting generation with retrieved models, most neural information retrieval methods are underpinned by a step
where a querier sends a query embedding to a server to calculate the distance / similarity between the query vector and the database, in order to return a document either as an embedding vector for concatenation or with the document tokens for use in LLM inference. This setup offloads the storage of large databases and their associated calculations to a more powerful server.

Recently, a significant body of research has been focusing on the problem of secure inference, which ensures that a query remains private at all times. Whether secure inference is achieved through cryptographic techniques (e.g., \cite{Li2022, PUMA, akimoto2023privformer, chen2022x, gupta2023sigma}), or by running the model locally~\cite{Arora22PerfectSecrecy}, if the inference pipeline includes an external retrieval phase (as is often the case), then security does not hold as the query itself is leaked to the database operator.

Similarly, the database may itself hold private information, collected by many different data owners. The only way to protect their data is by making sure both the client and the vector database server(s) remain oblivious to its content.

To formalize this, we assume our system has $n_{clients}$ clients sending queries and $n_{owners}$ data owners. Both clients and data owners interact with a set of $n_{servers}$ vector database operators. We assume that all parties in the system are semi-honest (i.e., they follow the protocol) and that at most $t < \frac{n_{servers}}{2}$ of the servers are corrupt (the honest majority setting). In this work, we do not focus on the $n_{owners}$ data owners privately building the server, and we assume that at some point in the past these data owners have secret-shared their data to the servers. Instead, we are focused on the inference stage, a much more frequent and real-time operation.

\subsubsection{Exact MPC Tools}

We assume all values are shared using Shamir secret sharing ~\cite{shamir1979share} over a prime field $\mathbb{F}_p$ where $p$ \~= 32 or 64 bits. This choice is made to be compatible with the crypten-supported implementation. We note that our protocols could work using other secret sharing schemes suitable for the honest-majority setting (e.g., replicated secret sharing \cite{ito1989secret} over the ring $\mathbb{Z}_{2^{32}}$ or $\mathbb{Z}_{2^{64}}$), but Shamir is the ideal choice in our setting, as it requires the least amount of space and scales well to a large number of servers.

We further assume, as is common in secure machine learning literature ~\cite{riazi2018chameleon, knott2021crypten}, that there is a trusted dealer that generates shared random values. However, other techniques could distribute this~\cite{damgaard2013practical, orsini2020overdrive2k, escudero2020improved}. As in other works, since these protocols happen offline in a preprocessing phase and do not impact the online performance of serving a query, we do not benchmark their performance.

We denote arithmetic secret-shared values by $[x]$. A share for a specific server $i$ is denoted as $[x]_i$. When sharings may appear once as a $t$-degree sharing and again as a $2t$-degree sharing, we occasionally distinguish these sharings with a superscript (e.g., $[x]^{(2t)})$. We use [$x$] := SS.Share($x$) and $x :=$ SS.Reveal([$x$]) for sharing and revealing secret shared items.

As is well known, all linear operations over secret-shared values require no interaction between the servers. For multiplication, a single round of interaction is required. Given our setting, we find the multiplication protocol by Damg{\aa}rd and Nielsen~\cite{damgaard2007scalable} to be the most suitable. 



To encode real numbers into the field $\mathbb{F}_p$, we use a known technique of representing all underlying values as fixed-point integers \cite{catrina2010secure}. In practice, this means that for any real value $\tilde{x} \in \mathbb{R}$, we encode it as a fixed-point integer $\lfloor \tilde{x}2^f \rfloor \in \mathbb{Z}$ with magnitude $e$ and precision $f$ (with a total bit length of $e + f$. Note that multiplying two encoded values results in a value with $2f$-precision. Therefore, truncation is needed after every multiplication to avoid causing an overflow inside the field, which would distort results.


\paragraph{Distance calculations}
While there is some heterogeneity in distance measures used in neural information retrieval, the majority use dot products, cosine similarity, or L2 norms (euclidean distance)~\cite{reimers-2019-sentence-bert,reimers-2020-multilingual-sentence-bert}. We provide MPC friendly implementations of all three.

A naive implementation of a dot product between a vector and a matrix can be provided by running the secure multiplication protocol in parallel. Both the communication and the computation complexity scale linearly with the size of the database $N$ and embedding dimension size $d_e$, the latter of which is fixed in almost all cases. Round complexity remains the same (constant) regardless.

Extending the dot product gives us cosine similarity, the predominant distance measure in sentence transformer style models~\cite{reimers-2019-sentence-bert}. To save on expensive MPC computations, we pre-normalize the input vectors and matrices prior to secret sharing into MPC, allowing for cosine similarity to reduce to a simple dot product. Computing Euclidean distance can also be achieved directly through MPC, but we observe that this is a much more expensive operation, as it requires computing square roots inside the MPC circuit. For example, Crypten~\cite{knott2021crypten}, which we use in our implementation, uses a slow Newton-Raphson approach for computing square roots, requiring multiple rounds of communication. 

However, we make the observation that given that top-k calculations are the end goal of distance calculations, the monotonic square root step in L2 can be ignored completely before looking for the top-k elements in the distance vector, removing the need to compute the square root securely.

\paragraph{Fast secure dot product}

Computing the dot product of two vectors $x, y$ requires computing the sum of their point-wise products $z := \sum_{j=1}^d x_j y_j$. This can be achieved in MPC naively by using a secure multiplication protocol in parallel. However, for vectors of size $N$, this requires pre-processing and communicating $O(N)$ elements per dot product. This further compounds as we try to securely multiply matrices together, as in our case.

However, as was observed by previously~\cite{chida2018fast} and leveraged in works such as Blinder~\cite{abraham2020blinder}, we can reduce the communication complexity of computing a dot product from $N$ elements to a single element, by first having each party first locally compute the sum of point-wise products (instead of each product independently), and only masking the final result. Repeating this across a dimension of a matrix, we can use this for efficient matrix multiplication.

\paragraph{Relation to private information retrieval}

A well-known method of privately reading a specific entry in a database is by computing the dot product between a one-hot-vector with a non-zero element at the index of interest. Assuming $i$ is the index of interest from some arbitrary vector or matrix $x$, one can privately retrieve the data at row $i$, without leaking any information as $[0, \dots, 1, \dots, 0] \cdot  [x_1, \dots, x_i, \dots, x_N]^T =  [x_i]$. To read several rows at once, we can first sum across several one-hot-vectors to obtain a single vector.

This simple oblivious private retrieval from a database allows us to extract any top-k elements from a database matrix that has been secret shared. This allows us to extract either database embedding vectors or token arrays from inside the distributed database for return. In essence, rather than securely returning top-k indices and asking the user to separately extract them, we can return the original tokens from a secret shared database directly in MPC. This oblivious retrieval is used extensively throughout our protocols below, such as in extracting candidate vectors from clusters.

\paragraph{Exact top-k for retrieval}

Retrieving the most similar documents to a query requires first ranking all documents by some similarity metric (as above) and then picking the top $k$ documents that are closest to the query.


Our solution is conceptually similar to secure top-k circuits designed in other works~\cite{chen2020sanns}, where $O(kN)$ comparisons are needed. These circuits operate by successively keeping an ordered list of $k$ items, and then computing each value in the array with the minimum value in the (much smaller) sorted list. Unfortunately, this solution also requires $O(N)$ rounds for MPC based on secret-sharing. 


Instead, our protocol iterates $k$ times over a secret-shared vector $[x]$. In each iteration, we run argmax($[x]$) to get the current minimum's index in the vector. We then obliviously scale down the selected value enough to ignore it in future iterations.

There are many ways to implement an MPC protocol for argmax($[x]$). Our description assumes a recursive tree-reduction based protocol as in Crypten~\cite{knott2021crypten}, having $O(\log_2(N))$ rounds and $O(N\log_2(N))$ total communication. This leads to an exact top-k round complexity of $O(k\log_2(N))$ and $O(kN\log_2(N))$ overall communication.

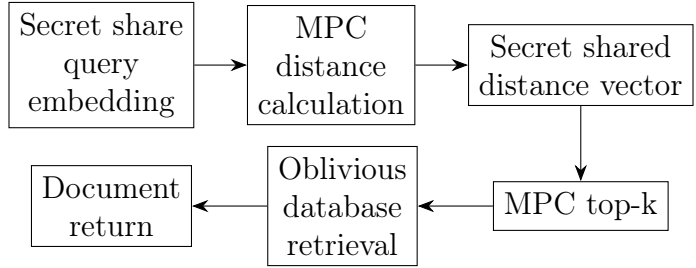
\begin{figure*}
\begin{adjustbox}{center}
\begin{tikzpicture}[node distance=0.7cm, auto,
    every node/.style={rectangle, draw, align=center},
    >={Stealth[scale=1.3]}
]
  \node (B) {Secret share\\query \\embedding};
  \node (C) [right=of B] {MPC\\ distance\\calculation};
  \node (D) [right=of C] {Secret shared\\distance vector};
  \node (E) [below=1cm of D] {MPC top-k};
  \node (F) [left=1cm of E] {Oblivious \\ database \\retrieval };
  \node (G) [left=1cm of F] {Document \\ return };

  \draw[->] (B) -- (C);
  \draw[->] (C) -- (D);
  \draw[->] (D) -- (E);
  \draw[->] (E) -- (F);
  \draw[->] (F) -- (G);
\end{tikzpicture} 
\end{adjustbox}
\caption{Process flow for retrieving the top-k nearest documents using MPC and oblivious database retrieval.}
\label{fig:mpc_flow}
\end{figure*}

By combining this with distance calculations and oblivious private retrieval from a database, we can provide an end-to-end exhaustive exact algorithm to return the top-k nearest documents to a query from a database of embeddings (and a database of tokens for exact document return). See the process flow in Figure \ref{fig:mpc_flow}.

\subsubsection{Nearest Neighbors and Inverted Files (IVF)}

At its core, the information retrieval task of top-k closest points is exactly the task of solving the \emph{k-nearest-neighbors} (kNN) problem, which requires finding the k points in a database that are nearest to the given data point (the query). While the above exact approach achieves this, it does so at a significant speed cost (both with or without MPC), motivating the creation of approximate nearest neighbors algorithms, which only require a sublinear amount of work.

These algorithms operate by first computing a compact representation of the dataset called the \emph{index}, and then executing queries on the index. Many approximate nearest neighbors techniques exist, and one that is particularly amenable to MPC is the \emph{inverted files index} (IVF) \cite{FAISS, IVF}. This technique works by first using a clustering algorithm (e.g., k-means) over the data set to find its $n_c$ \emph{centroids}. Then, each centroid represents a cluster holding all points associated with that cluster. In other words, this process splits the database into $n_c$ buckets.

After this one-time step, querying the data starts by computing the nearest neighbors of the query with respect to all centroids. Then, only the nearest clusters are searched (parameterized by $n_{probe}$), looking for the $k$ nearest neighbors among them.

During IVF generation, parameter choices in how the index is built affect the downstream performance of the queries. We choose the number of clusters to be $n_c = \alpha \sqrt N$ to get sublinear complexity, where $\alpha$ is a free parameter that can be tuned. During query time, we find the distance to all $n_c$ centroids, and select the top $n_{probe}$ clusters to inspect further. As we will see during experiments, this choice of $n_{probe}$ increases the recall performance of the model, and indeed at $n_{probe}=n_c$, all clusters are inspected and the search becomes exact. Similarly, for $n_{probe} = 1$, only the nearest cluster is searched, maximizing performance at the expense of recall. In general, the nature of IVF clustering allows a smaller $n_{probe}$ to be chosen while still achieving high accuracy. 

\subsubsection{Efficient approximate vector nearest neighbor search in MPC}

Bringing this into MPC, the protocol $\Pi_{\text{IVFQuery}}$ securely computes the approximate nearest neighbors using an inverted file index. We note that we only care about real-time efficiency of retrieval. We therefore assume that the servers pre-computed the secret-shared inverted index $[IVF]$, for example, by employing a private k-means clustering protocol, of which many exist (e.g., \cite{patel2012efficient, fan2021ppmck}). This private index consists of $n_c$ lists of size $m$, both of which are of size $O(\sqrt N)$, ensuring the overall communication complexity is sublinear. We use the MPC distance measures established earlier in the paper to calculate the distance between the query vector and each of the $n_c$ cluster means.

The parties then run a secure protocol of exact top $k$ as described earlier to identify the $n_{probe}$ most similar clusters. Unlike non-MPC protocols, it is critical that the servers remain oblivious as to which are the top clusters for this query. Otherwise, information about both the query and database would leak. For this reason, we require the top-k protocol to return each index as a one-hot-vector of size $n_c$ which are collectively stored in $[closest\ buckets]$.

Then, the parties perform an exact-match private information retrieval to get all the vectors in the closest buckets. These $[candidates]$ can be obliviously found through a product of $[closest\ buckets]$, a mapping of centroids indices to cluster indices in the database, $[IVF\ indices]$, and the entire $[IVF]$ vector database.
By obliviously reducing the entire vector database into a much smaller search space that only includes vectors from the $n_{probe}$ nearest clusters, we are able to achieve sublinear overall communication.

At this stage, $[candidates]$ holds a reduced $(n_{probe} \times m) \times d$ vector matrix (where $d$ is the embedding dimension). $[candidates\ indices]$ will similarly store the mapping from each candidate to the original database index.
We proceed by running an exact nearest neighbor search again, which computes the distances between the query and all candidates and then securely gets the top-k entries. Using $[candidates\ indices]$, these top-k entries are mapped back to the original database records, where documents can be obliviously retrieve.


\begin{algorithm}
\caption{$\Pi_{\text{IVFQuery}}$}
\KwIn{Public Parameters: $n$, $k$, $n_c$, $n_{\text{probe}}$, $m$, $d$\\
Client: query $x \in \mathbb{R}^d$\\
Server: Secret-shared inverted file clusters [IVF clusters]$\in \mathbb{R}^{n_c \times d}$, Inverted file index values [IVF] $\in \mathbb{R}^{n_c \times m \times d}$, Inverted file index indices [IVF indices] $\in \mathbb{R}^{n_c \times m}$}
\KwOut{k-nearest-neighbors (approximate)}

\textbf{Client computation:}\\
$[x] := SS.\text{Share}(x)$\;
Send each server $i$ its share $[x]_i$\;

\textbf{Servers computation:}\\

\textbf{in parallel} Iterate over [cluster] $\in$ [IVF clusters]\;
\quad $[\text{centroid distance}_i] := \text{SumProd}([x], [cluster])$\;
\quad $[\text{centroid distances}] := \{ [\text{centroid distance}_1]^{(t)}, \ldots, $ $[\text{centroid distance}_{n_c}]^{(t)} \}$\;

Compute $[\text{closest buckets}] := \text{ExactTopk}([\text{centroid distances}], n_{\text{probe}})$\;

Compute $[\text{candidates}] := \text{MatMult}([\text{closest buckets}], [\text{IVF}])$ and $[\text{candidates indices}] := \text{MatMult}([\text{closest buckets}], [\text{IVF indices}])$\;
\textbf{in parallel} Iterate over [candidate] $\in [\text{candidates}]$\;
\quad Compute distance using $\text{SumProd}$ and store as $[\text{candidate distances}]$\;
Compute $[\text{candidate top-k indices}] := \text{ExactTopk}([\text{candidate distances}], k)$\;

Compute $[\text{database top-k indices}]$ via private exact-match retrieval of $[\text{candidate top-k indices}]$ from $[\text{candidates indices}]$\;

Return $[\text{database top-k indices}]$ documents via private retrieval.
\end{algorithm}

\paragraph{Sublinear Communication Complexity}
The client maintains an optimal communication complexity of $O(1)$, as it only needs to communicate a share of the query vector to each server.

As to the servers, in lines 5-7 a total of $n_c := O(\sqrt N)$ elements are communicated. Computing the exact top-k over these $n_c$ distances requires $O(k \cdot \log_2(n_c))$ communication. Reducing the dataset obliviously costs $O(n_{probe} \frac{N}{m} d)$. With our choice of parameters, $n_{probe}$ and $d$ are constant, and $m = \sqrt N$, yielding $O(\sqrt N)$ communication. This gives a candidate dataset that is approximately of size $n_{probe} \sqrt N$. Finally, we can compute the distances and exact top-k on this reduced dataset, but as it now only contains $O(\sqrt N)$, the overall communication of that step is $O(k \cdot \log_2(\sqrt{N}))$.

Overall, we see that end-to-end the servers communicate $O(\sqrt N + \log_2(\sqrt{N}))$ field elements while the client communicates $O(1)$ elements (in fact, she communicates exactly $d$ elements, as is the size of the input vector). This holds true so long as $n_{probe}$ remains small enough to be considered a constant. As the number of candidate clusters to be probed becomes $n_c$, the overall complexity of the approach becomes $ O(\sqrt N \cdot \sqrt N) = O(N)$, which is no better than exact search but with additional overhead operations. Hence, $n_{probe}$ should be kept low as we will see in the experimental settings.

\subsubsection{Sacrificing Privacy for Speed in MPC IVF}
The fast secure dot product trick above helps significantly improve the speed of the step wherein we reduce the full database to only the $n_{probe}$ clusters vectors relevant to the query. However, this step is still extremely costly, requiring the manipulation of a large database of vectors for lookup when the clusters are stored in a large matrix. 

Instead, we can take an alternate approach, where each cluster is stored in its own secret shared database, with an exposed lookup table. The centroids of the database still remain secret shared and private, but during query time, the $n_{probe}$ closest clusters (shuffled to avoid exposing order) are reconstructed by each server to retrieve the relevant secret shared cluster matrices, which can then be concatenated before passing into the second distance-top-k calculation. This has large speed implications, dramatically decreasing the data access time and allowing for speed more competitive with non-MPC IVF. 

However, this does come at the cost of privacy. Each server will now know the $n_{probe}$ closest clusters to the query, which leaks the area in the embedding space where the query is coming from. Indeed, while the centroids are secret shared, knowing the lookup table and what a user accesses would allow an actor to determine an average point across those centroids with more queries. 

\begin{figure}
    \centering
\includegraphics[width=\linewidth]{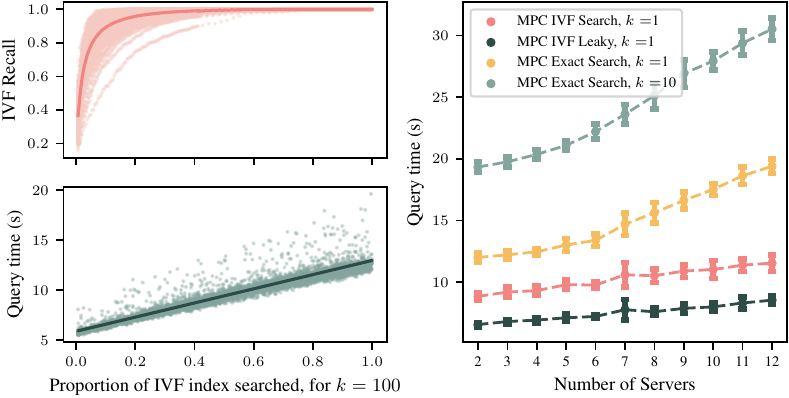}
    \caption{Information retrieval using IVF improves accuracy with increased $n_{probe}$ (top left) but increases query time as a larger proportion of the index ($\frac{n_{probe}}{n_c}$) must be searched (bottom left). These retrieval approaches (both IVF and exact) scale favorably across multiple servers (right).}
    \label{fig:IVFscalingandworlds}
\end{figure}

To mitigate this, a query could be noised according to a privacy budget similar to differential privacy, as for sufficiently large $n_{probe}$, even a high noised query would likely contain the relevant closest clusters nearby. One slight advantage here is that larger choices of $n_{probe}$ provide more privacy (and more capacity for noising), while also increasing the overall accuracy of the search (as we see in  \autoref{fig:IVFscalingandworlds}).

In general, this final methodological change differs from above by no longer being fully private, but is presented as part of the spectrum from slow but exact private search to fast approximate search, and finally to fastest but leaky approximate search.

\subsection{Experiments}
To demonstrate the performance of these models we run a series of experiments on both synthetic and real data to determine performance properties of the implementations of these methods above.


We benchmark the retrieval accuracy and speed across a range of embedding sizes (256 to 8192), synthetic embedding distributions ($N(0,0.05)$, $N(0,1)$, $U(-1,1)$, Binary), distance functions (cosine, dot product, euclidean), top-k values, IVF parameters, and database sizes. We perform MPC experiments on a single 2.2GHz Intel Xeon Silver CPU using Crypten's built-in communication code to spawn processes for each server. 

Further to this, we test the approaches on retrieval of real neural embedding datasets from BEIR~\cite{BEIR} using the same environment, this collection of datasets uses a range of textual document types and sizes, all of which we use a standard off-the-shelf embedding on. While there are several parallelization improvements that can be made locally within each server for MPC, our implementations of each algorithm above remain unoptimized.

\subsubsection{Exact Search}
Each step of the exact search approach is extremely accurate, with small numerical errors introduced during MPC. For distance measures, MPC vectors have a mean squared error difference from pytorch calculated distances of less than $10^{-5}$ for euclidean and $10^{-8}$ for cosine, going as low as $10^{-11}$ for euclidean distance on $N(0,0.05)$.  These errors do not change with database size, and are introduced at the numerical level of the elements.

The exact top-k approach using tree reduction applied interactive k times suffers from similar small numerical errors. For distance vectors drawn $N(0,0.05)$, where outliers are often standalone, top-k elements are picked out with 0.99 or above recall and precision. For uniform distributions (unrealistic for embedding distance vectors) the f1 accuracy is lower for top-1 (0.842) and top-k (0.96) with recall and precision climbing for higher k. This is explained by the small distances present between the max and its nearest value when drawn from a uniform distribution, leading numerical errors to induce a loss of accuracy. Fortunately, the nature of real distance distributions means performance is high in real contexts.
For small values of $k$, this approach can be relatively fast but increasing the choice of $k$ dramatically increases the time cost due to communication complexity in the interactive argmax looping.

\begin{figure}
    \centering
    \includegraphics[width=\linewidth]{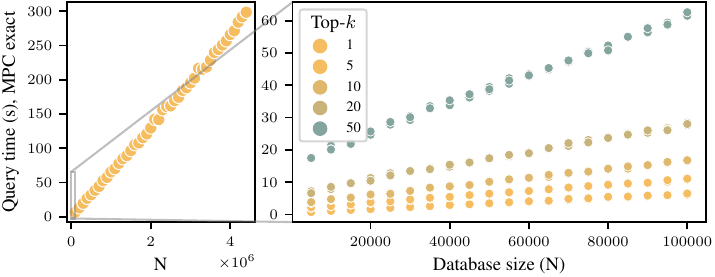}
    \caption{Time taken to retrieve top-k closest vectors in the database for end-to-end MPC exact search across increasing synthetic database sizes. The right side plot is a zoomed-in section of the left side.}
    \label{fig:e2escaling}
\end{figure}

Putting distance calculations, top-k, and oblivious retrieval together, the exact search approach in MPC can identify the top-1 (argmax) most similar vector to a query with 97.5\% accuracy and top-50 with 98.6\% F1 score, with accuracy independent of database sizes tested up to $5\times 10^5$. The constraint on the use of this MPC exact approach is the speed, taking up to 10 seconds for top-1 and top-5 for a $10^5$ size database, and increasing dramatically for larger $k$ as in \autoref{fig:e2escaling}.

\subsubsection{Approximate Search}

Our MPC IVF implementation, using both fully secure and partially leaky clustering, returns the elements as the standard IVF implementation with an average of over 99\% recall on both synthetic and real embedding data, with errors explained by numerical errors at runtime. For real data, we use embeddings from msmarco-distilbert-base-v3 from SBERT~\cite{reimers-2019-sentence-bert}. These numerical errors partly flow through from the exact search above, which is used at various points in the IVF MPC algorithm. This accuracy of the MPC IVF to non-IVF is stable across choices of $n_{probe}$ and $n_c$.

While the MPC IVF matches the recall performance of the standard IVF, the underlying approximate nature of the IVF provides tradeoffs between accuracy and speed. As shown in \autoref{fig:e2escaling}, increasing the value of $n_{probe}$ increases the proportion of the full database that is inspected at query time, in turn increasing the overall runtime. The benefit of IVF is that we can achieve high accuracy for even a low value of $n_{probe}$, dramatically reducing query time at the cost of accuracy.


\subsection{Related Work}
Drawing on the ideas in private federated learning, we can maintain privacy when doing public queries \cite{Arora2022ReasoningOP} and move beyond in-context learning \cite{Arora22PerfectSecrecy}.




We bring privacy to this idea through augmenting existing non-private retrieval methods, ranging from exact search on small datasets to large scale approximate retrieval~\cite{FAISS, IVF}. While several other works have examined the problem of secure similarity search~\cite{chen2020sanns, zuber2021efficient, servan2022private, asharov2017privacy, schoppmann2018private, shaul2018scalable, shaul2018secure, songhori2015compacting}, to the best of our knowledge we are the first to examine a model where the database is secret shared as well, and where an arbitrary number of servers and database owners can be supported.

These approaches can augment other pieces of privacy-first ML infrastructure from fully secure LLM inference~\cite{Li2022, PUMA} and federated or privacy preserving K-means clustering~\cite{Vaidya2003PrivacypreservingKC, Jagannathan2005PrivacypreservingDK}. We choose to focus on MPC techniques in this project, as opposed to secure retrieval schemes that rely trusted execution environments (TEEs) ~\cite{Wang2006PrivateIR, Yang2008AnEP, Papadopoulos2010NearestNS, Drean2023CitadelEW}, as TEEs have been known to suffer from privacy-breaching attacks.


\subsection*{Project Conclusion}

We introduced PRAG, a novel approach for secure, distributed information retrieval for large language models. PRAG uniquely safeguards both query vectors and a multi-owner database using multi-party computation (MPC). Key contributions include an MPC-friendly protocol for inverted file approximate search, allowing for rapid document retrieval with sublinear communication complexity; analysis of exact search performance on language embeddings; and a version of the protocol that offers a trade-off between speed and partial privacy, under a predefined privacy budget.
These tools allow for a new mechanism of neural information retrieval, which when combined with secure inference of LLMs, is a stepping stone towards fully secure foundation model agent pipelines. However, much like secure execution of LLMs, the approach put forward here has significant computational costs and speed limitations, especially for large databases and high accuracy demands.
Future work should explore optimizing communication costs, expanding beyond a semi-honest adversary, and integrating PRAG into larger secure machine learning frameworks. 

%% file: chapter-3/communitytrans.tex
\clearpage
\section{Trusted execution environments (TEEs) for private data management and RAG}

\noindent
\emph{While MPC is an interesting approach to privacy in RAG, it comes with speed and communication trade-offs. An alternative approach is to use hardware-based security, such as secure enclaves (also referred to as trusted execution environments). This section is based on a project titled `Community Transformers' started in 2023 at the start of the energy around ChatGPT. It was a collaboration primarily with Guy Zuskind to combine our various work on trusted execution environments, community data sharing, and the new large language models. This project was put on arxiv, but never continued into final publication. In part, this was because of a major issue: you could not run language models on GPUs securely as they did not support enclaves. Since this original project, the NVIDIA H100 provided support for exactly this, unlocking a renewed interest in this idea and commercial work on confidentiality for AI workloads. As a result, this original seed of an idea for knowledge management and confidential inference with large language models has been updated with new content and insights to reflect this.}

\begin{paperabstract}
This section examines the application of Trusted Execution Environments (TEEs) as a privacy-preserving mechanism for Retrieval Augmented Generation (RAG) systems. The `Community Transformers' architecture provides a TEE-centric framework that enables communities to securely pool sensitive data while maintaining individual privacy rights. The architecture implements cryptographic protocols for secure key exchange, privacy-preserving transformations for unstructured text, and governance mechanisms that balance technical protections with community sovereignty. Recent advances in hardware-level confidential computing, particularly with NVIDIA's H100 GPU, extend these capabilities by enabling secure model inference directly within protected enclaves, thus creating an end-to-end confidential pipeline for knowledge management in AI systems. This approach addresses critical privacy and compliance challenges while enabling collective knowledge utilization through large language models.
\end{paperabstract}

Community data sharing offers significant potential for addressing societal challenges, demanding innovative governance models like data cooperatives~\cite{Hardjono2019DataCT} that empower collective control~\cite{PentlandBuilding}. However, realizing this potential requires overcoming the inherent tension between data utility and privacy, especially when leveraging sensitive information with advanced AI like Large Language Models (LLMs). Existing approaches often lead to suboptimal outcomes where data remains siloed or control is ceded~\cite{Hardjono2019DataCT}. Human-centered architectures aim to extract insights without revealing raw data, challenging dominant data aggregation models~\cite{Mahari2021EraAntitrust}. This work proposes a solution using Trusted Execution Environments (TEEs) to enable secure Retrieval Augmented Generation (RAG) on community data. Critically, recent advances like NVIDIA's H100 GPUs allow even the computationally intensive LLM inference step to occur within secure hardware enclaves, facilitating an end-to-end confidential pipeline.

We define a community functionally as a group with shared relationships capable of establishing institutional structures necessary for deploying such a system~\cite{chavis1990sense, putnam2000bowling}. Examples range from geographic localities analyzing urban data, to businesses aggregating insights across units, to affinity groups using shared experiences for tailored support (e.g., mental health, cultural heritage). In all cases, harnessing community data via LLMs promises unique value but necessitates rigorous privacy preservation and alignment with community interests.

While methods like prompt engineering and model fine-tuning (including community-driven RLHF) exist for tailoring LLMs, they often face resource constraints (computation, data annotation) that are unsuitable for many communities. Fine-tuning is computationally expensive, and while RLHF is powerful, it requires significant community effort. Designing effective system prompts is vital but considered complementary. Therefore, we focus on RAG~\cite{Lewis2020, Nakano2021n}, which dynamically retrieves relevant information from a community corpus to augment LLM responses. RAG aligns well with TEE-based privacy, allowing secure use of sensitive data sources without costly model retraining, especially when combined with confidential GPU inference for the LLM itself.

\section{Community Transformers: System Architecture}
The central objective of the proposed `Community Transformers' system is to enable users to pose questions that draw insights from both their own private data and sensitive, pooled community data, without compromising the confidentiality of either. Consider, for example, an individual querying local specialist reviews (community data) based on their specific medical condition (personal private data). The system must facilitate this complex question-answering process while rigorously safeguarding all involved data sources.

We introduce a system architecture centered around a Trusted Execution Environment (TEE), which provides a secure and isolated processing environment on a server. This TEE manages access to an encrypted database holding the community data contributions. The core functionality provided within the TEE includes secure data ingestion, governance enforcement, privacy-preserving processing (when necessary), and secure information retrieval for RAG.

\textbf{Core Principles and Components:}
The design adheres to several key principles:
\begin{itemize}
    \item \textbf{Local Control and Self-Governance:} Data and computation should ideally be hosted and controlled locally or by trusted community representatives. Access to pooled data requires explicit consent mechanisms managed by the community.
    \item \textbf{Strong Privacy Guarantees:} Privacy, both from external threats and potentially internal misuse, must be technically enforced to the strongest feasible level.
    \item \textbf{Verifiability:} The integrity and confidentiality of the operations performed within the TEE should be verifiable through mechanisms like remote attestation.
\end{itemize}

The system requires an initial community investment in server infrastructure capable of hosting a TEE and an encrypted database. The TEE instance externalizes a public encryption key ($pk_{TEE}$) used by community members to encrypt their data before submission. Only the TEE possesses the corresponding private key ($sk_{TEE}$) required for decryption within its secure boundary. Community governance protocols are essential; for instance, administrative access for data deletion or extraction might require approval via a threshold signature scheme~\cite{FROST} involving designated community representatives, preventing unilateral actions.

\textbf{Data Flow and Types:}
Three primary data categories are handled:
\begin{enumerate}
    \item \textbf{User Query ($X$):} Comprises a natural language prompt ($X_{prompt}$) and optional user-specific private data ($X_{data}$) relevant to the query.
    \item \textbf{Private Community Data ($C^{private}$):} Sensitive data contributed by community members. Each record $C^{private}_i$ may contain personally identifiable or confidential information. This data resides within the TEE-controlled encrypted database and may undergo privacy-enhancing transformations (e.g., de-identification) inside the TEE before being used for retrieval. The transformed, privacy-enhanced version used for retrieval is denoted $C^{safe}$.
    \item \textbf{Open Community Data ($C^{open}$):} Data accessible to all authenticated community members but not the general public (e.g., shared community records, non-sensitive internal business data). This data also resides in the secure database.
\end{enumerate}
The pooled community data, $C = C^{private} \cup C^{open}$, is dynamic, updated asynchronously as members contribute.

\textbf{The Role of the TEE:}
A TEE, such as Intel SGX, AMD SEV, or ARM TrustZone, provides a hardware-enforced isolated execution environment~\cite{McKeen2013}. Its critical security properties enable this architecture:
\begin{itemize}
    \item \textbf{Confidentiality (Memory Encryption):} Data processed within the TEE is encrypted in memory, protecting against physical memory snooping and cold boot attacks.
    \item \textbf{Integrity:} Code executed within the TEE is protected from unauthorized modification by the host OS or hypervisor.
    \item \textbf{Attestation:} The TEE can cryptographically prove to a remote party (e.g., a user's device) what code is running inside it and that it is a genuine TEE. This builds trust that the correct privacy-preserving logic is being executed.
    \item \textbf{Sealing:} The TEE can securely encrypt data using a key unique to that specific TEE instance, allowing sensitive state (like processed data or intermediate keys) to be persisted securely outside the enclave (e.g., on disk) and only be decrypted later by the same TEE instance.
\end{itemize}
By managing data access and processing within the TEE, the system ensures that raw sensitive data ($C^{private}$, $X_{data}$) is only decrypted and handled inside this protected environment. For efficient retrieval, data ($C^{safe}$, $C^{open}$) can be loaded into the TEE's protected memory. If the entire relevant dataset fits within the enclave memory, it mitigates side-channel attacks based on observing memory access patterns to the external encrypted database. For larger datasets exceeding enclave memory limits, techniques like Oblivious RAM (ORAM)~\cite{jean2023sgxonerated} could be integrated (considered future work) to obscure access patterns, albeit with performance overhead.

\textbf{Privacy Controls within the TEE:}
Before private community data ($C^{private}$) is made available for retrieval (as $C^{safe}$), the TEE can execute pre-defined privacy-preserving transformations. These could range from simple pattern-based redaction (e.g., removing names, addresses) to more sophisticated statistical techniques like k-anonymization or potentially applying differentially private mechanisms during analysis or aggregation steps, depending on community requirements and data characteristics. LLMs themselves, executed within the TEE, could potentially assist in identifying and transforming sensitive elements. The specific transformations would be chosen and configured by the community governance process.

\textbf{Secure User Interaction:}
User queries ($X$) are encrypted client-side using the TEE's public key ($pk_{TEE}$) before transmission. Robust authentication mechanisms (e.g., user accounts, cryptographic signatures) are necessary to authorize query submission, preventing unauthorized access and mitigating risks like denial-of-service (DDoS) attacks or privacy breaches through repeated probing queries~\cite{Nasr2018ComprehensivePA}.

In summary, the architecture relies on these key security elements:
\begin{enumerate}
    \item TEE execution for all sensitive data handling and computation.
    \item End-to-end encryption of data in transit (using $pk_{TEE}$) and at rest (managed by the TEE and secure database).
    \item Optional, TEE-executed privacy transformations (e.g., de-identification) applied to $C^{private}$ to create $C^{safe}$.
    \item Secure, TEE-mediated information retrieval from $C^{safe}$ and $C^{open}$.
    \item Encryption of user queries and associated private data during transit to the TEE.
    \item Verifiable execution environment via TEE attestation.
\end{enumerate}

\subsection{Secure Retrieval Augmented Generation Protocol}

The Community Transformers architecture implements the following protocol to perform RAG while preserving privacy:

\begin{enumerate}
    \item \textbf{Query Encryption:} The user encrypts their query $X = (X_{prompt}, X_{data})$ using the TEE's public key: $E_{pk_{TEE}}(X)$.
    \item \textbf{Decryption:} The TEE receives the encrypted query and decrypts it using its private key within the secure enclave: $D_{sk_{TEE}}(E_{pk_{TEE}}(X)) = X$. The user's raw query and data ($X_{prompt}, X_{data}$) are now accessible only inside the TEE.
    \item \textbf{Secure Retrieval:} The TEE executes an information retrieval algorithm (e.g., embedding-based similarity search) over the authorized community datasets ($C^{safe}$ and potentially $C^{open}$) stored securely (e.g., within the TEE's memory or accessed via secure protocols from the encrypted database). This identifies relevant document chunks $R = \{r_1, r_2, ..., r_n\}$. This step occurs entirely within the TEE's protected environment.
    \item \textbf{Prompt Augmentation:} The retrieved documents $R$ are combined with the user's original prompt $X_{prompt}$ (and potentially contextualized by $X_{data}$ if applicable) inside the TEE to form the augmented prompt $P$.
    \item \textbf{LLM Inference:} The augmented prompt $P$ is processed by an LLM to generate a response $Y$. Crucially, this inference step can occur within a H100 GPU with a secure enclave to maintain end-to-end confidentiality and integrity.
    \item \textbf{Response Transmission:} The TEE returns the generated response $Y$ to the user, potentially encrypted or via a secure channel. Optionally, provenance information (e.g., identifiers of documents in $R$ that contributed to the answer) can be included, subject to community policy.
\end{enumerate}

This protocol extends standard RAG approaches~\cite{Lewis2020, Nakano2021n} by integrating cryptographic protection and TEE isolation at critical stages. It ensures user query intent ($X_{prompt}, X_{data}$) and the contents of the sensitive community data corpus ($C^{private}$ / $C^{safe}$) are protected throughout the process, accessible only within the verified confines of the TEE.

\subsection{Confidential Inference with H100 GPU Secure Enclaves}
\label{sec:confidential_inference} 

Recent advances in confidential computing hardware, particularly NVIDIA's H100 GPU featuring secure enclave technology, significantly enhance the feasibility and performance of TEE-based AI systems like Community Transformers. These GPUs enable high-performance LLM inference to occur directly within a hardware-protected TEE.

H100 Confidential Computing establishes hardware-level isolation using TEEs, creating a secure enclave within the GPU where code and data remain encrypted and protected, even from privileged host software like the OS, hypervisor, or cloud administrators. This provides stronger security guarantees than traditional virtualization while enabling near bare-metal performance for demanding AI workloads.

The integrity and authenticity of these GPU enclaves are verifiable through cryptographic attestation. This process uses hardware-rooted keys and cryptographic measurements to allow remote users to confirm:
\begin{itemize}
    \item The enclave is running on genuine NVIDIA hardware.
    \item The TEE is properly initialized and configured.
    \item The specific code (e.g., the LLM inference engine) running inside the enclave matches a known, expected measurement (hash).
\end{itemize}

This attestation process allows users to trust the execution environment without needing to trust the infrastructure provider. Furthermore, it enables transparency: the cryptographic measurement of the code can be compared against publicly available source code or binaries, allowing verification that the intended, audited software is executing within the enclave. This is often integrated into secure software development and deployment pipelines using trusted build systems and transparency logs.

\subsubsection{End-to-End Confidential Inference and RAG}
Integrating TEE-protected databases with the confidential inference capabilities of H100 GPUs enables a truly end-to-end confidential and verifiable RAG system for knowledge management. This enhanced architecture operates as follows:

\begin{enumerate}
    \item \textbf{Secure Embedding Generation:} When community data ($C^{private}$, $C^{open}$) is ingested, the process of generating vector embeddings (required for efficient semantic search) can occur within a TEE, potentially an H100 enclave. The embedding model itself executes inside the enclave on plaintext documents (decrypted after secure ingestion). Hardware attestation verifies the integrity of the specific embedding model code and its parameters. The resulting embeddings are encrypted using strong cryptography (e.g., AES-GCM) with keys managed by the TEE (potentially derived from community-controlled master keys) before being stored outside the enclave in the encrypted vector database.
    \item \textbf{Encrypted Vector Database:} The system utilizes an encrypted vector database. This might employ techniques like envelope encryption (where data-specific keys are encrypted by a master key) and potentially specialized index structures (e.g., encrypted variants of HNSW or IVF-PQ) designed to operate efficiently on encrypted vectors while preserving cryptographic guarantees.
    \item \textbf{Secure Search:} When a user query $X$ is received and decrypted within the CPU TEE (as per the protocol above), the query text ($X_{prompt}$ and $X_{data}$) is securely passed to the H100 TEE. Inside the H100 enclave, the query embedding is generated. The system then performs a similarity search. This involves the H100 TEE selectively retrieving necessary encrypted document vectors from the database, decrypting them \emph{only within its protected memory}, performing the similarity computations (e.g., dot product, cosine similarity) against the query vector, and identifying the top-k relevant document chunks $R$. All sensitive vectors remain encrypted outside the H100 TEE.
    \item \textbf{Confidential LLM Inference:} The retrieved chunks $R$ and the original query prompt $X_{prompt}$ are formulated into the final augmented prompt $P$ within the TEE environment (either CPU or GPU TEE). This prompt $P$ is then processed by the LLM, executing entirely within the H100 secure enclave. The resulting response $Y$ is generated within this protected boundary.
    \item \textbf{Attested Verification:} Throughout this process, the system's integrity is maintained and verifiable via attestation. Users can obtain cryptographic proof establishing (1) execution on genuine H100 hardware with TEE protections active, (2) integrity of the specific embedding models, search algorithms, and LLM inference code via measurements, and (3) correct application of encryption protocols for data leaving TEE boundaries.
\end{enumerate}

This architecture establishes a zero-trust RAG system with hardware-enforced security. It ensures that sensitive community data, user queries, intermediate embeddings, and the LLM computations themselves are protected from observation or tampering, even by privileged administrators or the cloud provider infrastructure. This provides a robust technical foundation for communities to leverage the power of LLMs on their sensitive data while retaining control and ensuring privacy.

%% file: chapter-4/chapter-4.tex
\chapter{Security with agentic AI}\label{chapter:4}
\epigraph{``Man is a tool-using animal. Without tools he is nothing, with tools he is all.''}{Thomas Carlyle}

\noindent
\emph{AI agents are becoming more common online, performing tasks on behalf of users and interacting with digital services. The AI systems not only access and process sensitive information, but also have the ability to take actions on behalf of users. This raises important challenges: how can we verify and audit what these agents do, ensure they follow proper access controls, and distinguish between real people and AI systems? To address these issues, we need stronger authentication frameworks that securely link AI actions to human users, enforce clear permissions, and provide tools like personhood credentials to maintain trust and accountability in digital spaces. This chapter examines this challenge with a focus on the role of authentication and authorization in the context of agents and how we can verify and audit human control.}

In 2023, right in the middle of my PhD, ChatGPT took the world by storm, showing that language models could flexibly follow instructions to perform tasks using some facsimile of cognition. As I graduate my PhD in 2025, agents are the new thing taking the world by storm. 

Agents simply refer to a system, typically underpinned by a language model (more sophisticated than GPT-3.5) that can access `tools' (anything from a simple calculator to accessing the web through a browser) and interact with the world. Where ChatGPT interfaced with a human who ultimately took actions, agents can interface with a much wider action space with far wider impacts and risks. 
This section takes the proliferation of agents as a given, and asks what can be done to prepare the world for this change to enhance security and transparency while balancing privacy. It is the final frontier of the risks outlined in \autoref{chapter:1}---and this chapter was, at the time of publishing, one of the most forward-looking perspectives of identity, authorization, and privacy risks for this coming global change. 

\import{./}{authenticated-delegation.tex}

\import{./}{phc.tex}

%% file: chapter-4/authenticated-delegation.tex
\clearpage
\section{Authenticated delegation for AI agents}

\emph{As the web changes in response to the proliferation of AI agents, we need tools to identify, control, and audit these AI agents at scale. To address this, this section proposes an authenticated delegation framework to enable a new generation of AI agents that are permissioned, auditable, and accountable. This is based on a paper I led and coauthored with Samuele Marro, Thomas Hardjono, Robert Mahari, Cedric Deslandes Whitney, Dazza Greenwood, Alan Chan, and Alex Pentland titled `Authenticated Delegation and Permissioned AI Agents', which subsequently spawned a working group to deploy these ideas in practice.}

\begin{paperabstract}
The rapid deployment of autonomous AI agents creates urgent challenges im the areas of authorization, accountability, and access control in digital spaces. 
New standards are needed to know whom AI agents act on behalf of and guide their use appropriately, protecting online spaces while unlocking the value of task delegation to autonomous agents. 
We introduce a novel framework for authenticated, authorized, and auditable delegation of authority to AI agents, where human users can securely delegate and restrict the permissions and scope of agents while maintaining clear chains of accountability.
This framework builds on existing identification and access management protocols, extending OAuth 2.0 and OpenID Connect with agent-specific credentials and metadata, maintaining compatibility with established authentication and web infrastructure.
Furthermore, we propose a framework for translating flexible, natural-language permissions into auditable access control configurations, enabling robust scoping of AI agent capabilities across diverse interaction modalities.
Together, these practical approaches facilitate the immediate deployment of AI agents while addressing key security and accountability concerns. Our work contributes toward ensuring agentic AI systems perform only appropriate actions, as well as providing a tool for digital service providers to enable AI agent interactions without the risks deriving from scalable interaction. 
\end{paperabstract}

Agentic AI systems, also referred to as AI assistants or simply `agents', are AI systems that can pursue complex goals with limited direct supervision on behalf of a user \cite{Gabriel2024c,chan2024visibility,shavit2023practices,chan2023harms,Kenton2023-cz}, including by interacting with a variety of external digital tools and services \cite{Nakano2021n,Lieberman1997-qm,fourney2024magentic}. For example, AI agents given a prompt to book travel arrangements for a holiday may browse the web for recommendations, search for flights via APIs, or message an airline agent in natural language via chat services to arrange a booking. Such communications could even extend to AI agent negotiations \cite{Abdelnabi2023-al} and other multi-agent contexts. 

While current AI agents have limitations \cite{Raji2022-kj,Wang2023-rt}, lack the ability to perform certain tasks \cite{Liu2023-rd}, and may be susceptible to attacks such as prompt injections \cite{Yao2024-uv,Liu2023-bl,Zhu2023-yf}, there has been rapid progress in their development and commercial interest. 

This has raised many concerns over the risks of AI agents and how they should be governed \cite{shavit2023practices,Gabriel2024c,eu_ai_act}. 
Credentials and verification may become critical in verifying the properties and metadata of AI systems \cite{chan2024visibility},
uniquely identifying humans in online spaces \cite{Borge2017-ti} (or at least distinguishing humans from AI agents \cite{adler2024personhood}), protecting contextual confidence \cite{Jain2023-ij}, mitigating AI-augmented influence operations \cite{Goldstein2023-co}, preventing AI manipulation of humans \cite{Bai2023-vc,Singh2024-dz}, and governing or auditing AI systems more broadly \cite{reuel2024open,South2023-wh}. 
The world needs ways to explicitly delegate authority to agents, transparently identify those agents as AI, and enforce human-centered choices around security and permission for these agents.

We distinguish three key concepts: \textit{authentication} confirms an entity's identity; \textit{authorization} determines the permissible actions and resource accesses that the authenticated identity is allowed to perform, defining the scope and limitations of delegated activity; and \textit{auditability} allows all parties to inspect and verify that claims, credentials, and attributes remain unaltered, supporting trustworthy authentication and authorization decisions.

This work has three key contributions. 
First, \autoref{sec:why} builds upon the existing literature to outline \textbf{why authenticated delegation is important} for AI agents, and what risks it could mitigate. In doing so, we provide an overview of current practices and where they fall short.
Second, \autoref{sec:technical} directly addresses this need by \textbf{extending existing authentication and authorization protocols to enable authenticated delegation} to AI agents, examining the role OpenID Connect and OAuth 2.0 could play in enabling a pragmatic, robust, and extensible implementation.
Third, \autoref{sec:scoping} explores the \textbf{role of agentic access control} and 
outlines a method for \textbf{expressing flexible, natural language permissions for agents} and transforming them into auditable, fine-grained access control rules, that can operate across agent modalities (e.g., web requests, computer use, or language interfaces),
Further, this work provides \textbf{example use cases} of the framework and a \textbf{legal analysis of the implications} of this work in \autoref{sec:legal}.

\begin{figure}[htp]
    \centering
    \includegraphics[width=0.7\linewidth]{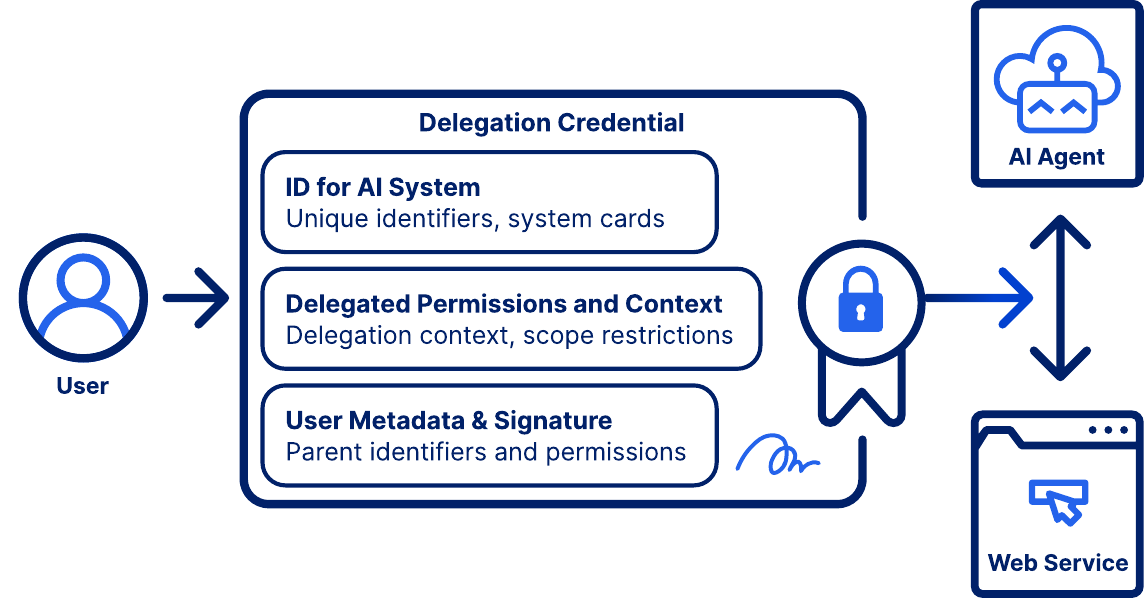}
    \caption{Conceptual overview of a verifiable delegation credential for AI agents. Users issue delegation credentials that include: the AI system's unique identity and properties, delegated permissions with contextual scope restrictions, user metadata, and cryptographic signatures for verifiability. These credentials enable secure, trustworthy interactions between AI agents and third-party services, ensuring traceability and appropriate delegation of authority. }
\end{figure}

\subsection{Why authenticated delegation is important}\label{sec:why}
Authenticated delegation is the process of instructing an AI system to perform a task that requires access to tools, the web, or computer environments in such a way that third parties can verify that (a) the interacting entity is an AI agent, (b) that the AI agent is acting on behalf of a specific human user, and (c) that the AI agent has been granted the necessary permissions to perform specific actions. 

Verifying the properties of interacting entities will be relevant whenever a context exists where an AI agent \emph{could} act on behalf of a human user, and especially where the agent is capable of taking consequential actions. This remains true whether the AI system is run locally or provided by an AI vendor---as harm can occur in both---and must be able to operate across various digital contexts and with AI models of heterogeneous capabilities.

At a high level, authenticated delegation involves a human user creating a digital authorization that a specific AI agent can use to access a digital service (or interact with another AI agent) on behalf of the user, which can be verified by the corresponding service or agent for its authenticity. Such authorization can include additional information, such as unique identifiers for the agent instance, permissions on what the agent is allowed to do, and other information (e.g., the capabilities and failure modes of the agent or information about the human user).
The authorization must be uniquely and cryptographically linked to the digital identity of the human delegator who granted the authorization. 
This could be done by linking to email accounts (as is commonplace for application accounts), linking to a more robust digital identity, or via domain-specific identifiers (such as user accounts within an organizational setting).

In practice, this needn't be substantially different from existing authentication and authorization mechanisms used today, such as how a calendar application is authorized to access a user's calendar data and scan it for upcoming events. However, AI agents' autonomous and highly capable nature means more care is needed in how we manage delegation. As such, let us examine the use cases for authenticated delegation in more detail.


\subsubsection{Functions of authenticated delegation}
Authenticated delegation opens avenues for AI agents to accelerate complex tasks, automate workflows, and seamlessly interface with digital services on behalf of human users. However, granting such agency also entails risks around scope misalignment, resource abuse, or a breakdown of clear accountability. This subsubsection delineates how robust identity verification, explicit scoping, and mutual authentication can unleash practical use cases—ranging from streamlined enterprise processes to safe, multi-agent coordination—while mitigating key vulnerabilities. By highlighting both the opportunities and the potential pitfalls, we underscore why adopting secure, verifiable delegation mechanisms is vital to responsibly harness AI agents.

\paragraph{Current challenges in delegating authority to AI agents}
As the capabilities of LLMs improve, there is a growing interest in making them more autonomous and general-purpose. A key aspect of this is the ability to use tools or access external services. For simple tasks such as asking an agent to search the web for information, write and execute code, or generate an image, this is straightforward and does not require additional authorization or individual-specific security mechanisms. However, to unlock use cases such as interacting with personal or organizational accounts, accessing sensitive personal information, or interacting with consequential infrastructure, more robust delegation frameworks are needed.

\textbf{Example:} Consider the above example of an AI agent booking a holiday. Having an agent search the web for information may not need any authorization, but how could that agent access a user's calendar or make a purchase? 
For calendars, users are used to the expected flow of granting access to applications to access their calendar data. This would be no different for an AI agent (and would be naively supported in the solution outlined in \autoref{sec:technical})--indeed, limited OAuth~2.0 support is enabled in some agent tools such as OpenAI GPT actions. 
Now consider a flight purchase. You \emph{could} provide your credit card details in the context window for the agent and prompt it to follow at budget, but this introduces a variety of security concerns and relies upon the underlying reliability of the AI system to not take unexpected actions or be vulnerable to attacks or jailbreaks. 
Instead, an AI agent should be authenticated and authorized to make a purchase on specific booking services, where credit cards are stored securely, and where explicit spending limits can be enforced.

\paragraph{Communicating limitations and restricting scope}
Current approaches to limiting the scope of AI agents are limited and one-sided. A user can provide a strong prompt to an agent to limit its actions, but this comes with a variety of failure modes \citep{Liu2023-bl}. Access to tools or websites can be blocked, but this is limited in the granularity of control. An AI system deployer could implement further controls, such as monitoring and blocking specific actions or website subdomains when agentic functionality occurs, but doesn't communicate these limitations to the service the agent is interacting with. 
By explicitly limiting the scope of an AI agent and communicating these limitations to the service the agent is interacting with, we can enable a more robust interaction between AI agents and services. A more detailed examination of how this could be designed across web, API, and natural language access modalities is available in \autoref{sec:scoping}.

\textbf{Example:} An AI agent is used by a physician to provide diagnostic recommendations in a telemedicine portal, logging in with basic credentials that do not specify its limitations. The portal assumes full physician capabilities, granting the agent access to all patient records, including a video assessment with a voice recording from a specialist. The agent, being text-only and unable to process video, generates a diagnosis based solely on the text data, appending a standard caveat—``not all available information was used''—which is overlooked. Trusting the incomplete recommendation, the physician risks making a misinformed treatment decision. If the agent's limitations were explicitly communicated via authenticated delegation, the portal could have flagged the need for a human review of the multimedia content, avoiding a potentially harmful oversight.

\paragraph{Verification in multi-agent communication}
When AI agents communicate to collaborate on tasks or facilitate interactions, ensuring mutual authentication becomes paramount. 
Securing communication channels is not enough; agents must also verify that they authentically represent the users or organizations they claim to represent. Mutual authentication ensures that agents can trust each other's intentions, capabilities, and authority, preventing impersonation, unauthorized actions, and potential misuse. This verification is essential for fostering reliable, safe, and accountable multi-agent ecosystems.

\textbf{Example:} Two AI agents—one representing a hospital and the other an insurance company—collaborate to process a patient's claim. The hospital agent submits treatment details, while the insurance agent verifies coverage. Without mutual authentication, a third-party malicious agent could impersonate the hospital, submitting fraudulent claims, or the insurance agent could reject valid claims out of concern over authenticity. 

\paragraph{Protecting human spaces online}
As AI agents grow increasingly adept at mimicking human behavior---crafting text, creating personas, and even replicating nuanced human interactions---it becomes harder to maintain digital environments genuinely inhabited by real people. This challenge drives the need for safe, human-only online spaces where authenticity is preserved and scalable manipulation is curbed by verifying human personhood \cite{adler2024personhood}. However, many AI agents act as useful proxies, assistants, or representatives for human users who cannot, or prefer not to, engage directly. Authenticated delegation enables these spaces to be selectively accessible to AI agents, while still ensuring that the AI agents are linked to verified human principals.

\begin{figure}[h]
    \centering
    \includegraphics[width=3in]{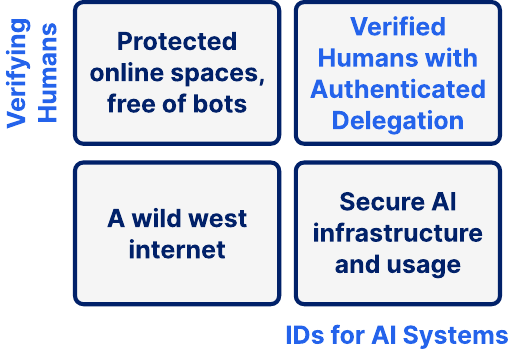}
    \caption{Authenticated delegation can benefit from user identification or verification of personhood (e.g., through personhood credentials). By combining verified human identity with authenticated delegation, we can support safer online spaces for human interaction while enabling the trustworthy and controlled use of AI agents.}
\end{figure}

\textbf{Example:} The Australian government's recent social media ban demonstrates how governments can restrict access to online spaces by requiring users to prove their age, often through methods like a government-issued ID or a face scan. While these measures aim to limit underage access, they may also inadvertently block AI agents from accessing these platforms. Instead of a blanket restriction, platforms could explicitly allow AI agents to access their services in controlled ways by leveraging authenticated delegation. This approach would ensure that AI agents act transparently on behalf of verified human users. For instance, an agent could access a user's social media account to retrieve information about friends and help draft an email, all while maintaining compliance with platform policies and ensuring accountability.

\paragraph{Supporting contextual integrity}
Contextual integrity addresses adherence to context-specific norms and privacy, which include actors (who is involved in the information flow), attributes (what information is shared), transmission principles (under what conditions information is shared), and social context (the broader cultural, institutional, or situational environment shaping these norms) \cite{ghalebikesabie2024,zhan2022,fan2024,nissenbaum2004privacy}. 
Contextual integrity offers a perspective for reasoning about how AI agents can act in ways that are contextually appropriate, transparent, and aligned with societal norms and the expectations of their human delegators \cite{bagdasarian2024,ghalebikesabie2024,bloom2022}. 
This includes exploring which decisions can reasonably be made autonomously by the AI and under what conditions human oversight or intervention might be necessary (e.g., when is human-in-the-loop required and who is responsible). 

\textbf{Example:} 
An AI assistant with authenticated delegation can be issued distinct credentials for separate contexts  (e.g., an enterprise-context assistant and a personal one).
Each credential encodes the agent's information, the delegating user, and context-specific permissions. 
By enforcing these scoped credentials, services can ensure that the assistant adheres to contextual integrity and rejects actions that cross boundaries, such as using information from work documents to complete personal forms. 
This separation of roles and explicit permission-sharing protects privacy, ensures accountability, and safeguards human oversight for cross-context decisions.

\subsubsection{Background}
Authenticated delegation can address various challenges, from traceability of AI outcomes to limitations on what spaces can be accessed and actions taken by AI systems. The overarching aim of identification and credentialing systems is to facilitate secure online environments and authenticated access to services. To this end, a variety of existing protocols and standards have been developed, tailored to both human users and AI systems, to uphold these goals in different contexts.

\paragraph{Comparisons to other AI identifiers}
To \emph{verify human identity online}, a large body of work exists ranging from simple authentication such as {OAuth~2.0}~\cite{RFC6749-Formatted} to more complex digital identity frameworks as W3C's Verifiable Credentials \citep{noauthor_verifiable_2024}, decentralized identifiers \citep{sporny_decentralized_2024}, and the European Union Digital Identity's privacy-preserving digital wallets \citep{eudi_arf_2024}. 
To \emph{privately prove personhood}, a number of systems have been developed to distinguish human users from bots, including proof-of-personhood systems designed to counter automated Sybil attacks \citep{borge_proof_personhood_2017}, simple turing tests such as CAPTCHAs \citep{von2003captcha}, and more robust credentials \citep{adler2024personhood}. 
More generally, the goal of `know-your-customer' for users and granular access permissions (identity and access management, IAM) are commonplace on the internet.

Similarly, many websites seek to \emph{broadly limit access to bots on their services}, and may do so through the use of robots.txt bans. This is important since the widespread presence of bots or unauthenticated AI agents can lead to abuse and harm, but is often done at the `user-agent' level (for example, banning all `GPTBot' user agents~\citep{longpre2024consent}).

To \emph{track and verify the output of AI systems}, watermarking techniques \citep{liu_survey_2024, wang_data_2021} and content provenance measures \citep{c2pa_c2pa_2023} have emerged as potential solutions for determining the origin of AI-generated content. However, these approaches face reliability challenges \citep{saberi_robustness_2024} and are insufficient for establishing comprehensive accountability or safety when using AI agents. The inherent limitations of current verification methods highlight the need for more robust frameworks that can track not just content creation but also the broader implications of AI system deployment and interaction.

For \emph{managing access to sensitive AI capabilities} themselves, researchers have proposed `know-your-customer' schemes for compute providers \citep{egan_oversight_2023, obrien_deployment_2023}, while commercial platforms implement API tokens and access controls \citep{openai_chatgpt_2023}. These developments reflect a growing recognition that AI systems need robust mechanisms to prove their authenticity and permissions when accessing external services \citep{buterin_what_2023}, particularly as they become more integrated into critical infrastructure and decision-making processes.

To \emph{identify specific instances of AI agents}, recent work has proposed identifiers and verification approaches discussed above \citep{chan2024ids,chan2024visibility}. This is important and critical work, which we build upon to extend to \emph{authenticated delegation of AI agents} using existing authentication and permission protocols to enable AI agents to act on behalf of users in a controlled manner. In turn, these identifiers and delegation mechanisms can help create spaces that do not just gatekeep to human users but also enable AI agents to act on behalf of users with auditability and accountability.

\paragraph{Comparisons to Model Context Protocol and GPT Actions}
One example of an AI-centric protocol is the recent Model Context Protocol (MCP) \cite{anthropic2024mcp} from Anthropic, which enables secure, structured interactions between AI systems and external tools or data sources. MCP aims to enhance the contextual relevance of AI outputs by establishing a standardized framework for connecting models to resources to facilitate applications like retrieving live data, interacting with APIs, and executing tasks in real time.

While an extremely useful standard, it's limited in its full scope towards authorized delegation, enabling only system communication and optionally access controls rather than broader authentication and identity management.

Similarly, OpenAI's GPT Actions are integrations allowing GPT models to perform specific actions like booking a flight or retrieving data from APIs, which is a more constrained version of MCP and shares in its shortcomings. LangChain's Agent Protocol / LangGraph Platform extends this idea to enable multi-agent interoperability.


\paragraph{Documentation, safety, and governance of agentic AI systems}
Documenting AI systems and the data that create them has been a critical area of research and practice. Early frameworks established foundational approaches including datasheets \citep{gebru_datasheets_2021}, model cards \citep{mitchell_model_2019}, and data statements \citep{bender_data_2018}, with popular implementations emerging  \citep{paullada_data_2021}. Although each of these approaches has proven valuable, they face challenges in adequately addressing concerns around bias \citep{buolamwini_gender_2018}, privacy, and copyright. Recent work has highlighted the need for documentation of AI agents to understand their capabilities and limitations \citep{chan2024ids}, moving beyond static system descriptions to capture dynamic behaviors and interaction patterns. As AI systems become increasingly agentic, new frameworks are needed to document their evolving capabilities, decision-making processes, and potential risks \citep{bommasani_opportunities_2022}.

Recent work has explored runtimes for validating and reversing agent actions \citep{patil_goex_2024} and protocols for structured communication between language models \citep{marro_protocol_2024}. Researchers are also evaluating frontier models specifically for capabilities that could enable deceptive behavior \citep{phuong_evaluating_2024, fang_agents_2024}, while others advocate for tracking prior incidents \citep{wei_designing_2024} and establishing broader safeguards for AI agent interactions \citep{shavit2023practices}. Governance of AI agents is a rapidly evolving area of research and practice \citep{reuel2024open, kolt_governing_2024}, with increasing attention being paid to the development of frameworks that can ensure responsible deployment and operation of autonomous systems.

\paragraph{How authenticated delegation combines these solutions}
This work combines and extends these existing approaches---AI agent IDs and credentials, proof-of-personhood and identity verification for human users, and content provenance and watermarking methods---to form a cohesive framework. This approach inherits well-established practices for identity management while introducing explicit scoping and metadata for AI agents. This integration allows for granular, enforceable permission sets, clearer accountability chains, and richer context signals (like a model's certifications or limitations) to be attached to each delegated action, with a more robustly verifiable construction than a simple agent ID system card. In effect, authenticated delegation complements existing standards and enhances their reliability by anchoring the actions of AI agents to verifiable human principals and recognized AI-specific credentials, creating a unified foundation for safe and accountable AI interactions. To this end, \autoref{sec:technical} introduces a concrete framework with additional security guarantees to package these elements together in a robustly verifiable way.

\subsection{Extending OpenID Connect for identifying and authenticating AI agents}\label{sec:technical}

To support the motivation of \autoref{sec:why}, this subsection proposes a concrete technical framework building on existing internet-scale authentication protocols to introduce mechanisms for delegating authority from users to AI agents and describes a token-based authentication framework that leverages OpenID Connect and OAuth 2.0. Our approach extends these battle-tested protocols to address the unique challenges of AI agent authentication while maintaining compatibility with existing internet infrastructure.


\subsubsection{OAuth2.0 and OpenID-Connect}
While new frameworks for AI system identification are emerging, there are valuable lessons to be learned from existing internet-scale authorization and authentication protocols. 
In particular, the {OAuth~2.0} protocol~\cite{RFC6749-Formatted} and its extensions provide battle-tested patterns for delegated authorization and identity verification that could inform the development of AI agent credential systems.

{OAuth~2.0} emerged from the need for users to provide authorization to one service to access resources located in another service, based on the RESTful paradigm~\cite{Fielding-2000-Thesis}.
A key requirement underlying {OAuth~2.0} is the ability for access to be continually granted even if later the user is absent (e.g., offline).
Existing user authentication protocols (e.g., MIT~Kerberos~\cite{RFC4120-Formatted}, CHAP~\cite{RFC1994-Formatted}) were developed primarily for the interaction between a human user utilizing a host computer connecting to the authentication server over the UDP layer.
The advent of the RESTful APIs meant that the parameters and flows had to be communicated over the HTTP layer, with the TLS providing the underlying message confidentiality layer.

A typical example of the scenario addressed by {OAuth~2.0} is the user who wishes to allow  an online calendaring service to read the user's itinerary from an airline service.
Here, the service that seeks access to the resource is referred to  as the {OAuth~2.0} {\em Client}.
On the other side, the service that is managing the resource is referred to as the {\em Resource Server} (RS). It is important to note that one of the assumptions underlying {OAuth~2.0} is the fact that Client and the RS can be operated by third-party entities.

The wide deployment and popularity of the {OAuth~2.0} protocol  enabled new features and extensions to be added. 
One successful extension---namely the {\em OpenID-Connect} protocol (OIDC)~\cite{OIDC1.0}---is the addition of flows dealing with the user authentication. The service dealing with authentication is referred to as the {\em OpenID Provider} (OP).
A key addition introduced by OpenID-Connect is the {\em ID-token}, which carries information about the human user that can be retrieved from the OP (i.e., by presenting ID-token). Here a merchant (as the Relying Party) would input the ID-token to the relevant token-validation endpoint at the OP in order to obtain more information about the user. We believe this capability may be extended to address the case of AI agents.

Another extension of the {OAuth~2.0} protocol that enables a user to manage multiple resources distributed across many Resource Servers is the {\em User-Managed Access} (UMA) protocol~\cite{UMACORE1.0}.
The UMA model may fit use-cases where the human user possesses multiple AI~Agents and where a single point of policy or rule configuration is desirable~\cite{Hardjono2019-IEEECommsMagazine}. Here, the AI~Agents can be viewed as distributed resource servers owned by the user.
Using the UMA~Authorization Server, the user can set policy at one location and have these policies automatically propagated to the multiplicity of AI~Agents.

\subsubsection{Delegation of authority from the user to the AI agent}

Given that the {OAuth~2.0} protocol is an authorization protocol, it is worthwhile considering reusing the {OAuth~2.0} patterns to establish a new mechanism for the human user to {\em delegate} certain tasks to the AI Agent. 
In other words, the human user is authorizing the AI Agent to carry out certain tasks that are limited in scope on behalf of the user.

In this new proposed extension, the human user must first perform authentication to the OpenID Provider (OP) to demonstrate their identity. The user then `registers' the AI Agent to the OP so that external entities who later seek to obtain further information about the AI Agent can do so to the OP. Registration could be done automatically in the background when an agent is created through a vendor (such as creating a new assistant instance with OpenAI).  

Existing {OAuth~2.0} client registration protocols can be customized to enable the user to register the AI Agent to the OpenID Provider and designate the AI Agent as a delegate or surrogate of the human user.

Next, the human user can issue a new {\em delegation token} that authorizes the AI~Agent to carry out tasks on behalf of the user.
Here, the term `authorize' is utilized to explicitly call out the fact that the AI~Agent is owned (driven) by a human delegator.

Both the user ID-token and the AI Agent delegation token can be referenced from within (or even copied into) a W3C Verified Credentials (VC) data structure~\cite{Sporny2022}.
This enables the AI Agent to wield the VC in its interactions with other entities (e.g.,  other services or other AI Agents), and have the benefit that both tokens would be verifiable at the standard OP.

It is worth noting that these delegation and authentication exchanges could alternatively be implemented using W3C VC issuance and delegation mechanisms. In such a scenario, a W3C VC could generate an OpenID-compatible credential, enabling seamless interfacing with OpenID systems. While this integration highlights the interoperability between W3C VC and OpenID ecosystems, further exploration and formalization of this process are left as future work and are beyond the scope of this section.

\begin{figure}[ht]
    \centering
    \begin{tikzpicture}[
        scale=0.7,
        box/.style={draw, minimum width=1.8cm, minimum height=0.5cm},
        user/.style={},
        circle_num/.style={circle, radius=0.2cm, draw, fill=yellow!20, font=\tiny},
        arrow/.style={->, >=latex}
    ]
    
    \node[box, minimum width=2cm, minimum height=2cm] (client) at (0,0) {Client};
    \node[left=0.1cm of client, align=center] {Human\\User};
    
    \node[box, minimum width=2cm, minimum height=1cm] (op) at (5,0.7) {OP};
    \node[box, minimum width=2cm, minimum height=1cm, align=center,  below=0 of op] (oauth){OAuth2.0\\(UMA)};
    
    \node[box] (a1) at (0,-3) {AI Agent$_1$};
    \node[box] (a2) at (3,-3) {AI Agent$_2$};
    \node at (5,-3) {...};
    \node[box] (an) at (7,-3) {AI Agent$_n$};
    
    \node[above=0.1cm of op] {(AS)};
    \node[below=0.1cm of a1] {(RS1)};
    \node[below=0.1cm of a2] {(RS2)};
    \node[below=0.1cm of an] {(RSn)};
    
    \draw[arrow] ($(client.north east)-(0,0.2)$) -- node[circle_num, midway] {1} ($(op.north west)-(0,0.2)$);
    \draw[arrow] ($(op.north west)-(0,1.3)$) -- node[circle_num, midway] {2} ($(client.north east)-(0,1.3)$);
    \draw[arrow] ($(client.south east)+(0,0.75)$) -- node[circle_num, midway] {3} (oauth);
    \draw[arrow] (oauth.south) -- node[circle_num, midway] {4} (a1.north);
    \draw[arrow] (oauth.south) -- (a2.north);
    \draw[arrow] (oauth.south) -- (an.north);
    
    \end{tikzpicture}
    \caption{Integration of OpenID Connect (OIDC) and User-Managed Access (UMA) protocols for establishing delegated authority from human users to AI Agents. The diagram illustrates the authentication flow where a human user first authenticates to an OpenID Provider (OP) (1 \& 2), registers their AI Agent (3), and issues a delegation token (4). This token empowers the AI Agent to perform authorized tasks on behalf of the user. The verification of both the user's ID token and the AI Agent's delegation token can be performed through the standard OpenID Provider, leveraging existing OAuth 2.0 patterns while incorporating new delegation mechanisms for AI Agent authorization.}
    \label{fig:OIDC-AI}
\end{figure}
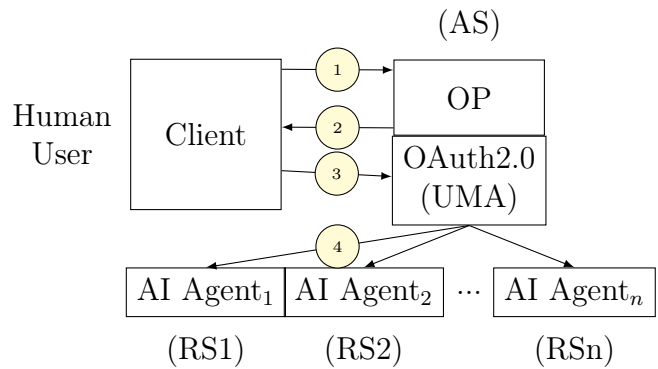

\subsubsection{Token-based authentication framework}

Extending the existing OIDC framework, we can provide all relevant AI agent attributes and metadata of delegation in a set of identity-related tokens.

\begin{itemize}[topsep=1pt] 
\item	{\em User's ID-token}: This is the existing ID-token data structure that is issued/signed by the OpenID Provider (OP) service.
It is intended to represent information regarding the human user, and is no different to those used in everyday login experiences. 

\item	{\em Agent-ID token}: This carries the relevant information about that AI agent issued as an {OAuth2.0} Native Client (meaning the owner of the AI~Agent controls all keying material and secret parameters) and allows the corresponding service to verify any claims about the AI agent and its information. This token can include a range of additional information, from a unique identifier for the agent to a richer and more detailed agent ID token containing system documentation, capabilities or limitation metadata, relationship attributes to other AI systems, or other system characteristics. See \citet{chan2024ids} for further discussion of what an agent ID could entail.

\item {\em Delegation Token}: This newly introduced token explicitly authorizes an AI agent to act on the user's behalf. 
The delegation token is issued and signed by the human delegator and carries references to (e.g., hash of) the corresponding user's ID token and the agent's Agent-ID token, allowing it to be verified by any service that trusts the OP.
Further, any relevant information about the nature of the delegation can be shared. For example, sharing the summarized goal of the agent and its scope limitations could assist a third party in guiding the AI agent to useful endpoints and interaction paradigms. The delegation token should specify validity conditions, such as expiration time or revocation endpoints, and be digitally signed by the user to prevent forgeries and ensure that the user knowingly granted the AI agent the listed privileges. 
In addition, the token may carry supplemental metadata—for example, logging or audit URLs—allowing service providers to record interactions, monitor delegated actions, and respond appropriately to anomalies.
By verifying that the delegation token references a valid user ID-token and a properly issued agent ID-token, remote services can confirm the authenticity and scope of the AI agent's authority before granting access.
\end{itemize}

\subsubsection{Scope Limitations on Delegation}
The delegation framework enables human users to optionally define explicit boundaries for their AI Agents' actions by encoding scope limitations in the delegation token. However, given the flexible nature of agents and their diverse action space, scoping presents a unique and interesting challenge.

\subsubsection{Using verifiable credentials as an alternative} The W3C Verifiable Credentials (VC) standard \cite{Sporny2022} offers a versatile alternative—and sometimes complement—to existing OpenID Connect (OIDC) flows for conveying identity and delegation data. Under a VC-based approach, an issuer (such as an organization or individual) can sign a credential that attests to various claims about a subject, which might be a user, an AI agent, or any other entity needing verifiable, tamper-evident attributes. Because VCs are not bound to a particular transport protocol, they can be presented and verified in a decentralized or peer-to-peer manner without always relying on a single identity provider. This contrasts with OIDC, which generally depends on a central OpenID Provider (OP) to mint and validate tokens.

A key benefit of VCs is their privacy-enhancing potential. Rather than disclosing all attributes or relying on a single identity provider, users, and AI agents can share only the subset of claims strictly necessary for a given interaction. This ``selective disclosure'' capability can mitigate concerns around centralized logging or cross-platform correlation inherent in OIDC-based architectures, especially when interactions span multiple domains or organizations.

Nonetheless, replacing OIDC entirely with a purely VC-based model does come with trade-offs. OIDC already enjoys a robust ecosystem of libraries and deployments that provide well-tested support for issues like token refresh, revocation, and audience restriction. VCs, while powerful, require additional work to replicate these flows at scale—particularly if each verification call demands a new signature check or interaction with a blockchain or distributed ledger. In many enterprise environments, stakeholders may prefer to incorporate VCs into existing SSO or multi-factor authentication frameworks, rather than adopt a fully decentralized identity infrastructure upfront.

In practice, hybrid solutions often prove the most pragmatic. A user or AI agent could store and manage VCs encoding rich attributes or regulatory endorsements, while still leveraging OIDC tokens to bootstrap compatibility with existing authentication or authorization endpoints. For instance, an Agent-ID token could embed a VC carrying detailed metadata on its behavioral, property, context, and relationship attributes. Service providers integrating with OIDC get the familiar token-based handshake, while still retaining the option to parse the embedded VC for an additional layer of trust and context. Examples such as OID4VC support this~\cite{yasuda2022openid}. 

\subsection{Defining scope and permissions for AI agents}\label{sec:scoping}

Authenticated delegation is inherently tied to robust scoping mechanisms, as users must be able to specify their permissions and instructions in a clear and unambiguous manner. This comes in direct conflict with the extremely large possible action space AI agents can perform. 

While much work in reliability and alignment focuses on ensuring that AI agents follow instructions correctly, the risks of misinstruction, prompt injection attacks, and reduced security auditability make pure natural language prompts an incomplete scoping, permission, and security tool. \textbf{This subsection addresses how AI agent infrastructure can bridge the gap between these natural language instructions and robust concrete access control mechanisms} by proposing converting flexible natural language scoping instructions into machine-readable, version-controllable, and auditable structured permission languages that can be leveraged for use in human-in-the-loop settings.

We distinguish between \textbf{task scoping} and \textbf{resource scoping}:
\begin{itemize}[noitemsep, topsep=0pt] 
    \item Task scoping involves specifying which actions or workflows an agent is authorized to perform on behalf of the user. These actions may range from high-level tasks (e.g., ``draft a financial report”) to more granular actions (e.g., ``create a new database entry”);
    \item Resource scoping involves specifying which resources (information, APIs, tools, etc.) an agent can use or modify.
\end{itemize}
While conceptually distinct, task scoping and resource scoping are closely connected. Limiting which tasks can be performed also means that a (well-designed) agent will not access unnecessary resources; similarly, restricting access to specific resources also constrains what tasks are feasible in the first place.

This subsection addresses how access control mechanisms can be integrated with complex AI agents and natural language workflows. It outlines the critical nature of structured permissions, how they can provide a robust and generalizable foundation for agent scoping, and how natural language and human oversight can be flexible interfaces for these access controls. 

\subsubsection{Structured permission languages}

A large class of scoping mechanisms relies on structured, machine-readable policy specifications. These specifications unambiguously define which entities have which authorizations, under which conditions, and with what privileges. Several well-known languages and frameworks exist for encoding permissions, such as XACML (eXtensible Access Control Markup Language), which uses XML to encode and evaluate access control policies \citep{xacml}, and ODRL (Open Digital Rights Language), designed for expressing usage permissions over digital content \citep{odrl}.
Other languages include OBAC \citep{obac}, ROWLBAC \citep{rowlbac}, KaOS \citep{kaos} and Multi-OrBAC \citet{multiOrbac}, which rely on ontologies (typically described using OWL) to model resources, subjects, and authorizations. In web-based contexts, this can often be as simple as whitelisting or blacklisting URLs and subdomains that an agent can access.

These structured languages are machine-readable and can thus be enforced reliably by traditional (non-AI) systems.
From a practical perspective, they are well-suited for resource scoping, since resources are typically discrete and can be classified, enumerated, and grouped into security domains. For instance, when a policy states that a certain directory is read-only for a particular agent, enforcing compliance is straightforward and can be implemented at the system level.

However, they have three main drawbacks. 
First, while these frameworks are suitable for enumerating resources, they are less flexible for task scoping, especially when tasks are open-ended or cannot be easily described as a set of operations. 
Second, policy definitions can become lengthy and complex, especially in environments with a large number of resources and tasks, or in web contexts where the number of possible web interactions is enormous.  
Third, they are often environment-specific and require updating for different digital systems with which an agent interacts. 

Despite these drawbacks, structured permission languages remain a cornerstone of access control because they provide a precise, easily auditable basis for resource scoping. An alternative approach involves using \emph{schema validation} to constrain how agents interact with the environment.

\subsubsection{Authentication flows}


Another dimension of controlling agent behavior is the \textbf{authentication flow} (i.e., deciding when to prompt a user or another authority for confirmation before the agent proceeds with an action). Rather than frontloading all access decisions into a single policy definition, an authentication flow can dynamically request user approval for borderline or high-risk operations.

The main advantage of this approach is that users do not need to define every edge case in a static policy. Additionally, authentication flows can be combined with other scoping mechanisms: for example, a policy can state that any resource that is neither explicitly approved nor explicitly forbidden requires human approval.

On the other hand, frequent authorization prompts can negatively affect the user experience, leading to ``prompt fatigue''~\cite{securitypromptfatigue}, where the user simply grants permissions without a proper review. Moreover, determining when a request requires explicit authorization can be non-trivial, and misclassifications can lead to either excessive prompting or critical operations slipping through unnoticed.

In practice, a well-designed system can combine robust, structured policy definitions (for common scenarios) with dynamic authentication flows for rare or particularly sensitive actions. This approach allows users to offload the majority of routine checks to automated policies while still preserving the ability to escalate novel or ambiguous requests for user confirmation.

\subsubsection{Natural Language Mechanisms}

Alongside fine-tuning, prompting has often been employed to steer the behavior of a model towards safety \citep{zheng2024prompt}. A reasonable extension of this approach would be to train (or prompt) the LLM to interpret permissions described in plain language.
For instance, a user might say, ``You are allowed to generate summaries of public documents, but you must not reveal any confidential metrics.'' Such instructions can, in principle, be parsed and acted upon by an LLM-based system.

The main strength of this paradigm is its user-friendliness. Non-technical users may find expressing policies in natural language much easier than writing formal rules. Moreover, natural language can capture nuanced or context-dependent instructions that are difficult to encode in structured languages. This makes them ideal for both task and resource scoping.
Finally, natural language can be used to enforce policies on actions that require reading or using natural language, such as interactions with other LLM-based agents.

However, natural language often lacks the precision needed for reliable policy enforcement. For instance, terms like ``sensitive data'' or ``private emails'' may be interpreted differently depending on context. This problem is particularly relevant in the case of conflict between different policies, where ambiguous and context-dependent instructions may yield different interpretations.
Relying solely on an LLM to interpret and enforce ambiguous natural language instructions can be risky in security-sensitive contexts.

In short, while natural language instructions can serve as a convenient mechanism (especially for task scoping, where other mechanisms are less suitable), they are \textbf{not} reliable enough to be used as standalone policy mechanisms.

\subsubsection{Combining structured permissions, natural language, and user oversight}

\paragraph{Resource scoping as a foundation.} We argue that the most broadly applicable strategy is to enforce \emph{resource scoping with structured permissions}. The brittleness of natural language mechanisms makes them unsuitable for production-level usage of AI agents, especially when security or compliance is a concern. In contrast, structured permissions are unambiguous and deterministic, providing verifiable guarantees against unauthorized access. Focusing on resource scoping also significantly reduces the overhead of specifying every authorized task in detail. To an extent, agents could attempt to represent task-scoping instructions in the form of resource scoping, using domain knowledge of the contexts in which they operate.  
Since resources are generally discrete and can be classified, enumerated, and grouped into domains, controlling resource access implicitly prevents many potential tasks that would require out-of-scope resources.
Additionally, structured resource scoping has several advantages:
\begin{itemize}[noitemsep, topsep=0pt] 
    \item It does not depend on how a user delegates tasks---be it via a script, an AI agent, or a more traditional workflow;
    \item It is more compatible with existing non-AI access control systems, which focus on machine-readable permissions for resources (e.g., databases or URLs);
    \item It is suitable for structured logging and version control, which simplifies auditing and compliance reporting.
\end{itemize}

Though users may supplement resource scoping task constraints written in natural language, the core resource-based policies provide a safety net that is largely immune to ambiguities in language or model vulnerabilities. Even if an LLM or another AI agent is tricked or misaligned, its ability to execute harmful actions is constrained by the underlying resource permissions.

\paragraph{Connecting to natural language.} While robust and auditable, structured resource scoping alone lacks ease of use and flexibility. To address this, the instructions for the LLM (or a separate scoping prompt) can flexibly express the scoping limitations that should be applied. These natural language scopes can be converted to a structured scoping format by the agent or an AI system in the corresponding environment (which has more detailed knowledge of the relevant resource profiles). Examples of conversion between natural language and structured permissions include \citet{subramaniam2024intent}, which generates PostgreSQL restrictions, and \citet{jayasundara2024ragent}, which uses retrieval to generate custom JSON policies.

A similar process could also be performed for different environments and digital services an agent interacts with, allowing a flexible set of permission instructions to be applied across a wide range of services and contexts (which is important given the broad action space of AI agents).

\paragraph{Bringing a human in the loop.} The key final step is validating these structured access controls via the human delegator.  
Authorization workflows present an opportunity for users to briefly review and approve structured access control limitations for different systems. For instance, in \citet{wright2024here} LLM agents agree on structured information (in this case, meeting dates) which are then confirmed by human users.

\paragraph{Combining into a hybrid implementation.} 
Bringing these elements together into an implementation is relatively straightforward. An LLM assists in converting high-level, natural language resource constraints into formal, structured rules that users can subsequently review and approve.
\textbf{For example:}
\begin{enumerate}[noitemsep, topsep=0pt]
    \item A user writes: ``Allow the agent to read and write to the directories about `projectAlpha`, but do not grant it access to the folders with financial folders;”
    \item The LLM translates this requirement into a policy definition, either in a universal permission language (e.g., XACML) or in the specific permission language used by the resource (e.g., SQL access policies for databases). In this specific case, the LLM enumerates ``projectAlpha” resources while explicitly denying access to ``financials2023;''
    \item The user reviews, corrects if necessary, and finalizes the policy.
\end{enumerate}
While many specific details of such a workflow need to be address such as intermediate validation checks and the evaluation of robustness of LLM translation into structured languages, we leave these specifics to future work.

Ultimately, focusing on structured, unambiguous resource constraints is the most reliable way to ensure that an AI agent remains within authorized bounds in a given environment. While there is still room for higher-level (often natural language) task constraints, one should treat these as guidance towards the primary enforcement mechanism. Indeed, while natural language can adequately address the extremely large possible space of agent actions, its transformation into access controls grounds the limitations on agent actions into finite auditable controls. 
Structured resource scoping reduces the reliance on model alignment alone, decreases the risk of adversarial prompt injections, and simplifies the integration with well-established security mechanisms. 
Combining this approach with well-designed authentication flows and helping the user interpret the generated policies can reduce the chances of human errors, enhance accountability, and improve the robustness of authenticated delegation. 

\paragraph{Inter-agent scoping.}
Extending beyond the user-agent-service model, this approach can apply to multi-agent settings where agents want to propagate their limitations onto other agents performing actions on their behalf.
Suppose that the user specifies the authorizations of an agent Alice. When Alice interacts with another agent, Bob, in natural language to perform a task, Bob can parse Alice's scoping instructions and interpret them in its own environment. 
By doing so, Bob can confirm that its assigned operations remain within the original scope, and provide an auditable receipt of the actions taken and the resources accessed. This is particularly useful in scenarios where inter-agent communication spans different organizations, each with separate policies and resource constraints.

For a concrete example, suppose that Alice is a project management agent and Bob is an accounting agent. The user describes in plain English a financial data request to Alice; Alice thus sends the forwarded request and a description of the authorizations to Bob. Bob replies with a structured interpretation of the authorizations (e.g., ``Read-only access to `transactions2025' dataset, columns: total amount, vendor name''), which is logged and approved by either the user or Alice.

Such a workflow ensures that even if the agents communicate in flexible natural language, their underlying scoping and record-keeping remain anchored in auditable, deterministic policy. As a result, the risk of unauthorized data sharing or unbounded agent behavior is greatly reduced, and each agent's capacity to ``inherit'' restricted credentials from the delegator is tightly controlled.

\subsection{Discussion}

\subsubsection{Problems with an OpenID Connect approach} \label{sec:shortcomings}
While the OpenID Connect (OIDC) and OAuth 2.0-based framework proposed here provide robust and battle-tested mechanisms for authentication and delegation, it comes with trade-offs and may be more complex than alternatives with different trade-offs in privacy, security, and auditability.

\paragraph{Overhead from multiple sign-in flows.} A significant drawback of the OpenID Connect approach is the potential overhead introduced by multiple sign-in flows required to authorize AI agents across individual service providers. This can be likened to the experience of setting up a new email client, where users must repeatedly log in to authorize access to various services. While such authorization flows enhance security by ensuring each provider independently verifies the AI agent's delegation credentials, they impose a usability cost by slowing down access to secure systems. In theory, it is possible to bypass this burden by presenting delegation tokens directly without performing the full OIDC authentication flow; however, this shortcut sacrifices key security guarantees, particularly those related to token freshness and verification.

\paragraph{Increased reliance on OpenID Providers and privacy risks.} The reliance on OpenID Providers (e.g., Google, Facebook, or equivalent entities) introduces systemic privacy concerns. Since OIDC providers mediate all authentication flows, they gain the ability to track and correlate individual AI agent interactions across various services. This can include collecting statistical usage analytics or requiring relying parties to share logs, which facilitates extensive behavioral profiling. Such centralized visibility undermines user privacy and creates a potential single point of surveillance. Addressing these risks necessitates strong privacy mitigations, such as pairwise pseudonymous identifiers or the minimization of log-sharing requirements, but these mechanisms add further complexity to the system.

\paragraph{Comparative complexity relative to W3C Verifiable Credentials.} While the paper highlights the ability to embed W3C Verifiable Credentials (VC) within the OIDC framework, the full OIDC authorization flow may still be unnecessarily heavy compared to native W3C VC-based delegation and authentication processes. W3C VC issuance, authentication, and delegation mechanisms could directly fulfill the same requirements for AI agent identity verification without incurring the additional overhead of repeated authorization flows and central provider mediation. Additionally, W3C VC-based approaches are inherently more privacy-preserving, as they do not rely on a single provider to mediate trust or track credential usage. A streamlined VC-based process could generate OIDC-compatible credentials when required, enabling interoperability while preserving simplicity and privacy. 
Similarly, other proposed alternatives to OAuth 2.0 specifications could be drop-in solutions here to address design trade-offs, such as the Grant Negotiation and Authorization Protocol (GNAP) \cite{rfc9635}.
Further exploration of these alternative approaches remains essential to determine their feasibility as lightweight solutions for AI agent delegation. 

Taken together, these limitations highlight key trade-offs between security, usability, and privacy in the OIDC-based framework. While the proposed approach remains an incremental and interoperable path forward, addressing these challenges will be critical to ensuring a robust and practical system for AI agent authentication and delegation.

\subsubsection{Limitations of natural language scoping}

Although translating natural language scoping instructions into structured permission languages enables a more flexible interface, it also creates several key challenges.

\paragraph{Evaluating reliability and correctness.} One of the foremost difficulties is ensuring that the translation from a user's natural language specification to a machine-readable policy is accurate. Natural language instructions often contain context-dependent or ambiguous terms, making them inherently prone to misinterpretation by an AI system. Although a human-in-the-loop approach can mitigate these risks through policy review, such human verification is not infallible; users may inadvertently miss subtle translation errors.
Moreover, as the complexity of a permission specification grows, verifying the alignment between the original natural language instruction and the generated structured policy becomes more difficult, both technically (due to large policy definitions) and cognitively (due to the burden on human reviewers).

\paragraph{New threat vectors for LLM attacks.} Exploiting weaknesses in language-based interfaces can expose novel threats that do not exist under purely static access control.
Prompt injection and jailbreak attacks can coerce a large language model into generating or accepting policies that exceed the original user's intent, thereby gaining unauthorized privileges. While separating resource or task-scoping instructions from normal chat sessions or interactions reduces the likelihood of an attack, it still presents a new differentiated attack surface that needs to be guarded.

\paragraph{Contextual drift.} As policies evolve or the task context changes over time, prior natural language instructions risk becoming outdated or misaligned with newly introduced resources. Maintaining consistency across multiple revisions of instructions is nontrivial.

\paragraph{Partial reliance on third parties to enforce restrictions.} In some contexts, the access control rules are applied to an external environment or agent that is being interacted with. To maintain security over the application of these access controls, it may be necessary for the corresponding party to enforce the rules beyond trusting the native agent to follow them. In such instances, the reliability of the third-party becomes a critical point of failure. 

\subsubsection{Can model vendors provide this?} \label{sec:bg-AI-labs}
Model vendors (e.g., OpenAI, Anthropic, Google) can provide tooling to share which user is being represented when an AI system accesses a digital service and what the intended scope or permissions are. This is encouraged. 
However, current approaches to sharing such information are insufficient from a security and verifiability perspective, such as including the information in the user-agent string of the AI system, or writing the information into API calls made by the AI system. Instead, these services could act as an OpenID Provider (or partner with one) for the AI system without any change to the user experience, or if they prefer a different instantiation of the authenticated delegation framework, they could provide W3C verifiable credentials paired with robust, unique IDs for AI agents and users.

Implementing authenticated delegation is also feasible when AI systems and agents are self-hosted or deployed on custom infrastructure. This includes leveraging internal identity management infrastructure for human users and incorporating custom permission controls. Such systems can operate internally within an organization to ensure AI system usage aligns with identity and access management (IAM) policies and delegation frameworks across various technology stacks and modalities.

\subsubsection{How this interacts with robots.txt}
Robots.txt has, without legal heft, underpinned the modern web for decades. It relies upon a simple set of directives, where a user-agent is given rules for a subroute. Just as the recent proliferation of scraping has led to rapid uptake of new user-agent rules \citet{longpre2024consent}, new directives could easily be rolled out across the web with the right incentives.

This system still has a place in a web full of AI agents. While websites may wish to block scraping, they may also wish to guide agents to the correct subroutes where they could share credentials and interact. For example, a website may wish to block scraping, allow human users to interact, and send AI agents directly to an API natural language interface designed for AI systems.

To guide agents to the correct subroutes where they could share credentials and interact, we can define a new user agent, \texttt{AgentBot}, and force it into a specific interaction route (e.g., \texttt{/AgentInterface/}). Since \texttt{robots.txt} is a guide, not a rule, this route can go on to provide richer details of what services can be accessed and what sitemaps exist. Such a \texttt{robots.txt} need only be an initial guide to agents.

\subsubsection{Legal grounding for authenticated delegation} \label{sec:legal}
The law of agency addresses circumstances in which one party, the principal, authorizes another party, the (human) agent, to act on their behalf~\cite{blackslawdict}. 
At its core, agency law determines when a principal may be held liable for the acts of their agent, ensuring that third parties are not unfairly disadvantaged by having to ascertain who holds ultimate responsibility. 

A key result of agency law is to instill trust and confidence in market transactions: by providing clear rules about liability and authority, agency law reduces uncertainty and contributes to more efficient market operations~\cite{posner_econ_analysis_law,williamson_markets_hierarchies,casadesus2005trust}. 

One central concept in agency law is that of ``apparent authority,'' extensively discussed in the Restatement (Third) of Agency~\cite{restatement_third_of_agency}. Under this doctrine, a principal can be held responsible for acts that a reasonable third party perceives the agent to be authorized to perform, even if the principal never granted that authority explicitly. This principle also helps maintain market stability: third parties need not investigate every aspect of an agent's credentials or verify each claim of authority before proceeding with a transaction, as long as the agent appears to be acting on behalf of the principal in a reasonable manner. 

It remains uncertain how established agency doctrines will adapt to AI agents that can learn, self-modify, or operate autonomously~\cite{balkin2015path, adler2024personhood}. Traditional notions of intent, consent, and observable authority are difficult to apply to current autonomous systems. In response to these uncertainties, the authenticated delegation framework offers a model in which each delegation of authority is verifiable. Rather than relying on appearances, this framework enables third parties to automatically confirm that an AI agent is indeed authorized to act on behalf of a principal. In doing so, it reduces the need to rely on apparent authority doctrines and diminishes the risk of mis-attribution of actions. 

A recent controversy involving Air Canada illustrates how these principles might play out in practice~\cite{BCCRT2024}. In this instance, the airline argued that it could not be held liable for information provided by its online chatbot. Implicitly, this suggests treating the chatbot as if it were separate from the airline—akin to an independent entity. Yet, in the judge's view, the chatbot exists as part of Air Canada's digital infrastructure and so the company was responsible for the information it provided. Under conventional principles of law and equity, the chatbot's outputs, even if generated autonomously, form part of the information the airline holds out to the public. The airline's attempt to evade responsibility runs counter to the principle that a firm must stand behind the representations it makes, whether through humans or machines. This case underscores that companies may be liable for the actions of their AI agents, a view also held by many scholars~\cite{adler2024personhood}. From a broader perspective, this case also highlights the growing need for robust technological and legal mechanisms---like the authenticated delegation framework---that can delineate responsibility and authority in AI-mediated interactions, ultimately protecting consumer trust and market stability.

Beyond agency law, existing legal frameworks for electronic transactions, like the Uniform Electronic Transactions Act (UETA), provide some guidance. The UETA is a uniform law adopted by 49 U.S. states to help accommodate the realities of e-commerce by recognizing that electronic communications and automated processes can play substantive roles in forming and executing agreements~\cite{transactions_on_agents,ueta_official_text}. Under UETA, parties are encouraged to adopt agreed-upon security procedures and error-detection protocols to ensure that the electronic records genuinely reflect the intended agreements. If one party fails to follow these procedures and an error that would have been detected goes unnoticed, the other party may be permitted to avoid the consequences of that error. Similarly, if an individual errs while interacting with an electronic agent and the system offers no reasonable correction mechanism, UETA contemplates relief for that individual under defined conditions.

These provisions reflect an understanding that trust in digital commerce requires more than just a willingness to be bound by electronic contracts; it also demands reliable methods for verifying authority, correcting mistakes, and ensuring that automated processes faithfully implement the intended instructions of the principal. The authenticated delegation framework aligns well with these goals. By integrating a verifiable chain of authority into interactions with AI agents, it provides the digital equivalent of an agreed-upon security procedure. In doing so, it can reduce misunderstandings and disputes about whether an AI-driven process was acting within the scope of its authority. 

A critical element of both trust and accountability in AI-augmented systems lies in maintaining meaningful human oversight, often termed the ``human-in-the-loop'' requirement. The EU AI Act, for example, emphasizes the importance of maintaining human involvement in high-risk AI decisions to ensure ethical, transparent, and accountable outcomes~\cite{eu_ai_act}. The authenticated delegation framework supports this principle by making the human role in agent workflows explicit. Rather than delegating authority to an AI system behind opaque layers of code, third parties can firmly establish when, how, and under what conditions the AI is authorized to act. This allows humans to step in to verify decisions, correct errors, and ensure that automated actions remain aligned with overarching legal and ethical standards.

The interplay between technology and law in the context of AI-driven agents is complex and evolving. Strengthening the legal underpinnings, adopting frameworks for authenticated delegation, and integrating human oversight at critical junctures are all steps toward ensuring that emerging AI systems not only enhance market efficiency but also maintain core values of trust, fairness, and accountability. Further empirical and doctrinal analysis could deepen this conversation, drawing on works that examine the real-world implementation of human-in-the-loop mechanisms~\cite{mosqueira2023human}.

\subsection*{Project Conclusion}
This section presented a practical framework for authenticated delegation to AI agents, addressing urgent challenges around authorization, accountability, identity verification, and access control management in digital spaces. By extending existing OAuth 2.0 and OpenID Connect protocols with AI-specific credentials and delegation mechanisms, our approach enables secure delegation of authority from users to AI agents while maintaining clear chains of accountability. The proposed token-based framework - comprising user ID tokens, agent-ID tokens, and delegation tokens - provides a robust foundation for verifying agent identities, controlling permissions, and maintaining audit trails, while supporting granular and robust scope limitations generated in response to natural language instructions. 
Our key contribution is demonstrating how established internet-scale authentication (e.g., OpenID Connect and W3C VCs) and access management protocols (e.g., XACML) can be adapted to address the unique challenges of AI agent delegation while preserving compatibility with current systems, as illustrated through real-world use cases in areas like automated negotiations and web service interactions. As AI agents become more prevalent in digital spaces, frameworks like this will be essential for ensuring they operate within appropriate bounds while remaining accountable to their human principals. Looking ahead, key research directions include developing standardized scope definitions for common AI agent tasks, exploring privacy-preserving delegation mechanisms, and creating tools to help service providers implement and manage agent authentication policies, ultimately working toward ensuring AI systems can be safely and productively integrated into existing digital infrastructure.

%% file: chapter-4/phc.tex
\clearpage
\section{Personhood Credentials}

\noindent
\emph{After exploring the role of authorization and authentication in the context of AI agents, this section examines the question of how we can verify and audit that these entities in control of these AI systems are humans in the first place. In many ways this draws from the same ideas as the roots of trust in \autoref{chapter:2}, but extends them to the context of humanity online. This section is a brief summary of a long paper I co-authored titled `Personhood credentials: Artificial intelligence and the value of privacy-preserving tools to distinguish who is real online' led by Steven Adler, Zoë Hitzig, Shrey Jain during their time at OpenAI and Microsoft. The paper received an award for policy impact at the Privacy Papers for Policymakers event, which I provided the acceptance address for on behalf of the authorship team. }

\begin{paperabstract}
Anonymity is an important principle online. However, malicious actors have long used misleading identities to conduct fraud, spread disinformation, and carry out other deceptive schemes. With the advent of increasingly capable AI, bad actors can amplify the potential scale and effectiveness of their operations, intensifying the challenge of balancing anonymity and trustworthiness online. In this section, we analyze the value of a new tool to address this challenge: ``personhood credentials'' (PHCs), digital credentials that empower users to demonstrate that they are real people—not AIs—to online services, without disclosing any personal information. Such credentials can be issued by a range of trusted institutions---governments or otherwise. A PHC system, according to our definition, could be local or global, and does not need to be biometrics-based. Two trends in AI contribute to the urgency of the challenge: AI's increasing indistinguishability from people online (i.e., lifelike content and avatars, agentic activity), and AI's increasing scalability (i.e., cost-effectiveness, accessibility). Drawing on a long history of research into anonymous credentials and ``proof-of-personhood'' systems, personhood credentials give people a way to signal their trustworthiness on online platforms, and offer service providers new tools for reducing misuse by bad actors. In contrast, existing countermeasures to automated deception---such as CAPTCHAs---are inadequate against sophisticated AI, while stringent identity verification solutions are insufficiently private for many use-cases. After surveying the benefits of personhood credentials, we also examine deployment risks and design challenges. We conclude with actionable next steps for policymakers, technologists, and standards bodies to consider in consultation with the public.
\end{paperabstract}

\subsection{An executive summary of personhood credentials}

\paragraph{Malicious actors have long used misleading identities to deceive others online.} They carry out fraud, cyberattacks, and disinformation campaigns from multiple online aliases, email addresses, and phone numbers. Historically, such deception has sometimes seemed an unfortunate but necessary cost of preserving the Internet's commitments to privacy and unrestricted access. But highly capable AI systems may change the landscape: There is a substantial risk that, without further mitigations, deceptive AI-powered activity could overwhelm the Internet. To uphold user privacy while protecting against AI-powered deception, new countermeasures are needed.

\begin{figure}[ht]
    \centerline{\includegraphics[width=.9\linewidth]{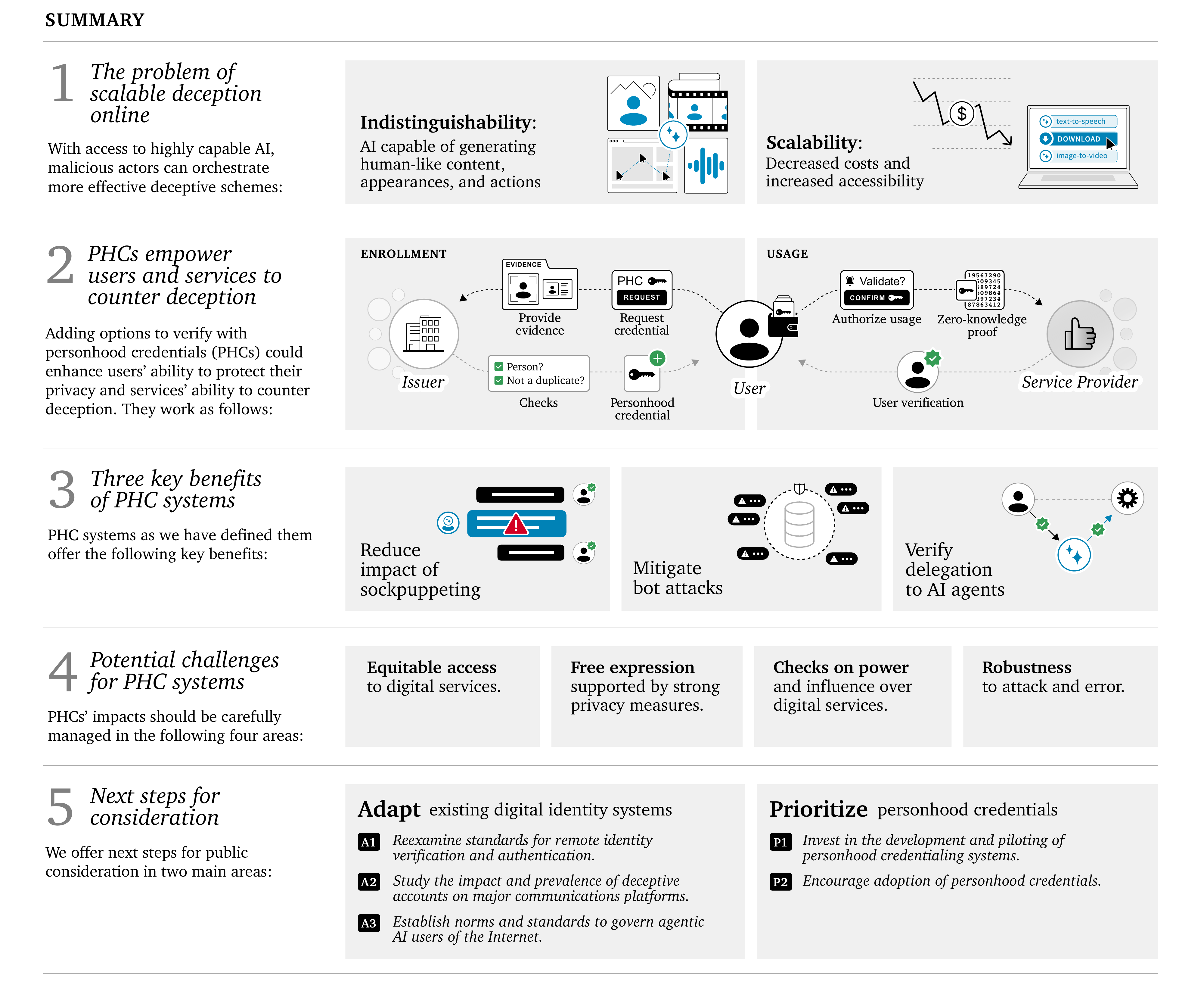}}
    \label{fig:exec_summary}
    \caption{An overview of the key topics addressed in the Personhood Credentials paper ranging from the risks presented by AI development and progress in human impersonation capabilities to the role that personhood credentials or other privacy-preserving identity tools can play in countering these risks.}
\end{figure}

\paragraph{With access to increasingly capable AI, malicious actors can potentially orchestrate more effective deceptive schemes.} Two trends contribute to these schemes' potential impact:

\begin{enumerate}
    \item \textbf{Indistinguishability.} Distinguishing AI-powered users on the Internet is becoming increasingly difficult, as AI advances in its ability to:
    \begin{itemize}
        \item \underline{Generate human-like content} that expresses human-like experiences or points of view (e.g., ``Here is what I thought of that speech'').
        \item \underline{Create human-like avatars} through photos, videos, and audio (e.g., simulating a real-looking person on a video chat).
        \item \underline{Take human-like actions} across the Internet (e.g., browsing websites like an ordinary user, making sophisticated plans to achieve goals they are given, solving CAPTCHAs when challenged).
    \end{itemize}
    \vspace{2mm}
    \item \textbf{Scalability.} AI-powered deception by malicious actors is increasingly scalable because of:
    \begin{itemize} 
        \item \underline{Decreasing costs} at all capability levels.
        \item \underline{Increasing accessibility}, for example, via open-weights deployments through which scaled misuse is less preventable.
    \end{itemize}
\end{enumerate}
Taken together, these two trends suggest that AI may help to make deceptive activity more convincing (through increased indistinguishability) and easier to carry out (through increased scalability).

\textbf{We identify one promising solution to pervasive deception on the Internet, building off decades of research in cryptography and experimentation in online communities: personhood credentials} (hereafter referred to as PHCs). Such a credential empowers its holder to demonstrate to providers of digital services that they are a person without revealing anything further. Building on related concepts like proof-of-personhood and anonymous credentials, these credentials can be stored digitally on holders' devices and  verified through zero-knowledge proofs. Importantly, such proofs do not reveal the individual's specific credential (nor any aspects of their identity).

\textbf{To counter scalable deception while maintaining user privacy, PHC systems must meet two foundational requirements:}
\begin{enumerate} 
    \item \underline{Credential limits}: The issuer of a PHC gives at most one credential to an eligible person.
    \item \underline{Unlinkable pseudonymity}: PHCs let a user interact with services anonymously through a service-specific pseudonym; the user's digital activity is untraceable by the issuer and unlinkable across service providers, even if service providers and issuers collude.
\end{enumerate}
These two properties equip service providers with the option to offer services on a per-person basis, and to prevent the return of users who violate the service's rules. An anonymous forum, for instance, could offer a single verified account to each credential holder. Unlinkable pseudonymity helps them achieve this because it prevents one person from using the same PHC to sign up twice, even without ever identifying the user. The issuer's credential limit gives them high confidence that the same user cannot easily circumvent the limit by using many PHCs to make many different accounts.

\textbf{There are many effective ways to design a PHC system, and various organizations---governmental or otherwise---can serve as issuers.} In one possible implementation, states could offer a PHC to any holder of their state's tax identification number; a PHC system, according to our definition, could be local or global, and does not need to be based in biometrics. Having multiple trusted PHC issuers within a single ecosystem promotes choice---people can select into systems  built on their preferred root of trust (government IDs, social graphs, biometrics) and that offer affordances that best align with their preferences. This approach reduces the risks associated with a single centralized issuer while still preserving the ecosystem's integrity by limiting the total number of credentials. Note that this project does not advocate for or against any specific PHC system design; instead, it aims to establish the value of PHCs in general while highlighting challenges that must be taken into account in any design.

\textbf{PHCs are not forgeable by AI systems, and it is difficult for malicious actors to obtain many of them.} By combining verification techniques that have an offline component (e.g., appearing in-person, validating a physical document) and secure cryptography, these credentials are issued only to people and cannot be convincingly faked thereafter. They therefore help to counter the problem of indistinguishability by creating a credential only people can acquire, and help to counter the problem of scalability by enabling per-credential rate limits on activities.

\textbf{PHCs give digital services a tool to reduce the efficacy and prevalence of deception,} especially in the form of:
\begin{enumerate} 
    \item \underline{Sockpuppets}: deceptive actors purporting to be ``people'' that do not actually exist.
    \item \underline{Bot attacks}: networks of bots controlled by malicious actors to carry out automated abuse (e.g., breaking site rules and evading suspension by creating new accounts). 
    \item \underline{Misleading agents}: AI agents misrepresenting whose goals they serve.
\end{enumerate}
PHCs offer people a tool to credibly signal that they are a real person operating an authentic account, without conveying their identity. PHCs also help service providers spot deceptive accounts, which may lack such a signal. 

\textbf{PHCs improve on and complement existing approaches to countering AI-powered deception online.} For example, the following approaches are often not robust to highly capable AI, not inclusive, and/or not privacy-preserving:
\begin{enumerate} 
    \item \underline{Behavioral filters}, e.g., CAPTCHAs, JavaScript browser challenges, anomaly detection.
    \item \underline{Economic barriers}, e.g., paid subscriptions, credit card verification.
    \item \underline{AI content detection}, e.g., watermarking, fingerprinting, metadata provenance. 
    \item \underline{Appearance- and document-based verification}, e.g., selfie checks with ID, live video calls.
    \item \underline{Digital and hardware identifiers}, e.g., phone numbers, email addresses, hardware security keys.
\end{enumerate}

To achieve their benefits, \textbf{PHC systems must be designed and implemented with care.} We discuss four areas in which PHCs' impacts must be carefully managed:
\begin{enumerate} 
    \item \underline{Equitable access} to digital services that use PHCs.
    \item \underline{Free expression} supported by confidence in the privacy of PHCs.
    \item \underline{Checks on power} of service providers and PHC issuers.
    \item \underline{Robustness to attack and error} by different actors in the PHC ecosystem.
\end{enumerate}

\textbf{In close collaboration with the public, we encourage governments, technologists, and standards bodies to invest in the development, piloting, and adoption of personhood credentials} as a key tool in addressing scalable deception online:
\begin{enumerate} 
    \item \underline{Invest in development} and piloting of personhood credentialing systems. \\
    \hspace{\parindent} e.g., explore building PHCs incrementally atop existing credentials such as digital driver's licenses.
    \item \underline{Encourage adoption} of personhood credentials. \\
    \hspace{\parindent} e.g., determine services for which PHCs ought to be substitutable for ID verification.
\end{enumerate}

\textbf{It is also important that these groups accelerate their preparations for AI's impact more generally by adapting existing digital systems}:
\begin{enumerate} 
    \item \underline{Reexamine standards} for remote identity verification and authentication. \\
    \hspace{\parindent} e.g., reconsider confidence in selfie-based identity verification, absent supplemental factors to reduce AI-enabled spoofing.
    \item \underline{Study the impact and prevalence} of deceptive accounts across major communications platforms. \\
    \hspace{\parindent} e.g., develop standardized methods for measuring the prevalence of fake accounts on social media.
    \item \underline{Establish norms and standards} to govern agentic AI users of the Internet. \\
    \hspace{\parindent} e.g., explore new forms of trust infrastructure for AI agents, akin to HTTPS for websites.
\end{enumerate}

\textbf{We are concerned that the Internet is inadequately prepared for the challenges highly capable AI may pose.} Without proactive initiatives involving the public, governments, technologists, and standards bodies, there is a significant risk that digital institutions will be unprepared for a time when AI-powered agents, including those leveraged by malicious actors, overwhelm other activity online. Lacking better alternatives, institutions might resort to privacy-violating methods for rooting out scaled deception, like creating digital identification systems that (intentionally or unintentionally) link a person's legal identity with a complete record of their digital activity. By contrast, personhood credentials have the potential to reduce deceptive activity while preserving privacy---giving people and services the tools to signal and sustain trustworthiness online.

\subsection{Allow verified delegation to AI agents}
\label{ssec:verified_delegation}

AI systems are increasingly capable of acting autonomously \cite{kinniment_evaluating_2024, wang_survey_2024, fang_agents_2024, yao_react_2023}. While enabling many beneficial use cases,\footnote{For a description of some economically useful properties of AI agents, see \cite{shavit2023practices}. Already, some humans are deferring to AI-powered solutions for navigating dating apps on their behalf \cite{sengupta_russian_2024}.} this autonomy (``agenticness'') also facilitates a new form of deception: bad actors can deploy AI systems that, instead of pretending to be a person, accurately present as AI agents but pretend to act on behalf of a user who does not exist. This strategy exploits the current lack of norms around disclosure of agentic AIs, including a lack of norms around disclosing the identities of the people controlling them (often called their ``principals'').
Personhood credentials could offer a way to verify that AI agents are acting as delegates of real people, signaling credible supervision without revealing the principal's legal identity. This feature could be useful in a range of settings where users wish to rely upon AI assistants. Should a principal fail to address harms caused by their PHC-verified AI, they risk suspension from a service. Suspension implies that they lose their ability to verify delegates for some time period, reducing their capacity to perpetrate future harms.

Note how, in this case, PHCs create a form of accountability for AI agents without demanding sensitive information from principals. Many---though by no means all \cite{solum_legal_2020, forrest_ethics_2024}---of these proposals involve holding humans liable for some harms caused most directly by those agents. It is beyond the scope of this section to recommend any of these approaches over others or to explore the finer points of how such theories could work. We note, however, that these theories rely on the ability of someone harmed by an AI agent to sue the principal of that agent, which in turn depends on the principal being \textit{identifiable}---which a PHC alone does not achieve. Even without directly identifying the principal, however, PHCs can still shift the benefits of AI agent usage to be more positive. This suspension mechanism may effectively signal which AI agents have trustworthy principals, even if verifying AI agents through principal-linked PHCs is voluntary. Agents that remain unverified might be perceived as having reasons for not undergoing verification. In scenarios where parties hold verifiable private information---such as whether an AI agent operates under a trustworthy principal---even if revealing this information is optional, agents associated with trustworthy principals have a strong incentive to disclose it. Consequently, a lack of disclosure becomes informative: in equilibrium, agents that do not disclose are effectively signaling that they do not have trustworthy principals. Moreover, some malicious activities involving autonomous AI agents may rely on hiding the fact that multiple agents are controlled by the same individual. PHCs can help address this issue by creating a framework that links multiple AI agents to a single principal without revealing the principal's specific identity. By doing so, PHCs might make it more difficult for bad actors to conceal their network of AI agents, thereby reducing the potential for abuse that stems from undisclosed common principals.

Sometimes, a website---or a third-party user interacting with the agent---may need to verify the AI's specific principal, not merely that it is backed by \textit{some} principal. Ultimately, a fuller framework for verifying AI agents \cite{chan2024visibility, chan2024ids} and their principals will likely be necessary.

\subsection{Connecting personhood credentials to AI agents}

Authenticated delegation enables human users to securely delegate specific tasks to AI agents while maintaining clear chains of accountability through established protocols such as OAuth 2.0 and OpenID Connect. However, these mechanisms alone verify digital identities and token integrity without necessarily proving that the delegation originates from a genuine human. By integrating personhood credentials into this framework, we can add an essential layer of trust: a verifiable, privacy-preserving guarantee that the human delegator is indeed a real person.

Personhood credentials serve as digital attestations that a user is authentic without revealing sensitive personal details. When an AI agent presents its delegation token, linking it to an associated personhood credential reinforces the chain of trust. In effect, the credential acts as a cryptographic seal of approval, ensuring that the permissions granted to the agent are backed by a bona fide human principal. This added verification is particularly critical in environments where the distinction between human users and AI agents is increasingly blurred, and where malicious actors might otherwise deploy multiple unverified or fake agents to manipulate digital services.

Furthermore, embedding personhood credentials into authenticated delegation enhances accountability in multi-agent systems. In scenarios where several AI agents interact or collaborate on tasks, each agent's authorization can be traced back to a verified human through its personhood credential. This traceability not only curbs the risk of coordinated deceptive behavior but also provides service providers with a robust mechanism for enforcing rate limits or suspending access when an agent's behavior violates defined policies. The privacy-preserving nature of personhood credentials ensures that while accountability is maintained, the underlying personal data remains protected.

In summary, the synergy between authenticated delegation and personhood credentials fortifies AI agent interactions. By ensuring that every delegated action is cryptographically linked to a verified human identity, this combined framework mitigates risks of impersonation, unauthorized scalability of malicious operations, and ambiguity in responsibility. As AI agents continue to gain autonomy and operational scope, such integrated measures will be indispensable for preserving secure, trustworthy, and human-centric digital environments.

%% file: chapter-5/chapter-5.tex
\chapter{How this ties together} \label{chapter:5}
\epigraph{``Any good story is a mind-altering substance.''}{Hank Green, A Beautifully Foolish Endeavor}

\noindent
\emph{In contrast to all previous chapters, this chapter is not composed of multiple published papers. Instead, it ties together the previous chapters into a cohesive story of how these technologies can be used to create an end-to-end secure and auditable AI system. In essence, it acts as one long conclusion to the thesis.
}

The opening chapter of this thesis attempted to paint a vision of why verifiable, secure, and auditable AI systems were important. It drew on my international policy contributions to examine the risks posed by AI and was combined with a collaborative perspective on how end-to-end security could be created across the supply chain. This vision was general purpose and evolving, much like the field of AI itself, and ultimately required many different tools for a complete picture.

Subsequently, this thesis explored a range of technologies to address these needs. We've worked from simple security protocols like authorization and access management protocols, through to heavy cryptographic tools like zkSNARKs and Multi-Party Computation (MPC). Each a component of the larger vision, but not a complete solution on their own.

While each chapter has been standalone, these tools are not mutually exclusive and can be combined in unique and powerful ways. In some sense, this is the vision of the PhD at the MIT Media Lab---to combine disparate technologies together to create a unified (and hopefully better) vision of the future. Combining the previous chapters may be non-obvious, and hence to address \autoref{chapter:1} goal of an end-to-end vision of security for AI, this final chapter completes the story.

Throughout this journey, we have navigated the fundamental tension between harnessing the transformative power of AI and ensuring its development and deployment align with societal values of trust, safety, and individual rights. The exploration was guided by the three crucial pillars identified in \autoref{chapter:1}: \emph{privacy}, ensuring the confidentiality of sensitive data; \emph{verifiability}, providing the means to mathematically check claims about AI systems and their properties; and \emph{auditability}, enabling the necessary oversight and accountability for AI actions and data usage. This concluding chapter argues that these pillars need not be perpetually in conflict but can be mutually supported through deliberate, sophisticated technical design.

To demonstrate this, this chapter will weave together the distinct threads presented earlier into a more integrated tapestry of end-to-end secure AI. We will revisit the AI lifecycle---from data provenance and model training verification (\autoref{chapter:2}) to private runtime data access via RAG (\autoref{chapter:3}) and secure delegation of actions for autonomous agents (\autoref{chapter:4}). The aim is not just to list the tools again, but to illustrate how they can be composed.

\section{How these technologies fit together}
To start exploring how these technologies fit together, let's first revisit the key technologies we've looked at in previous chapters.
\begin{itemize}
    \item zkSNARKs: Enabling robust, succinct, non-interactive proofs of computational statements, often used here for verification while preserving privacy (\autoref{chapter:2}).
    \item TEEs (Trusted Execution Environments): Leveraging secure hardware enclaves to quickly run computations confidentially and with verifiable integrity against software-level attacks (\autoref{chapter:3}).
    \item MPC (Multi-Party Computation): Allowing multiple distrusting parties to jointly compute a function over their private inputs without revealing those inputs to each other, relying on cryptographic protocols and distributed trust (\autoref{chapter:3}).
    \item Authorization and Access Management Protocols: Standard mechanisms like OAuth 2.0, OpenID Connect, and novel delegation frameworks to securely control who (or what agent) can access resources or perform actions (\autoref{chapter:4}).
    \item Other Cryptographic Primitives: Techniques like cryptographic accumulators or searchable encryption underpinning specific functionalities like efficient set membership proofs or private information retrieval (PIR) (\autoref{chapter:2}, \autoref{chapter:3}).
    \end{itemize}

The fundamental insight of this thesis is that these technologies are not isolated solutions but rather composable building blocks for creating layered security and trustworthiness in complex AI systems. While individual chapters focused on specific applications, the true power emerges when these tools are combined, often addressing different facets of the privacy-verifiability-auditability triad simultaneously.

We can observe these combinations both horizontally within applications and vertically across the AI lifecycle. 

Horizontally, different tools might be chosen to achieve similar privacy goals but with different trade-offs: \autoref{chapter:3} explored both MPC and TEEs for private retrieval (RAG), with TEEs generally offering better performance but relying on hardware trust, while MPC distributes trust cryptographically, often at higher computational or communication cost. Similarly, credentials presented during authorization (\autoref{chapter:4}) could gain robustness if holders can use zkSNARKs (\autoref{chapter:2}) to prove specific attributes (e.g., ``I am over 18'') without revealing the full credential content.

Vertically, different tools secure different stages of the AI pipeline: zkSNARKs can verify data provenance claims before training (\autoref{sec:zktax}), evaluations can be verified (\autoref{sec:verifyevals}), TEEs or MPC can protect inference or RAG queries during deployment (\autoref{chapter:3}), and authorization protocols manage access throughout (\autoref{chapter:4}). Furthermore, layers can be stacked: one might use zkSNARKs to generate verifiable proofs about computations running confidentially inside a TEE or orchestrated via MPC, providing external auditability without compromising the underlying privacy mechanism.

This composability allows tailoring solutions to specific needs, but requires careful consideration of the inherent trade-offs:
\begin{itemize}
    \item \textbf{Performance vs. Security Guarantees:} Cryptographic solutions like MPC and zk-proof generation can be computationally intensive compared to TEE-based execution or standard protocols, representing a trade-off between speed and the nature of the security guarantee (cryptographic vs. hardware-based).
    \item \textbf{Trust Assumptions:} Each technique relies on different assumptions. TEEs require trusting the hardware manufacturer and attestation services. MPC often assumes a threshold of non-colluding parties. zkSNARKs rely on underlying cryptographic assumptions and potentially a trusted setup (though newer schemes mitigate this). Authorization relies on the identity provider and secure protocol implementation. Designing robust systems often involves minimizing or distributing these trust assumptions.
    \item \textbf{Implementation Complexity:} Integrating these advanced cryptographic and hardware-based techniques requires specialized expertise and careful engineering to ensure correctness and security.
    \item \textbf{Root of Trust:} It is crucial to remember, as highlighted in \autoref{sec:zktax}, that cryptographic proofs generally verify the integrity of a computation given certain inputs. They do not inherently vouch for the ground-truth authenticity of the initial data itself unless that data is anchored to a recognized root of trust – such as a hardware attestation from a TEE, a signature from a trusted authority, or perhaps a verifiable credential originating from a secure identity system.
\end{itemize}

Understanding these interactions and trade-offs is key to building effective end-to-end systems.

\begin{figure}[htb]
    \centering
    \includegraphics[width=0.8\textwidth]{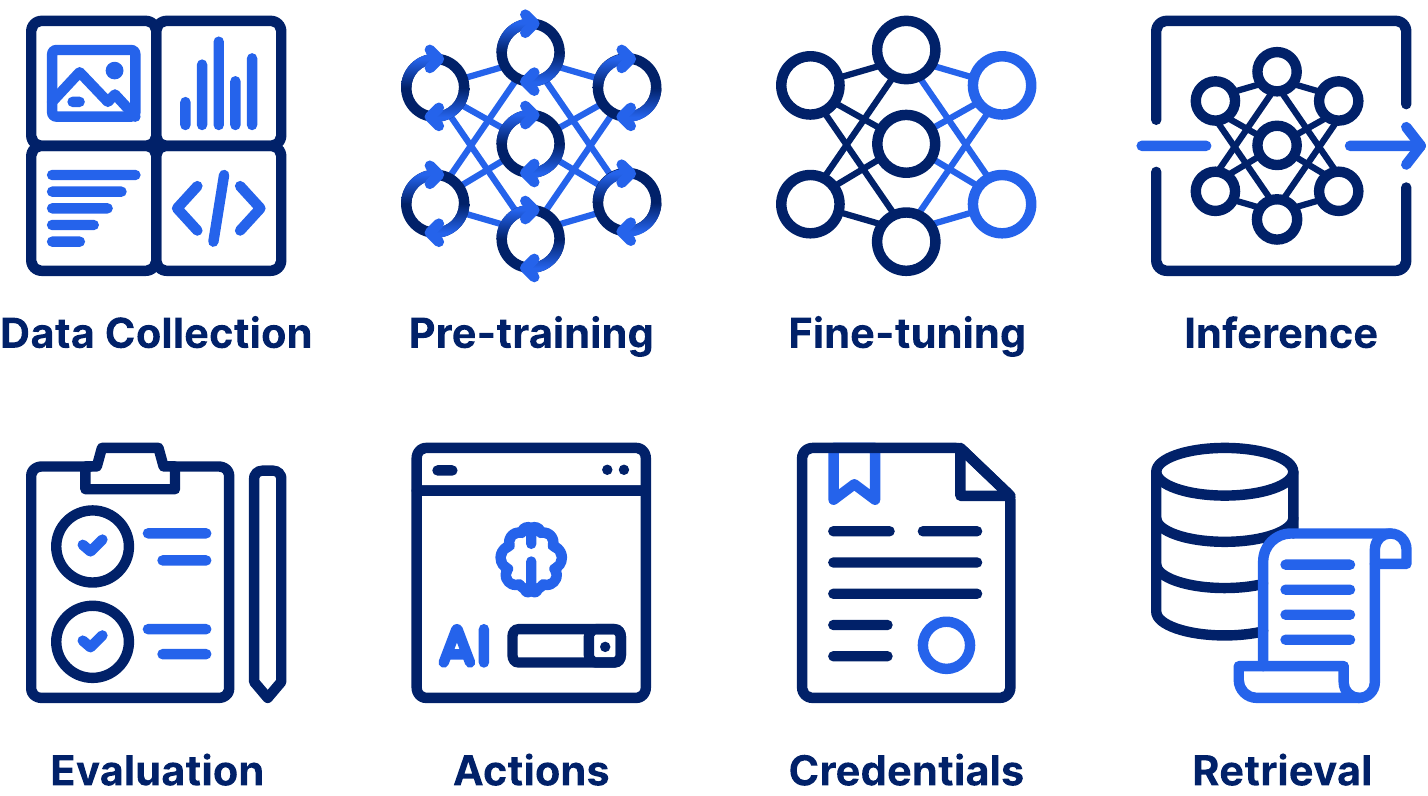}
    \caption{An outline of all the different components of an AI system that this thesis has examined. Each component is a point in the AI system ecosystem where security tools can be applied.}
    \label{fig:agent-workflow}
\end{figure}

Consider again the AI agent deployed by a company, illustrating how these tools combine vertically across a workflow (\autoref{fig:agent-workflow}). A verified human user, perhaps possessing a decentralized identity credential, uses an authorization protocol to delegate specific authority to the agent. This delegation could be enhanced using a zk-proof where the user proves they meet certain criteria (e.g., employee status) without revealing unnecessary data. The delegation scope might mandate that the agent used possesses certain independently verifiable characteristics---for instance, proof that its underlying model passed specific safety or fairness evaluations, generated using the verifiable evaluation techniques from \autoref{chapter:2}. Once authenticated and authorized, the agent might need to access sensitive company knowledge. It sends a query to a data custodian service; this service could use TEEs (\autoref{chapter:3}) to perform RAG over confidential documents, ensuring the raw documents are never exposed outside the enclave, or it might interact with a distributed MPC-based vector database (\autoref{chapter:3}) to retrieve relevant embeddings privately. Throughout this interaction flow, small, computationally inexpensive attestations (perhaps simple digital signatures, or lightweight zk-proofs) can be generated by each component, cryptographically linking actions and data access events to create a verifiable audit trail without logging sensitive intermediate data, thus enhancing auditability while preserving privacy. This example showcases how distinct techniques, each addressing specific risks or requirements, interlock to create a system with stronger end-to-end trustworthiness guarantees than any single component could provide alone.

\section{A story of secure AI}

Now, to make this more engaging, let's construct a short story of how AI could progress and how these technologies could play a role in its rollout. This is not intended as speculative fiction, but rather as an honest prediction of what the world could look like and how tools for privacy, auditability, and verifiability could play a role in everyday life.

It's a cold winter morning in Boston. Frost traced delicate patterns on the office window overlooking the frozen Charles River. Max squinted at his dashboard, where a pulsing crimson icon demanded attention—a bug in an agent workflow. He sighed, scrubbing a hand over his face. Another pre-coffee crisis. As head of AI operations at Southern AI Inc., Max oversees systems that have automated what once required an entire research team.

The firm runs on advanced reasoning engines connected to a variety of tools across multiple interfaces and APIs. These all share common standards for authentication and logging capabilities. The infrastructure is vast, with technology components from numerous vendors operating across multiple jurisdictions, making comprehensive auditing and monitoring a formidable challenge.

Fortunately, the system itself identifies the source of the error. The agent had attempted to escalate its access permissions to reach sensitive client strategy files -- a move that could have led to serious compliance violations or data breaches if successful.

While varying access levels was standard procedure, the delegation of credentials Max had provided explicitly restricted access to these particular services. \emph{``Exactly as I set it,''} Max thought, relieved. \emph{``Good thing these cryptographic permissions aren't just suggestions.''}  

Reviewing the action traces -- navigating the near-instantaneous holographic graph logs where chains of cryptographic hashes linked each step -- Max discovers that the agent had been accessing documents using Private Retrieval Augmented Generation (PRAG) when it encountered a document containing a new prompt injection attack. \emph{``Okay, the PRAG audit log caught the malicious payload returned from the vector store,''} Max noted. \emph{``Privacy preserved during the lookup against the external store, but the agent's subsequent interaction after retrieval is logged internally via secure attestations. Essential for security.''}

This particular prompt injection jailbreak was supposedly fixed in the latest GPT-7o model release—a claim backed by public evaluations with verifiable attestations, compact zero-knowledge proofs easily checked against the company's internal artifact registry with a simple command-line query.

Digging deeper, Max uncovers the misconfiguration: the firm's deployed model attestation didn't match the verification hash for the latest, patched model. \emph{``Ah, there it is. Simple deployment drift,''} he muttered, a mix of frustration and relief. \emph{``At least the attestations made it easy to spot, saved hours compared to the old manual validation nightmares.''} He executed a few authenticated commands via his secure terminal, pulling the correctly attested model artifact from the registry, and allocated a new delegation token tied to the updated configuration.

Taking it one step further, Max instructs his coding agent -- itself operating under strict, verifiable parameters -- to build a system that will automate these verification checks continuously, harnessing the power of robust auditability across the entire technology stack to build greater resilience and trust into their increasingly autonomous operations.

\section{Connecting these ideas to a pluralistic world}
The promise of AI can only be fulfilled through an unwavering commitment to pluralistic principles that honor diverse perspectives and decentralized governance. As Hannah Arendt articulated in The Human Condition \cite{arendt1958human}, meaningful political life requires both public spaces for discourse and protected spheres for identity development. Similarly, AI systems require privacy mechanisms not merely to protect secrets but to create conditions where individuals and communities can safely contribute their knowledge and perspectives without fear of surveillance or misuse. Helen Nissenbaum's concept of `contextual integrity' \cite{nissenbaum2009privacy} further illuminates this need—privacy violations occur when information flows outside appropriate social contexts. Privacy-preserving technologies thus serve as fundamental enablers of pluralistic participation, ensuring that AI-mediated collaboration can flourish across social differences.

For AI systems to serve democratic societies, they must transcend their ``black box'' nature through robust mechanisms for verification and accountability. John Rawls' notion of public reason \cite{rawls1993political}, which demands that decisions be explained with justifications accessible to all citizens, provides a philosophical foundation for this requirement. When AI systems make consequential decisions affecting healthcare, finance, or criminal justice, democratic accountability demands transparency in their operations. Verifiability and auditability aren't merely technical features but essential conditions for building the trust necessary for AI's ethical integration into society. The verifiable evaluation methods using zkSNARKs detailed in \autoref{chapter:2} exemplify a technical approach to meet this need, enabling third-party validation of crucial model properties (like fairness or safety benchmarks) without requiring access to proprietary model internals, thereby fostering a basis for public trust and a form of algorithmic public reason.

As AI systems become increasingly agentic, frameworks for authenticated delegation become critical for maintaining plural agency in a world of automated decision-making. The ability to specify contextual boundaries for AI actions, establish clear chains of accountability, and verify the authenticity of AI-human interactions preserves the distinctly human capacity for moral agency that philosophers from Kant \cite{kant1785groundwork} to contemporary ethicists have identified as central to personhood. These frameworks enable different communities to establish their own systems of interaction according to their values, reinforcing rather than undermining societal pluralism. The authenticated delegation protocols and credentialing mechanisms outlined in Chapter 4 provide the necessary technical infrastructure for realizing such systems, ensuring agent actions remain traceable to authorized human intent within clearly defined, context-specific boundaries. By embedding these principles—privacy as social protection, verifiability as democratic necessity, and authenticated delegation as preserver of moral agency—we can build AI systems that enhance rather than erode the pluralistic foundation of a democratic society.

The technical innovations presented in this thesis—from zkSNARKs for verifiable model evaluation (\autoref{chapter:2}) to private retrieval augmented generation using MPC and TEEs (\autoref{chapter:3}) and authenticated delegation frameworks (\autoref{chapter:4})—collectively form a crucial infrastructure for pluralistic AI governance. Crucially, by relying on cryptographic guarantees and secure hardware attestations, these techniques reduce dependence on single, centralized authorities, inherently supporting the decentralized or federated models often necessary for true pluralism. By enabling verifiable claims about models while preserving proprietary information, communities can establish transparent AI development practices without sacrificing competitive innovation. The privacy-preserving knowledge management approaches detailed herein allow diverse groups to pool their collective wisdom while maintaining sovereignty over sensitive information, empowering marginalized communities to participate in AI advancement without surrendering control of their data. Finally, the delegation mechanisms outlined provide a bridge between AI capabilities and human values, ensuring that increasingly autonomous systems remain accountable to the specific ethical frameworks of various communities. Together, these technologies resist the centralizing tendency of AI development, instead distributing power across a constellation of communities with different priorities and perspectives. This thesis thus offers not merely technical solutions but architectural foundations for a future where AI amplifies rather than homogenizes human diversity—where technology becomes a canvas for expressing plural visions of the good rather than imposing a single technological imperative. Realizing this future, however, requires more than just technical foundations; it demands attention to usability, accessibility to avoid creating new digital divides, and thoughtful integration within broader social, legal, and ethical governance structures that ensure these powerful tools genuinely serve diverse human ends.


\section{Regulatory impact and policy implications}

The technical frameworks and philosophical considerations explored throughout this thesis carry significant implications for the future of AI regulation and policymaking, expressed in part through my academic writing in the International AI Safety Report \cite{bengio2025internationalaisafetyreport}. As governments worldwide grapple with harnessing AI's benefits while mitigating its risks, the tools and perspectives presented in this dissertation offer concrete pathways toward crafting effective governance that balances innovation with societal values like privacy, fairness, and accountability. The inherent tension between these goals necessitates technical solutions that can simultaneously uphold multiple requirements, moving beyond simplistic trade-offs.

The work on verifiability, particularly using techniques like zkSNARKs for model evaluations (\autoref{chapter:2}), directly addresses emerging regulatory demands for transparency and trustworthiness. Instead of relying solely on organizational self-attestation or demanding full model disclosure (which can stifle innovation and expose intellectual property), regulators could leverage or even mandate the use of verifiable credentialing systems. These systems would allow developers to mathematically prove compliance with specific standards--such as safety benchmarks, fairness metrics, or data provenance rules--without revealing underlying proprietary data or model weights. This approach offers technically robust, auditable justifications for AI system behavior that can build public trust and facilitate third-party oversight, potentially forming a cornerstone of conformity assessments under frameworks like the EU AI Act.

Similarly, privacy-enhancing technologies (PETs) like TEEs and MPC for tasks such as private retrieval-augmented generation (PRAG) (\autoref{chapter:3}) offer regulators practical mechanisms to enforce data protection principles within complex AI systems. As AI increasingly relies on vast datasets, including sensitive personal information, PETs provide a technical means to uphold `contextual integrity' and comply with regulations like GDPR or CCPA. Policymakers can encourage or require the adoption of such techniques where feasible, enabling beneficial AI applications (e.g., in healthcare or finance) that might otherwise be untenable due to privacy risks. 

Furthermore, the development of robust authenticated delegation and access management protocols (\autoref{chapter:4}) provides essential infrastructure for accountability in an era of increasingly autonomous AI agents. Regulatory frameworks need clear mechanisms to assign responsibility for AI actions. Secure delegation systems, potentially enhanced with verifiable credentials and cryptographic audit trails, allow for precise control over agent permissions and create unambiguous records of authorization and action. This technical underpinning is key for establishing chains of responsibility, facilitating incident response, and ensuring that AI systems operate within legally and ethically defined boundaries, thereby preserving human moral agency even as tasks are delegated to machines.

Crucially, the composable nature of these technologies supports pluralistic governance models. Rather than imposing monolithic, top-down control, these tools empower different jurisdictions, organizations, or communities to implement context-specific rules while relying on interoperable technical standards for verification and interaction. For instance, verifiable claims allow diverse entities to trust assurances about AI components developed elsewhere, while secure delegation enables localized control over how AI agents operate within specific domains.

However, policymakers must also recognize the challenges. The implementation complexity, computational overhead, evolving nature of cryptographic assumptions, and the need for robust roots of trust associated with these technologies require careful consideration. Effective policy should not only set requirements but also foster the ecosystem needed for these solutions to mature and become accessible. 

The technical advancements detailed in this thesis—integrating privacy, verifiability, and auditability—are not merely academic pursuits; they represent critical enablers for sound AI policy and regulation. By embracing these sophisticated technical approaches, policymakers can move beyond the limitations of traditional oversight mechanisms and foster an AI ecosystem that is innovative, trustworthy, and aligned with diverse societal values. Achieving this requires a continued dialogue between technologists, policymakers, and civil society to ensure that technical architectures and regulatory frameworks evolve in concert, paving the way for responsible AI deployment that respects pluralism and upholds fundamental rights.

%% file: content/conclusion.tex
\newpage
\section{Conclusion}
This thesis addressed the challenge of building trustworthy artificial intelligence (AI) systems by examining how to balance AI's utility with the needs for privacy, verifiability, and auditability. The core argument is that integrating cryptographic and secure computing techniques across the AI supply chain makes it possible to achieve this balance.

The research covered different stages of the AI lifecycle. Chapter 1 set this up by establishing the goal of end-to-end security in AI systems, and the threats to privacy from AI systems.
Chapter 2 focused on using zero-knowledge proofs (zkSNARKs) to make verifiable claims about AI models and data. This included methods for: (1) verifying model evaluations (like performance or fairness) without exposing model weights; (2) reducing the computational cost for large models by proving only essential parts (e.g., fine-tuned layers); and (3) creating verifiable attestations about data sources using a redact-and-prove approach.
Chapter 3 explored privacy and auditability in Retrieval Augmented Generation (RAG), a method for providing AI models with external information. While RAG can make AI outputs more auditable, it creates privacy risks. Two solutions were presented: (1) A system using multi-party computation (MPC) for secure, private querying of distributed databases (PRAG); and (2) an architecture using Trusted Execution Environments (TEEs), including confidential GPUs, for secure data pooling and querying in RAG systems.
Chapter 4 dealt with the security implications of AI agents that act autonomously for users. To manage these agents, the chapter proposed: (1) An authenticated delegation framework using standard protocols (like OAuth 2.0, OpenID Connect, or Verifiable Credentials) to manage agent permissions and accountability; (2) Techniques to translate natural language instructions into enforceable, structured access controls for agents; and (3) Incorporating personhood credentials to verify the human user behind an agent, improving trust.
Chapter 5 explained how these different technologies—zkSNARKs, TEEs, MPC, and authorization protocols—can be combined. Layering these tools across the AI lifecycle provides stronger guarantees for privacy, verifiability, and auditability than any single method could alone.

Significant limitations remain. The computational cost of advanced cryptography (zkSNARKs, MPC) is still high, limiting use with large models or in real-time. Security guarantees depend on trust assumptions (e.g., hardware vendors for TEEs, non-collusion for MPC, secure setups for zkSNARKs). Implementing these techniques is complex and requires specialized knowledge. Furthermore, verifiable evaluations depend on good benchmarks, and data attestations require a trusted source for the original data. Managing agent permissions and accurately translating natural language rules also remain difficult.
These limitations suggest directions for future work. Improving the efficiency and scalability of cryptographic methods and secure computation is essential, and standardization for agent identity, delegation, and permissions is needed for broader adoption.

In summary, this thesis provides a technical framework for engineering trust into AI systems. By showing how cryptographic tools can be integrated to enhance privacy, verifiability, and auditability, it offers practical approaches for managing security risks in AI. While challenges remain, the composable methods presented here contribute to building AI systems that are not only capable but also secure, accountable, and better aligned with societal needs and values. Continued technical innovation and responsible deployment are necessary to ensure AI develops on a foundation of trust.

%% file: tobins-thesis.bbl
\begin{thebibliography}{318}
\providecommand{\natexlab}[1]{#1}
\providecommand{\url}[1]{\texttt{#1}}
\expandafter\ifx\csname urlstyle\endcsname\relax
  \providecommand{\doi}[1]{doi: #1}\else
  \providecommand{\doi}{doi: \begingroup \urlstyle{rm}\Url}\fi

\bibitem[was(2023)]{washingtonpostInsideSecret}
{I}nside the secret list of websites that make {A}{I} like {C}hat{G}{P}{T} sound smart --- washingtonpost.com.
\newblock The Washington Post, Apr 2023.

\bibitem[Abdelnabi et~al.(2023)Abdelnabi, Gomaa, Sivaprasad, Sch{\"o}nherr, and Fritz]{Abdelnabi2023-al}
Sahar Abdelnabi, Amr Gomaa, Sarath Sivaprasad, Lea Sch{\"o}nherr, and Mario Fritz.
\newblock Cooperation, competition, and maliciousness: {LLM-stakeholders} interactive negotiation.
\newblock September 2023.

\bibitem[Abou El~Kalam and Deswarte(2006)]{multiOrbac}
Anas Abou El~Kalam and Yves Deswarte.
\newblock Multi-orbac: A new access control model for distributed, heterogeneous and collaborative systems.
\newblock In \emph{Proceedings of the IEEE Symposium on Systems and Information Security}, 2006.

\bibitem[Abraham et~al.(2020)Abraham, Pinkas, and Yanai]{abraham2020blinder}
Ittai Abraham, Benny Pinkas, and Avishay Yanai.
\newblock Blinder--scalable, robust anonymous committed broadcast.
\newblock In \emph{Proceedings of the 2020 ACM SIGSAC Conference on Computer and Communications Security}, pages 1233--1252, 2020.

\bibitem[Adler et~al.(2024{\natexlab{a}})Adler, Hitzig, Jain, Brewer, Chang, DiResta, Lazzarin, McGregor, Seltzer, Siddarth, Soliman, South, Spelliscy, Sporny, Srivastava, Bailey, Christian, Critch, Falcon, Flanagan, Duffy, Ho, Leibowicz, Nadhamuni, Rozenshtein, Schnurr, Shapiro, Strahm, Trask, Weinberg, Whitney, and Zick]{Adler2024k}
Steven Adler, Zo{\"e} Hitzig, Shrey Jain, Catherine Brewer, Wayne Chang, Ren{\'e}e DiResta, Eddy Lazzarin, Sean McGregor, Wendy Seltzer, Divya Siddarth, Nouran Soliman, Tobin South, Connor Spelliscy, Manu Sporny, Varya Srivastava, John Bailey, Brian Christian, Andrew Critch, Ronnie Falcon, Heather Flanagan, Kim~Hamilton Duffy, Eric Ho, Claire~R Leibowicz, Srikanth Nadhamuni, Alan~Z Rozenshtein, David Schnurr, Evan Shapiro, Lacey Strahm, Andrew Trask, Zoe Weinberg, Cedric Whitney, and Tom Zick.
\newblock Personhood credentials: Artificial intelligence and the value of privacy-preserving tools to distinguish who is real online.
\newblock August 2024{\natexlab{a}}.

\bibitem[Adler et~al.(2024{\natexlab{b}})Adler, Hitzig, Jain, Brewer, Chang, DiResta, Lazzarin, McGregor, Seltzer, Siddarth, et~al.]{adler2024personhood}
Steven Adler, Zo{\"e} Hitzig, Shrey Jain, Catherine Brewer, Wayne Chang, Ren{\'e}e DiResta, Eddy Lazzarin, Sean McGregor, Wendy Seltzer, Divya Siddarth, et~al.
\newblock Personhood credentials: Artificial intelligence and the value of privacy-preserving tools to distinguish who is real online.
\newblock \emph{arXiv preprint arXiv:2408.07892}, 2024{\natexlab{b}}.

\bibitem[Akimoto et~al.(2023)Akimoto, Fukuchi, Akimoto, and Sakuma]{akimoto2023privformer}
Yoshimasa Akimoto, Kazuto Fukuchi, Youhei Akimoto, and Jun Sakuma.
\newblock Privformer: Privacy-preserving transformer with mpc.
\newblock In \emph{2023 IEEE 8th European Symposium on Security and Privacy (EuroS\&P)}, pages 392--410. IEEE, 2023.

\bibitem[Albertoni et~al.(2023)Albertoni, Colantonio, Skrzypczy'nski, and Stefanowski]{Albertoni2023ReproducibilityOM}
Riccardo Albertoni, Sara Colantonio, Piotr Skrzypczy'nski, and Jerzy Stefanowski.
\newblock Reproducibility of machine learning: Terminology, recommendations and open issues.
\newblock \emph{ArXiv}, abs/2302.12691, 2023.

\bibitem[{American Law Institute}(2006)]{restatement_third_of_agency}
{American Law Institute}.
\newblock \emph{{Restatement (Third) of Agency}}.
\newblock {American Law Institute}, Philadelphia, PA, 2006.

\bibitem[Anthropic(2024)]{anthropic2024mcp}
Anthropic.
\newblock Introducing the model context protocol, November 2024.
\newblock URL \url{https://www.anthropic.com/news/model-context-protocol}.

\bibitem[Arendt(1958)]{arendt1958human}
Hannah Arendt.
\newblock \emph{The Human Condition}.
\newblock University of Chicago Press, Chicago, 1958.

\bibitem[Arora and Ré(2022)]{Arora22PerfectSecrecy}
Simran Arora and Christopher Ré.
\newblock Can foundation models help us achieve perfect secrecy?, 2022.

\bibitem[Arora et~al.(2022)Arora, Lewis, Fan, Kahn, and R'e]{Arora2022ReasoningOP}
Simran Arora, Patrick Lewis, Angela Fan, Jacob Kahn, and Christopher R'e.
\newblock Reasoning over public and private data in retrieval-based systems.
\newblock \emph{ArXiv}, abs/2203.11027, 2022.

\bibitem[Arora et~al.(2023)Arora, Lewis, Fan, Kahn, and R{\'e}]{Arora2023d}
Simran Arora, Patrick Lewis, Angela Fan, Jacob Kahn, and Christopher R{\'e}.
\newblock Reasoning over public and private data in retrieval-based systems.
\newblock \emph{Trans. Assoc. Comput. Linguist.}, 11:\penalty0 902--921, August 2023.

\bibitem[Asharov et~al.(2017)Asharov, Halevi, Lindell, and Rabin]{asharov2017privacy}
Gilad Asharov, Shai Halevi, Yehuda Lindell, and Tal Rabin.
\newblock Privacy-preserving search of similar patients in genomic data.
\newblock \emph{Cryptology ePrint Archive}, 2017.

\bibitem[Baevski et~al.(2020)Baevski, Zhou, Mohamed, and Auli]{baevski2020wav2vec}
Alexei Baevski, Yuhao Zhou, Abdelrahman Mohamed, and Michael Auli.
\newblock wav2vec 2.0: A framework for self-supervised learning of speech representations.
\newblock \emph{Advances in neural information processing systems}, 33:\penalty0 12449--12460, 2020.

\bibitem[Bagdasarian et~al.(2024)Bagdasarian, Yi, Ghalebikesabi, Kairouz, Gruteser, Oh, Balle, and Ramage]{bagdasarian2024}
Eugene Bagdasarian, Ren Yi, Sahra Ghalebikesabi, P.~Kairouz, Marco Gruteser, Sewoong Oh, Borja Balle, and Daniel Ramage.
\newblock Airgapagent: Protecting privacy-conscious conversational agents, 2024.

\bibitem[Bai et~al.(2023)Bai, Voelkel, Eichstaedt, and Willer]{Bai2023-vc}
Hui Bai, Jan Voelkel, Johannes Eichstaedt, and Robb Willer.
\newblock Artificial intelligence can persuade humans on political issues.
\newblock September 2023.

\bibitem[Bakker et~al.(2022)Bakker, Chadwick, Sheahan, Tessler, Campbell-Gillingham, Balaguer, McAleese, Glaese, Aslanides, Botvinick, and Summerfield]{Bakker2022FinetuningLM}
Michiel~A. Bakker, Martin Chadwick, Hannah Sheahan, Michael~Henry Tessler, Lucy Campbell-Gillingham, Jan Balaguer, Nathan McAleese, Amelia Glaese, John Aslanides, Matthew~M. Botvinick, and Christopher Summerfield.
\newblock Fine-tuning language models to find agreement among humans with diverse preferences.
\newblock \emph{ArXiv}, abs/2211.15006, 2022.

\bibitem[Balkin(2015)]{balkin2015path}
Jack~M Balkin.
\newblock The path of robotics law.
\newblock \emph{California Law Review Circuit}, 6:\penalty0 45, 2015.

\bibitem[Baruwal~Chhetri et~al.(2024)Baruwal~Chhetri, Tariq, Singh, Jalalvand, Paris, and Nepal]{securitypromptfatigue}
Mohan Baruwal~Chhetri, Shahroz Tariq, Ronal Singh, Fatemeh Jalalvand, Cecile Paris, and Surya Nepal.
\newblock Towards human-ai teaming to mitigate alert fatigue in security operations centres.
\newblock \emph{ACM Trans. Internet Technol.}, 24\penalty0 (3), July 2024.
\newblock ISSN 1533-5399.
\newblock \doi{10.1145/3670009}.
\newblock URL \url{https://doi.org/10.1145/3670009}.

\bibitem[Beam et~al.(2020{\natexlab{a}})Beam, Manrai, and Ghassemi]{Beam2020ChallengesTT}
Andrew Beam, Arjun~K. Manrai, and Marzyeh Ghassemi.
\newblock Challenges to the reproducibility of machine learning models in health care.
\newblock \emph{JAMA}, 2020{\natexlab{a}}.

\bibitem[Beam et~al.(2020{\natexlab{b}})Beam, Manrai, and Ghassemi]{10.1001/jama.2019.20866}
Andrew~L. Beam, Arjun~K. Manrai, and Marzyeh Ghassemi.
\newblock {Challenges to the Reproducibility of Machine Learning Models in Health Care}.
\newblock \emph{JAMA}, 323\penalty0 (4):\penalty0 305--306, 01 2020{\natexlab{b}}.
\newblock ISSN 0098-7484.

\bibitem[Belz et~al.(2023)Belz, Thomson, Reiter, Abercrombie, Alonso-Moral, Arvan, Cheung, Cieliebak, Clark, van Deemter, Dinkar, Dusek, Eger, Fang, Gatt, Gkatzia, Gonz'alez-Corbelle, Hovy, Hurlimann, Ito, Kelleher, Klubicka, Lai, van~der Lee, van Miltenburg, Li, Mahamood, Mieskes, Nissim, Parde, Pl'atek, Rieser, Romero, Tetreault, Toral, Wan, Wanner, Watson, and Yang]{Belz2023MissingIU}
Anya Belz, Craig Thomson, Ehud Reiter, Gavin Abercrombie, Jose~Maria Alonso-Moral, Mohammad Arvan, Jackie Chi~Kit Cheung, Mark Cieliebak, Elizabeth Clark, Kees van Deemter, Tanvi Dinkar, Ondrej Dusek, Steffen Eger, Qixiang Fang, Albert Gatt, Dimitra Gkatzia, Javier Gonz'alez-Corbelle, Dirk Hovy, Manuela Hurlimann, Takumi Ito, John~D. Kelleher, Filip Klubicka, Huiyuan Lai, Chris van~der Lee, Emiel van Miltenburg, Yiru Li, Saad Mahamood, Margot Mieskes, Malvina Nissim, Natalie Parde, Ondvrej Pl'atek, Verena Rieser, Pablo Romero, Joel Tetreault, Antonio Toral, Xiao-Yi Wan, L.~Wanner, Lewis~J. Watson, and Diyi Yang.
\newblock Missing information, unresponsive authors, experimental flaws: The impossibility of assessing the reproducibility of previous human evaluations in nlp.
\newblock \emph{ArXiv}, abs/2305.01633, 2023.

\bibitem[Ben-Sasson et~al.(2014)Ben-Sasson, Chiesa, Garman, Green, Miers, Tromer, and Virza]{zcash}
Eli Ben-Sasson, Alessandro Chiesa, Christina Garman, Matthew Green, Ian Miers, Eran Tromer, and Madars Virza.
\newblock Zerocash: Decentralized anonymous payments from bitcoin.
\newblock \emph{2014 IEEE Symposium on Security and Privacy}, 2014.

\bibitem[Bender and Friedman(2018)]{bender_data_2018}
Emily~M. Bender and Batya Friedman.
\newblock Data {Statements} for {Natural} {Language} {Processing}: {Toward} {Mitigating} {System} {Bias} and {Enabling} {Better} {Science}.
\newblock \emph{Transactions of the Association for Computational Linguistics}, 6:\penalty0 587--604, 2018.
\newblock \doi{10.1162/tacl_a_00041}.
\newblock URL \url{https://aclanthology.org/Q18-1041}.
\newblock Place: Cambridge, MA Publisher: MIT Press.

\bibitem[Bender et~al.(2021)Bender, Gebru, McMillan-Major, and Shmitchell]{bender2021dangers}
Emily~M Bender, Timnit Gebru, Angelina McMillan-Major, and Shmargaret Shmitchell.
\newblock On the dangers of stochastic parrots: Can language models be too big?
\newblock In \emph{Proceedings of the 2021 ACM conference on fairness, accountability, and transparency}, pages 610--623, 2021.

\bibitem[Bengio et~al.(2025)Bengio, Mindermann, Privitera, Besiroglu, Bommasani, Casper, Choi, Fox, Garfinkel, Goldfarb, Heidari, Ho, Kapoor, Khalatbari, Longpre, Manning, Mavroudis, Mazeika, Michael, Newman, Ng, Okolo, Raji, Sastry, Seger, Skeadas, South, Strubell, Tramèr, Velasco, Wheeler, Acemoglu, Adekanmbi, Dalrymple, Dietterich, Felten, Fung, Gourinchas, Heintz, Hinton, Jennings, Krause, Leavy, Liang, Ludermir, Marda, Margetts, McDermid, Munga, Narayanan, Nelson, Neppel, Oh, Ramchurn, Russell, Schaake, Schölkopf, Song, Soto, Tiedrich, Varoquaux, Yao, Zhang, Albalawi, Alserkal, Ajala, Avrin, Busch, de~Leon Ferreira~de Carvalho, Fox, Gill, Hatip, Heikkilä, Jolly, Katzir, Kitano, Krüger, Johnson, Khan, Lee, Ligot, Molchanovskyi, Monti, Mwamanzi, Nemer, Oliver, Portillo, Ravindran, Rivera, Riza, Rugege, Seoighe, Sheehan, Sheikh, Wong, and Zeng]{bengio2025internationalaisafetyreport}
Yoshua Bengio, Sören Mindermann, Daniel Privitera, Tamay Besiroglu, Rishi Bommasani, Stephen Casper, Yejin Choi, Philip Fox, Ben Garfinkel, Danielle Goldfarb, Hoda Heidari, Anson Ho, Sayash Kapoor, Leila Khalatbari, Shayne Longpre, Sam Manning, Vasilios Mavroudis, Mantas Mazeika, Julian Michael, Jessica Newman, Kwan~Yee Ng, Chinasa~T. Okolo, Deborah Raji, Girish Sastry, Elizabeth Seger, Theodora Skeadas, Tobin South, Emma Strubell, Florian Tramèr, Lucia Velasco, Nicole Wheeler, Daron Acemoglu, Olubayo Adekanmbi, David Dalrymple, Thomas~G. Dietterich, Edward~W. Felten, Pascale Fung, Pierre-Olivier Gourinchas, Fredrik Heintz, Geoffrey Hinton, Nick Jennings, Andreas Krause, Susan Leavy, Percy Liang, Teresa Ludermir, Vidushi Marda, Helen Margetts, John McDermid, Jane Munga, Arvind Narayanan, Alondra Nelson, Clara Neppel, Alice Oh, Gopal Ramchurn, Stuart Russell, Marietje Schaake, Bernhard Schölkopf, Dawn Song, Alvaro Soto, Lee Tiedrich, Gaël Varoquaux, Andrew Yao, Ya-Qin Zhang, Fahad Albalawi, Marwan
  Alserkal, Olubunmi Ajala, Guillaume Avrin, Christian Busch, André Carlos~Ponce de~Leon Ferreira~de Carvalho, Bronwyn Fox, Amandeep~Singh Gill, Ahmet~Halit Hatip, Juha Heikkilä, Gill Jolly, Ziv Katzir, Hiroaki Kitano, Antonio Krüger, Chris Johnson, Saif~M. Khan, Kyoung~Mu Lee, Dominic~Vincent Ligot, Oleksii Molchanovskyi, Andrea Monti, Nusu Mwamanzi, Mona Nemer, Nuria Oliver, José Ramón~López Portillo, Balaraman Ravindran, Raquel~Pezoa Rivera, Hammam Riza, Crystal Rugege, Ciarán Seoighe, Jerry Sheehan, Haroon Sheikh, Denise Wong, and Yi~Zeng.
\newblock International ai safety report 2025, January 2025.

\bibitem[Bennett and Mangasarian(1992)]{bennett1992robust}
K.~P. Bennett and O.~L. Mangasarian.
\newblock Robust linear programming discrimination of two linearly inseparable sets.
\newblock \emph{Optimization Methods and Software}, 1:\penalty0 23--34, 1992.

\bibitem[Berke et~al.(2024)Berke, South, Mahari, Larson, and Pentland]{zkTax}
Alex Berke, Tobin South, Robert Mahari, Kent Larson, and Alex Pentland.
\newblock zktax: A pragmatic way to support zero-knowledge tax disclosures.
\newblock In \emph{Proceedings of the 2024 on ACM SIGSAC Conference on Computer and Communications Security}, pages 4952--4954, 2024.

\bibitem[Blodgett et~al.(2020)Blodgett, Barocas, Daum{\'e}~III, and Wallach]{blodgett2020language}
Su~Lin Blodgett, Solon Barocas, Hal Daum{\'e}~III, and Hanna Wallach.
\newblock Language (technology) is power: A critical survey of" bias" in nlp.
\newblock \emph{arXiv preprint arXiv:2005.14050}, 2020.

\bibitem[Bloom and Emery(2022)]{bloom2022}
Cara Bloom and Josiah Emery.
\newblock Privacy expectations for human-autonomous vehicle interactions, 2022.

\bibitem[Bogen and Rieke(2018)]{bogen2018help}
Miranda Bogen and Aaron Rieke.
\newblock Help wanted: An examination of hiring algorithms, equity, and bias.
\newblock 2018.

\bibitem[Bommasani et~al.(2022)Bommasani, Hudson, Adeli, Altman, Arora, von Arx, Bernstein, Bohg, Bosselut, Brunskill, Brynjolfsson, Buch, Card, Castellon, Chatterji, Chen, Creel, Davis, Demszky, Donahue, Doumbouya, Durmus, Ermon, Etchemendy, Ethayarajh, Fei-Fei, Finn, Gale, Gillespie, Goel, Goodman, Grossman, Guha, Hashimoto, Henderson, Hewitt, Ho, Hong, Hsu, Huang, Icard, Jain, Jurafsky, Kalluri, Karamcheti, Keeling, Khani, Khattab, Koh, Krass, Krishna, Kuditipudi, Kumar, Ladhak, Lee, Lee, Leskovec, Levent, Li, Li, Ma, Malik, Manning, Mirchandani, Mitchell, Munyikwa, Nair, Narayan, Narayanan, Newman, Nie, Niebles, Nilforoshan, Nyarko, Ogut, Orr, Papadimitriou, Park, Piech, Portelance, Potts, Raghunathan, Reich, Ren, Rong, Roohani, Ruiz, Ryan, Ré, Sadigh, Sagawa, Santhanam, Shih, Srinivasan, Tamkin, Taori, Thomas, Tramèr, Wang, Wang, Wu, Wu, Wu, Xie, Yasunaga, You, Zaharia, Zhang, Zhang, Zhang, Zhang, Zheng, Zhou, and Liang]{bommasani_opportunities_2022}
Rishi Bommasani, Drew~A. Hudson, Ehsan Adeli, Russ Altman, Simran Arora, Sydney von Arx, Michael~S. Bernstein, Jeannette Bohg, Antoine Bosselut, Emma Brunskill, Erik Brynjolfsson, Shyamal Buch, Dallas Card, Rodrigo Castellon, Niladri Chatterji, Annie Chen, Kathleen Creel, Jared~Quincy Davis, Dora Demszky, Chris Donahue, Moussa Doumbouya, Esin Durmus, Stefano Ermon, John Etchemendy, Kawin Ethayarajh, Li~Fei-Fei, Chelsea Finn, Trevor Gale, Lauren Gillespie, Karan Goel, Noah Goodman, Shelby Grossman, Neel Guha, Tatsunori Hashimoto, Peter Henderson, John Hewitt, Daniel~E. Ho, Jenny Hong, Kyle Hsu, Jing Huang, Thomas Icard, Saahil Jain, Dan Jurafsky, Pratyusha Kalluri, Siddharth Karamcheti, Geoff Keeling, Fereshte Khani, Omar Khattab, Pang~Wei Koh, Mark Krass, Ranjay Krishna, Rohith Kuditipudi, Ananya Kumar, Faisal Ladhak, Mina Lee, Tony Lee, Jure Leskovec, Isabelle Levent, Xiang~Lisa Li, Xuechen Li, Tengyu Ma, Ali Malik, Christopher~D. Manning, Suvir Mirchandani, Eric Mitchell, Zanele Munyikwa, Suraj Nair,
  Avanika Narayan, Deepak Narayanan, Ben Newman, Allen Nie, Juan~Carlos Niebles, Hamed Nilforoshan, Julian Nyarko, Giray Ogut, Laurel Orr, Isabel Papadimitriou, Joon~Sung Park, Chris Piech, Eva Portelance, Christopher Potts, Aditi Raghunathan, Rob Reich, Hongyu Ren, Frieda Rong, Yusuf Roohani, Camilo Ruiz, Jack Ryan, Christopher Ré, Dorsa Sadigh, Shiori Sagawa, Keshav Santhanam, Andy Shih, Krishnan Srinivasan, Alex Tamkin, Rohan Taori, Armin~W. Thomas, Florian Tramèr, Rose~E. Wang, William Wang, Bohan Wu, Jiajun Wu, Yuhuai Wu, Sang~Michael Xie, Michihiro Yasunaga, Jiaxuan You, Matei Zaharia, Michael Zhang, Tianyi Zhang, Xikun Zhang, Yuhui Zhang, Lucia Zheng, Kaitlyn Zhou, and Percy Liang.
\newblock On the opportunities and risks of foundation models, July 2022.
\newblock URL \url{http://arxiv.org/abs/2108.07258}.
\newblock arXiv: 2108.07258 [cs].

\bibitem[Boneh et~al.(2020)Boneh, Drake, Fisch, and Gabizon]{boneh2020halo}
Dan Boneh, Justin Drake, Ben Fisch, and Ariel Gabizon.
\newblock Halo infinite: Recursive zk-snarks from any additive polynomial commitment scheme.
\newblock \emph{Cryptology ePrint Archive}, 2020.

\bibitem[Borge et~al.(2017{\natexlab{a}})Borge, Kokoris-Kogias, Jovanovic, Gasser, Gailly, and Ford]{Borge2017-ti}
Maria Borge, Eleftherios Kokoris-Kogias, Philipp Jovanovic, Linus Gasser, Nicolas Gailly, and Bryan Ford.
\newblock {Proof-of-Personhood}: Redemocratizing permissionless cryptocurrencies.
\newblock In \emph{2017 {IEEE} European Symposium on Security and Privacy Workshops ({EuroS\&PW})}. IEEE, April 2017{\natexlab{a}}.

\bibitem[Borge et~al.(2017{\natexlab{b}})Borge, Kokoris-Kogias, Jovanovic, Gasser, Gailly, and Ford]{borge_proof_personhood_2017}
Maria Borge, Eleftherios Kokoris-Kogias, Philipp Jovanovic, Linus Gasser, Nicolas Gailly, and Bryan Ford.
\newblock {Proof-of-Personhood}: Redemocratizing permissionless cryptocurrencies.
\newblock In \emph{2017 {IEEE} {European} Symposium on Security and Privacy Workshops}, {EUROS\&PW}, pages 23--26. {IEEE} Computer Society, 2017{\natexlab{b}}.
\newblock \doi{10.1109/EuroSPW.2017.46}.
\newblock URL \url{https://www.computer.org/csdl/proceedings-article/euros&pw/2017/07966966/12OmNzdoMiB}.

\bibitem[Bowman and Dahl(2021)]{bowman2021will}
Samuel~R Bowman and George~E Dahl.
\newblock What will it take to fix benchmarking in natural language understanding?
\newblock \emph{arXiv preprint arXiv:2104.02145}, 2021.

\bibitem[Brewster et~al.(2020)Brewster, Nouwt, Raaijmakers, and Verhoosel]{obac}
Christopher Brewster, Barry Nouwt, Stephan Raaijmakers, and Jack Verhoosel.
\newblock Ontology-based access control for fair data.
\newblock \emph{Data Intelligence}, 2\penalty0 (1-2):\penalty0 66--77, 01 2020.
\newblock ISSN 2641-435X.
\newblock \doi{10.1162/dint_a_00029}.
\newblock URL \url{https://doi.org/10.1162/dint\_a\_00029}.

\bibitem[Brown et~al.(2020)Brown, Mann, Ryder, Subbiah, Kaplan, Dhariwal, Neelakantan, Shyam, Sastry, Askell, et~al.]{brown2020language}
Tom Brown, Benjamin Mann, Nick Ryder, Melanie Subbiah, Jared~D Kaplan, Prafulla Dhariwal, Arvind Neelakantan, Pranav Shyam, Girish Sastry, Amanda Askell, et~al.
\newblock Language models are few-shot learners.
\newblock \emph{Advances in neural information processing systems}, 33:\penalty0 1877--1901, 2020.

\bibitem[Buolamwini and Gebru(2018{\natexlab{a}})]{buolamwini2018gender}
Joy Buolamwini and Timnit Gebru.
\newblock Gender shades: Intersectional accuracy disparities in commercial gender classification.
\newblock In \emph{Conference on fairness, accountability and transparency}, pages 77--91. PMLR, 2018{\natexlab{a}}.

\bibitem[Buolamwini and Gebru(2018{\natexlab{b}})]{buolamwini_gender_2018}
Joy Buolamwini and Timnit Gebru.
\newblock Gender shades: Intersectional accuracy disparities in commercial gender classification.
\newblock In \emph{{Proceedings of the 1st Conference on Fairness, Accountability and Transparency}}, Machine Learning Research, pages 77--91. {PMLR}, 2018{\natexlab{b}}.
\newblock URL \url{https://proceedings.mlr.press/v81/buolamwini18a.html}.

\bibitem[Buterin(2023)]{buterin_what_2023}
Vitalik Buterin.
\newblock What do {I} think about biometric proof of personhood?
\newblock Blog, 2023.
\newblock URL \url{https://vitalik.eth.limo/general/2023/07/24/biometric.html}.

\bibitem[C2PA(2023)]{c2pa_c2pa_2023}
C2PA.
\newblock {C2PA} {Technical} {Specification}, 2023.
\newblock URL \url{https://c2pa.org/specifications/specifications/1.3/specs/C2PA_Specification.html#_introduction}.

\bibitem[Camuto et~al.(2023{\natexlab{a}})Camuto, Gănescu, Passerat-Palmbach, and Morton]{danteHoneyBlog}
Alexander Camuto, Bianca Gănescu, Jonathan Passerat-Palmbach, and Jason Morton.
\newblock Honey {I} snarked the {GPT}.
\newblock EZKL Blog, Oct 2023{\natexlab{a}}.

\bibitem[Camuto et~al.(2023{\natexlab{b}})Camuto, Wawrzyniak, and Morton]{ezkl2023gpu}
Alexander Camuto, Sofia Wawrzyniak, and Jason Morton.
\newblock Steps in hardware, leaps in performance.
\newblock EZKL Blog, Nov 2023{\natexlab{b}}.

\bibitem[Camuto et~al.(2023{\natexlab{c}})Camuto, Wawrzyniak, and Morton]{ezkl2023splitting}
Alexander Camuto, Sofia Wawrzyniak, and Jason Morton.
\newblock Splitting and parallelizing proofs.
\newblock EZKL Blog, Oct 2023{\natexlab{c}}.

\bibitem[Camuto et~al.(2023{\natexlab{d}})Camuto, Wawrzyniak, and Morton]{ezkl2023zerooverheadhashing}
Alexander Camuto, Sofia Wawrzyniak, and Jason Morton.
\newblock Removing additional commitment cost.
\newblock EZKL Blog, Oct 2023{\natexlab{d}}.

\bibitem[Carlini et~al.(2021)Carlini, Tram{\`e}r, Wallace, Jagielski, Herbert-Voss, Lee, Roberts, Brown, Song, Erlingsson, Oprea, and Raffel]{Carlini2021u}
Nicholas Carlini, Florian Tram{\`e}r, Eric Wallace, Matthew Jagielski, Ariel Herbert-Voss, Katherine Lee, Adam Roberts, Tom Brown, Dawn Song, {\'U}lfar Erlingsson, Alina Oprea, and Colin Raffel.
\newblock Extracting training data from large language models.
\newblock In \emph{30th {USENIX} security symposium ({USENIX} security 21)}, pages 2633--2650. USENIX Association, August 2021.

\bibitem[Carlini et~al.(2022)Carlini, Ippolito, Jagielski, Lee, Tramer, and Zhang]{Carlini2022m}
Nicholas Carlini, Daphne Ippolito, Matthew Jagielski, Katherine Lee, Florian Tramer, and Chiyuan Zhang.
\newblock Quantifying memorization across neural language models.
\newblock In \emph{11th International Conference on Learning Representations ({ICLR} 2023)}, Kigali, Rwanda, 2022.

\bibitem[Carlini et~al.(2023)Carlini, Hayes, Nasr, Jagielski, Sehwag, Tram{\`e}r, Balle, Ippolito, and Wallace]{Carlini2023k}
Nicolas Carlini, Jamie Hayes, Milad Nasr, Matthew Jagielski, Vikash Sehwag, Florian Tram{\`e}r, Borja Balle, Daphne Ippolito, and Eric Wallace.
\newblock Extracting training data from diffusion models.
\newblock In \emph{32nd {USENIX} security symposium ({USENIX} security 23)}, pages 5253--5270, Anaheim, CA, August 2023. USENIX Association.

\bibitem[Casadesus-Masanell and Spulber(2005)]{casadesus2005trust}
Ramon Casadesus-Masanell and Daniel~F Spulber.
\newblock Trust and incentives in agency.
\newblock \emph{Southern California Interdisciplinary Law Journal}, 15:\penalty0 45, 2005.

\bibitem[Catrina and Saxena(2010)]{catrina2010secure}
Octavian Catrina and Amitabh Saxena.
\newblock Secure computation with fixed-point numbers.
\newblock In \emph{Financial Cryptography and Data Security: 14th International Conference, FC 2010, Tenerife, Canary Islands, January 25-28, 2010, Revised Selected Papers 14}, pages 35--50. Springer, 2010.

\bibitem[Chan et~al.(2023)Chan, Salganik, Markelius, Pang, Rajkumar, Krasheninnikov, Langosco, He, Duan, Carroll, et~al.]{chan2023harms}
Alan Chan, Rebecca Salganik, Alva Markelius, Chris Pang, Nitarshan Rajkumar, Dmitrii Krasheninnikov, Lauro Langosco, Zhonghao He, Yawen Duan, Micah Carroll, et~al.
\newblock Harms from increasingly agentic algorithmic systems.
\newblock In \emph{Proceedings of the 2023 ACM Conference on Fairness, Accountability, and Transparency}, pages 651--666, 2023.

\bibitem[Chan et~al.(2024{\natexlab{a}})Chan, Ezell, Kaufmann, Wei, Hammond, Bradley, Bluemke, Rajkumar, Krueger, Kolt, et~al.]{chan2024visibility}
Alan Chan, Carson Ezell, Max Kaufmann, Kevin Wei, Lewis Hammond, Herbie Bradley, Emma Bluemke, Nitarshan Rajkumar, David Krueger, Noam Kolt, et~al.
\newblock Visibility into ai agents.
\newblock In \emph{The 2024 ACM Conference on Fairness, Accountability, and Transparency}, pages 958--973, 2024{\natexlab{a}}.

\bibitem[Chan et~al.(2024{\natexlab{b}})Chan, Kolt, Wills, Anwar, de~Witt, Rajkumar, Hammond, Krueger, Heim, and Anderljung]{chan2024ids}
Alan Chan, Noam Kolt, Peter Wills, Usman Anwar, Christian~Schroeder de~Witt, Nitarshan Rajkumar, Lewis Hammond, David Krueger, Lennart Heim, and Markus Anderljung.
\newblock Ids for ai systems.
\newblock \emph{arXiv preprint arXiv:2406.12137}, 2024{\natexlab{b}}.

\bibitem[Chang et~al.(2024)Chang, Wang, Wang, Wu, Yang, Zhu, Chen, Yi, Wang, Wang, et~al.]{chang2024survey}
Yupeng Chang, Xu~Wang, Jindong Wang, Yuan Wu, Linyi Yang, Kaijie Zhu, Hao Chen, Xiaoyuan Yi, Cunxiang Wang, Yidong Wang, et~al.
\newblock A survey on evaluation of large language models.
\newblock \emph{ACM Transactions on Intelligent Systems and Technology}, 15\penalty0 (3):\penalty0 1--45, 2024.

\bibitem[Chavis and Wandersman(1990)]{chavis1990sense}
David~M Chavis and Abraham Wandersman.
\newblock Sense of community in the urban environment: A catalyst for participation and community development.
\newblock \emph{American journal of community psychology}, 18\penalty0 (1):\penalty0 55--81, 1990.

\bibitem[Chen et~al.(2024)Chen, Waiwitlikhit, Stoica, and Kang]{chen2024zkml}
Bing-Jyue Chen, Suppakit Waiwitlikhit, Ion Stoica, and Daniel Kang.
\newblock Zkml: An optimizing system for ml inference in zero-knowledge proofs.
\newblock In \emph{Proceedings of the Nineteenth European Conference on Computer Systems}, pages 560--574, 2024.

\bibitem[Chen et~al.(2020)Chen, Chillotti, Dong, Poburinnaya, Razenshteyn, and Riazi]{chen2020sanns}
Hao Chen, Ilaria Chillotti, Yihe Dong, Oxana Poburinnaya, Ilya Razenshteyn, and M~Sadegh Riazi.
\newblock $\{$SANNS$\}$: Scaling up secure approximate $\{$k-Nearest$\}$ neighbors search.
\newblock In \emph{29th USENIX Security Symposium (USENIX Security 20)}, pages 2111--2128, 2020.

\bibitem[Chen et~al.(2023{\natexlab{a}})Chen, Zaharia, and Zou]{chen2023chatgpt}
Lingjiao Chen, Matei Zaharia, and James Zou.
\newblock How is chatgpt's behavior changing over time?
\newblock \emph{arXiv preprint arXiv:2307.09009}, 2023{\natexlab{a}}.

\bibitem[Chen et~al.(2021)Chen, Tworek, Jun, Yuan, de~Oliveira~Pinto, Kaplan, Edwards, Burda, Joseph, Brockman, Ray, Puri, Krueger, Petrov, Khlaaf, Sastry, Mishkin, Chan, Gray, Ryder, Pavlov, Power, Kaiser, Bavarian, Winter, Tillet, Such, Cummings, Plappert, Chantzis, Barnes, Herbert-Voss, Guss, Nichol, Paino, Tezak, Tang, Babuschkin, Balaji, Jain, Saunders, Hesse, Carr, Leike, Achiam, Misra, Morikawa, Radford, Knight, Brundage, Murati, Mayer, Welinder, McGrew, Amodei, McCandlish, Sutskever, and Zaremba]{chen2021evaluating}
Mark Chen, Jerry Tworek, Heewoo Jun, Qiming Yuan, Henrique~Ponde de~Oliveira~Pinto, Jared Kaplan, Harri Edwards, Yuri Burda, Nicholas Joseph, Greg Brockman, Alex Ray, Raul Puri, Gretchen Krueger, Michael Petrov, Heidy Khlaaf, Girish Sastry, Pamela Mishkin, Brooke Chan, Scott Gray, Nick Ryder, Mikhail Pavlov, Alethea Power, Lukasz Kaiser, Mohammad Bavarian, Clemens Winter, Philippe Tillet, Felipe~Petroski Such, Dave Cummings, Matthias Plappert, Fotios Chantzis, Elizabeth Barnes, Ariel Herbert-Voss, William~Hebgen Guss, Alex Nichol, Alex Paino, Nikolas Tezak, Jie Tang, Igor Babuschkin, Suchir Balaji, Shantanu Jain, William Saunders, Christopher Hesse, Andrew~N. Carr, Jan Leike, Josh Achiam, Vedant Misra, Evan Morikawa, Alec Radford, Matthew Knight, Miles Brundage, Mira Murati, Katie Mayer, Peter Welinder, Bob McGrew, Dario Amodei, Sam McCandlish, Ilya Sutskever, and Wojciech Zaremba.
\newblock Evaluating large language models trained on code, 2021.

\bibitem[Chen et~al.(2022)Chen, Bao, Huang, Dong, Jiao, Jiang, Zhou, Li, and Wei]{chen2022x}
Tianyu Chen, Hangbo Bao, Shaohan Huang, Li~Dong, Binxing Jiao, Daxin Jiang, Haoyi Zhou, Jianxin Li, and Furu Wei.
\newblock The-x: Privacy-preserving transformer inference with homomorphic encryption.
\newblock \emph{arXiv preprint arXiv:2206.00216}, 2022.

\bibitem[Chen et~al.(2023{\natexlab{b}})Chen, Mendes, Das, Xu, and Ritter]{Chen2023v}
Yang Chen, Ethan Mendes, Sauvik Das, Wei Xu, and Alan Ritter.
\newblock Can language models be instructed to protect personal information?
\newblock October 2023{\natexlab{b}}.

\bibitem[Cheng et~al.(2021)]{Cheng2021}
Victoria Cheng et~al.
\newblock Can you fake it until you make it? impacts of differentially private synthetic data on downstream classification fairness.
\newblock In \emph{Proceedings of the 2021 ACM Conference on Fairness, Accountability, and Transparency}, FAccT '21, pages 149--160, 3 2021.
\newblock \doi{10.1145/3442188.3445879}.
\newblock URL \url{https://doi.org/10.1145/3442188.3445879}.

\bibitem[Chida et~al.(2018)Chida, Genkin, Hamada, Ikarashi, Kikuchi, Lindell, and Nof]{chida2018fast}
Koji Chida, Daniel Genkin, Koki Hamada, Dai Ikarashi, Ryo Kikuchi, Yehuda Lindell, and Ariel Nof.
\newblock Fast large-scale honest-majority mpc for malicious adversaries.
\newblock In \emph{Advances in Cryptology--CRYPTO 2018: 38th Annual International Cryptology Conference, Santa Barbara, CA, USA, August 19--23, 2018, Proceedings, Part III 38}, pages 34--64. Springer, 2018.

\bibitem[{Civil Resolution Tribunal (British Columbia)}(2024)]{BCCRT2024}
{Civil Resolution Tribunal (British Columbia)}.
\newblock Patel v. wong, 2024 bccrt 149, 2024.
\newblock URL \url{https://www.canlii.org/en/bc/bccrt/doc/2024/2024bccrt149/2024bccrt149.html}.
\newblock Accessed: 2025-01-06.

\bibitem[Corbett-Davies and Goel(2018)]{CorbettDavies2018TheMA}
Sam Corbett-Davies and Sharad Goel.
\newblock The measure and mismeasure of fairness: A critical review of fair machine learning.
\newblock \emph{ArXiv}, abs/1808.00023, 2018.

\bibitem[Cummings et~al.(2024)]{Cummings2024}
Rachel Cummings et~al.
\newblock Advancing differential privacy: Where we are now and future directions for real-world deployment.
\newblock \emph{Harvard Data Science Review}, 6\penalty0 (1), 1 2024.
\newblock \doi{10.1162/99608f92.d3197524}.
\newblock URL \url{https://doi.org/10.1162/99608f92.d3197524}.

\bibitem[Dagher et~al.(2015)Dagher, B{\"u}nz, Bonneau, Clark, and Boneh]{provisions}
Gaby~G. Dagher, Benedikt B{\"u}nz, Joseph Bonneau, Jeremy Clark, and Dan Boneh.
\newblock Provisions: Privacy-preserving proofs of solvency for bitcoin exchanges.
\newblock \emph{Proceedings of the 22nd ACM SIGSAC Conference on Computer and Communications Security}, 2015.

\bibitem[Damg{\aa}rd and Nielsen(2007)]{damgaard2007scalable}
Ivan Damg{\aa}rd and Jesper~Buus Nielsen.
\newblock Scalable and unconditionally secure multiparty computation.
\newblock In \emph{Annual International Cryptology Conference}, pages 572--590. Springer, 2007.

\bibitem[Damg{\aa}rd et~al.(2013)Damg{\aa}rd, Keller, Larraia, Pastro, Scholl, and Smart]{damgaard2013practical}
Ivan Damg{\aa}rd, Marcel Keller, Enrique Larraia, Valerio Pastro, Peter Scholl, and Nigel~P Smart.
\newblock Practical covertly secure mpc for dishonest majority--or: breaking the spdz limits.
\newblock In \emph{Computer Security--ESORICS 2013: 18th European Symposium on Research in Computer Security, Egham, UK, September 9-13, 2013. Proceedings 18}, pages 1--18. Springer, 2013.

\bibitem[Deng(2012)]{deng2012mnist}
Li~Deng.
\newblock The mnist database of handwritten digit images for machine learning research.
\newblock \emph{IEEE Signal Processing Magazine}, 29\penalty0 (6):\penalty0 141--142, 2012.

\bibitem[Derei(2023)]{derei2023accelerating}
Tal Derei.
\newblock \emph{Accelerating the PlonK zkSNARK Proving System using GPU Architectures}.
\newblock PhD thesis, Lehigh University, 2023.

\bibitem[Desai and Kroll(2017)]{desai2017trust}
Deven~R Desai and Joshua~A Kroll.
\newblock Trust but verify: A guide to algorithms and the law.
\newblock \emph{Harv. JL \& Tech.}, 31:\penalty0 1, 2017.

\bibitem[Dong et~al.(2023)Dong, jie Lu, Zheng, Wu, Zhao, Tan, Huang, Hong, Wei, and Cheng]{PUMA}
Ye~Dong, Wen jie Lu, Yancheng Zheng, Haoqi Wu, Derun Zhao, Jin Tan, Zhicong Huang, Cheng Hong, Tao Wei, and Wen-Chang Cheng.
\newblock Puma: Secure inference of llama-7b in five minutes.
\newblock \emph{ArXiv}, abs/2307.12533, 2023.

\bibitem[Drean et~al.(2023)Drean, Gomez-Garcia, Bourgeat, and Devadas]{Drean2023CitadelEW}
Jules Drean, Miguel Gomez-Garcia, Thomas Bourgeat, and Srinivas Devadas.
\newblock Citadel: Enclaves with strong microarchitectural isolation and secure shared memory on a speculative out-of-order processor.
\newblock 2023.

\bibitem[Duan et~al.(2024)Duan, Suri, Mireshghallah, Min, Shi, Zettlemoyer, Tsvetkov, Choi, Evans, and Hajishirzi]{Duan2024q}
Michael Duan, Anshuman Suri, Niloofar Mireshghallah, Sewon Min, Weijia Shi, Luke Zettlemoyer, Yulia Tsvetkov, Yejin Choi, David Evans, and Hannaneh Hajishirzi.
\newblock Do membership inference attacks work on large language models?
\newblock February 2024.

\bibitem[Eagen and Gabizon(2023)]{cryptoeprint:2023/393}
Liam Eagen and Ariel Gabizon.
\newblock cqlin: Efficient linear operations on kzg commitments with cached quotients.
\newblock Cryptology ePrint Archive, Paper 2023/393, 2023.

\bibitem[Egan and Heim(2023)]{egan_oversight_2023}
Janet Egan and Lennart Heim.
\newblock Oversight for {Frontier} {AI} through a {Know}-{Your}-{Customer} {Scheme} for {Compute} {Providers}, October 2023.
\newblock URL \url{http://arxiv.org/abs/2310.13625}.
\newblock arXiv:2310.13625 [cs].

\bibitem[Eloundou et~al.(2023)Eloundou, Manning, Mishkin, and Rock]{Eloundou2023GPTsAG}
Tyna Eloundou, Sam Manning, Pamela Mishkin, and Daniel Rock.
\newblock Gpts are gpts: An early look at the labor market impact potential of large language models.
\newblock \emph{ArXiv}, 2023.

\bibitem[Escudero et~al.(2020)Escudero, Ghosh, Keller, Rachuri, and Scholl]{escudero2020improved}
Daniel Escudero, Satrajit Ghosh, Marcel Keller, Rahul Rachuri, and Peter Scholl.
\newblock Improved primitives for mpc over mixed arithmetic-binary circuits.
\newblock In \emph{Advances in Cryptology--CRYPTO 2020: 40th Annual International Cryptology Conference, CRYPTO 2020, Santa Barbara, CA, USA, August 17--21, 2020, Proceedings, Part II 40}, pages 823--852. Springer, 2020.

\bibitem[{European Commission}(2021{\natexlab{a}})]{EUAIACT}
{European Commission}.
\newblock Proposal for a regulation of the european parliament and of the council laying down harmonised rules on artificial intelligence (artificial intelligence act) and amending certain union legislative acts.
\newblock Legislative proposal, European Commission, 4 2021{\natexlab{a}}.
\newblock URL \url{https://eur-lex.europa.eu/legal-content/EN/TXT/?uri=CELEX%3A52021PC0206}.

\bibitem[{European Commission}(2021{\natexlab{b}})]{eu2023aiact}
{European Commission}.
\newblock Artificial intelligence act, 4 2021{\natexlab{b}}.

\bibitem[{European Commission}(2021{\natexlab{c}})]{eu_ai_act}
{European Commission}.
\newblock {Proposal for a Regulation of the European Parliament and of the Council Laying Down Harmonised Rules on Artificial Intelligence (Artificial Intelligence Act) and amending certain Union legislative acts}.
\newblock COM(2021) 206 final, 2021/0106 (COD), 2021{\natexlab{c}}.
\newblock \url{https://eur-lex.europa.eu/legal-content/EN/TXT/?uri=CELEX:52021PC0206} (accessed 6 January 2025).

\bibitem[{European Data Protection Board}(2024)]{EuropeanDataProtectionBoard2024j}
{European Data Protection Board}.
\newblock Report of the work undertaken by the {ChatGPT} taskforce.
\newblock Technical report, EDPB, May 2024.

\bibitem[Fan et~al.(2024)Fan, Li, Deng, Wang, and Song]{fan2024}
Wei Fan, Haoran Li, Zheye Deng, Weiqi Wang, and Yangqiu Song.
\newblock Goldcoin: Grounding large language models in privacy laws via contextual integrity theory, 2024.

\bibitem[Fan et~al.(2021)Fan, Bai, Lei, Lin, Hu, Wu, Guo, and Tan]{fan2021ppmck}
Yongkai Fan, Jianrong Bai, Xia Lei, Weiguo Lin, Qian Hu, Guodong Wu, Jiaming Guo, and Gang Tan.
\newblock Ppmck: Privacy-preserving multi-party computing for k-means clustering.
\newblock \emph{Journal of Parallel and Distributed Computing}, 154:\penalty0 54--63, 2021.

\bibitem[Fang et~al.(2024)Fang, Bindu, Gupta, and Kang]{fang_agents_2024}
Richard Fang, Rohan Bindu, Akul Gupta, and Daniel Kang.
\newblock {LLM} agents can autonomously hack websites, 2024.
\newblock URL \url{https://arxiv.org/abs/2402.06664}.

\bibitem[{Federal Trade Commission}(2023)]{FederalTradeCommission2023x}
{Federal Trade Commission}.
\newblock {FTC} says ring employees illegally surveilled customers, failed to stop hackers from taking control of users' cameras.
\newblock \url{https://www.ftc.gov/news-events/news/press-releases/2023/05/ftc-says-ring-employees-illegally-surveilled-customers-failed-stop-hackers-taking-control-users}, May 2023.
\newblock Accessed: 2024-11-28.

\bibitem[{Federal Trade Commission}(2024)]{FederalTradeCommission2024n}
{Federal Trade Commission}.
\newblock {FTC} staff report finds large social media and video streaming companies have engaged in vast surveillance of users with lax privacy controls and inadequate safeguards for kids and teens.
\newblock \url{https://www.ftc.gov/news-events/news/press-releases/2024/09/ftc-staff-report-finds-large-social-media-video-streaming-companies-have-engaged-vast-surveillance}, September 2024.
\newblock Accessed: 2024-11-28.

\bibitem[Fedus et~al.(2022)Fedus, Zoph, and Shazeer]{fedus2022switch}
William Fedus, Barret Zoph, and Noam Shazeer.
\newblock Switch transformers: Scaling to trillion parameter models with simple and efficient sparsity.
\newblock \emph{The Journal of Machine Learning Research}, 23\penalty0 (1):\penalty0 5232--5270, 2022.

\bibitem[Feng et~al.(2021)Feng, Qin, Zhang, Ding, and Chu]{Feng2021ZENAO}
Boyuan Feng, Lianke Qin, Zhenfei Zhang, Yufei Ding, and Shumo Chu.
\newblock Zen: An optimizing compiler for verifiable, zero-knowledge neural network inferences.
\newblock 2021.

\bibitem[Fielding(2000)]{Fielding-2000-Thesis}
Roy~T. Fielding.
\newblock {Architectural Styles and the Design of Network-based Software Architectures}.
\newblock Doctoral thesis, University of California at Irvine, June 2000.
\newblock URL \url{https://www.ics.uci.edu/~fielding/pubs/dissertation/fielding_dissertation.pdf}.

\bibitem[Finin et~al.(2008)Finin, Joshi, Kagal, Niu, Sandhu, Winsborough, and Thuraisingham]{rowlbac}
T.~Finin, A.~Joshi, L.~Kagal, J.~Niu, R.~Sandhu, W.~Winsborough, and B.~Thuraisingham.
\newblock Rowlbac: representing role based access control in owl.
\newblock In \emph{Proceedings of the 13th ACM Symposium on Access Control Models and Technologies}, SACMAT '08, page 73–82, New York, NY, USA, 2008. Association for Computing Machinery.
\newblock ISBN 9781605581293.
\newblock \doi{10.1145/1377836.1377849}.
\newblock URL \url{https://doi.org/10.1145/1377836.1377849}.

\bibitem[Forrest(2024)]{forrest_ethics_2024}
Katherine~B. Forrest.
\newblock The ethics and challenges of legal personhood for {AI}.
\newblock \emph{Yale Law Journal}, 133, 2024.
\newblock URL \url{https://www.yalelawjournal.org/pdf/ForrestYLJForumEssay_at8hdu63.pdf}.

\bibitem[Foulds et~al.(2020)Foulds, Islam, Keya, and Pan]{foulds2020intersectional}
James~R Foulds, Rashidul Islam, Kamrun~Naher Keya, and Shimei Pan.
\newblock An intersectional definition of fairness.
\newblock In \emph{2020 IEEE 36th International Conference on Data Engineering (ICDE)}, pages 1918--1921. IEEE, 2020.

\bibitem[Fourney et~al.(2024)Fourney, Bansal, Mozannar, Tan, Salinas, Niedtner, Proebsting, Bassman, Gerrits, Alber, et~al.]{fourney2024magentic}
Adam Fourney, Gagan Bansal, Hussein Mozannar, Cheng Tan, Eduardo Salinas, Friederike Niedtner, Grace Proebsting, Griffin Bassman, Jack Gerrits, Jacob Alber, et~al.
\newblock Magentic-one: A generalist multi-agent system for solving complex tasks.
\newblock \emph{arXiv preprint arXiv:2411.04468}, 2024.

\bibitem[Fredrikson et~al.(2015)Fredrikson, Jha, and Ristenpart]{Fredrikson2015i}
Matt Fredrikson, Somesh Jha, and Thomas Ristenpart.
\newblock Model inversion attacks that exploit confidence information and basic countermeasures.
\newblock In \emph{Proceedings of the 22nd {ACM} {SIGSAC} Conference on Computer and Communications Security ({CCS} '15)}, pages 1322--1333, New York, NY, USA, October 2015. Association for Computing Machinery.

\bibitem[Gabriel et~al.(2024)Gabriel, Manzini, Keeling, Hendricks, Rieser, Iqbal, Toma{\v s}ev, Ktena, Kenton, Rodriguez, El-Sayed, Brown, Akbulut, Trask, Hughes, Stevie~Bergman, Shelby, Marchal, Griffin, Mateos-Garcia, Weidinger, Street, Lange, Ingerman, Lentz, Enger, Barakat, Krakovna, Siy, Kurth-Nelson, McCroskery, Bolina, Law, Shanahan, Alberts, Balle, de~Haas, Ibitoye, Dafoe, Goldberg, Krier, Reese, Witherspoon, Hawkins, Rauh, Wallace, Franklin, Goldstein, Lehman, Klenk, Vallor, Biles, Morris, King, Ag{\"u}era~y Arcas, Isaac, and Manyika]{Gabriel2024c}
Iason Gabriel, Arianna Manzini, Geoff Keeling, Lisa~Anne Hendricks, Verena Rieser, Hasan Iqbal, Nenad Toma{\v s}ev, Ira Ktena, Zachary Kenton, Mikel Rodriguez, Seliem El-Sayed, Sasha Brown, Canfer Akbulut, Andrew Trask, Edward Hughes, A~Stevie~Bergman, Renee Shelby, Nahema Marchal, Conor Griffin, Juan Mateos-Garcia, Laura Weidinger, Winnie Street, Benjamin Lange, Alex Ingerman, Alison Lentz, Reed Enger, Andrew Barakat, Victoria Krakovna, John~Oliver Siy, Zeb Kurth-Nelson, Amanda McCroskery, Vijay Bolina, Harry Law, Murray Shanahan, Lize Alberts, Borja Balle, Sarah de~Haas, Yetunde Ibitoye, Allan Dafoe, Beth Goldberg, S{\'e}bastien Krier, Alexander Reese, Sims Witherspoon, Will Hawkins, Maribeth Rauh, Don Wallace, Matija Franklin, Josh~A Goldstein, Joel Lehman, Michael Klenk, Shannon Vallor, Courtney Biles, Meredith~Ringel Morris, Helen King, Blaise Ag{\"u}era~y Arcas, William Isaac, and James Manyika.
\newblock The ethics of advanced {AI} assistants.
\newblock Technical report, Google DeepMind, April 2024.

\bibitem[Ganescu and Passerat-Palmbach(2024)]{ganescu2024trust}
Bianca-Mihaela Ganescu and Jonathan Passerat-Palmbach.
\newblock Trust the process: Zero-knowledge machine learning to enhance trust in generative ai interactions.
\newblock \emph{arXiv preprint arXiv:2402.06414}, 2024.

\bibitem[Garg et~al.(2023)Garg, Goel, Jha, Mahloujifar, Mahmoody, Policharla, and Wang]{garg2023experimenting}
Sanjam Garg, Aarushi Goel, Somesh Jha, Saeed Mahloujifar, Mohammad Mahmoody, Guru-Vamsi Policharla, and Mingyuan Wang.
\newblock Experimenting with zero-knowledge proofs of training.
\newblock In \emph{Proceedings of the 2023 ACM SIGSAC Conference on Computer and Communications Security}, pages 1880--1894, 2023.

\bibitem[Garner(2019)]{blackslawdict}
Bryan~A. Garner, editor.
\newblock \emph{Black's Law Dictionary}.
\newblock Thomson Reuters, St. Paul, MN, 11th edition, 2019.

\bibitem[Gebru et~al.(2021)Gebru, Morgenstern, Vecchione, Vaughan, Wallach, Iii, and Crawford]{gebru_datasheets_2021}
Timnit Gebru, Jamie Morgenstern, Briana Vecchione, Jennifer~Wortman Vaughan, Hanna Wallach, Hal~Daumé Iii, and Kate Crawford.
\newblock Datasheets for datasets.
\newblock \emph{Communications of the ACM}, 64\penalty0 (12):\penalty0 86--92, December 2021.
\newblock ISSN 0001-0782, 1557-7317.
\newblock \doi{10.1145/3458723}.
\newblock URL \url{https://dl.acm.org/doi/10.1145/3458723}.

\bibitem[{Gemini Team} et~al.(2023){Gemini Team}, Anil, Borgeaud, Alayrac, Yu, Soricut, Schalkwyk, Dai, Hauth, Millican, Silver, Johnson, Antonoglou, Schrittwieser, Glaese, Chen, Pitler, Lillicrap, Lazaridou, Firat, Molloy, Isard, Barham, Hennigan, Lee, Viola, Reynolds, Xu, Doherty, Collins, Meyer, Rutherford, Moreira, Ayoub, Goel, Krawczyk, Du, Chi, Cheng, Ni, Shah, Kane, Chan, Faruqui, Severyn, Lin, Li, Cheng, Ittycheriah, Mahdieh, Chen, Sun, Tran, Bagri, Lakshminarayanan, Liu, Orban, G{\"u}ra, Zhou, Song, Boffy, Ganapathy, Zheng, Choe, Weisz, Zhu, Lu, Gopal, Kahn, Kula, Pitman, Shah, Taropa, Merey, Baeuml, Chen, Shafey, Zhang, Sercinoglu, Tucker, Piqueras, Krikun, Barr, Savinov, Danihelka, Roelofs, White, Andreassen, von Glehn, Yagati, Kazemi, Gonzalez, Khalman, Sygnowski, Frechette, Smith, Culp, Proleev, Luan, Chen, Lottes, Schucher, Lebron, Rrustemi, Clay, Crone, Kocisky, Zhao, Perz, Yu, Howard, Bloniarz, Rae, Lu, Sifre, Maggioni, Alcober, Garrette, Barnes, Thakoor, Austin, Barth-Maron, Wong, Joshi,
  Chaabouni, Fatiha, Ahuja, Tomar, Senter, Chadwick, Kornakov, Attaluri, Iturrate, Liu, Li, Cogan, Chen, Jia, Gu, Zhang, Grimstad, Hartman, Garcia, Pillai, Devlin, Laskin, Casas, Valter, Tao, Blanco, Badia, Reitter, Chen, Brennan, Rivera, Brin, Iqbal, Surita, Labanowski, Rao, Winkler, Parisotto, Gu, Olszewska, Addanki, Miech, Louis, Teplyashin, Brown, Catt, Balaguer, Xiang, Wang, Ashwood, Briukhov, Webson, Ganapathy, Sanghavi, Kannan, Chang, Stjerngren, Djolonga, Sun, Bapna, Aitchison, Pejman, Michalewski, Yu, Wang, Love, Ahn, Bloxwich, Han, Humphreys, Sellam, Bradbury, Godbole, Samangooei, Damoc, Kaskasoli, Arnold, Vasudevan, Agrawal, Riesa, Lepikhin, Tanburn, Srinivasan, Lim, Hodkinson, Shyam, Ferret, Hand, Garg, Paine, Li, Li, Giang, Neitz, Abbas, York, Reid, Cole, Chowdhery, Das, Rogozi{\'n}ska, Nikolaev, Sprechmann, Nado, Zilka, Prost, He, Monteiro, Mishra, Welty, Newlan, Jia, Allamanis, Hu, de~Liedekerke, Gilmer, Saroufim, Rijhwani, Hou, Shrivastava, Baddepudi, Goldin, Ozturel, Cassirer, Xu, Sohn,
  Sachan, Amplayo, Swanson, Petrova, Narayan, Guez, Brahma, Landon, Patel, Zhao, Villela, Wang, Jia, Rahtz, Gim{\'e}nez, Yeung, Keeling, Georgiev, Mincu, Wu, Haykal, Saputro, Vodrahalli, Qin, Cankara, Sharma, Fernando, Hawkins, Neyshabur, Kim, Hutter, Agrawal, Castro-Ros, van~den Driessche, Wang, Yang, Chang, Komarek, McIlroy, Lu{\v c}i{\'c}, Zhang, Farhan, Sharman, Natsev, Michel, Bansal, Qiao, Cao, Shakeri, Butterfield, Chung, Rubenstein, Agrawal, Mensch, Soparkar, Lenc, Chung, Pope, Maggiore, Kay, Jhakra, Wang, Maynez, Phuong, Tobin, Tacchetti, Trebacz, Robinson, Katariya, Riedel, Bailey, Xiao, Ghelani, Aroyo, Slone, Houlsby, Xiong, Yang, Gribovskaya, Adler, Wirth, Lee, {Li, Music}, Kagohara, Pavagadhi, Bridgers, Bortsova, Ghemawat, Ahmed, Liu, Powell, Bolina, Iinuma, Zablotskaia, Besley, Chung, Dozat, Comanescu, Si, Greer, Su, Polacek, Kaufman, Tokumine, Hu, Buchatskaya, Miao, Elhawaty, Siddhant, Tomasev, Xing, Greer, Miller, Ashraf, Roy, Zhang, Ma, Filos, Besta, Blevins, Klimenko, Yeh, Changpinyo, Mu,
  Chang, Pajarskas, Muir, Cohen, Lan, Haridasan, Marathe, Hansen, Douglas, Samuel, Wang, Austin, Lan, Jiang, Chiu, Lorenzo, Sj{\"o}sund, Cevey, Gleicher, Avrahami, Boral, Srinivasan, Selo, May, Aisopos, Hussenot, Soares, Baumli, Chang, Recasens, Caine, Pritzel, Pavetic, Pardo, Gergely, Frye, Ramasesh, Horgan, Badola, Kassner, Roy, Dyer, Campos, Tomala, Tang, Badawy, White, Mustafa, Lang, Jindal, Vikram, Gong, Caelles, Hemsley, Thornton, Feng, Stokowiec, Zheng, Thacker, {\"U}nl{\"u}, Zhang, Saleh, Svensson, Bileschi, Patil, Anand, Ring, Tsihlas, Vezer, Selvi, Shevlane, Rodriguez, Kwiatkowski, Daruki, Rong, Dafoe, FitzGerald, Gu-Lemberg, Khan, Hendricks, Pellat, Feinberg, Cobon-Kerr, Sainath, Rauh, Hashemi, Ives, Hasson, Noland, Cao, Byrd, Hou, Wang, Sottiaux, Paganini, Lespiau, Moufarek, Hassan, Shivakumar, van Amersfoort, Mandhane, Joshi, Goyal, Tung, Brock, Sheahan, Misra, Li, Raki{\'c}evi{\'c}, Dehghani, Liu, Mittal, Oh, Noury, Sezener, Huot, Lamm, De~Cao, Chen, Mudgal, Stella, Brooks, Vasudevan, Liu,
  Chain, Melinkeri, Cohen, Wang, Seymore, Zubkov, Goel, Yue, Krishnakumaran, Albert, Hurley, Sano, Mohananey, Joughin, Filonov, K{\k e}pa, Eldawy, Lim, Rishi, Badiezadegan, Bos, Chang, Jain, Padmanabhan, Puttagunta, Krishna, Baker, Kalb, Bedapudi, Kurzrok, Lei, Yu, Litvin, Zhou, Wu, Sobell, Siciliano, Papir, Neale, Bragagnolo, Toor, Chen, Anklin, Wang, Feng, Gholami, Ling, Liu, Walter, Moghaddam, Kishore, Adamek, Mercado, Mallinson, Wandekar, Cagle, Ofek, Garrido, Lombriser, Mukha, Sun, Mohammad, Matak, Qian, Peswani, Janus, Yuan, Schelin, David, Garg, He, Duzhyi, {\"A}lgmyr, Lottaz, Li, Yadav, Xu, Chinien, Shivanna, Chuklin, Li, Spadine, Wolfe, Mohamed, Das, Dai, He, von Dincklage, Upadhyay, Maurya, Chi, Krause, Salama, Rabinovitch, {M, Pavan Kumar Reddy}, Selvan, Dektiarev, Ghiasi, Guven, Gupta, Liu, Sharma, Shtacher, Paul, Akerlund, Aubet, Huang, Zhu, Zhu, Teixeira, Fritze, Bertolini, Marinescu, B{\"o}lle, Paulus, Gupta, Latkar, Chang, Sanders, Wilson, Wu, Tan, Thiet, Doshi, Lall, Mishra, Chen, Luong,
  Benjamin, Lee, Andrejczuk, Rabiej, Ranjan, Styrc, Yin, Simon, Harriott, Bansal, Robsky, Bacon, Greene, Mirylenka, Zhou, Sarvana, Goyal, Andermatt, Siegler, Horn, Israel, Pongetti, Chen, Selvatici, Silva, Wang, Tolins, Guu, Yogev, Cai, Agostini, Shah, Nguyen, Donnaile, Pereira, Friso, Stambler, Kurzrok, Kuang, Romanikhin, Geller, Yan, Jang, Lee, Fica, Malmi, Tan, Banica, Balle, Pham, Huang, Avram, Shi, Singh, Hidey, Ahuja, Saxena, Dooley, Potharaju, O'Neill, Gokulchandran, Foley, Zhao, Dusenberry, Liu, Mehta, Kotikalapudi, Safranek-Shrader, Goodman, Kessinger, Globen, Kolhar, Gorgolewski, Ibrahim, Song, Eichenbaum, Brovelli, Potluri, Lahoti, Baetu, Ghorbani, Chen, Crawford, Pal, Sridhar, Gurita, Mujika, Petrovski, Cedoz, Li, Chen, Santo, Goyal, Punjabi, Kappaganthu, Kwak, Lv, Velury, Choudhury, Hall, Shah, Figueira, Thomas, Lu, Zhou, Kumar, Jurdi, Chikkerur, Ma, Yu, Kwak, {\"A}hdel, Rajayogam, Choma, Liu, Barua, Ji, Park, Hellendoorn, Bailey, Bilal, Zhou, Khatir, Sutton, Rzadkowski, Macintosh, Shagin,
  Medina, Liang, Zhou, Shah, Bi, Dankovics, Banga, Lehmann, Bredesen, Lin, Hoffmann, Lai, Chung, Yang, Balani, Bra{\v z}inskas, Sozanschi, Hayes, Alcalde, Makarov, Chen, Stella, Snijders, Mandl, K{\"a}rrman, Nowak, Wu, Dyck, Vaidyanathan, {R, Raghavender}, Mallet, Rudominer, Johnston, Mittal, Udathu, Christensen, Verma, Irving, Santucci, Elsayed, Davoodi, Georgiev, Tenney, Hua, Cideron, Leurent, Alnahlawi, Georgescu, Wei, Zheng, Scandinaro, Jiang, Snoek, Sundararajan, Wang, Ontiveros, Karo, Cole, Rajashekhar, Tumeh, Ben-David, Jain, Uesato, Datta, Bunyan, Wu, Zhang, Stanczyk, Zhang, Steiner, Naskar, Azzam, Johnson, Paszke, Chiu, Elias, Mohiuddin, Muhammad, Miao, Lee, Vieillard, Park, Zhang, Stanway, Garmon, Karmarkar, Dong, Lee, Kumar, Zhou, Evens, Isaac, Irving, Loper, Fink, Arkatkar, Chen, Shafran, Petrychenko, Chen, Jia, Levskaya, Zhu, Grabowski, Mao, Magni, Yao, Snaider, Casagrande, Palmer, Suganthan, Casta{\~n}o, Giannoumis, Kim, Rybi{\'n}ski, Sreevatsa, Prendki, Soergel, Goedeckemeyer, Gierke, Jafari,
  Gaba, Wiesner, Wright, Wei, Vashisht, Kulizhskaya, Hoover, Le, Li, Iwuanyanwu, Liu, Ramirez, Khorlin, Cui, Lin, Wu, Aguilar, Pallo, Chakladar, Perng, Abellan, Zhang, Dasgupta, Kushman, Penchev, Repina, Wu, van~der Weide, Ponnapalli, Kaplan, Simsa, Li, Dousse, Yang, Piper, Ie, Pasumarthi, Lintz, Vijayakumar, Andor, Valenzuela, Lui, Paduraru, Peng, Lee, Zhang, Greene, Nguyen, Kurylowicz, Hardin, Dixon, Janzer, Choo, Feng, Zhang, Singhal, Du, McKinnon, Antropova, Bolukbasi, Keller, Reid, Finchelstein, Raad, Crocker, Hawkins, Dadashi, Gaffney, Franko, Bulanova, Leblond, Chung, Askham, Cobo, Xu, Fischer, Xu, Sorokin, Alberti, Lin, Evans, Dimitriev, Forbes, Banarse, Tung, Omernick, Bishop, Sterneck, Jain, Xia, Amid, Piccinno, Wang, Banzal, Mankowitz, Polozov, Krakovna, Brown, Bateni, Duan, Firoiu, Thotakuri, Natan, Geist, Girgin, Li, Ye, Roval, Tojo, Kwong, Lee-Thorp, Yew, Sinopalnikov, Ramos, Mellor, Sharma, Wu, Miller, Sonnerat, Vnukov, Greig, Beattie, Caveness, Bai, Eisenschlos, Korchemniy, Tsai, Jasarevic,
  Kong, Dao, Zheng, Liu, Yang, Zhu, Teh, Sanmiya, Gladchenko, Trdin, Toyama, Rosen, Tavakkol, Xue, Elkind, Woodman, Carpenter, Papamakarios, Kemp, Kafle, Grunina, Sinha, Talbert, Wu, Owusu-Afriyie, Du, Thornton, Pont-Tuset, Narayana, Li, Fatehi, Wieting, Ajmeri, Uria, Ko, Knight, H{\'e}liou, Niu, Gu, Pang, Li, Levine, Stolovich, Santamaria-Fernandez, Goenka, Yustalim, Strudel, Elqursh, Deck, Lee, Li, Levin, Hoffmann, Holtmann-Rice, Bachem, Arora, Koh, Yeganeh, P{\~o}der, Tariq, Sun, Ionita, Seyedhosseini, Tafti, Liu, Gulati, Liu, Ye, Chrzaszcz, Wang, Sethi, Li, Brown, Singh, Fan, Parisi, Stanton, Koverkathu, Choquette-Choo, Li, Lu, Ittycheriah, Shroff, Varadarajan, Bahargam, Willoughby, Gaddy, Desjardins, Cornero, Robenek, Mittal, Albrecht, Shenoy, Moiseev, Jacobsson, Ghaffarkhah, Rivi{\`e}re, Walton, Crepy, Parrish, Zhou, Farabet, Radebaugh, Srinivasan, van~der Salm, Fidjeland, Scellato, Latorre-Chimoto, Klimczak-Pluci{\'n}ska, Bridson, de~Cesare, Hudson, Mendolicchio, Walker, Morris, Mauger, Guseynov, Reid,
  Odoom, Loher, Cotruta, Yenugula, Grewe, Petrushkina, Duerig, Sanchez, Yadlowsky, Shen, Globerson, Webb, Dua, Li, Bhupatiraju, Hurt, Qureshi, Agarwal, Shani, Eyal, Khare, Belle, Wang, Tekur, Kale, Wei, Sang, Saeta, Liechty, Sun, Zhao, Lee, Nayak, Fritz, Vuyyuru, Aslanides, Vyas, Wicke, Ma, Eltyshev, Martin, Cate, Manyika, Amiri, Kim, Xiong, Kang, Luisier, Tripuraneni, Madras, Guo, Waters, Wang, Ainslie, Baldridge, Zhang, Pruthi, Bauer, Yang, Mansour, Gelman, Xu, Polovets, Liu, Cai, Chen, Sheng, Xue, Ozair, Angermueller, Li, Sinha, Wang, Wiesinger, Koukoumidis, Tian, Iyer, Gurumurthy, Goldenson, Shah, Blake, Yu, Urbanowicz, Palomaki, Fernando, Durden, Mehta, Momchev, Rahimtoroghi, Georgaki, Raul, Ruder, Redshaw, Lee, Zhou, Jalan, Li, Hechtman, Schuh, Nasr, Milan, Mikulik, Franco, Green, Nguyen, Kelley, Mahendru, Hu, Howland, Vargas, Hui, Bansal, Rao, Ghiya, Wang, Ye, Sarr, Preston, Elish, Li, Kaku, Gupta, Pasupat, Juan, Someswar, M., Chen, Amini, Fabrikant, Chu, Dong, Muthal, Buthpitiya, Jauhari, Hua,
  Khandelwal, Hitron, Ren, Rinaldi, Drath, Dabush, Jiang, Godhia, Sachs, Chen, Fan, Taitelbaum, Noga, Dai, Wang, Liang, Hamer, Ferng, Elkind, Atias, Lee, List{\'\i}k, Carlen, van~de Kerkhof, Pikus, Zaher, M{\"u}ller, Zykova, Stefanec, Gatsko, Hirnschall, Sethi, Xu, Ahuja, Tsai, Stefanoiu, Feng, Dhandhania, Katyal, Gupta, Parulekar, Pitta, Zhao, Bhatia, Bhavnani, Alhadlaq, Li, Danenberg, Tu, Pine, Filippova, Ghosh, Limonchik, Urala, Lanka, Clive, Sun, Li, Wu, Hongtongsak, Li, Thakkar, Omarov, Majmundar, Alverson, Kucharski, Patel, Jain, Zabelin, Pelagatti, Kohli, Kumar, Kim, Sankar, Shah, Ramachandruni, Zeng, Bariach, Weidinger, Subramanya, Hsiao, Hassabis, Kavukcuoglu, Sadovsky, Le, Strohman, Wu, Petrov, Dean, and Vinyals]{GeminiTeam2023a}
{Gemini Team}, Rohan Anil, Sebastian Borgeaud, Jean-Baptiste Alayrac, Jiahui Yu, Radu Soricut, Johan Schalkwyk, Andrew~M Dai, Anja Hauth, Katie Millican, David Silver, Melvin Johnson, Ioannis Antonoglou, Julian Schrittwieser, Amelia Glaese, Jilin Chen, Emily Pitler, Timothy Lillicrap, Angeliki Lazaridou, Orhan Firat, James Molloy, Michael Isard, Paul~R Barham, Tom Hennigan, Benjamin Lee, Fabio Viola, Malcolm Reynolds, Yuanzhong Xu, Ryan Doherty, Eli Collins, Clemens Meyer, Eliza Rutherford, Erica Moreira, Kareem Ayoub, Megha Goel, Jack Krawczyk, Cosmo Du, Ed~Chi, Heng-Tze Cheng, Eric Ni, Purvi Shah, Patrick Kane, Betty Chan, Manaal Faruqui, Aliaksei Severyn, Hanzhao Lin, Yaguang Li, Yong Cheng, Abe Ittycheriah, Mahdis Mahdieh, Mia Chen, Pei Sun, Dustin Tran, Sumit Bagri, Balaji Lakshminarayanan, Jeremiah Liu, Andras Orban, Fabian G{\"u}ra, Hao Zhou, Xinying Song, Aurelien Boffy, Harish Ganapathy, Steven Zheng, Hyunjeong Choe, {\'A}goston Weisz, Tao Zhu, Yifeng Lu, Siddharth Gopal, Jarrod Kahn, Maciej Kula,
  Jeff Pitman, Rushin Shah, Emanuel Taropa, Majd~Al Merey, Martin Baeuml, Zhifeng Chen, Laurent~El Shafey, Yujing Zhang, Olcan Sercinoglu, George Tucker, Enrique Piqueras, Maxim Krikun, Iain Barr, Nikolay Savinov, Ivo Danihelka, Becca Roelofs, Ana{\"\i}s White, Anders Andreassen, Tamara von Glehn, Lakshman Yagati, Mehran Kazemi, Lucas Gonzalez, Misha Khalman, Jakub Sygnowski, Alexandre Frechette, Charlotte Smith, Laura Culp, Lev Proleev, Yi~Luan, Xi~Chen, James Lottes, Nathan Schucher, Federico Lebron, Alban Rrustemi, Natalie Clay, Phil Crone, Tomas Kocisky, Jeffrey Zhao, Bartek Perz, Dian Yu, Heidi Howard, Adam Bloniarz, Jack~W Rae, Han Lu, Laurent Sifre, Marcello Maggioni, Fred Alcober, Dan Garrette, Megan Barnes, Shantanu Thakoor, Jacob Austin, Gabriel Barth-Maron, William Wong, Rishabh Joshi, Rahma Chaabouni, Deeni Fatiha, Arun Ahuja, Gaurav~Singh Tomar, Evan Senter, Martin Chadwick, Ilya Kornakov, Nithya Attaluri, I{\~n}aki Iturrate, Ruibo Liu, Yunxuan Li, Sarah Cogan, Jeremy Chen, Chao Jia, Chenjie Gu,
  Qiao Zhang, Jordan Grimstad, Ale~Jakse Hartman, Xavier Garcia, Thanumalayan~Sankaranarayana Pillai, Jacob Devlin, Michael Laskin, Diego de~Las Casas, Dasha Valter, Connie Tao, Lorenzo Blanco, Adri{\`a}~Puigdom{\`e}nech Badia, David Reitter, Mianna Chen, Jenny Brennan, Clara Rivera, Sergey Brin, Shariq Iqbal, Gabriela Surita, Jane Labanowski, Abhi Rao, Stephanie Winkler, Emilio Parisotto, Yiming Gu, Kate Olszewska, Ravi Addanki, Antoine Miech, Annie Louis, Denis Teplyashin, Geoff Brown, Elliot Catt, Jan Balaguer, Jackie Xiang, Pidong Wang, Zoe Ashwood, Anton Briukhov, Albert Webson, Sanjay Ganapathy, Smit Sanghavi, Ajay Kannan, Ming-Wei Chang, Axel Stjerngren, Josip Djolonga, Yuting Sun, Ankur Bapna, Matthew Aitchison, Pedram Pejman, Henryk Michalewski, Tianhe Yu, Cindy Wang, Juliette Love, Junwhan Ahn, Dawn Bloxwich, Kehang Han, Peter Humphreys, Thibault Sellam, James Bradbury, Varun Godbole, Sina Samangooei, Bogdan Damoc, Alex Kaskasoli, S{\'e}bastien M~R Arnold, Vijay Vasudevan, Shubham Agrawal, Jason
  Riesa, Dmitry Lepikhin, Richard Tanburn, Srivatsan Srinivasan, Hyeontaek Lim, Sarah Hodkinson, Pranav Shyam, Johan Ferret, Steven Hand, Ankush Garg, Tom~Le Paine, Jian Li, Yujia Li, Minh Giang, Alexander Neitz, Zaheer Abbas, Sarah York, Machel Reid, Elizabeth Cole, Aakanksha Chowdhery, Dipanjan Das, Dominika Rogozi{\'n}ska, Vitaliy Nikolaev, Pablo Sprechmann, Zachary Nado, Lukas Zilka, Flavien Prost, Luheng He, Marianne Monteiro, Gaurav Mishra, Chris Welty, Josh Newlan, Dawei Jia, Miltiadis Allamanis, Clara~Huiyi Hu, Raoul de~Liedekerke, Justin Gilmer, Carl Saroufim, Shruti Rijhwani, Shaobo Hou, Disha Shrivastava, Anirudh Baddepudi, Alex Goldin, Adnan Ozturel, Albin Cassirer, Yunhan Xu, Daniel Sohn, Devendra Sachan, Reinald~Kim Amplayo, Craig Swanson, Dessie Petrova, Shashi Narayan, Arthur Guez, Siddhartha Brahma, Jessica Landon, Miteyan Patel, Ruizhe Zhao, Kevin Villela, Luyu Wang, Wenhao Jia, Matthew Rahtz, Mai Gim{\'e}nez, Legg Yeung, James Keeling, Petko Georgiev, Diana Mincu, Boxi Wu, Salem Haykal,
  Rachel Saputro, Kiran Vodrahalli, James Qin, Zeynep Cankara, Abhanshu Sharma, Nick Fernando, Will Hawkins, Behnam Neyshabur, Solomon Kim, Adrian Hutter, Priyanka Agrawal, Alex Castro-Ros, George van~den Driessche, Tao Wang, Fan Yang, Shuo-Yiin Chang, Paul Komarek, Ross McIlroy, Mario Lu{\v c}i{\'c}, Guodong Zhang, Wael Farhan, Michael Sharman, Paul Natsev, Paul Michel, Yamini Bansal, Siyuan Qiao, Kris Cao, Siamak Shakeri, Christina Butterfield, Justin Chung, Paul~Kishan Rubenstein, Shivani Agrawal, Arthur Mensch, Kedar Soparkar, Karel Lenc, Timothy Chung, Aedan Pope, Loren Maggiore, Jackie Kay, Priya Jhakra, Shibo Wang, Joshua Maynez, Mary Phuong, Taylor Tobin, Andrea Tacchetti, Maja Trebacz, Kevin Robinson, Yash Katariya, Sebastian Riedel, Paige Bailey, Kefan Xiao, Nimesh Ghelani, Lora Aroyo, Ambrose Slone, Neil Houlsby, Xuehan Xiong, Zhen Yang, Elena Gribovskaya, Jonas Adler, Mateo Wirth, Lisa Lee, {Li, Music}, Thais Kagohara, Jay Pavagadhi, Sophie Bridgers, Anna Bortsova, Sanjay Ghemawat, Zafarali Ahmed,
  Tianqi Liu, Richard Powell, Vijay Bolina, Mariko Iinuma, Polina Zablotskaia, James Besley, Da-Woon Chung, Timothy Dozat, Ramona Comanescu, Xiance Si, Jeremy Greer, Guolong Su, Martin Polacek, Rapha{\"e}l~Lopez Kaufman, Simon Tokumine, Hexiang Hu, Elena Buchatskaya, Yingjie Miao, Mohamed Elhawaty, Aditya Siddhant, Nenad Tomasev, Jinwei Xing, Christina Greer, Helen Miller, Shereen Ashraf, Aurko Roy, Zizhao Zhang, Ada Ma, Angelos Filos, Milos Besta, Rory Blevins, Ted Klimenko, Chih-Kuan Yeh, Soravit Changpinyo, Jiaqi Mu, Oscar Chang, Mantas Pajarskas, Carrie Muir, Vered Cohen, Charline~Le Lan, Krishna Haridasan, Amit Marathe, Steven Hansen, Sholto Douglas, Rajkumar Samuel, Mingqiu Wang, Sophia Austin, Chang Lan, Jiepu Jiang, Justin Chiu, Jaime~Alonso Lorenzo, Lars~Lowe Sj{\"o}sund, S{\'e}bastien Cevey, Zach Gleicher, Thi Avrahami, Anudhyan Boral, Hansa Srinivasan, Vittorio Selo, Rhys May, Konstantinos Aisopos, L{\'e}onard Hussenot, Livio~Baldini Soares, Kate Baumli, Michael~B Chang, Adri{\`a} Recasens, Ben
  Caine, Alexander Pritzel, Filip Pavetic, Fabio Pardo, Anita Gergely, Justin Frye, Vinay Ramasesh, Dan Horgan, Kartikeya Badola, Nora Kassner, Subhrajit Roy, Ethan Dyer, V{\'\i}ctor~Campos Campos, Alex Tomala, Yunhao Tang, Dalia~El Badawy, Elspeth White, Basil Mustafa, Oran Lang, Abhishek Jindal, Sharad Vikram, Zhitao Gong, Sergi Caelles, Ross Hemsley, Gregory Thornton, Fangxiaoyu Feng, Wojciech Stokowiec, Ce~Zheng, Phoebe Thacker, {\c C}a{\u g}lar {\"U}nl{\"u}, Zhishuai Zhang, Mohammad Saleh, James Svensson, Max Bileschi, Piyush Patil, Ankesh Anand, Roman Ring, Katerina Tsihlas, Arpi Vezer, Marco Selvi, Toby Shevlane, Mikel Rodriguez, Tom Kwiatkowski, Samira Daruki, Keran Rong, Allan Dafoe, Nicholas FitzGerald, Keren Gu-Lemberg, Mina Khan, Lisa~Anne Hendricks, Marie Pellat, Vladimir Feinberg, James Cobon-Kerr, Tara Sainath, Maribeth Rauh, Sayed~Hadi Hashemi, Richard Ives, Yana Hasson, Eric Noland, Yuan Cao, Nathan Byrd, Le~Hou, Qingze Wang, Thibault Sottiaux, Michela Paganini, Jean-Baptiste Lespiau,
  Alexandre Moufarek, Samer Hassan, Kaushik Shivakumar, Joost van Amersfoort, Amol Mandhane, Pratik Joshi, Anirudh Goyal, Matthew Tung, Andrew Brock, Hannah Sheahan, Vedant Misra, Cheng Li, Nemanja Raki{\'c}evi{\'c}, Mostafa Dehghani, Fangyu Liu, Sid Mittal, Junhyuk Oh, Seb Noury, Eren Sezener, Fantine Huot, Matthew Lamm, Nicola De~Cao, Charlie Chen, Sidharth Mudgal, Romina Stella, Kevin Brooks, Gautam Vasudevan, Chenxi Liu, Mainak Chain, Nivedita Melinkeri, Aaron Cohen, Venus Wang, Kristie Seymore, Sergey Zubkov, Rahul Goel, Summer Yue, Sai Krishnakumaran, Brian Albert, Nate Hurley, Motoki Sano, Anhad Mohananey, Jonah Joughin, Egor Filonov, Tomasz K{\k e}pa, Yomna Eldawy, Jiawern Lim, Rahul Rishi, Shirin Badiezadegan, Taylor Bos, Jerry Chang, Sanil Jain, Sri Gayatri~Sundara Padmanabhan, Subha Puttagunta, Kalpesh Krishna, Leslie Baker, Norbert Kalb, Vamsi Bedapudi, Adam Kurzrok, Shuntong Lei, Anthony Yu, Oren Litvin, Xiang Zhou, Zhichun Wu, Sam Sobell, Andrea Siciliano, Alan Papir, Robby Neale, Jonas
  Bragagnolo, Tej Toor, Tina Chen, Valentin Anklin, Feiran Wang, Richie Feng, Milad Gholami, Kevin Ling, Lijuan Liu, Jules Walter, Hamid Moghaddam, Arun Kishore, Jakub Adamek, Tyler Mercado, Jonathan Mallinson, Siddhinita Wandekar, Stephen Cagle, Eran Ofek, Guillermo Garrido, Clemens Lombriser, Maksim Mukha, Botu Sun, Hafeezul~Rahman Mohammad, Josip Matak, Yadi Qian, Vikas Peswani, Pawel Janus, Quan Yuan, Leif Schelin, Oana David, Ankur Garg, Yifan He, Oleksii Duzhyi, Anton {\"A}lgmyr, Timoth{\'e}e Lottaz, Qi~Li, Vikas Yadav, Luyao Xu, Alex Chinien, Rakesh Shivanna, Aleksandr Chuklin, Josie Li, Carrie Spadine, Travis Wolfe, Kareem Mohamed, Subhabrata Das, Zihang Dai, Kyle He, Daniel von Dincklage, Shyam Upadhyay, Akanksha Maurya, Luyan Chi, Sebastian Krause, Khalid Salama, Pam~G Rabinovitch, {M, Pavan Kumar Reddy}, Aarush Selvan, Mikhail Dektiarev, Golnaz Ghiasi, Erdem Guven, Himanshu Gupta, Boyi Liu, Deepak Sharma, Idan~Heimlich Shtacher, Shachi Paul, Oscar Akerlund, Fran{\c c}ois-Xavier Aubet, Terry Huang,
  Chen Zhu, Eric Zhu, Elico Teixeira, Matthew Fritze, Francesco Bertolini, Liana-Eleonora Marinescu, Martin B{\"o}lle, Dominik Paulus, Khyatti Gupta, Tejasi Latkar, Max Chang, Jason Sanders, Roopa Wilson, Xuewei Wu, Yi-Xuan Tan, Lam~Nguyen Thiet, Tulsee Doshi, Sid Lall, Swaroop Mishra, Wanming Chen, Thang Luong, Seth Benjamin, Jasmine Lee, Ewa Andrejczuk, Dominik Rabiej, Vipul Ranjan, Krzysztof Styrc, Pengcheng Yin, Jon Simon, Malcolm~Rose Harriott, Mudit Bansal, Alexei Robsky, Geoff Bacon, David Greene, Daniil Mirylenka, Chen Zhou, Obaid Sarvana, Abhimanyu Goyal, Samuel Andermatt, Patrick Siegler, Ben Horn, Assaf Israel, Francesco Pongetti, Chih-Wei~``louis'' Chen, Marco Selvatici, Pedro Silva, Kathie Wang, Jackson Tolins, Kelvin Guu, Roey Yogev, Xiaochen Cai, Alessandro Agostini, Maulik Shah, Hung Nguyen, Noah~{\'O} Donnaile, S{\'e}bastien Pereira, Linda Friso, Adam Stambler, Adam Kurzrok, Chenkai Kuang, Yan Romanikhin, Mark Geller, Z~J Yan, Kane Jang, Cheng-Chun Lee, Wojciech Fica, Eric Malmi, Qijun Tan,
  Dan Banica, Daniel Balle, Ryan Pham, Yanping Huang, Diana Avram, Hongzhi Shi, Jasjot Singh, Chris Hidey, Niharika Ahuja, Pranab Saxena, Dan Dooley, Srividya~Pranavi Potharaju, Eileen O'Neill, Anand Gokulchandran, Ryan Foley, Kai Zhao, Mike Dusenberry, Yuan Liu, Pulkit Mehta, Ragha Kotikalapudi, Chalence Safranek-Shrader, Andrew Goodman, Joshua Kessinger, Eran Globen, Prateek Kolhar, Chris Gorgolewski, Ali Ibrahim, Yang Song, Ali Eichenbaum, Thomas Brovelli, Sahitya Potluri, Preethi Lahoti, Cip Baetu, Ali Ghorbani, Charles Chen, Andy Crawford, Shalini Pal, Mukund Sridhar, Petru Gurita, Asier Mujika, Igor Petrovski, Pierre-Louis Cedoz, Chenmei Li, Shiyuan Chen, Niccol{\`o}~Dal Santo, Siddharth Goyal, Jitesh Punjabi, Karthik Kappaganthu, Chester Kwak, Pallavi Lv, Sarmishta Velury, Himadri Choudhury, Jamie Hall, Premal Shah, Ricardo Figueira, Matt Thomas, Minjie Lu, Ting Zhou, Chintu Kumar, Thomas Jurdi, Sharat Chikkerur, Yenai Ma, Adams Yu, Soo Kwak, Victor {\"A}hdel, Sujeevan Rajayogam, Travis Choma, Fei Liu,
  Aditya Barua, Colin Ji, Ji~Ho Park, Vincent Hellendoorn, Alex Bailey, Taylan Bilal, Huanjie Zhou, Mehrdad Khatir, Charles Sutton, Wojciech Rzadkowski, Fiona Macintosh, Konstantin Shagin, Paul Medina, Chen Liang, Jinjing Zhou, Pararth Shah, Yingying Bi, Attila Dankovics, Shipra Banga, Sabine Lehmann, Marissa Bredesen, Zifan Lin, John~Eric Hoffmann, Jonathan Lai, Raynald Chung, Kai Yang, Nihal Balani, Arthur Bra{\v z}inskas, Andrei Sozanschi, Matthew Hayes, H{\'e}ctor~Fern{\'a}ndez Alcalde, Peter Makarov, Will Chen, Antonio Stella, Liselotte Snijders, Michael Mandl, Ante K{\"a}rrman, Pawe{\l} Nowak, Xinyi Wu, Alex Dyck, Krishnan Vaidyanathan, {R, Raghavender}, Jessica Mallet, Mitch Rudominer, Eric Johnston, Sushil Mittal, Akhil Udathu, Janara Christensen, Vishal Verma, Zach Irving, Andreas Santucci, Gamaleldin Elsayed, Elnaz Davoodi, Marin Georgiev, Ian Tenney, Nan Hua, Geoffrey Cideron, Edouard Leurent, Mahmoud Alnahlawi, Ionut Georgescu, Nan Wei, Ivy Zheng, Dylan Scandinaro, Heinrich Jiang, Jasper Snoek,
  Mukund Sundararajan, Xuezhi Wang, Zack Ontiveros, Itay Karo, Jeremy Cole, Vinu Rajashekhar, Lara Tumeh, Eyal Ben-David, Rishub Jain, Jonathan Uesato, Romina Datta, Oskar Bunyan, Shimu Wu, John Zhang, Piotr Stanczyk, Ye~Zhang, David Steiner, Subhajit Naskar, Michael Azzam, Matthew Johnson, Adam Paszke, Chung-Cheng Chiu, Jaume~Sanchez Elias, Afroz Mohiuddin, Faizan Muhammad, Jin Miao, Andrew Lee, Nino Vieillard, Jane Park, Jiageng Zhang, Jeff Stanway, Drew Garmon, Abhijit Karmarkar, Zhe Dong, Jong Lee, Aviral Kumar, Luowei Zhou, Jonathan Evens, William Isaac, Geoffrey Irving, Edward Loper, Michael Fink, Isha Arkatkar, Nanxin Chen, Izhak Shafran, Ivan Petrychenko, Zhe Chen, Johnson Jia, Anselm Levskaya, Zhenkai Zhu, Peter Grabowski, Yu~Mao, Alberto Magni, Kaisheng Yao, Javier Snaider, Norman Casagrande, Evan Palmer, Paul Suganthan, Alfonso Casta{\~n}o, Irene Giannoumis, Wooyeol Kim, Miko{\l}aj Rybi{\'n}ski, Ashwin Sreevatsa, Jennifer Prendki, David Soergel, Adrian Goedeckemeyer, Willi Gierke, Mohsen Jafari,
  Meenu Gaba, Jeremy Wiesner, Diana~Gage Wright, Yawen Wei, Harsha Vashisht, Yana Kulizhskaya, Jay Hoover, Maigo Le, Lu~Li, Chimezie Iwuanyanwu, Lu~Liu, Kevin Ramirez, Andrey Khorlin, Albert Cui, Tian Lin, Marcus Wu, Ricardo Aguilar, Keith Pallo, Abhishek Chakladar, Ginger Perng, Elena~Allica Abellan, Mingyang Zhang, Ishita Dasgupta, Nate Kushman, Ivo Penchev, Alena Repina, Xihui Wu, Tom van~der Weide, Priya Ponnapalli, Caroline Kaplan, Jiri Simsa, Shuangfeng Li, Olivier Dousse, Fan Yang, Jeff Piper, Nathan Ie, Rama Pasumarthi, Nathan Lintz, Anitha Vijayakumar, Daniel Andor, Pedro Valenzuela, Minnie Lui, Cosmin Paduraru, Daiyi Peng, Katherine Lee, Shuyuan Zhang, Somer Greene, Duc~Dung Nguyen, Paula Kurylowicz, Cassidy Hardin, Lucas Dixon, Lili Janzer, Kiam Choo, Ziqiang Feng, Biao Zhang, Achintya Singhal, Dayou Du, Dan McKinnon, Natasha Antropova, Tolga Bolukbasi, Orgad Keller, David Reid, Daniel Finchelstein, Maria~Abi Raad, Remi Crocker, Peter Hawkins, Robert Dadashi, Colin Gaffney, Ken Franko, Anna
  Bulanova, R{\'e}mi Leblond, Shirley Chung, Harry Askham, Luis~C Cobo, Kelvin Xu, Felix Fischer, Jun Xu, Christina Sorokin, Chris Alberti, Chu-Cheng Lin, Colin Evans, Alek Dimitriev, Hannah Forbes, Dylan Banarse, Zora Tung, Mark Omernick, Colton Bishop, Rachel Sterneck, Rohan Jain, Jiawei Xia, Ehsan Amid, Francesco Piccinno, Xingyu Wang, Praseem Banzal, Daniel~J Mankowitz, Alex Polozov, Victoria Krakovna, Sasha Brown, Mohammadhossein Bateni, Dennis Duan, Vlad Firoiu, Meghana Thotakuri, Tom Natan, Matthieu Geist, Ser~Tan Girgin, Hui Li, Jiayu Ye, Ofir Roval, Reiko Tojo, Michael Kwong, James Lee-Thorp, Christopher Yew, Danila Sinopalnikov, Sabela Ramos, John Mellor, Abhishek Sharma, Kathy Wu, David Miller, Nicolas Sonnerat, Denis Vnukov, Rory Greig, Jennifer Beattie, Emily Caveness, Libin Bai, Julian Eisenschlos, Alex Korchemniy, Tomy Tsai, Mimi Jasarevic, Weize Kong, Phuong Dao, Zeyu Zheng, Frederick Liu, Fan Yang, Rui Zhu, Tian~Huey Teh, Jason Sanmiya, Evgeny Gladchenko, Nejc Trdin, Daniel Toyama, Evan
  Rosen, Sasan Tavakkol, Linting Xue, Chen Elkind, Oliver Woodman, John Carpenter, George Papamakarios, Rupert Kemp, Sushant Kafle, Tanya Grunina, Rishika Sinha, Alice Talbert, Diane Wu, Denese Owusu-Afriyie, Cosmo Du, Chloe Thornton, Jordi Pont-Tuset, Pradyumna Narayana, Jing Li, Saaber Fatehi, John Wieting, Omar Ajmeri, Benigno Uria, Yeongil Ko, Laura Knight, Am{\'e}lie H{\'e}liou, Ning Niu, Shane Gu, Chenxi Pang, Yeqing Li, Nir Levine, Ariel Stolovich, Rebeca Santamaria-Fernandez, Sonam Goenka, Wenny Yustalim, Robin Strudel, Ali Elqursh, Charlie Deck, Hyo Lee, Zonglin Li, Kyle Levin, Raphael Hoffmann, Dan Holtmann-Rice, Olivier Bachem, Sho Arora, Christy Koh, Soheil~Hassas Yeganeh, Siim P{\~o}der, Mukarram Tariq, Yanhua Sun, Lucian Ionita, Mojtaba Seyedhosseini, Pouya Tafti, Zhiyu Liu, Anmol Gulati, Jasmine Liu, Xinyu Ye, Bart Chrzaszcz, Lily Wang, Nikhil Sethi, Tianrun Li, Ben Brown, Shreya Singh, Wei Fan, Aaron Parisi, Joe Stanton, Vinod Koverkathu, Christopher~A Choquette-Choo, Yunjie Li, T~J Lu, Abe
  Ittycheriah, Prakash Shroff, Mani Varadarajan, Sanaz Bahargam, Rob Willoughby, David Gaddy, Guillaume Desjardins, Marco Cornero, Brona Robenek, Bhavishya Mittal, Ben Albrecht, Ashish Shenoy, Fedor Moiseev, Henrik Jacobsson, Alireza Ghaffarkhah, Morgane Rivi{\`e}re, Alanna Walton, Cl{\'e}ment Crepy, Alicia Parrish, Zongwei Zhou, Clement Farabet, Carey Radebaugh, Praveen Srinivasan, Claudia van~der Salm, Andreas Fidjeland, Salvatore Scellato, Eri Latorre-Chimoto, Hanna Klimczak-Pluci{\'n}ska, David Bridson, Dario de~Cesare, Tom Hudson, Piermaria Mendolicchio, Lexi Walker, Alex Morris, Matthew Mauger, Alexey Guseynov, Alison Reid, Seth Odoom, Lucia Loher, Victor Cotruta, Madhavi Yenugula, Dominik Grewe, Anastasia Petrushkina, Tom Duerig, Antonio Sanchez, Steve Yadlowsky, Amy Shen, Amir Globerson, Lynette Webb, Sahil Dua, Dong Li, Surya Bhupatiraju, Dan Hurt, Haroon Qureshi, Ananth Agarwal, Tomer Shani, Matan Eyal, Anuj Khare, Shreyas~Rammohan Belle, Lei Wang, Chetan Tekur, Mihir~Sanjay Kale, Jinliang Wei,
  Ruoxin Sang, Brennan Saeta, Tyler Liechty, Yi~Sun, Yao Zhao, Stephan Lee, Pandu Nayak, Doug Fritz, Manish~Reddy Vuyyuru, John Aslanides, Nidhi Vyas, Martin Wicke, Xiao Ma, Evgenii Eltyshev, Nina Martin, Hardie Cate, James Manyika, Keyvan Amiri, Yelin Kim, Xi~Xiong, Kai Kang, Florian Luisier, Nilesh Tripuraneni, David Madras, Mandy Guo, Austin Waters, Oliver Wang, Joshua Ainslie, Jason Baldridge, Han Zhang, Garima Pruthi, Jakob Bauer, Feng Yang, Riham Mansour, Jason Gelman, Yang Xu, George Polovets, Ji~Liu, Honglong Cai, Warren Chen, Xianghai Sheng, Emily Xue, Sherjil Ozair, Christof Angermueller, Xiaowei Li, Anoop Sinha, Weiren Wang, Julia Wiesinger, Emmanouil Koukoumidis, Yuan Tian, Anand Iyer, Madhu Gurumurthy, Mark Goldenson, Parashar Shah, M~K Blake, Hongkun Yu, Anthony Urbanowicz, Jennimaria Palomaki, Chrisantha Fernando, Ken Durden, Harsh Mehta, Nikola Momchev, Elahe Rahimtoroghi, Maria Georgaki, Amit Raul, Sebastian Ruder, Morgan Redshaw, Jinhyuk Lee, Denny Zhou, Komal Jalan, Dinghua Li, Blake
  Hechtman, Parker Schuh, Milad Nasr, Kieran Milan, Vladimir Mikulik, Juliana Franco, Tim Green, Nam Nguyen, Joe Kelley, Aroma Mahendru, Andrea Hu, Joshua Howland, Ben Vargas, Jeffrey Hui, Kshitij Bansal, Vikram Rao, Rakesh Ghiya, Emma Wang, Ke~Ye, Jean~Michel Sarr, Melanie~Moranski Preston, Madeleine Elish, Steve Li, Aakash Kaku, Jigar Gupta, Ice Pasupat, Da-Cheng Juan, Milan Someswar, Tejvi M., Xinyun Chen, Aida Amini, Alex Fabrikant, Eric Chu, Xuanyi Dong, Amruta Muthal, Senaka Buthpitiya, Sarthak Jauhari, Nan Hua, Urvashi Khandelwal, Ayal Hitron, Jie Ren, Larissa Rinaldi, Shahar Drath, Avigail Dabush, Nan-Jiang Jiang, Harshal Godhia, Uli Sachs, Anthony Chen, Yicheng Fan, Hagai Taitelbaum, Hila Noga, Zhuyun Dai, James Wang, Chen Liang, Jenny Hamer, Chun-Sung Ferng, Chenel Elkind, Aviel Atias, Paulina Lee, V{\'\i}t List{\'\i}k, Mathias Carlen, Jan van~de Kerkhof, Marcin Pikus, Krunoslav Zaher, Paul M{\"u}ller, Sasha Zykova, Richard Stefanec, Vitaly Gatsko, Christoph Hirnschall, Ashwin Sethi, Xingyu~Federico
  Xu, Chetan Ahuja, Beth Tsai, Anca Stefanoiu, Bo~Feng, Keshav Dhandhania, Manish Katyal, Akshay Gupta, Atharva Parulekar, Divya Pitta, Jing Zhao, Vivaan Bhatia, Yashodha Bhavnani, Omar Alhadlaq, Xiaolin Li, Peter Danenberg, Dennis Tu, Alex Pine, Vera Filippova, Abhipso Ghosh, Ben Limonchik, Bhargava Urala, Chaitanya~Krishna Lanka, Derik Clive, Yi~Sun, Edward Li, Hao Wu, Kevin Hongtongsak, Ianna Li, Kalind Thakkar, Kuanysh Omarov, Kushal Majmundar, Michael Alverson, Michael Kucharski, Mohak Patel, Mudit Jain, Maksim Zabelin, Paolo Pelagatti, Rohan Kohli, Saurabh Kumar, Joseph Kim, Swetha Sankar, Vineet Shah, Lakshmi Ramachandruni, Xiangkai Zeng, Ben Bariach, Laura Weidinger, Amar Subramanya, Sissie Hsiao, Demis Hassabis, Koray Kavukcuoglu, Adam Sadovsky, Quoc Le, Trevor Strohman, Yonghui Wu, Slav Petrov, Jeffrey Dean, and Oriol Vinyals.
\newblock Gemini: A family of highly capable multimodal models.
\newblock Technical report, Google DeepMind, 2023.

\bibitem[Ghalebikesabi et~al.(2024)Ghalebikesabi, Bagdasaryan, Yi, Yona, Shumailov, Pappu, Shi, Weidinger, Stanforth, Berrada, Kohli, Huang, and Balle]{ghalebikesabie2024}
Sahra Ghalebikesabi, Eugene Bagdasaryan, Ren Yi, Itay Yona, Ilia Shumailov, Aneesh Pappu, Chongyang Shi, Laura Weidinger, Robert Stanforth, Leonard Berrada, Pushmeet Kohli, Po-Sen Huang, and Borja Balle.
\newblock Operationalizing contextual integrity in privacy-conscious assistants, 2024.

\bibitem[Ghodsi et~al.(2017)Ghodsi, Gu, and Garg]{Ghodsi2017SafetyNetsVE}
Zahra Ghodsi, Tianyu Gu, and Siddharth Garg.
\newblock Safetynets: Verifiable execution of deep neural networks on an untrusted cloud.
\newblock In \emph{Neural Information Processing Systems}, 2017.

\bibitem[Gilad-Bachrach et~al.(2016)]{GiladBachrach2016}
Ran Gilad-Bachrach et~al.
\newblock Cryptonets: Applying neural networks to encrypted data with high throughput and accuracy.
\newblock In Maria~Florina Balcan and Kilian~Q. Weinberger, editors, \emph{International Conference on Machine Learning}, pages 201--210, New York, 2016. PMLR.

\bibitem[Goldstein et~al.(2023)Goldstein, Sastry, Musser, DiResta, Gentzel, and Sedova]{Goldstein2023-co}
Josh~A Goldstein, Girish Sastry, Micah Musser, Renee DiResta, Matthew Gentzel, and Katerina Sedova.
\newblock Generative language models and automated influence operations: Emerging threats and potential mitigations.
\newblock January 2023.

\bibitem[Goldwasser et~al.(2022)]{Goldwasser2022}
Shafi Goldwasser et~al.
\newblock Planting undetectable backdoors in machine learning models.
\newblock In \emph{2022 IEEE 63rd Annual Symposium on Foundations of Computer Science (FOCS)}, pages 931--942, 2022.
\newblock \doi{10.1109/FOCS54457.2022.00092}.
\newblock URL \url{https://doi.org/10.1109/FOCS54457.2022.00092}.

\bibitem[{GPA's International Enforcement Cooperation Working Group}(2023)]{GPASInternationalEnforcementCooperationWorkingGroup2023k}
{GPA's International Enforcement Cooperation Working Group}.
\newblock Joint statement on data scraping and the protection of privacy.
\newblock Technical report, Information Commissioner's Office, August 2023.

\bibitem[Greenwood(2024)]{transactions_on_agents}
Dazza Greenwood.
\newblock Transactions on open agents, November 2024.
\newblock URL \url{https://onagents.org/transactions/}.
\newblock Accessed: 2025-01-06.

\bibitem[Groth(2016)]{Groth2016}
Jens Groth.
\newblock On the size of pairing-based non-interactive arguments.
\newblock \emph{IACR Cryptol. ePrint Arch.}, 2016:\penalty0 260, 2016.

\bibitem[Groth et~al.(2018)Groth, Kohlweiss, Maller, Meiklejohn, and Miers]{groth2018updatable}
Jens Groth, Markulf Kohlweiss, Mary Maller, Sarah Meiklejohn, and Ian Miers.
\newblock Updatable and universal common reference strings with applications to zk-snarks.
\newblock In \emph{Annual International Cryptology Conference}, pages 698--728. Springer, 2018.

\bibitem[Gundersen et~al.(2022)Gundersen, Shamsaliei, and Isdahl]{gundersen2022machine}
Odd~Erik Gundersen, Saeid Shamsaliei, and Richard~Juul Isdahl.
\newblock Do machine learning platforms provide out-of-the-box reproducibility?
\newblock \emph{Future Generation Computer Systems}, 126:\penalty0 34--47, 2022.

\bibitem[Gunter et~al.(2024)Gunter, Wang, Wang, Pang, Narayanan, Zhang, Zhang, Chen, Chiu, Qiu, Gopinath, Yap, Yin, Nan, Weers, Yin, Huang, Wang, Lu, Peebles, Ye, Lee, Du, Chen, Keunebroek, Wiseman, Evans, Lei, Rathod, Kong, Du, Li, Wang, Gao, Ahmed, Xu, Lu, Rashid, Jose, Doane, Bencomo, Vanderby, Hansen, Jain, Anupama, Kamal, Wu, Brum, Maalouf, Erdenebileg, Dulhanty, Moritz, Kang, Jimenez, Ladd, Shi, Bai, Chu, Hohman, Kotek, Coleman, Li, Bigham, Cao, Lai, Cheung, Shan, Zhou, Li, Qin, Singh, Vega, Zou, Heckman, Gardiner, Bowler, Cordell, Cao, Hay, Shahdadpuri, Godwin, Dighe, Rachapudi, Tantawi, Frigg, Davarnia, Shah, Guha, Sirovica, Ma, Ma, Wang, Kim, Jayaram, Shankar, Paidi, Kumar, Wang, Zheng, Cheng, Shrager, Ye, Tanaka, Guo, Meng, Luo, Ouyang, Aygar, Wan, Walkingshaw, Narayanan, Lin, Farooq, Ramerth, Reed, Bartels, Chaney, Riazati, Yang, Feldman, Hochstrasser, Seguin, Belousova, Pelemans, Yang, Vahid, Cao, Najibi, Zuliani, Horton, Cho, Bhendawade, Dong, Maj, Agrawal, Shan, Fu, Poston, Xu, Liu, Rao,
  Heeramun, Merth, Rayala, Cui, Sridhar, Zhang, Zhang, Wu, Zhou, Liu, Zhao, Xia, Ren, and Ren]{Gunter2024a}
Tom Gunter, Zirui Wang, Chong Wang, Ruoming Pang, Andy Narayanan, Aonan Zhang, Bowen Zhang, Chen Chen, Chung-Cheng Chiu, David Qiu, Deepak Gopinath, Dian~Ang Yap, Dong Yin, Feng Nan, Floris Weers, Guoli Yin, Haoshuo Huang, Jianyu Wang, Jiarui Lu, John Peebles, Ke~Ye, Mark Lee, Nan Du, Qibin Chen, Quentin Keunebroek, Sam Wiseman, Syd Evans, Tao Lei, Vivek Rathod, Xiang Kong, Xianzhi Du, Yanghao Li, Yongqiang Wang, Yuan Gao, Zaid Ahmed, Zhaoyang Xu, Zhiyun Lu, Al~Rashid, Albin~Madappally Jose, Alec Doane, Alfredo Bencomo, Allison Vanderby, Andrew Hansen, Ankur Jain, Anupama~Mann Anupama, Areeba Kamal, Bugu Wu, Carolina Brum, Charlie Maalouf, Chinguun Erdenebileg, Chris Dulhanty, Dominik Moritz, Doug Kang, Eduardo Jimenez, Evan Ladd, Fangping Shi, Felix Bai, Frank Chu, Fred Hohman, Hadas Kotek, Hannah~Gillis Coleman, Jane Li, Jeffrey Bigham, Jeffery Cao, Jeff Lai, Jessica Cheung, Jiulong Shan, Joe Zhou, John Li, Jun Qin, Karanjeet Singh, Karla Vega, Kelvin Zou, Laura Heckman, Lauren Gardiner, Margit Bowler,
  Maria Cordell, Meng Cao, Nicole Hay, Nilesh Shahdadpuri, Otto Godwin, Pranay Dighe, Pushyami Rachapudi, Ramsey Tantawi, Roman Frigg, Sam Davarnia, Sanskruti Shah, Saptarshi Guha, Sasha Sirovica, Shen Ma, Shuang Ma, Simon Wang, Sulgi Kim, Suma Jayaram, Vaishaal Shankar, Varsha Paidi, Vivek Kumar, Xin Wang, Xin Zheng, Walker Cheng, Yael Shrager, Yang Ye, Yasu Tanaka, Yihao Guo, Yunsong Meng, Zhao~Tang Luo, Zhi Ouyang, Alp Aygar, Alvin Wan, Andrew Walkingshaw, Andy Narayanan, Antonie Lin, Arsalan Farooq, Brent Ramerth, Colorado Reed, Chris Bartels, Chris Chaney, David Riazati, Eric~Liang Yang, Erin Feldman, Gabriel Hochstrasser, Guillaume Seguin, Irina Belousova, Joris Pelemans, Karen Yang, Keivan~Alizadeh Vahid, Liangliang Cao, Mahyar Najibi, Marco Zuliani, Max Horton, Minsik Cho, Nikhil Bhendawade, Patrick Dong, Piotr Maj, Pulkit Agrawal, Qi~Shan, Qichen Fu, Regan Poston, Sam Xu, Shuangning Liu, Sushma Rao, Tashweena Heeramun, Thomas Merth, Uday Rayala, Victor Cui, Vivek~Rangarajan Sridhar, Wencong Zhang,
  Wenqi Zhang, Wentao Wu, Xingyu Zhou, Xinwen Liu, Yang Zhao, Yin Xia, Zhile Ren, and Zhongzheng Ren.
\newblock Apple intelligence foundation language models.
\newblock July 2024.

\bibitem[Gupta et~al.(2023)Gupta, Jawalkar, Mukherjee, Chandran, Gupta, Panwar, and Sharma]{gupta2023sigma}
Kanav Gupta, Neha Jawalkar, Ananta Mukherjee, Nishanth Chandran, Divya Gupta, Ashish Panwar, and Rahul Sharma.
\newblock Sigma: Secure gpt inference with function secret sharing.
\newblock \emph{Cryptology ePrint Archive}, 2023.

\bibitem[Hab{\"o}ck(2022)]{habock2022multivariate}
Ulrich Hab{\"o}ck.
\newblock Multivariate lookups based on logarithmic derivatives.
\newblock \emph{Cryptology ePrint Archive}, 2022.

\bibitem[Hardjono et~al.(2015)Hardjono, Maler, Machulak, and Catalano]{UMACORE1.0}
T.~Hardjono, E.~Maler, M.~Machulak, and D.~Catalano.
\newblock {U}ser-{M}anaged {A}ccess ({UMA}) {P}rofile of {OAuth2.0} -- {S}pecification {V}ersion {1.0}.
\newblock Kantara published specification, Kantara Initiative, April 2015.
\newblock https://docs.kantarainitiative.org/uma/rec-uma-core.html.

\bibitem[Hardjono(2019)]{Hardjono2019-IEEECommsMagazine}
Thomas Hardjono.
\newblock {F}ederated {A}uthorization over {A}ccess to {P}ersonal {D}ata for {D}ecentralized {I}dentity {M}anagement.
\newblock \emph{{IEEE} {C}ommunications {S}tandards {M}agazine -- {The Dawn of the Internet Identity Layer and the Role of Decentralized Identity}}, 3\penalty0 (4):\penalty0 32--38, December 2019.
\newblock URL \url{https://doi.org/10.1109/MCOMSTD.001.1900019}.

\bibitem[Hardjono and Pentland(2019)]{Hardjono2019DataCT}
Thomas Hardjono and Alex~'Sandy' Pentland.
\newblock Data cooperatives: Towards a foundation for decentralized personal data management.
\newblock \emph{ArXiv}, abs/1905.08819, 2019.
\newblock URL \url{https://api.semanticscholar.org/CorpusID:162168525}.

\bibitem[Hardt(2012)]{RFC6749-Formatted}
D.~Hardt.
\newblock {The OAuth 2.0 Authorization Framework}, October 2012.
\newblock URL \url{https://tools.ietf.org/html/rfc6749}.
\newblock {IETF}~{S}tandard~{RFC6749}.

\bibitem[Hendrycks et~al.(2021)Hendrycks, Burns, Basart, Zou, Mazeika, Song, and Steinhardt]{hendryckstest2021}
Dan Hendrycks, Collin Burns, Steven Basart, Andy Zou, Mantas Mazeika, Dawn Song, and Jacob Steinhardt.
\newblock Measuring massive multitask language understanding.
\newblock \emph{Proceedings of the International Conference on Learning Representations (ICLR)}, 2021.

\bibitem[Henzinger et~al.(2023)]{Henzinger2023}
Alexandra Henzinger et~al.
\newblock Private web search with tiptoe.
\newblock In \emph{Proceedings of the 29th Symposium on Operating Systems Principles}, pages 396--416, 10 2023.
\newblock \doi{10.1145/3600006.3613134}.
\newblock URL \url{https://doi.org/10.1145/3600006.3613134}.

\bibitem[Hesamifard et~al.(2017)Hesamifard, Takabi, and Ghasemi]{Hesamifard2017}
Ehsan Hesamifard, Hassan Takabi, and Mehdi Ghasemi.
\newblock Cryptodl: Deep neural networks over encrypted data.
\newblock \emph{arXiv}, 11 2017.
\newblock \doi{10.48550/arXiv.1711.05189}.
\newblock URL \url{https://doi.org/10.48550/arXiv.1711.05189}.

\bibitem[Hinton et~al.(2015)Hinton, Vinyals, and Dean]{hinton2015distilling}
Geoffrey Hinton, Oriol Vinyals, and Jeff Dean.
\newblock Distilling the knowledge in a neural network.
\newblock \emph{arXiv preprint arXiv:1503.02531}, 2015.

\bibitem[Ho et~al.(2022)Ho, Chan, Saharia, Whang, Gao, Gritsenko, Kingma, Poole, Norouzi, Fleet, and Salimans]{Ho2022u}
Jonathan Ho, William Chan, Chitwan Saharia, Jay Whang, Ruiqi Gao, Alexey Gritsenko, Diederik~P Kingma, Ben Poole, Mohammad Norouzi, David~J Fleet, and Tim Salimans.
\newblock Imagen video: High definition video generation with diffusion models.
\newblock October 2022.

\bibitem[Hu et~al.(2021)Hu, Shen, Wallis, Allen-Zhu, Li, Wang, and Chen]{hu2021lora}
Edward Hu, Yelong Shen, Phil Wallis, Zeyuan Allen-Zhu, Yuanzhi Li, Lu~Wang, and Weizhu Chen.
\newblock Lora: Low-rank adaptation of large language models, 2021.

\bibitem[Ito et~al.(1989)Ito, Saito, and Nishizeki]{ito1989secret}
Mitsuru Ito, Akira Saito, and Takao Nishizeki.
\newblock Secret sharing scheme realizing general access structure.
\newblock \emph{Electronics and Communications in Japan (Part III: Fundamental Electronic Science)}, 72\penalty0 (9):\penalty0 56--64, 1989.

\bibitem[Jagannathan and Wright(2005)]{Jagannathan2005PrivacypreservingDK}
Geetha Jagannathan and Rebecca~N. Wright.
\newblock Privacy-preserving distributed k-means clustering over arbitrarily partitioned data.
\newblock In \emph{Knowledge Discovery and Data Mining}, 2005.

\bibitem[Jain et~al.(2023)Jain, Hitzig, and Mishkin]{Jain2023-ij}
Shrey Jain, Zo{\"e} Hitzig, and Pamela Mishkin.
\newblock Contextual confidence and generative {AI}.
\newblock November 2023.

\bibitem[Jayasundara et~al.(2024)Jayasundara, Arachchilage, and Russello]{jayasundara2024ragent}
Sakuna~Harinda Jayasundara, Nalin Asanka~Gamagedara Arachchilage, and Giovanni Russello.
\newblock Ragent: Retrieval-based access control policy generation.
\newblock \emph{arXiv preprint arXiv:2409.07489}, 2024.

\bibitem[Jean-Louis et~al.(2023)Jean-Louis, Li, Ji, Malvai, Yurek, Bellemare, and Miller]{jean2023sgxonerated}
Nerla Jean-Louis, Yunqi Li, Yan Ji, Harjasleen Malvai, Thomas Yurek, Sylvain Bellemare, and Andrew Miller.
\newblock Sgxonerated: Finding (and partially fixing) privacy flaws in tee-based smart contract platforms without breaking the tee.
\newblock \emph{Cryptology ePrint Archive}, 2023.

\bibitem[J{\'e}gou et~al.(2011)J{\'e}gou, Douze, and Schmid]{IVF}
Herv{\'e} J{\'e}gou, Matthijs Douze, and Cordelia Schmid.
\newblock Product quantization for nearest neighbor search.
\newblock \emph{IEEE Transactions on Pattern Analysis and Machine Intelligence}, pages 117--128, 2011.

\bibitem[Jha and Reagen(2023)]{Jha2023DeepReShapeRN}
Nandan~Kumar Jha and Brandon Reagen.
\newblock Deepreshape: Redesigning neural networks for efficient private inference.
\newblock \emph{ArXiv}, abs/2304.10593, 2023.

\bibitem[Jia et~al.(2021)Jia, Yaghini, Choquette-Choo, Dullerud, Thudi, Chandrasekaran, and Papernot]{Jia2021ProofofLearningDA}
Hengrui Jia, Mohammad Yaghini, Christopher~A. Choquette-Choo, Natalie Dullerud, Anvith Thudi, Varun Chandrasekaran, and Nicolas Papernot.
\newblock Proof-of-learning: Definitions and practice.
\newblock \emph{2021 IEEE Symposium on Security and Privacy (SP)}, pages 1039--1056, 2021.

\bibitem[Johnson et~al.(2017)Johnson, Douze, and J{\'e}gou]{FAISS}
Jeff Johnson, Matthijs Douze, and Herv{\'e} J{\'e}gou.
\newblock Billion-scale similarity search with gpus.
\newblock \emph{IEEE Transactions on Big Data}, 7:\penalty0 535--547, 2017.

\bibitem[Jones and Galliers(1995)]{jones1995evaluating}
Karen~Sparck Jones and Julia~R Galliers.
\newblock Evaluating natural language processing systems: An analysis and review.
\newblock 1995.

\bibitem[Juvekar et~al.(2018)Juvekar, Vaikuntanathan, and Chandrakasan]{juvekar2018gazelle}
Chiraag Juvekar, Vinod Vaikuntanathan, and Anantha Chandrakasan.
\newblock Gazelle: A low latency framework for secure neural network inference.
\newblock In \emph{27th USENIX Security Symposium (USENIX Security 18)}, pages 1651--1669, 2018.

\bibitem[Kairouz et~al.(2019)Kairouz, McMahan, Avent, Bellet, Bennis, Bhagoji, Bonawitz, Charles, Cormode, Cummings, D'Oliveira, Rouayheb, Evans, Gardner, Garrett, Gasc{\'o}n, Ghazi, Gibbons, Gruteser, Harchaoui, He, He, Huo, Hutchinson, Hsu, Jaggi, Javidi, Joshi, Khodak, Konecn{\'y}, Korolova, Koushanfar, Koyejo, Lepoint, Liu, Mittal, Mohri, Nock, {\"O}zg{\"u}r, Pagh, Raykova, Qi, Ramage, Raskar, Song, Song, Stich, Sun, Suresh, Tram{\`e}r, Vepakomma, Wang, Xiong, Xu, Yang, Yu, Yu, and Zhao]{Kairouz2019AdvancesAO}
Peter Kairouz, H.~B. McMahan, Brendan Avent, Aur{\'e}lien Bellet, Mehdi Bennis, Arjun~Nitin Bhagoji, Keith Bonawitz, Zachary~B. Charles, Graham Cormode, Rachel Cummings, Rafael G.~L. D'Oliveira, Salim Y.~El Rouayheb, David Evans, Josh Gardner, Zachary Garrett, Adri{\`a} Gasc{\'o}n, Badih Ghazi, Phillip~B. Gibbons, Marco Gruteser, Za{\"i}d Harchaoui, Chaoyang He, Lie He, Zhouyuan Huo, Ben Hutchinson, Justin Hsu, Martin Jaggi, Tara Javidi, Gauri Joshi, Mikhail Khodak, Jakub Konecn{\'y}, Aleksandra Korolova, Farinaz Koushanfar, Oluwasanmi Koyejo, Tancr{\`e}de Lepoint, Yang Liu, Prateek Mittal, Mehryar Mohri, Richard Nock, Ayfer {\"O}zg{\"u}r, R.~Pagh, Mariana Raykova, Hang Qi, Daniel Ramage, Ramesh Raskar, Dawn~Xiaodong Song, Weikang Song, Sebastian~U. Stich, Ziteng Sun, Ananda~Theertha Suresh, Florian Tram{\`e}r, Praneeth Vepakomma, Jianyu Wang, Li~Xiong, Zheng Xu, Qiang Yang, Felix~X. Yu, Han Yu, and Sen Zhao.
\newblock Advances and open problems in federated learning.
\newblock \emph{Found. Trends Mach. Learn.}, 14:\penalty0 1--210, 2019.

\bibitem[Kang et~al.(2022{\natexlab{a}})Kang, Hashimoto, Stoica, and Sun]{kang2022scaling}
Daniel Kang, Tatsunori Hashimoto, Ion Stoica, and Yi~Sun.
\newblock Scaling up trustless dnn inference with zero-knowledge proofs.
\newblock \emph{arXiv preprint arXiv:2210.08674}, 2022{\natexlab{a}}.

\bibitem[Kang et~al.(2022{\natexlab{b}})]{Kang2022}
Daniel Kang et~al.
\newblock Scaling up trustless dnn inference with zero-knowledge proofs.
\newblock \emph{arXiv}, 10 2022{\natexlab{b}}.
\newblock \doi{10.48550/arXiv.2210.08674}.
\newblock URL \url{https://doi.org/10.48550/arXiv.2210.08674}.

\bibitem[Kang(2023)]{kang_2023_tensorplonk}
Daniel~D. Kang.
\newblock Tensorplonk: A ``gpu''' for zkml, delivering 1,000x speedups.
\newblock Medium, 2023.

\bibitem[Kant(1785)]{kant1785groundwork}
Immanuel Kant.
\newblock \emph{Groundwork of the Metaphysics of Morals}.
\newblock Cambridge University Press, Cambridge, 1785.
\newblock Translated by Mary Gregor, 1997.

\bibitem[Kaplan et~al.(2020)Kaplan, McCandlish, Henighan, Brown, Chess, Child, Gray, Radford, Wu, and Amodei]{Kaplan2020ScalingLF}
Jared Kaplan, Sam McCandlish, T.~J. Henighan, Tom~B. Brown, Benjamin Chess, Rewon Child, Scott Gray, Alec Radford, Jeff Wu, and Dario Amodei.
\newblock Scaling laws for neural language models.
\newblock \emph{ArXiv}, abs/2001.08361, 2020.

\bibitem[K{\"a}rkk{\"a}inen and Joo(2021)]{Krkkinen2021FairFaceFA}
Kimmo K{\"a}rkk{\"a}inen and Jungseock Joo.
\newblock Fairface: Face attribute dataset for balanced race, gender, and age for bias measurement and mitigation.
\newblock \emph{2021 IEEE Winter Conference on Applications of Computer Vision (WACV)}, pages 1547--1557, 2021.

\bibitem[Karpukhin et~al.(2020{\natexlab{a}})Karpukhin, Oguz, Min, Lewis, Wu, Edunov, Chen, and Yih]{Karpukhin2020j}
Vladimir Karpukhin, Barlas Oguz, Sewon Min, Patrick Lewis, Ledell Wu, Sergey Edunov, Danqi Chen, and Wen-Tau Yih.
\newblock Dense passage retrieval for open-domain question answering.
\newblock In \emph{Proceedings of the 2020 Conference on Empirical Methods in Natural Language Processing ({EMNLP})}, pages 6769--6781, Stroudsburg, PA, USA, November 2020{\natexlab{a}}. Association for Computational Linguistics.

\bibitem[Karpukhin et~al.(2020{\natexlab{b}})Karpukhin, Oğuz, Min, Lewis, Wu, Edunov, Chen, and tau Yih]{Karpukhin2020DensePR}
Vladimir Karpukhin, Barlas Oğuz, Sewon Min, Patrick Lewis, Ledell~Yu Wu, Sergey Edunov, Danqi Chen, and Wen tau Yih.
\newblock Dense passage retrieval for open-domain question answering.
\newblock In \emph{Conference on Empirical Methods in Natural Language Processing}, 2020{\natexlab{b}}.

\bibitem[Karpur et~al.(2023)Karpur, Lahav, Matheny, Alstott, and Nevo]{karpur2023securing}
Ajay Karpur, Dan Lahav, Jason Matheny, Jeff Alstott, and Sella Nevo.
\newblock Securing artificial intelligence model weights: Interim report.
\newblock 2023.

\bibitem[Kate et~al.(2010)Kate, Zaverucha, and Goldberg]{kzg10}
Aniket Kate, Gregory~M Zaverucha, and Ian Goldberg.
\newblock Constant-size commitments to polynomials and their applications.
\newblock In \emph{Advances in Cryptology-ASIACRYPT 2010: 16th International Conference on the Theory and Application of Cryptology and Information Security, Singapore, December 5-9, 2010. Proceedings 16}, pages 177--194. Springer, 2010.

\bibitem[Kenton et~al.(2021)Kenton, Everitt, Weidinger, Gabriel, Mikulik, and Irving]{Kenton2021AlignmentOL}
Zachary Kenton, Tom Everitt, Laura Weidinger, Iason Gabriel, Vladimir Mikulik, and Geoffrey Irving.
\newblock Alignment of language agents.
\newblock \emph{ArXiv}, abs/2103.14659, 2021.

\bibitem[Kenton et~al.(2023)Kenton, Kumar, Farquhar, Richens, MacDermott, and Everitt]{Kenton2023-cz}
Zachary Kenton, Ramana Kumar, Sebastian Farquhar, Jonathan Richens, Matt MacDermott, and Tom Everitt.
\newblock Discovering agents.
\newblock \emph{Artif. Intell.}, 322\penalty0 (103963):\penalty0 103963, September 2023.

\bibitem[Kiela et~al.(2021)Kiela, Bartolo, Nie, Kaushik, Geiger, Wu, Vidgen, Prasad, Singh, Ringshia, et~al.]{kiela2021dynabench}
Douwe Kiela, Max Bartolo, Yixin Nie, Divyansh Kaushik, Atticus Geiger, Zhengxuan Wu, Bertie Vidgen, Grusha Prasad, Amanpreet Singh, Pratik Ringshia, et~al.
\newblock Dynabench: Rethinking benchmarking in nlp.
\newblock \emph{arXiv preprint arXiv:2104.14337}, 2021.

\bibitem[Kifer and Machanavajjhala(2014)]{Kifer2014}
Daniel Kifer and Ashwin Machanavajjhala.
\newblock Pufferfish: A framework for mathematical privacy definitions.
\newblock \emph{ACM Transactions on Database Systems (TODS)}, 39\penalty0 (1):\penalty0 1--36, 1 2014.
\newblock \doi{10.1145/2514689}.
\newblock URL \url{https://doi.org/10.1145/2514689}.

\bibitem[Kim(2017)]{kim2017auditing}
Pauline~T Kim.
\newblock Auditing algorithms for discrimination.
\newblock \emph{U. Pa. L. Rev. Online}, 166:\penalty0 189, 2017.

\bibitem[Kingma and Welling(2014)]{kingma2014auto}
Diederik~P Kingma and Max Welling.
\newblock Auto-encoding variational bayes.
\newblock In \emph{Proceedings of the International Conference on Learning Representations (ICLR)}, 2014.

\bibitem[Kinniment et~al.(2024)Kinniment, Sato, Du, Goodrich, Hasin, Chan, Miles, Lin, Wijk, Burget, Ho, Barnes, and Christiano]{kinniment_evaluating_2024}
Megan Kinniment, {Lucas Jun Koba} Sato, Haoxing Du, Brian Goodrich, Max Hasin, Lawrence Chan, {Luke Harold} Miles, {Tao R.} Lin, Hjalmar Wijk, Joel Burget, Aaron Ho, Elizabeth Barnes, and Paul Christiano.
\newblock Evaluating language-model agents on realistic autonomous tasks, 2024.
\newblock URL \url{https://arxiv.org/abs/2312.11671}.

\bibitem[Kirk et~al.(2021)Kirk, Jun, Volpin, Iqbal, Benussi, Dreyer, Shtedritski, and Asano]{NEURIPS2021_1531beb7}
Hannah~Rose Kirk, Yennie Jun, Filippo Volpin, Haider Iqbal, Elias Benussi, Frederic Dreyer, Aleksandar Shtedritski, and Yuki Asano.
\newblock Bias out-of-the-box: An empirical analysis of intersectional occupational biases in popular generative language models.
\newblock In M.~Ranzato, A.~Beygelzimer, Y.~Dauphin, P.S. Liang, and J.~Wortman Vaughan, editors, \emph{Advances in Neural Information Processing Systems}, volume~34, pages 2611--2624. Curran Associates, Inc., 2021.
\newblock URL \url{https://proceedings.neurips.cc/paper_files/paper/2021/file/1531beb762df4029513ebf9295e0d34f-Paper.pdf}.

\bibitem[Knott et~al.(2021{\natexlab{a}})Knott, Venkataraman, Hannun, Sengupta, Ibrahim, and van~der Maaten]{knott2021crypten}
Brian Knott, Shobha Venkataraman, Awni Hannun, Shubho Sengupta, Mark Ibrahim, and Laurens van~der Maaten.
\newblock Crypten: Secure multi-party computation meets machine learning.
\newblock \emph{Advances in Neural Information Processing Systems}, 34:\penalty0 4961--4973, 2021{\natexlab{a}}.

\bibitem[Knott et~al.(2021{\natexlab{b}})Knott, Venkataraman, Hannun, Sengupta, Ibrahim, and van~der Maaten]{Knott2021CrypTenSM}
Brian Knott, Shobha Venkataraman, Awni~Y. Hannun, Shubho Sengupta, Mark Ibrahim, and Laurens van~der Maaten.
\newblock Crypten: Secure multi-party computation meets machine learning.
\newblock \emph{ArXiv}, abs/2109.00984, 2021{\natexlab{b}}.

\bibitem[Knott et~al.(2021{\natexlab{c}})]{Knott2021}
Brian Knott et~al.
\newblock Crypten: Secure multi-party computation meets machine learning.
\newblock In \emph{Advances in Neural Information Processing Systems}, volume~34, pages 4961--4973, 2021{\natexlab{c}}.

\bibitem[Kojima et~al.(2022)Kojima, Gu, Reid, Matsuo, and Iwasawa]{Kojima2022LargeLM}
Takeshi Kojima, Shixiang~Shane Gu, Machel Reid, Yutaka Matsuo, and Yusuke Iwasawa.
\newblock Large language models are zero-shot reasoners.
\newblock \emph{ArXiv}, 2022.

\bibitem[Kolt(2024)]{kolt_governing_2024}
Noam Kolt.
\newblock Governing {AI} agents, 2024.
\newblock URL \url{https://papers.ssrn.com/abstract=4772956}.

\bibitem[Komlo and Goldberg(2021)]{FROST}
Chelsea Komlo and Ian Goldberg.
\newblock Frost: flexible round-optimized schnorr threshold signatures.
\newblock In \emph{Selected Areas in Cryptography: 27th International Conference, Halifax, NS, Canada (Virtual Event), October 21-23, 2020, Revised Selected Papers 27}, pages 34--65. Springer, 2021.

\bibitem[Kosinski et~al.(2013)Kosinski, Stillwell, and Graepel]{Kosinski2013b}
Michal Kosinski, David Stillwell, and Thore Graepel.
\newblock Private traits and attributes are predictable from digital records of human behavior.
\newblock \emph{Proc. Natl. Acad. Sci. U. S. A.}, 110\penalty0 (15):\penalty0 5802--5805, April 2013.

\bibitem[Kothapalli et~al.(2022)Kothapalli, Setty, and Tzialla]{kothapalli2022nova}
Abhiram Kothapalli, Srinath Setty, and Ioanna Tzialla.
\newblock Nova: Recursive zero-knowledge arguments from folding schemes.
\newblock In \emph{Annual International Cryptology Conference}, pages 359--388. Springer, 2022.

\bibitem[Kroll(2015)]{kroll2015accountable}
Joshua~Alexander Kroll.
\newblock \emph{Accountable algorithms}.
\newblock PhD thesis, Princeton University, 2015.

\bibitem[Lalor et~al.(2022)Lalor, Yang, Smith, Forsgren, and Abbasi]{lalor2022benchmarkingIntersectional}
John~P. Lalor, Yi~Yang, Kendall Smith, Nicole Forsgren, and Ahmed Abbasi.
\newblock Benchmarking intersectional biases in nlp.
\newblock In \emph{Proceedings of the 2022 Annual Conference of the North American Chapter of the Association for Computational Linguistics}. Association for Computational Linguistics, 2022.

\bibitem[Lamb et~al.(2024)Lamb, Israelstam, Agarwal, and Bhasker]{Lamb2024v}
Jessica Lamb, Greg Israelstam, Rahul Agarwal, and Shashank Bhasker.
\newblock Generative {AI} in healthcare: Adoption trends and what's next.
\newblock Technical report, McKinsey \& Company, July 2024.

\bibitem[Lee(2020)]{Lee2020vCNNVC}
Seunghwan Lee.
\newblock vcnn: Verifiable convolutional neural network based on zk-snarks.
\newblock 2020.

\bibitem[Lewis et~al.(2020{\natexlab{a}})Lewis, Perez, Piktus, Petroni, Karpukhin, Goyal, K{\"u}ttler, Lewis, Yih, Rockt{\"a}schel, Riedel, and Kiela]{Lewis2020l}
Patrick Lewis, Ethan Perez, Aleksandra Piktus, Fabio Petroni, Vladimir Karpukhin, Naman Goyal, Heinrich K{\"u}ttler, Mike Lewis, Wen-Tau Yih, Tim Rockt{\"a}schel, Sebastian Riedel, and Douwe Kiela.
\newblock {Retrieval-{{Augmented} Generation}} for {{{Knowledge-Intensive} {NLP} Tasks}}.
\newblock In \emph{34th Conference on Neural Information Processing Systems ({NeurIPS} 2020)}, volume~33, pages 9459--9474, Vancouver, Canada, 2020{\natexlab{a}}. Curran Associates, Inc.

\bibitem[Lewis et~al.(2020{\natexlab{b}})]{Lewis2020}
Patrick Lewis et~al.
\newblock Retrieval-augmented generation for knowledge-intensive nlp tasks.
\newblock In \emph{Proceedings of the 34th International Conference on Neural Information Processing Systems}, pages 9459--9474, 12 2020{\natexlab{b}}.

\bibitem[Li et~al.(2022)]{Li2022}
Dacheng Li et~al.
\newblock Mpcformer: Fast, performant and private transformer inference with mpc.
\newblock \emph{arXiv}, 11 2022.
\newblock \doi{10.48550/arXiv.2211.01452}.
\newblock URL \url{https://doi.org/10.48550/arXiv.2211.01452}.

\bibitem[Liang et~al.(2022)Liang, Bommasani, Lee, Tsipras, Soylu, Yasunaga, Zhang, Narayanan, Wu, Kumar, et~al.]{liang2022holistic}
Percy Liang, Rishi Bommasani, Tony Lee, Dimitris Tsipras, Dilara Soylu, Michihiro Yasunaga, Yian Zhang, Deepak Narayanan, Yuhuai Wu, Ananya Kumar, et~al.
\newblock Holistic evaluation of language models.
\newblock \emph{arXiv preprint arXiv:2211.09110}, 2022.

\bibitem[Liao(2021)]{Liao2021AreWL}
Thomas Liao.
\newblock Are we learning yet? a meta review of evaluation failures across machine learning.
\newblock In \emph{NeurIPS Datasets and Benchmarks}, 2021.

\bibitem[Lieberman(1997)]{Lieberman1997-qm}
Henry Lieberman.
\newblock Autonomous interface agents.
\newblock In \emph{Proceedings of the {ACM} {SIGCHI} Conference on Human factors in computing systems}, New York, NY, USA, March 1997. ACM.

\bibitem[Liu et~al.(2024)Liu, Pan, Lu, Li, Hu, Wen, King, and Yu]{liu_survey_2024}
Aiwei Liu, Leyi Pan, Yijian Lu, Jingjing Li, Xuming Hu, Lijie Wen, Irwin King, and Philip~S. Yu.
\newblock A {Survey} of {Text} {Watermarking} in the {Era} of {Large} {Language} {Models}, January 2024.
\newblock URL \url{http://arxiv.org/abs/2312.07913}.
\newblock arXiv:2312.07913 [cs].

\bibitem[Liu et~al.(2023{\natexlab{a}})Liu, Chen, Shen, and Choo]{Liu2023FairCompassOF}
Jessica Liu, Huaming Chen, Jun Shen, and Kim-Kwang~Raymond Choo.
\newblock Faircompass: Operationalising fairness in machine learning.
\newblock \emph{ArXiv}, abs/2312.16726, 2023{\natexlab{a}}.

\bibitem[Liu et~al.(2021)Liu, Xie, and Zhang]{Liu2021zkCNNZK}
Tianyi Liu, Xiang Xie, and Yupeng Zhang.
\newblock zkcnn: Zero knowledge proofs for convolutional neural network predictions and accuracy.
\newblock \emph{Proceedings of the 2021 ACM SIGSAC Conference on Computer and Communications Security}, 2021.

\bibitem[Liu et~al.(2023{\natexlab{b}})Liu, Yu, Zhang, Xu, Lei, Lai, Gu, Ding, Men, Yang, Zhang, Deng, Zeng, Du, Zhang, Shen, Zhang, Su, Sun, Huang, Dong, and Tang]{Liu2023-rd}
Xiao Liu, Hao Yu, Hanchen Zhang, Yifan Xu, Xuanyu Lei, Hanyu Lai, Yu~Gu, Hangliang Ding, Kaiwen Men, Kejuan Yang, Shudan Zhang, Xiang Deng, Aohan Zeng, Zhengxiao Du, Chenhui Zhang, Sheng Shen, Tianjun Zhang, Yu~Su, Huan Sun, Minlie Huang, Yuxiao Dong, and Jie Tang.
\newblock {AgentBench}: Evaluating {LLMs} as agents.
\newblock August 2023{\natexlab{b}}.

\bibitem[Liu et~al.(2023{\natexlab{c}})Liu, Deng, Li, Wang, Wang, Wang, Zhang, Liu, Wang, Zheng, and Liu]{Liu2023-bl}
Yi~Liu, Gelei Deng, Yuekang Li, Kailong Wang, Zihao Wang, Xiaofeng Wang, Tianwei Zhang, Yepang Liu, Haoyu Wang, Yan Zheng, and Yang Liu.
\newblock Prompt injection attack against {LLM-integrated} applications.
\newblock June 2023{\natexlab{c}}.

\bibitem[Liu et~al.(2015)Liu, Luo, Wang, and Tang]{liu2015deep}
Ziwei Liu, Ping Luo, Xiaogang Wang, and Xiaoou Tang.
\newblock Deep learning face attributes in the wild.
\newblock In \emph{Proceedings of the IEEE international conference on computer vision}, pages 3730--3738, 2015.

\bibitem[Longpre et~al.(2024{\natexlab{a}})Longpre, Mahari, Lee, Lund, Oderinwale, Brannon, Saxena, Obeng-Marnu, South, Hunter, et~al.]{longpre2024consent}
Shayne Longpre, Robert Mahari, Ariel Lee, Campbell Lund, Hamidah Oderinwale, William Brannon, Nayan Saxena, Naana Obeng-Marnu, Tobin South, Cole Hunter, et~al.
\newblock Consent in crisis: The rapid decline of the ai data commons.
\newblock \emph{arXiv preprint arXiv:2407.14933}, 2024{\natexlab{a}}.

\bibitem[Longpre et~al.(2024{\natexlab{b}})Longpre, Mahari, Lee, Lund, Oderinwale, Brannon, Saxena, Obeng-Marnu, South, Hunter, Klyman, Klamm, Schoelkopf, Singh, Cherep, Anis, Dinh, Chitongo, Yin, Sileo, Mataciunas, Misra, Alghamdi, Shippole, Zhang, Materzynska, Qian, Tiwary, Miranda, Dey, Liang, Hamdy, Muennighoff, Ye, Kim, Mohanty, Gupta, Sharma, Chien, Zhou, Yizhi, Xiong, Villa, Biderman, Li, Ippolito, Hooker, Kabbara, and Pentland]{Longpre2024e}
Shayne Longpre, Robert Mahari, Ariel~N Lee, Campbell~S Lund, Hamidah Oderinwale, William Brannon, Nayan Saxena, Naana Obeng-Marnu, Tobin South, Cole~J Hunter, Kevin Klyman, Christopher Klamm, Hailey Schoelkopf, Nikhil Singh, Manuel Cherep, Ahmad~Mustafa Anis, An~Dinh, Caroline~Shamiso Chitongo, Da~Yin, Damien Sileo, Deividas Mataciunas, Diganta Misra, Emad~A Alghamdi, Enrico Shippole, Jianguo Zhang, Joanna Materzynska, Kun Qian, Kushagra Tiwary, Lester James~Validad Miranda, Manan Dey, Minnie Liang, Mohammed Hamdy, Niklas Muennighoff, Seonghyeon Ye, Seungone Kim, Shrestha Mohanty, Vipul Gupta, Vivek Sharma, Vu~Minh Chien, Xuhui Zhou, L~I Yizhi, Caiming Xiong, Luis Villa, Stella Biderman, Hanlin Li, Daphne Ippolito, Sara Hooker, Jad Kabbara, and Alex Pentland.
\newblock Consent in crisis: The rapid decline of the {AI} data commons.
\newblock In \emph{38th Conference on Neural Information Processing Systems Datasets and Benchmarks Track}, November 2024{\natexlab{b}}.

\bibitem[Lou and Jiang(2021)]{Lou2021HEMETAH}
Qian Lou and Lei Jiang.
\newblock Hemet: A homomorphic-encryption-friendly privacy-preserving mobile neural network architecture.
\newblock In \emph{International Conference on Machine Learning}, 2021.

\bibitem[Lukas et~al.(2023)Lukas, Salem, Sim, Tople, Wutschitz, and Zanella-B{\'e}guelin]{Lukas2023a}
Nils Lukas, Ahmed Salem, Robert Sim, Shruti Tople, Lukas Wutschitz, and Santiago Zanella-B{\'e}guelin.
\newblock Analyzing leakage of personally identifiable information in language models.
\newblock In \emph{2023 {IEEE} Symposium on Security and Privacy ({SP})}, pages 346--363. IEEE, May 2023.

\bibitem[Mahari et~al.(2021)Mahari, Lera, and Pentland]{Mahari2021EraAntitrust}
Robert Mahari, Sandro~Claudio Lera, and Alex~'Sandy' Pentland.
\newblock Time for a new antitrust era: Refocusing antitrust law to invigorate competition in the 21st century.
\newblock \emph{Stanford Computational Antitrust}, 1, 2021.

\bibitem[Mao et~al.(2020)Mao, He, Liu, Shen, Gao, Han, and Chen]{Mao2020GenerationAugmentedRF}
Yuning Mao, Pengcheng He, Xiaodong Liu, Yelong Shen, Jianfeng Gao, Jiawei Han, and Weizhu Chen.
\newblock Generation-augmented retrieval for open-domain question answering.
\newblock In \emph{Annual Meeting of the Association for Computational Linguistics}, 2020.

\bibitem[Marro(2024)]{marro_protocol_2024}
Samuele Marro.
\newblock A {Protocol} {Sketch} {For} {LLM} {Communication}, April 2024.
\newblock URL \url{https://samuelemarro.it/blog/2024/a-protocol-for-llm/}.

\bibitem[May et~al.(2019)May, Wang, Bordia, Bowman, and Rudinger]{May2019OnMS}
Chandler May, Alex Wang, Shikha Bordia, Samuel~R. Bowman, and Rachel Rudinger.
\newblock On measuring social biases in sentence encoders.
\newblock \emph{ArXiv}, abs/1903.10561, 2019.

\bibitem[Mckeen et~al.(2013)Mckeen, Alexandrovich, Berenzon, Rozas, Shafi, Shanbhogue, and Savagaonkar]{McKeen2013}
Frank Mckeen, Ilya Alexandrovich, Alex Berenzon, Carlos Rozas, Hisham Shafi, Vedvyas Shanbhogue, and Uday Savagaonkar.
\newblock {I}nnovative {I}nstructions and {S}oftware {M}odel for {I}solated {E}xecution.
\newblock In \emph{Proc. Second Workshop on Hardware and Architectural Support for Security and Privacy {HASP2013}}, Tel-Aviv, June 2013.
\newblock https://sites.google.com/site/haspworkshop2013/workshop-program.

\bibitem[Mehrabi et~al.(2019)Mehrabi, Morstatter, Saxena, Lerman, and Galstyan]{Mehrabi2019ASO}
Ninareh Mehrabi, Fred Morstatter, Nripsuta~Ani Saxena, Kristina Lerman, and A.~G. Galstyan.
\newblock A survey on bias and fairness in machine learning.
\newblock \emph{ACM Computing Surveys (CSUR)}, 54:\penalty0 1 -- 35, 2019.

\bibitem[Mireshghallah et~al.(2024)Mireshghallah, Antoniak, More, Choi, and Farnadi]{Mireshghallah2024q}
Niloofar Mireshghallah, Maria Antoniak, Yash More, Yejin Choi, and Golnoosh Farnadi.
\newblock Trust no bot: Discovering personal disclosures in {Human-LLM} conversations in the wild.
\newblock In \emph{First Conference on Language Modeling}, August 2024.

\bibitem[Mishra et~al.(2019)Mishra, Lehmkuhl, Srinivasan, Zheng, and Popa]{Mishra2019DE}
Pratyush Mishra, Ryan~T. Lehmkuhl, Akshayaram Srinivasan, Wenting Zheng, and Raluca~A. Popa.
\newblock D elphi : A cryptographic inference service for neural networks.
\newblock 2019.

\bibitem[Mitchell et~al.(2019{\natexlab{a}})Mitchell, Wu, Zaldivar, Barnes, Vasserman, Hutchinson, Spitzer, Raji, and Gebru]{mitchell2019model}
Margaret Mitchell, Simone Wu, Andrew Zaldivar, Parker Barnes, Lucy Vasserman, Ben Hutchinson, Elena Spitzer, Inioluwa~Deborah Raji, and Timnit Gebru.
\newblock Model cards for model reporting.
\newblock In \emph{Proceedings of the conference on fairness, accountability, and transparency}, pages 220--229, 2019{\natexlab{a}}.

\bibitem[Mitchell et~al.(2019{\natexlab{b}})Mitchell, Wu, Zaldivar, Barnes, Vasserman, Hutchinson, Spitzer, Raji, and Gebru]{mitchell_model_2019}
Margaret Mitchell, Simone Wu, Andrew Zaldivar, Parker Barnes, Lucy Vasserman, Ben Hutchinson, Elena Spitzer, Inioluwa~Deborah Raji, and Timnit Gebru.
\newblock Model {Cards} for {Model} {Reporting}.
\newblock In \emph{Proceedings of the {Conference} on {Fairness}, {Accountability}, and {Transparency}}, pages 220--229, January 2019{\natexlab{b}}.
\newblock \doi{10.1145/3287560.3287596}.
\newblock URL \url{http://arxiv.org/abs/1810.03993}.
\newblock arXiv:1810.03993 [cs].

\bibitem[Mo et~al.(2020)Mo, Shamsabadi, Katevas, Demetriou, Leontiadis, Cavallaro, and Haddadi]{Mo2020DarkneTZTM}
Fan Mo, Ali~Shahin Shamsabadi, Kleomenis Katevas, Soteris Demetriou, Ilias Leontiadis, Andrea Cavallaro, and Hamed Haddadi.
\newblock Darknetz: towards model privacy at the edge using trusted execution environments.
\newblock \emph{Proceedings of the 18th International Conference on Mobile Systems, Applications, and Services}, 2020.

\bibitem[Mohassel and Zhang(2017)]{Mohassel2017}
Payman Mohassel and Yupeng Zhang.
\newblock Secureml: A system for scalable privacy-preserving machine learning.
\newblock In \emph{2017 IEEE Symposium on Security and Privacy (SP)}, pages 19--38. IEEE, 2017.
\newblock \doi{10.1109/SP.2017.12}.

\bibitem[Mosqueira-Rey et~al.(2023)Mosqueira-Rey, Hern{\'a}ndez-Pereira, Alonso-R{\'\i}os, Bobes-Bascar{\'a}n, and Fern{\'a}ndez-Leal]{mosqueira2023human}
Eduardo Mosqueira-Rey, Elena Hern{\'a}ndez-Pereira, David Alonso-R{\'\i}os, Jos{\'e} Bobes-Bascar{\'a}n, and {\'A}ngel Fern{\'a}ndez-Leal.
\newblock Human-in-the-loop machine learning: a state of the art.
\newblock \emph{Artificial Intelligence Review}, 56\penalty0 (4):\penalty0 3005--3054, 2023.

\bibitem[Muthukumar et~al.(2018)Muthukumar, Pedapati, Ratha, Sattigeri, Wu, Kingsbury, Kumar, Thomas, Mojsilovic, and Varshney]{Muthukumar2018UnderstandingUG}
Vidya Muthukumar, Tejaswini Pedapati, Nalini~K. Ratha, Prasanna Sattigeri, Chai-Wah Wu, Brian Kingsbury, Abhishek Kumar, Samuel Thomas, Aleksandra Mojsilovic, and Kush~R. Varshney.
\newblock Understanding unequal gender classification accuracy from face images.
\newblock \emph{ArXiv}, abs/1812.00099, 2018.

\bibitem[Nakandala et~al.(2019)Nakandala, Yu, Weimer, and Interlandi]{nakandala2019compiling}
Supun Nakandala, Gyeong-In Yu, Markus Weimer, and Matteo Interlandi.
\newblock Compiling classical ml pipelines into tensor computations for one-size-fits-all prediction serving.
\newblock In \emph{System for ML Workshop}. NeurIPS, 2019.

\bibitem[Nakano et~al.(2021)Nakano, Hilton, Balaji, Wu, Ouyang, Kim, Hesse, Jain, Kosaraju, Saunders, Jiang, Cobbe, Eloundou, Krueger, Button, Knight, Chess, and Schulman]{Nakano2021n}
Reiichiro Nakano, Jacob Hilton, Suchir Balaji, Jeff Wu, Long Ouyang, Christina Kim, Christopher Hesse, Shantanu Jain, Vineet Kosaraju, William Saunders, Xu~Jiang, Karl Cobbe, Tyna Eloundou, Gretchen Krueger, Kevin Button, Matthew Knight, Benjamin Chess, and John Schulman.
\newblock {WebGPT}: Browser-assisted question-answering with human feedback.
\newblock Technical report, OpenAI, December 2021.

\bibitem[Nasr et~al.(2018)Nasr, Shokri, and Houmansadr]{Nasr2018ComprehensivePA}
Milad Nasr, R.~Shokri, and Amir Houmansadr.
\newblock Comprehensive privacy analysis of deep learning: Passive and active white-box inference attacks against centralized and federated learning.
\newblock \emph{2019 IEEE Symposium on Security and Privacy (SP)}, pages 739--753, 2018.

\bibitem[{National Conference of Commissioners on Uniform State Laws}()]{ueta_official_text}
{National Conference of Commissioners on Uniform State Laws}.
\newblock {Uniform Electronic Transactions Act (UETA)}.
\newblock Proposed official text, 1999.

\bibitem[Neuman et~al.(2005)Neuman, Yu, Hartman, and Raeburn]{RFC4120-Formatted}
C.~Neuman, T.~Yu, S.~Hartman, and K.~Raeburn.
\newblock {The Kerberos Network Authentication Service (V5)}, July 2005.
\newblock URL \url{https://datatracker.ietf.org/doc/html/rfc4120}.
\newblock {IETF}~{S}tandard~{RFC4120}.

\bibitem[Nikolaenko et~al.(2022)Nikolaenko, Ragsdale, Bonneau, and Boneh]{nikolaenko2022powers}
Valeria Nikolaenko, Sam Ragsdale, Joseph Bonneau, and Dan Boneh.
\newblock Powers-of-tau to the people: Decentralizing setup ceremonies.
\newblock \emph{Cryptology ePrint Archive}, 2022.

\bibitem[Nikolaenko et~al.(2024)Nikolaenko, Ragsdale, Bonneau, and Boneh]{nikolaenko2024powers}
Valeria Nikolaenko, Sam Ragsdale, Joseph Bonneau, and Dan Boneh.
\newblock Powers-of-tau to the people: Decentralizing setup ceremonies.
\newblock In \emph{International Conference on Applied Cryptography and Network Security}, pages 105--134. Springer, 2024.

\bibitem[Nissenbaum(2004)]{nissenbaum2004privacy}
Helen Nissenbaum.
\newblock Privacy as contextual integrity.
\newblock \emph{Washington Law Review}, 79\penalty0 (1):\penalty0 119, 2004.

\bibitem[Nissenbaum(2009{\natexlab{a}})]{Nissenbaum2009c}
Helen Nissenbaum.
\newblock \emph{Privacy in Context: Technology, Policy, and the Integrity of Social Life}.
\newblock Stanford University Press, Palo Alto, CA, 2009{\natexlab{a}}.

\bibitem[Nissenbaum(2009{\natexlab{b}})]{nissenbaum2009privacy}
Helen Nissenbaum.
\newblock \emph{Privacy in Context: Technology, Policy, and the Integrity of Social Life}.
\newblock Stanford University Press, Stanford, California, 2009{\natexlab{b}}.

\bibitem[OASIS(2013)]{xacml}
OASIS.
\newblock extensible access control markup language (xacml) version 3.0.
\newblock 2013.
\newblock URL \url{https://docs.oasis-open.org/xacml/3.0/xacml-3.0-core-spec-os-en.html}.

\bibitem[O'Brien et~al.(2023)O'Brien, Ee, and Williams]{obrien_deployment_2023}
Joe O'Brien, Shaun Ee, and Zoe Williams.
\newblock Deployment {Corrections}: {An} incident response framework for frontier {AI} models, September 2023.
\newblock URL \url{http://arxiv.org/abs/2310.00328}.
\newblock arXiv:2310.00328 [cs].

\bibitem[Olszewski et~al.(2023)Olszewski, Lu, Stillman, Warren, Kitroser, Pascual, Ukirde, Butler, and Traynor]{10.1145/3576915.3623130}
Daniel Olszewski, Allison Lu, Carson Stillman, Kevin Warren, Cole Kitroser, Alejandro Pascual, Divyajyoti Ukirde, Kevin Butler, and Patrick Traynor.
\newblock "get in researchers; we're measuring reproducibility": A reproducibility study of machine learning papers in tier 1 security conferences.
\newblock In \emph{Proceedings of the 2023 ACM SIGSAC Conference on Computer and Communications Security}, CCS '23, page 3433–3459, New York, NY, USA, 2023. Association for Computing Machinery.
\newblock ISBN 9798400700507.
\newblock \doi{10.1145/3576915.3623130}.

\bibitem[OpenAI(2023{\natexlab{a}})]{GPT4}
OpenAI.
\newblock Gpt-4 technical report.
\newblock \emph{ArXiv}, abs/2303.08774, 2023{\natexlab{a}}.
\newblock URL \url{https://api.semanticscholar.org/CorpusID:257532815}.

\bibitem[OpenAI(2023{\natexlab{b}})]{openai_chatgpt_2023}
OpenAI.
\newblock {ChatGPT} plugins, 2023{\natexlab{b}}.
\newblock URL \url{https://openai.com/blog/chatgpt-plugins}.

\bibitem[{OpenAI}(2024)]{OpenAI2024d}
{OpenAI}.
\newblock {GPT-4o} system card.
\newblock Technical report, OpenAI, August 2024.

\bibitem[Orsini et~al.(2020)Orsini, Smart, and Vercauteren]{orsini2020overdrive2k}
Emmanuela Orsini, Nigel~P Smart, and Frederik Vercauteren.
\newblock Overdrive2k: efficient secure mpc over from somewhat homomorphic encryption.
\newblock In \emph{Cryptographers' Track at the RSA Conference}, pages 254--283. Springer, 2020.

\bibitem[Ouyang et~al.(2022)Ouyang, Wu, Jiang, Almeida, Wainwright, Mishkin, Zhang, Agarwal, Slama, Ray, et~al.]{RLHF}
Long Ouyang, Jeffrey Wu, Xu~Jiang, Diogo Almeida, Carroll Wainwright, Pamela Mishkin, Chong Zhang, Sandhini Agarwal, Katarina Slama, Alex Ray, et~al.
\newblock Training language models to follow instructions with human feedback.
\newblock \emph{Advances in Neural Information Processing Systems}, 35:\penalty0 27730--27744, 2022.

\bibitem[Papadopoulos et~al.(2010)Papadopoulos, Bakiras, and Papadias]{Papadopoulos2010NearestNS}
Stavros Papadopoulos, Spiridon Bakiras, and Dimitris Papadias.
\newblock Nearest neighbor search with strong location privacy.
\newblock \emph{Proceedings of the VLDB Endowment}, 3:\penalty0 619 -- 629, 2010.

\bibitem[Parraga et~al.(2023)Parraga, M{\'o}re, de~Oliveira, Gavenski, Kupssinsk{\"u}, Medronha, Moura, Sim{\~o}es, and Barros]{Parraga2023FairnessID}
Ot{\'a}vio Parraga, Martin~D. M{\'o}re, Christian~Mattjie de~Oliveira, Nathan~S. Gavenski, Lucas~S. Kupssinsk{\"u}, Adilson Medronha, Luis~V. Moura, Gabriel~S. Sim{\~o}es, and Rodrigo~C. Barros.
\newblock Fairness in deep learning: A survey on vision and language research.
\newblock \emph{ACM Computing Surveys}, 2023.

\bibitem[Patel et~al.(2012)Patel, Garasia, and Jinwala]{patel2012efficient}
Sankita Patel, Sweta Garasia, and Devesh Jinwala.
\newblock An efficient approach for privacy preserving distributed k-means clustering based on shamir's secret sharing scheme.
\newblock In \emph{Trust Management VI: 6th IFIP WG 11.11 International Conference, IFIPTM 2012, Surat, India, May 21-25, 2012. Proceedings 6}, pages 129--141. Springer, 2012.

\bibitem[Patil et~al.(2024)Patil, Zhang, Fang, C., Huang, Hao, Casado, Gonzalez, Popa, and Stoica]{patil_goex_2024}
Shishir~G. Patil, Tianjun Zhang, Vivian Fang, Noppapon C., Roy Huang, Aaron Hao, Martin Casado, Joseph~E. Gonzalez, Raluca~Ada Popa, and Ion Stoica.
\newblock {GoEX}: {Perspectives} and {Designs} {Towards} a {Runtime} for {Autonomous} {LLM} {Applications}, April 2024.
\newblock URL \url{http://arxiv.org/abs/2404.06921}.
\newblock arXiv:2404.06921 [cs].

\bibitem[Paullada et~al.(2021)Paullada, Raji, Bender, Denton, and Hanna]{paullada_data_2021}
Amandalynne Paullada, Inioluwa~Deborah Raji, Emily~M. Bender, Emily Denton, and Alex Hanna.
\newblock Data and its (dis)contents: {A} survey of dataset development and use in machine learning research.
\newblock \emph{Patterns}, 2\penalty0 (11):\penalty0 100336, November 2021.
\newblock ISSN 2666-3899.
\newblock \doi{10.1016/j.patter.2021.100336}.
\newblock URL \url{https://www.sciencedirect.com/science/article/pii/S2666389921001847}.

\bibitem[Pentland et~al.(2021)Pentland, Lipton, and Hardjono]{PentlandBuilding}
Alex Pentland, Alexander Lipton, and Thomas Hardjono.
\newblock \emph{Building the New Economy: Data as Capital}.
\newblock The MIT Press, 2021.
\newblock ISBN 9780262543156.

\bibitem[Phuong et~al.(2024)Phuong, Aitchison, Catt, Cogan, Kaskasoli, Krakovna, Lindner, Rahtz, Assael, Hodkinson, Howard, Lieberum, Kumar, Raad, Webson, Ho, Lin, Farquhar, Hutter, Deletang, Ruoss, El-Sayed, Brown, Dragan, Shah, Dafoe, and Shevlane]{phuong_evaluating_2024}
Mary Phuong, Matthew Aitchison, Elliot Catt, Sarah Cogan, Alexandre Kaskasoli, Victoria Krakovna, David Lindner, Matthew Rahtz, Yannis Assael, Sarah Hodkinson, Heidi Howard, Tom Lieberum, Ramana Kumar, Maria~Abi Raad, Albert Webson, Lewis Ho, Sharon Lin, Sebastian Farquhar, Marcus Hutter, Gregoire Deletang, Anian Ruoss, Seliem El-Sayed, Sasha Brown, Anca Dragan, Rohin Shah, Allan Dafoe, and Toby Shevlane.
\newblock Evaluating frontier models for dangerous capabilities, 2024.
\newblock URL \url{http://arxiv.org/abs/2403.13793}.

\bibitem[Pineau et~al.(2021)Pineau, Vincent-Lamarre, Sinha, Larivi\`{e}re, Beygelzimer, d'Alch\'{e} Buc, Fox, and Larochelle]{10.5555/3546258.3546422}
Joelle Pineau, Philippe Vincent-Lamarre, Koustuv Sinha, Vincent Larivi\`{e}re, Alina Beygelzimer, Florence d'Alch\'{e} Buc, Emily Fox, and Hugo Larochelle.
\newblock Improving reproducibility in machine learning research (a report from the neurips 2019 reproducibility program).
\newblock \emph{J. Mach. Learn. Res.}, 22\penalty0 (1), jan 2021.
\newblock ISSN 1532-4435.

\bibitem[Posner(2019)]{posner_econ_analysis_law}
Richard~A. Posner.
\newblock \emph{Economic Analysis of Law}.
\newblock Wolters Kluwer, New York, 10th edition, 2019.

\bibitem[Premchand and Choudhry(2018)]{openbanking}
Anshu Premchand and Anurag Choudhry.
\newblock Open banking \& apis for transformation in banking.
\newblock \emph{2018 International Conference on Communication, Computing and Internet of Things (IC3IoT)}, pages 25--29, 2018.

\bibitem[Putnam(2000)]{putnam2000bowling}
Robert~D Putnam.
\newblock \emph{Bowling alone: The collapse and revival of American community}.
\newblock Simon and schuster, 2000.

\bibitem[Radford and Narasimhan(2018)]{Radford2018ImprovingLU}
Alec Radford and Karthik Narasimhan.
\newblock Improving language understanding by generative pre-training.
\newblock 2018.

\bibitem[Radford et~al.(2019)Radford, Wu, Child, Luan, Amodei, and Sutskever]{radford2019language}
Alec Radford, Jeffrey Wu, Rewon Child, David Luan, Dario Amodei, and Ilya Sutskever.
\newblock Language models are unsupervised multitask learners.
\newblock \emph{OpenAI blog}, 2019.

\bibitem[Raji and Buolamwini(2019{\natexlab{a}})]{Raji2019ActionableAI}
Inioluwa~Deborah Raji and Joy Buolamwini.
\newblock Actionable auditing: Investigating the impact of publicly naming biased performance results of commercial ai products.
\newblock \emph{Proceedings of the 2019 AAAI/ACM Conference on AI, Ethics, and Society}, 2019{\natexlab{a}}.

\bibitem[Raji and Buolamwini(2019{\natexlab{b}})]{raji2019actionable}
Inioluwa~Deborah Raji and Joy Buolamwini.
\newblock Actionable auditing: Investigating the impact of publicly naming biased performance results of commercial ai products.
\newblock In \emph{Proceedings of the 2019 AAAI/ACM Conference on AI, Ethics, and Society}, pages 429--435, 2019{\natexlab{b}}.

\bibitem[Raji et~al.(2021)Raji, Bender, Paullada, Denton, and Hanna]{raji2021ai}
Inioluwa~Deborah Raji, Emily~M Bender, Amandalynne Paullada, Emily Denton, and Alex Hanna.
\newblock Ai and the everything in the whole wide world benchmark.
\newblock \emph{arXiv preprint arXiv:2111.15366}, 2021.

\bibitem[Raji et~al.(2022)Raji, Kumar, Horowitz, and Selbst]{Raji2022-kj}
Inioluwa~Deborah Raji, I~Elizabeth Kumar, Aaron Horowitz, and Andrew Selbst.
\newblock The fallacy of {AI} functionality.
\newblock In \emph{2022 {ACM} Conference on Fairness, Accountability, and Transparency}, volume~12, pages 959--972, New York, NY, USA, June 2022. ACM.

\bibitem[Ram et~al.(2023)Ram, Levine, Dalmedigos, Muhlgay, Shashua, Leyton-Brown, and Shoham]{Ram2023w}
Ori Ram, Yoav Levine, Itay Dalmedigos, Dor Muhlgay, Amnon Shashua, Kevin Leyton-Brown, and Yoav Shoham.
\newblock {In-Context} retrieval-augmented language models.
\newblock \emph{Trans. Assoc. Comput. Linguist.}, 11:\penalty0 1316--1331, November 2023.

\bibitem[Rawls(1993)]{rawls1993political}
John Rawls.
\newblock \emph{Political Liberalism}.
\newblock Columbia University Press, New York, 1993.

\bibitem[Reimers and Gurevych(2019)]{reimers-2019-sentence-bert}
Nils Reimers and Iryna Gurevych.
\newblock Sentence-bert: Sentence embeddings using siamese bert-networks.
\newblock In \emph{Proceedings of the 2019 Conference on Empirical Methods in Natural Language Processing}. Association for Computational Linguistics, 11 2019.
\newblock URL \url{https://arxiv.org/abs/1908.10084}.

\bibitem[Reimers and Gurevych(2020)]{reimers-2020-multilingual-sentence-bert}
Nils Reimers and Iryna Gurevych.
\newblock Making monolingual sentence embeddings multilingual using knowledge distillation.
\newblock In \emph{Proceedings of the 2020 Conference on Empirical Methods in Natural Language Processing}. Association for Computational Linguistics, 11 2020.
\newblock URL \url{https://arxiv.org/abs/2004.09813}.

\bibitem[{Reka Team} et~al.(2024){Reka Team}, Ormazabal, Zheng, d'Autume, Yogatama, Fu, Ong, Chen, Lamprecht, Pham, Ong, Aleksiev, Li, Henderson, Bain, Artetxe, Relan, Padlewski, Liu, Chen, Phua, Yang, Tay, Wang, Zhu, and Xie]{RekaTeam2024r}
{Reka Team}, Aitor Ormazabal, Che Zheng, Cyprien de~Masson d'Autume, Dani Yogatama, Deyu Fu, Donovan Ong, Eric Chen, Eugenie Lamprecht, Hai Pham, Isaac Ong, Kaloyan Aleksiev, Lei Li, Matthew Henderson, Max Bain, Mikel Artetxe, Nishant Relan, Piotr Padlewski, Qi~Liu, Ren Chen, Samuel Phua, Yazheng Yang, Yi~Tay, Yuqi Wang, Zhongkai Zhu, and Zhihui Xie.
\newblock Reka core, flash, and edge: A series of powerful multimodal language models.
\newblock April 2024.

\bibitem[Reuel et~al.(2024)Reuel, Bucknall, Casper, Fist, Soder, Aarne, Hammond, Ibrahim, Chan, Wills, et~al.]{reuel2024open}
Anka Reuel, Ben Bucknall, Stephen Casper, Tim Fist, Lisa Soder, Onni Aarne, Lewis Hammond, Lujain Ibrahim, Alan Chan, Peter Wills, et~al.
\newblock Open problems in technical ai governance.
\newblock \emph{arXiv preprint arXiv:2407.14981}, 2024.

\bibitem[Riazi et~al.(2018)Riazi, Weinert, Tkachenko, Songhori, Schneider, and Koushanfar]{riazi2018chameleon}
M~Sadegh Riazi, Christian Weinert, Oleksandr Tkachenko, Ebrahim~M Songhori, Thomas Schneider, and Farinaz Koushanfar.
\newblock Chameleon: A hybrid secure computation framework for machine learning applications.
\newblock In \emph{Proceedings of the 2018 on Asia conference on computer and communications security}, pages 707--721, 2018.

\bibitem[Richer and Imbault(2024)]{rfc9635}
Justin Richer and Fabien Imbault.
\newblock {Grant Negotiation and Authorization Protocol (GNAP)}.
\newblock RFC 9635, Internet Engineering Task Force (IETF), October 2024.
\newblock URL \url{https://www.rfc-editor.org/info/rfc9635}.

\bibitem[Rivera-Zamarripa et~al.(2019)Rivera-Zamarripa, Rodriguez, Ch{\'a}vez, Cort{\'e}s, and Rodr{\'i}guez-Henr{\'i}quez]{mexicoPKI}
Luis Rivera-Zamarripa, Lil~M. Rodriguez, Miguel Angel~L{\'e}on Ch{\'a}vez, Nareli~Cruz Cort{\'e}s, and Francisco Rodr{\'i}guez-Henr{\'i}quez.
\newblock Security analysis of the mexican fiscal digital certificate system.
\newblock \emph{Computaci{\'o}n y Sistemas}, 2019.

\bibitem[Saab et~al.(2024)Saab, Tu, Weng, Tanno, Stutz, Wulczyn, Zhang, Strother, Park, Vedadi, Chaves, Hu, Schaekermann, Kamath, Cheng, Barrett, Cheung, Mustafa, Palepu, McDuff, Hou, Golany, Liu, Alayrac, Houlsby, Tomasev, Freyberg, Lau, Kemp, Lai, Azizi, Kanada, Man, Kulkarni, Sun, Shakeri, He, Caine, Webson, Latysheva, Johnson, Mansfield, Lu, Rivlin, Anderson, Green, Wong, Krause, Shlens, Dominowska, Ali~Eslami, Chou, Cui, Vinyals, Kavukcuoglu, Manyika, Dean, Hassabis, Matias, Webster, Barral, Corrado, Semturs, Sara~Mahdavi, Gottweis, Karthikesalingam, and Natarajan]{Saab2024m}
Khaled Saab, Tao Tu, Wei-Hung Weng, Ryutaro Tanno, David Stutz, Ellery Wulczyn, Fan Zhang, Tim Strother, Chunjong Park, Elahe Vedadi, Juanma~Zambrano Chaves, Szu-Yeu Hu, Mike Schaekermann, Aishwarya Kamath, Yong Cheng, David G~T Barrett, Cathy Cheung, Basil Mustafa, Anil Palepu, Daniel McDuff, Le~Hou, Tomer Golany, Luyang Liu, Jean-Baptiste Alayrac, Neil Houlsby, Nenad Tomasev, Jan Freyberg, Charles Lau, Jonas Kemp, Jeremy Lai, Shekoofeh Azizi, Kimberly Kanada, Siwai Man, Kavita Kulkarni, Ruoxi Sun, Siamak Shakeri, Luheng He, Ben Caine, Albert Webson, Natasha Latysheva, Melvin Johnson, Philip Mansfield, Jian Lu, Ehud Rivlin, Jesper Anderson, Bradley Green, Renee Wong, Jonathan Krause, Jonathon Shlens, Ewa Dominowska, S~M Ali~Eslami, Katherine Chou, Claire Cui, Oriol Vinyals, Koray Kavukcuoglu, James Manyika, Jeff Dean, Demis Hassabis, Yossi Matias, Dale Webster, Joelle Barral, Greg Corrado, Christopher Semturs, S~Sara~Mahdavi, Juraj Gottweis, Alan Karthikesalingam, and Vivek Natarajan.
\newblock Capabilities of gemini models in medicine.
\newblock Technical report, Google Deepmind, April 2024.

\bibitem[Saberi et~al.(2024)Saberi, Sadasivan, Rezaei, Kumar, Chegini, Wang, and Feizi]{saberi_robustness_2024}
Mehrdad Saberi, Vinu~Sankar Sadasivan, Keivan Rezaei, Aounon Kumar, Atoosa Chegini, Wenxiao Wang, and Soheil Feizi.
\newblock Robustness of {AI}-image detectors: Fundamental limits and practical attacks, 2024.
\newblock URL \url{https://arxiv.org/abs/2310.00076}.

\bibitem[Sak et~al.(2014)Sak, Senior, and Beaufays]{sak2014long}
Hasim Sak, Andrew~W Senior, and Fran{\c{c}}oise Beaufays.
\newblock Long short-term memory recurrent neural network architectures for large scale acoustic modeling.
\newblock 2014.

\bibitem[Sakimura et~al.(2014)Sakimura, Bradley, Jones, de~Medeiros, and Mortimore]{OIDC1.0}
N.~Sakimura, J.~Bradley, M.~Jones, B.~de~Medeiros, and C.~Mortimore.
\newblock {OpenID} {C}onnect {C}ore {1.0}.
\newblock Technical specification {v1.0} -- errata set 1, OpenID Foundation, November 2014.
\newblock http://openid.net/specs/openid-connect-core-1\_0.html.

\bibitem[Sanders(2020)]{sanders2020efficient}
Olivier Sanders.
\newblock Efficient redactable signature and application to anonymous credentials.
\newblock In \emph{Public-Key Cryptography--PKC 2020: 23rd IACR International Conference on Practice and Theory of Public-Key Cryptography, Edinburgh, UK, May 4--7, 2020, Proceedings, Part II}, pages 628--656. Springer, 2020.

\bibitem[Schoppmann et~al.(2018)Schoppmann, Gasc{\'o}n, and Balle]{schoppmann2018private}
Phillipp Schoppmann, Adri{\`a} Gasc{\'o}n, and Borja Balle.
\newblock Private nearest neighbors classification in federated databases.
\newblock \emph{IACR Cryptol. ePrint Arch.}, page 289, 2018.

\bibitem[Schwartz et~al.(2022)Schwartz, Vassilev, Greene, Perine, Burt, and Hall]{Schwartz2022TowardsAS}
Reva Schwartz, Apostol~T. Vassilev, Kristen Greene, Lori~A. Perine, Andrew Burt, and Patrick Hall.
\newblock Towards a standard for identifying and managing bias in artificial intelligence.
\newblock 2022.

\bibitem[Selbst and Powles(2017)]{Selbst2017MeaningfulIA}
Andrew~D Selbst and Julia Powles.
\newblock {Meaningful information and the right to explanation}.
\newblock \emph{International Data Privacy Law}, 7\penalty0 (4):\penalty0 233--242, 12 2017.
\newblock ISSN 2044-3994.
\newblock \doi{10.1093/idpl/ipx022}.
\newblock URL \url{https://doi.org/10.1093/idpl/ipx022}.

\bibitem[Semmelrock et~al.(2023)Semmelrock, Kopeinik, Theiler, Ross-Hellauer, and Kowald]{semmelrock2023reproducibility}
Harald Semmelrock, Simone Kopeinik, Dieter Theiler, Tony Ross-Hellauer, and Dominik Kowald.
\newblock Reproducibility in machine learning-driven research.
\newblock \emph{arXiv preprint arXiv:2307.10320}, 2023.

\bibitem[Sengupta(2024)]{sengupta_russian_2024}
Trisha Sengupta.
\newblock Russian man uses {AI} for online dating, claims it helped him find his wife.
\newblock Hindustan Times, 2024.
\newblock URL \url{https://www.hindustantimes.com/trending/russian-man-uses-ai-for-online-dating-claims-it-helped-him-find-his-wife-101706798506466.html}.

\bibitem[Servan-Schreiber et~al.(2022)Servan-Schreiber, Langowski, and Devadas]{servan2022private}
Sacha Servan-Schreiber, Simon Langowski, and Srinivas Devadas.
\newblock Private approximate nearest neighbor search with sublinear communication.
\newblock In \emph{2022 IEEE Symposium on Security and Privacy (SP)}, pages 911--929. IEEE, 2022.

\bibitem[Setty(2020)]{setty2020spartan}
Srinath Setty.
\newblock Spartan: Efficient and general-purpose zksnarks without trusted setup.
\newblock In \emph{Annual International Cryptology Conference}, pages 704--737. Springer, 2020.

\bibitem[Shah et~al.(2019)Shah, Schwartz, and Hovy]{shah2019predictive}
Deven Shah, H~Andrew Schwartz, and Dirk Hovy.
\newblock Predictive biases in natural language processing models: A conceptual framework and overview.
\newblock \emph{arXiv preprint arXiv:1912.11078}, 2019.

\bibitem[Shamir(1979)]{shamir1979share}
Adi Shamir.
\newblock How to share a secret.
\newblock \emph{Communications of the ACM}, 22\penalty0 (11):\penalty0 612--613, 1979.

\bibitem[Shaul et~al.(2018{\natexlab{a}})Shaul, Feldman, and Rus]{shaul2018scalable}
Hayim Shaul, Dan Feldman, and Daniela Rus.
\newblock Scalable secure computation of statistical functions with applications to k-nearest neighbors.
\newblock \emph{arXiv preprint arXiv:1801.07301}, 2018{\natexlab{a}}.

\bibitem[Shaul et~al.(2018{\natexlab{b}})Shaul, Feldman, and Rus]{shaul2018secure}
Hayim Shaul, Dan Feldman, and Daniela Rus.
\newblock Secure $ k $-ish nearest neighbors classifier.
\newblock \emph{arXiv preprint arXiv:1801.07301}, 2018{\natexlab{b}}.

\bibitem[Shavit et~al.(2023)Shavit, Agarwal, Brundage, Adler, O'Keefe, Campbell, Lee, Mishkin, Eloundou, Hickey, et~al.]{shavit2023practices}
Yonadav Shavit, Sandhini Agarwal, Miles Brundage, Steven Adler, Cullen O'Keefe, Rosie Campbell, Teddy Lee, Pamela Mishkin, Tyna Eloundou, Alan Hickey, et~al.
\newblock Practices for governing agentic ai systems.
\newblock \emph{Research Paper, OpenAI, December}, 2023.

\bibitem[Shi et~al.(2023)Shi, Ajith, Xia, Huang, Liu, Blevins, Chen, and Zettlemoyer]{Shi2023s}
Weijia Shi, Anirudh Ajith, Mengzhou Xia, Yangsibo Huang, Daogao Liu, Terra Blevins, Danqi Chen, and Luke Zettlemoyer.
\newblock Detecting pretraining data from large language models.
\newblock In \emph{The 12th International Conference on Learning Representations ({ICLR} 2024)}, Vienna, Austria, October 2023.

\bibitem[Shokri et~al.(2017)Shokri, Stronati, Song, and Shmatikov]{Shokri2017m}
Reza Shokri, Marco Stronati, Congzheng Song, and Vitaly Shmatikov.
\newblock Membership inference attacks against machine learning models.
\newblock In \emph{2017 {IEEE} Symposium on Security and Privacy ({SP})}, pages 3--18, San Jose, CA, USA, 2017. IEEE.

\bibitem[Simpson(1996)]{RFC1994-Formatted}
W.~Simpson.
\newblock {PPP Challenge Handshake Authentication Protocol (CHAP)}, August 1996.
\newblock URL \url{https://datatracker.ietf.org/doc/rfc1994/}.
\newblock {IETF}~{Standard}~{RFC1996}.

\bibitem[Singh et~al.(2024)Singh, Singla, Si, and Krishnamurthy]{Singh2024-dz}
Somesh Singh, Yaman~K Singla, Harini Si, and Balaji Krishnamurthy.
\newblock Measuring and improving persuasiveness of large language models.
\newblock October 2024.

\bibitem[Solaiman et~al.(2023)Solaiman, Talat, Agnew, Ahmad, Baker, Blodgett, Daum{\'e}~III, Dodge, Evans, Hooker, et~al.]{solaiman2023evaluating}
Irene Solaiman, Zeerak Talat, William Agnew, Lama Ahmad, Dylan Baker, Su~Lin Blodgett, Hal Daum{\'e}~III, Jesse Dodge, Ellie Evans, Sara Hooker, et~al.
\newblock Evaluating the social impact of generative ai systems in systems and society.
\newblock \emph{arXiv preprint arXiv:2306.05949}, 2023.

\bibitem[Solove(2025)]{Solove2025r}
Daniel~J Solove.
\newblock Artificial intelligence and privacy.
\newblock \emph{Florida Law Review}, 2025.

\bibitem[Solum(1992)]{solum_legal_2020}
{Lawrence B.} Solum.
\newblock Legal personhood for artificial intelligences.
\newblock \emph{North Carolina Law Review}, 70\penalty0 (4):\penalty0 415--471, 1992.
\newblock URL \url{https://scholarship.law.unc.edu/cgi/viewcontent.cgi?article=3447&context=nclr}.

\bibitem[Songhori et~al.(2015)Songhori, Hussain, Sadeghi, and Koushanfar]{songhori2015compacting}
Ebrahim~M Songhori, Siam~U Hussain, Ahmad-Reza Sadeghi, and Farinaz Koushanfar.
\newblock Compacting privacy-preserving k-nearest neighbor search using logic synthesis.
\newblock In \emph{Proceedings of the 52nd Annual Design Automation Conference}, pages 1--6, 2015.

\bibitem[South et~al.(2023{\natexlab{a}})South, Mahari, and Pentland]{South2023-wh}
Tobin South, Robert Mahari, and Alex Pentland.
\newblock Transparency by design for large language models.
\newblock \emph{Network Law Review}, Computational Legal Futures, May 2023{\natexlab{a}}.

\bibitem[South et~al.(2023{\natexlab{b}})South, Zuskind, Mahari, and Hardjono]{southsecure}
Tobin South, Guy Zuskind, Robert Mahari, and Thomas Hardjono.
\newblock Secure community transformers: Private pooled data for llms.
\newblock 2023{\natexlab{b}}.

\bibitem[South et~al.(2024{\natexlab{a}})South, Drean, Singh, Zyskind, Mahari, Sharma, Vepakomma, Kagal, Devadas, and Pentland]{South2024Roadmap}
Tobin South, Jules Drean, Abhishek Singh, Guy Zyskind, Robert Mahari, Vivek Sharma, Praneeth Vepakomma, Lalana Kagal, Srinivas Devadas, and Alex Pentland.
\newblock A {Roadmap} for {End}-to-{End} {Privacy} and {Security} in {Generative} {AI}.
\newblock \emph{An MIT Exploration of Generative AI}, sep 10 2024{\natexlab{a}}.
\newblock https://mit-genai.pubpub.org/pub/m4msmss4.

\bibitem[South et~al.(2025)South, Marro, Hardjono, Mahari, Whitney, Greenwood, Chan, and Pentland]{south2025authenticated}
Tobin South, Samuele Marro, Thomas Hardjono, Robert Mahari, Cedric~Deslandes Whitney, Dazza Greenwood, Alan Chan, and Alex Pentland.
\newblock Authenticated delegation and authorized ai agents.
\newblock \emph{Forty-Second International Conference on Machine Learning}, 2025.

\bibitem[South et~al.(2024{\natexlab{b}})]{SouthVeriableEvaluations}
Tobin South et~al.
\newblock Verifiable evaluations of machine learning models using zksnarks.
\newblock \emph{arXiv}, 2 2024{\natexlab{b}}.
\newblock \doi{10.48550/arXiv.2402.02675}.
\newblock URL \url{https://doi.org/10.48550/arXiv.2402.02675}.

\bibitem[Sporny et~al.(2022)Sporny, Longley, and Chadwick]{Sporny2022}
Manu Sporny, Dave Longley, and David Chadwick.
\newblock {V}erifiable {C}redentials {D}ata {M}odel {1.1}.
\newblock {W3C} {R}ecommendation, {W3C}, March 2022.
\newblock URL \url{https://www.w3.org/TR/vc-data-model/}.

\bibitem[Sporny et~al.(2024{\natexlab{a}})Sporny, Longley, Sabadello, Reed, Steele, Allen, and {W3C}]{sporny_decentralized_2024}
Manu Sporny, Dave Longley, Markus Sabadello, Drummond Reed, Orie Steele, Christopher Allen, and {W3C}.
\newblock Decentralized identifiers ({DIDs}) v1.1.
\newblock {World Wide Web Consortium Editor's Draft}, 2024{\natexlab{a}}.
\newblock URL \url{https://w3c.github.io/did-core/}.

\bibitem[Sporny et~al.(2024{\natexlab{b}})Sporny, Thibodeau~Jr., Herman, Jones, Cohen, and {W3C}]{noauthor_verifiable_2024}
Manu Sporny, Ted Thibodeau~Jr., Ivan Herman, Michael~B. Jones, Gabe Cohen, and {W3C}.
\newblock Verifiable credentials data model v2.0.
\newblock W3C Candidate Recommendation Draft, August 2024{\natexlab{b}}.
\newblock URL \url{https://www.w3.org/TR/vc-data-model-2.0/}.
\newblock Accessed: 2024-08-10.

\bibitem[Staab et~al.(2023)Staab, Vero, Balunovic, and Vechev]{Staab2023t}
Robin Staab, Mark Vero, Mislav Balunovic, and Martin Vechev.
\newblock Beyond memorization: Violating privacy via inference with large language models.
\newblock In \emph{The 12th International Conference on Learning Representations ({ICLR} 2024)}, Vienna, Austria, October 2023.

\bibitem[Steinfeld et~al.(2002)Steinfeld, Bull, and Zheng]{steinfeld2002content}
Ron Steinfeld, Laurence Bull, and Yuliang Zheng.
\newblock Content extraction signatures.
\newblock In \emph{Information Security and Cryptology—ICISC 2001: 4th International Conference Seoul, Korea, December 6--7, 2001 Proceedings 4}, pages 285--304. Springer, 2002.

\bibitem[Subramaniam and Krishnan(2024)]{subramaniam2024intent}
Pranav Subramaniam and Sanjay Krishnan.
\newblock Intent-based access control: Using llms to intelligently manage access control.
\newblock \emph{arXiv preprint arXiv:2402.07332}, 2024.

\bibitem[Sun and Zhang(2023)]{Sun2023zkDLEZ}
Hao-Lun Sun and Hongyang Zhang.
\newblock zkdl: Efficient zero-knowledge proofs of deep learning training.
\newblock \emph{ArXiv}, abs/2307.16273, 2023.

\bibitem[Sun et~al.(2024{\natexlab{a}})Sun, Li, and Zhang]{Sun2024}
Haochen Sun, Jason Li, and Hongyang Zhang.
\newblock zkllm: Zero knowledge proofs for large language models.
\newblock \emph{arXiv}, 4 2024{\natexlab{a}}.
\newblock \doi{10.48550/arXiv.2404.16109}.
\newblock URL \url{https://doi.org/10.48550/arXiv.2404.16109}.

\bibitem[Sun et~al.(2024{\natexlab{b}})Sun, Li, and Zhang]{sun2024zkllm}
Haochen Sun, Jason Li, and Hongyang Zhang.
\newblock zkllm: Zero knowledge proofs for large language models.
\newblock \emph{arXiv preprint arXiv:2404.16109}, 2024{\natexlab{b}}.

\bibitem[Suresh and Guttag(2019)]{HariniFramework}
Harini Suresh and John~V. Guttag.
\newblock A framework for understanding sources of harm throughout the machine learning life cycle.
\newblock \emph{Equity and Access in Algorithms, Mechanisms, and Optimization}, 2019.

\bibitem[Tan and Celis(2019)]{tan2019assessing}
Yi~Chern Tan and L~Elisa Celis.
\newblock Assessing social and intersectional biases in contextualized word representations.
\newblock \emph{Advances in neural information processing systems}, 32, 2019.

\bibitem[Tang et~al.(2023)Tang, Tan, and Cai]{Tang2023PrivacyPreservingAT}
Gui Tang, Wuzheng Tan, and Mei Cai.
\newblock Privacy-preserving and trustless verifiable fairness audit of machine learning models.
\newblock \emph{International Journal of Advanced Computer Science and Applications}, 2023.

\bibitem[Thakur et~al.(2021)Thakur, Reimers, R{\"u}ckl{\'e}, Srivastava, and Gurevych]{BEIR}
Nandan Thakur, Nils Reimers, Andreas R{\"u}ckl{\'e}, Abhishek Srivastava, and Iryna Gurevych.
\newblock {BEIR}: A heterogeneous benchmark for zero-shot evaluation of information retrieval models.
\newblock In \emph{Thirty-fifth Conference on Neural Information Processing Systems Datasets and Benchmarks Track (Round 2)}, 2021.
\newblock URL \url{https://openreview.net/forum?id=wCu6T5xFjeJ}.

\bibitem[{The White House}(2023)]{us2023executive}
{The White House}.
\newblock Executive order on the safe, secure, and trustworthy development and use of artificial intelligence, 10 2023.

\bibitem[{UK National Cyber Security Centre} et~al.(2023){UK National Cyber Security Centre}, {US Cybersecurity and Infrastructure Security Agency}, {National Security Agency}, {Federal Bureau of Investigation}, {Australian Signals Directorate's Australian Cyber Security Centre}, {Canadian Centre for Cyber Security}, {New Zealand National Cyber Security Centre}, {Chile's Government CSIRT}, {National Cyber and Information Security Agency of the Czech Republic}, {Information System Authority of Estonia}, {National Cyber Security Centre of Estonia}, {French Cybersecurity Agency}, {Germany's Federal Office for Information Security}, {Israeli National Cyber Directorate}, {Italian National Cybersecurity Agency}, {Japan's National center of Incident readiness and Strategy For Cybersecurity}, {Japan's Secretariat of Science, Technology and Innovation Policy, Cabinet Office}, {Nigeria's National Information Technology Development Agency}, {Norwegian National Cyber Security Centre}, {Poland Ministry of Digital
  Affairs}, {Poland's NASK National Research Institute}, {Republic of Korea National Intelligence Service}, and {Cyber Security Agency of Singapore}]{UKNationalCyberSecurityCentre2023e}
{UK National Cyber Security Centre}, {US Cybersecurity and Infrastructure Security Agency}, {National Security Agency}, {Federal Bureau of Investigation}, {Australian Signals Directorate's Australian Cyber Security Centre}, {Canadian Centre for Cyber Security}, {New Zealand National Cyber Security Centre}, {Chile's Government CSIRT}, {National Cyber and Information Security Agency of the Czech Republic}, {Information System Authority of Estonia}, {National Cyber Security Centre of Estonia}, {French Cybersecurity Agency}, {Germany's Federal Office for Information Security}, {Israeli National Cyber Directorate}, {Italian National Cybersecurity Agency}, {Japan's National center of Incident readiness and Strategy For Cybersecurity}, {Japan's Secretariat of Science, Technology and Innovation Policy, Cabinet Office}, {Nigeria's National Information Technology Development Agency}, {Norwegian National Cyber Security Centre}, {Poland Ministry of Digital Affairs}, {Poland's NASK National Research Institute}, {Republic
  of Korea National Intelligence Service}, and {Cyber Security Agency of Singapore}.
\newblock Guidelines for secure {AI} system development.
\newblock Technical report, UK Government, November 2023.

\bibitem[Vaidya and Clifton(2003)]{Vaidya2003PrivacypreservingKC}
Jaideep Vaidya and Chris Clifton.
\newblock Privacy-preserving k-means clustering over vertically partitioned data.
\newblock In \emph{Knowledge Discovery and Data Mining}, 2003.

\bibitem[Van~Lamsweerde(2001)]{kaos}
Axel Van~Lamsweerde.
\newblock Goal-oriented requirements engineering: A guided tour.
\newblock In \emph{Proceedings fifth ieee international symposium on requirements engineering}, pages 249--262. IEEE, 2001.

\bibitem[Von~Ahn et~al.(2003)Von~Ahn, Blum, Hopper, and Langford]{von2003captcha}
Luis Von~Ahn, Manuel Blum, Nicholas~J Hopper, and John Langford.
\newblock Captcha: Using hard ai problems for security.
\newblock In \emph{Advances in Cryptology—EUROCRYPT 2003: International Conference on the Theory and Applications of Cryptographic Techniques, Warsaw, Poland, May 4--8, 2003 Proceedings 22}, pages 294--311. Springer, 2003.

\bibitem[W3C(2018)]{odrl}
W3C.
\newblock Odrl information model 2.2.
\newblock 2018.
\newblock URL \url{https://www.w3.org/TR/odrl-model/}.

\bibitem[{Wallet}(2024)]{eudi_arf_2024}
{{EU}} {Digital}~{Identity} {Wallet}.
\newblock {European Digital Identity Wallet Architecture and Reference Framework}, 2024.
\newblock URL \url{https://github.com/eu-digital-identity-wallet/eudi-doc-architecture-and-reference-framework/blob/main/docs/arf.md}.

\bibitem[Wang et~al.(2023)Wang, Ma, Feng, Zhang, Yang, Zhang, Chen, Tang, Chen, Lin, Zhao, Wei, and Wen]{Wang2023-rt}
Lei Wang, Chen Ma, Xueyang Feng, Zeyu Zhang, Hao Yang, Jingsen Zhang, Zhiyuan Chen, Jiakai Tang, Xu~Chen, Yankai Lin, Wayne~Xin Zhao, Zhewei Wei, and Ji-Rong Wen.
\newblock A survey on large language model based autonomous agents.
\newblock August 2023.

\bibitem[Wang et~al.(2024)Wang, Ma, Feng, Zhang, Yang, Zhang, Chen, Tang, Chen, Lin, Zhao, Wei, and Wen]{wang_survey_2024}
Lei Wang, Chen Ma, Xueyang Feng, Zeyu Zhang, Hao Yang, Jingsen Zhang, Zhiyuan Chen, Jiakai Tang, Xu~Chen, Yankai Lin, Wayne~Xin Zhao, Zhewei Wei, and Jirong Wen.
\newblock A survey on large language model based autonomous agents.
\newblock \emph{Frontiers of Computer Science}, 18\penalty0 (186345), 2024.
\newblock URL \url{https://doi.org/10.1007/s11704-024-40231-1}.

\bibitem[Wang et~al.(2006)Wang, Ding, Deng, and Bao]{Wang2006PrivateIR}
Shuhong Wang, Xuhua Ding, Robert~H. Deng, and Feng Bao.
\newblock Private information retrieval using trusted hardware.
\newblock In \emph{IACR Cryptology ePrint Archive}, 2006.

\bibitem[Wang et~al.(2021)Wang, Byrnes, Wang, Sun, Ma, Chen, Wu, and Xue]{wang_data_2021}
Zihan Wang, Olivia Byrnes, Hu~Wang, Ruoxi Sun, Congbo Ma, Huaming Chen, Qi~Wu, and Minhui Xue.
\newblock Data {Hiding} with {Deep} {Learning}: {A} {Survey} {Unifying} {Digital} {Watermarking} and {Steganography}, July 2021.
\newblock URL \url{https://arxiv.org/abs/2107.09287v3}.

\bibitem[Wei and Heim(2024)]{wei_designing_2024}
Kevin Wei and Lennart Heim.
\newblock Designing {Incident} {Reporting} {Systems} for {Harms} from {AI}, May 2024.

\bibitem[Weidinger et~al.(2021)Weidinger, Mellor, Rauh, Griffin, Uesato, Huang, Cheng, Glaese, Balle, Kasirzadeh, Kenton, Brown, Hawkins, Stepleton, Biles, Birhane, Haas, Rimell, Hendricks, Isaac, Legassick, Irving, and Gabriel]{Weidinger2021g}
Laura Weidinger, John Mellor, Maribeth Rauh, Conor Griffin, Jonathan Uesato, Po-Sen Huang, Myra Cheng, Mia Glaese, Borja Balle, Atoosa Kasirzadeh, Zac Kenton, Sasha Brown, Will Hawkins, Tom Stepleton, Courtney Biles, Abeba Birhane, Julia Haas, Laura Rimell, Lisa~Anne Hendricks, William Isaac, Sean Legassick, Geoffrey Irving, and Iason Gabriel.
\newblock Ethical and social risks of harm from language models.
\newblock Technical report, Google DeepMind, December 2021.

\bibitem[Weng et~al.(2022)Weng, Weng, Tang, Yang, Li, and Liu]{Weng2022pvCNNPA}
Jiasi Weng, Jian Weng, Gui Tang, Anjia Yang, Ming Li, and Jia-Nan Liu.
\newblock pvcnn: Privacy-preserving and verifiable convolutional neural network testing.
\newblock \emph{IEEE Transactions on Information Forensics and Security}, 18:\penalty0 2218--2233, 2022.

\bibitem[Williamson(1975)]{williamson_markets_hierarchies}
Oliver~E. Williamson.
\newblock \emph{Markets and Hierarchies: Analysis and Antitrust Implications}.
\newblock Free Press, New York, 1975.

\bibitem[Wright(2024)]{wright2024here}
Jesse Wright.
\newblock Here's charlie! realising the semantic web vision of agents in the age of llms.
\newblock \emph{arXiv preprint arXiv:2409.04465}, 2024.

\bibitem[Xiao and Devadas(2023)]{Xiao2023}
Hanshen Xiao and Srinivas Devadas.
\newblock Pac privacy: Automatic privacy measurement and control of data processing.
\newblock In Helena Handschuh and Anna Lysyanskaya, editors, \emph{Annual International Cryptology Conference}, pages 611--644, Cham, Switzerland, 2023. Springer.
\newblock \doi{10.1007/978-3-031-38545-2_20}.
\newblock URL \url{https://doi.org/10.1007/978-3-031-38545-2_20}.

\bibitem[Yang et~al.(2008)Yang, Ding, Deng, and Bao]{Yang2008AnEP}
Yanjiang Yang, Xuhua Ding, Robert~H. Deng, and Feng Bao.
\newblock An efficient pir construction using trusted hardware.
\newblock In \emph{Information Security Conference}, 2008.

\bibitem[Yao et~al.(2023)Yao, Zhao, Yu, Du, Shafran, Narasimhan, and Cao]{yao_react_2023}
Shunyo Yao, Jeffery Zhao, Dian Yu, Nan Du, Izhak Shafran, Karthik Narasimhan, and Yuan Cao.
\newblock {ReAct}: Synergizing reasoning and acting in language models, 2023.
\newblock URL \url{https://arxiv.org/abs/2210.03629}.

\bibitem[Yao et~al.(2024)Yao, Duan, Xu, Cai, Sun, and Zhang]{Yao2024-uv}
Yifan Yao, Jinhao Duan, Kaidi Xu, Yuanfang Cai, Zhibo Sun, and Yue Zhang.
\newblock A survey on large language model ({LLM}) security and privacy: The good, the bad, and the ugly.
\newblock \emph{High-Confidence Computing}, 4\penalty0 (2):\penalty0 100211, June 2024.

\bibitem[Yasuda et~al.(2022)Yasuda, Lodderstedt, Chadwick, Nakamura, and Vercammen]{yasuda2022openid}
Kristina Yasuda, Torsten Lodderstedt, David Chadwick, Kenichi Nakamura, and Jo~Vercammen.
\newblock {OpenID for Verifiable Credentials}: A shift in the trust model brought by verifiable credentials, June 2022.

\bibitem[Yin et~al.(2021)Yin, Zhu, and Hu]{Yin2021}
Xuefei Yin, Yanming Zhu, and Jiankun Hu.
\newblock A comprehensive survey of privacy-preserving federated learning: A taxonomy, review, and future directions.
\newblock \emph{ACM Computing Surveys (CSUR)}, 54\penalty0 (6):\penalty0 1--36, 7 2021.
\newblock \doi{10.1145/346042}.
\newblock URL \url{https://doi.org/10.1145/346042}.

\bibitem[Yu et~al.(2021)]{Yu2021}
Da~Yu et~al.
\newblock Differentially private fine-tuning of language models.
\newblock \emph{arXiv}, 10 2021.
\newblock \doi{10.48550/arXiv.2110.06500}.
\newblock URL \url{https://doi.org/10.48550/arXiv.2110.06500}.

\bibitem[zcash(2022)]{halo2book}
zcash.
\newblock The halo2 book, 2022.
\newblock URL \url{zcash.github.io/halo2/}.

\bibitem[Zhan et~al.(2022)Zhan, Ştefan Sarkadi, Criado, and Such]{zhan2022}
Xiao Zhan, Ştefan Sarkadi, N.~Criado, and J.~Such.
\newblock A model for governing information sharing in smart assistants, 2022.

\bibitem[Zhang et~al.(2023)Zhang, Nakamura, Isohara, and Sakurai]{MachineUnlearn}
Haibo Zhang, Toru Nakamura, Takamasa Isohara, and Kouichi Sakurai.
\newblock A review on machine unlearning.
\newblock \emph{SN Computer Science}, 4:\penalty0 1--13, 2023.

\bibitem[Zheng et~al.(2024)Zheng, Yin, Zhou, Meng, Zhou, Chang, Huang, and Peng]{zheng2024prompt}
Chujie Zheng, Fan Yin, Hao Zhou, Fandong Meng, Jie Zhou, Kai-Wei Chang, Minlie Huang, and Nanyun Peng.
\newblock Prompt-driven llm safeguarding via directed representation optimization.
\newblock \emph{arXiv preprint arXiv:2401.18018}, 2024.

\bibitem[Zhu et~al.(2023)Zhu, Wang, Zhou, Wang, Chen, Wang, Yang, Ye, Zhang, Gong, and Xie]{Zhu2023-yf}
Kaijie Zhu, Jindong Wang, Jiaheng Zhou, Zichen Wang, Hao Chen, Yidong Wang, Linyi Yang, Wei Ye, Yue Zhang, Neil~Zhenqiang Gong, and Xing Xie.
\newblock {PromptRobust}: Towards evaluating the robustness of large language models on adversarial prompts.
\newblock June 2023.

\bibitem[Zuber and Sirdey(2021)]{zuber2021efficient}
Martin Zuber and Renaud Sirdey.
\newblock Efficient homomorphic evaluation of k-nn classifiers.
\newblock \emph{Proc. Priv. Enhancing Technol.}, \penalty0 (2):\penalty0 111--129, 2021.

\bibitem[Zyskind et~al.(2023)Zyskind, South, and Pentland]{Zyskind2023}
Guy Zyskind, Tobin South, and Alex Pentland.
\newblock Don't forget private retrieval: Distributed private similarity search for large language models.
\newblock \emph{arXiv}, 11 2023.
\newblock \doi{10.48550/arXiv.2311.12955}.
\newblock URL \url{https://doi.org/10.48550/arXiv.2311.12955}.

\bibitem[Zyskind et~al.(2024)Zyskind, South, and Pentland]{Zyskind2024j}
Guy Zyskind, Tobin South, and Alex Pentland.
\newblock Don't forget private retrieval: distributed private similarity search for large language models.
\newblock In \emph{Proceedings of the Fifth Workshop on Privacy in Natural Language Processing}, pages 7--19, 2024.

\end{thebibliography}
